\documentclass[11pt]{revtex4}
\usepackage{amsmath}
\usepackage{amsfonts}
\usepackage{amssymb}
\usepackage{amsxtra}
\newcommand\spm{\mathrel{\text{\framebox[0.9\width]{$\pm$}}}}
\newcommand\smp{\mathrel{\text{\framebox[0.9\width]{$\mp$}}}}
\newcommand\cpm{\mathrel{\text{\textcircled{\makebox{$\pm$}}}}}
\newcommand\cmp{\mathrel{\text{\textcircled{\makebox{$\mp$}}}}}
\newcommand\sminus{\boxminus}
\usepackage{supertabular}
\usepackage[pdftex]{graphicx}
\usepackage{tikz}
\usetikzlibrary{arrows,snakes}
\usepackage{subfig}
\usepackage{amsxtra} 
\begin{document}
\title{ 
 New way of second quantized theory of fermions with either Clifford or Grassmann coordinates
and  {\it spin-charge-family} theory. 
}
\author{N.S. Manko\v c Bor\v stnik${}^1$ and H.B.F. Nielsen${}^2$\\
${}^1$
University of Ljubljana, Slovenia\\
${}^2$Niels Bohr Institute, Denmark
Copenhagen, DK-2100
}
\begin{abstract} 
Fermions with the internal degrees of freedom described in Clifford space carry in any dimension 
a half integer spin. There are two kinds of spins in Clifford space. The spin-charge-family theory 
(\cite{norma92,norma93,IARD2016,n2014matterantimatter,nd2017,n2012scalars,JMP2013,%
normaJMP2015,nh2017}, and the references therein), assuming even d=(13+1), uses one kind 
of spins to describe in d=(3+1) spins and charges of quarks and leptons and antiquarks and 
antileptons, while the other kind is used to describe families. \\
In this work the new way of second quantization, suggested by the spin-charge-family theory, 
is presented. It is shown that the creation and annihilation operators of $1$-fermion states, 
written as products of nilpotents and projectors of an odd Clifford character, fulfill the 
anticommutation relations as required in the second quantization procedure for fermions, what 
means that $1$-fermion states are in Clifford space already second quantized, and that the 
creation operators for n-fermion second quantized vectors are products of one fermion creation 
operators, operating on the empty vacuum state. There is no need in this theory for the negative 
energy states fulfilled with fermions. \\
It is demonstrated that also in Grassmann space there exist the creation and annihilation operators 
of an odd Grassmann character, generating "fermions", which fulfill as well the anticommutation 
relations for fermions, representing correspondingly the second quantized $1$-"fermion" states. 
However, while the internal spins determined by the generators of the Lorentz group of the Clifford 
objects of both kinds are half integer, the internal spins determined by the Grassmann objects are 
integer. Grassmann space offers no families. \\
We discuss the new second quantization procedure of the fields in both spaces. For the Grassmann 
case we present the action, the basic states, the solution of the “Weyl” equation for free massless 
"fermions" and the discrete symmetry operators. 
A short overview of the achievements of the spin-charge-family theory is done, and open problems 
of this theory still waiting to be solved are presented. We compare the Grassmann and the Clifford 
case in order to better understand to how many open questions in physics of elementary fermion 
and boson fields and in cosmology the spin-charge-family theory is able to answer.
\end{abstract}
%



\maketitle

\section{Introduction}
\label{introduction} 

More than 50 years ago the {\it standard model} offered an elegant new step in understanding 
elementary fermion and boson fields by postulating:\\
{\bf i.} Massless family members of 
coloured quarks and colouress leptons, the left handed members as the weak charged doublets 
and the weak chargeless right hand members, the left handed quarks distinguishing in the hyper 
charge from the left handed leptons, each right handed member having a different hyper charge.
All fermion charges are in the fundamental representation of the corresponding groups. 
Antifermions 
carry the corresponding anticharges. The  existence of massless families to  each family member
is as well postulated. There is no right handed neutrino, since it would carry none of the so far 
observed charges, and correspondingly there is also no left handed antineutrino.\\ 
 {\bf ii.} The existence of the massless vector gauge fields to the observed charges of quarks 
and leptons, carrying charges in the corresponding adjoint representations.\\
 {\bf iii.}  The existence of a massive scalar Higgs, gaining  at some 
step of the expanding universe the nonzero vacuum expectation value,  causing masses of 
fermions and heavy bosons and the Yukawa couplings. The Higgs carry a half integer weak and 
hyper charge.\\
{\bf iv.} Fermion and boson fields can be (second) quantized.

The {\it standard model} assumptions have in the literature several explanations, mostly with 
many new not explained assumptions. The most successful seems to be the grand unifying 
theories~\cite{Geor,FritzMin,PatiSal,GeorGlas,Cho,ChoFreu,Zee,SalStra,DaeSalStra,Mec,%
HorPalCraSch,Asaka,ChaSla,Jackiw,Ant,Ramond,Horawa}, if postulating in addition the family 
group and the corresponding gauge scalar fields.

The {\it spin-charge-family} theory, the project of N.S.M.B.~\cite%
{norma92,norma93,IARD2016,n2014matterantimatter,nd2017,n2012scalars,JMP2013,%
normaJMP2015,nh2017,nh2018}, is offering the explanation for all the assumptions of the
{\it standard model}, unifying not only charges, but also charges and spins and families, 
explaining the appearance of families, of the vector gauge fields, of the scalar field and the
Yukawa couplings, offering the explanation for the matter-antimatter asymmetry, making 
several predictions. This theory also offers the explanation for the appearance of creation 
and annihilation operators, which in the Dirac theory~\cite{Dirac} is just assumed.

 The {\it spin-charge-family} theory is a kind of the Kaluza-Klein like theories~\cite{KaluzaKlein,%
Witten,Duff,App,SapTin,Wetterich,zelenaknjiga,mil,nh2017} due to the assumption that in 
$d\ge5$ (in the {\it spin-charge-family} theory $d\ge (13+1)$) fermions interact with the
gravity only. Correspondingly this theory shares with the 
Kaluza-Klein like theories their weak points, at least: {\bf i.} Not yet solved the quantization 
problem of gravitational fields. {\bf ii.} Breaking spontaneously of the starting symmetry which 
would at low energies manifest the observed almost massless fermions~\cite{Witten}.  
Concerning this second point we proved on the toy model of $d=(5+1)$ that the break of 
symmetry can lead to (almost) massless fermions~\cite{NHD,ND012,familiesNDproc,NHD2011}. 
It remains to study how does appear the spontaneous  breaking of the starting 
symmetry in $d=(13+1)$, first with the appearance of the condensate of two right handed
neutrinos, Table~\ref{Table con.}, Ref.~\cite{n2014matterantimatter}, and then when scalar 
fields with space index $(7,8)$ obtain nonzero vacuum expectation values. (This second point 
is common to all the unifying theories.)

Since in $d=(3+1)$-dimensional space --- at low energies --- the gauge gravitational fields 
manifest as the observed vector gauge fields~\cite{nd2017},  which can be quantized in the usual  
way, quantization procedure of gravity can
wait to be made. The author is in mean time trying to find out (together with the collaborators)  
how far can the {\it spin-charge-family} theory --- starting in $d=(13+1)$-dimensional space with
a simple and "elegant" action, Eq.~(\ref{wholeaction}) ---  reproduce in $d=(3+1)$ the observed 
properties of  quarks and leptons~\cite{IARD2016,n2014matterantimatter,nd2017,n2012scalars,%
JMP2013,normaJMP2015,nh2017,nh2018}, the observed gauge fields, the assumed scalar field, 
the appearance 
of the dark matter and of the matter-antimatter asymmetry, as well as the other
open questions, connecting elementary fermion and boson fields and cosmology. The work done so 
far seems very promising.

Let us in what follows and  in Subsect.~\ref{SCFT} overview shortly the starting assumptions 
and so far achievements of  the {\it spin-charge-family} theory, and discuss as well open problems.

The recognition that there are  in Grassmann space two kinds of the Clifford algebra objects~%
\cite{norma93} ($\gamma^a$ and $\tilde{\gamma}^a$) offered in the {\it spin-charge-family} 
theory the explanation for the origin of families~\cite{nh02,nh03,DKhn,norma92,norma93}, 
Table~\ref{Table III.}. 

The assumption made in the {\it spin-charge-family} theory that  the dimension of space
 is $\ge (13+1)$  enables  the explanation for by the {\it standard model} assumed spins 
and charges  of quarks and leptons~\cite{pikan2003,pikan2006}, explaining as well the 
miraculous cancellation of triangle anomalies~\cite{nh2017,normaJMP2015,n2014matterantimatter} 
in the {\it standard model}, however, without relating handedness and charges "by hand"
as it is needed in $SO(10)$~\cite{Alvarez,
Bilal,AlvarezBondiaMartin}.

 Since there are in $SO(13 +1)$ additional quantum numbers to those assumed by the 
{\it standard model},  the theory predicts that right handed neutrinos and left handed 
antineutrinos, carrying nonzero additional quantum numbers --- $\tau^{23}$ and $\tau^{4}$ 
instead of $Y$ in the {\it standard model}, $Y =\tau^{23}+\tau^{4}$ in 
Table~\ref{Table so13+1.}, Eqs.~(\ref{so1+3}, \ref{so42}, \ref{so64}, \ref{YQY'Q'andtilde}) --- 
are regular members of families of quarks and leptons~\cite{pikan2003,pikan2006, IARD2016,%
normaJMP2015}. This prediction is common also to $SO(10)$~\cite{Alvarez,
Bilal,AlvarezBondiaMartin}.

In the {\it spin-charge-family} theory spins and charges are described  by the superposition of 
$S^{ab}$ $(=\frac{i}{4}\,(\gamma^a \gamma^b - \gamma^b \gamma^a )$,
 Eq.~(\ref{twocliffordsab})), with 
$\gamma^a$ belonging to the first kind of the Clifford algebra objects and with $S^{mn}, 
(m,n)=(0,1,2,3)$,  describing spins and handedness of quarks and leptons (Eq.~(\ref{so1+3})), 
and $S^{st}, (s,t)=(5,6,\cdots,14)$, describing their charges, Table~\ref{Table so13+1.},
Eqs.~(\ref{so42}, \ref{so64}) and Refs.~\cite{norma93,nh02,DKhn,pikan2006}.
  
Family quantum numbers 
are determined by the second kind of the Clifford algebra objects, by the superposition of 
$\tilde{S}^{ab}$ $(=\frac{i}{4}\,(\gamma^a \gamma^b - \gamma^b \gamma^a )$),
Eq.~(\ref{twocliffordsab}), Table~\ref{Table III.}~%
\cite{norma93,nh03}.

The vector gauge fields, assumed in the {\it standard model} as the gauge fields of the
corresponding fermion charges, are  in the {\it spin-charge-family} theory explainable as 
the superposition of the gauge fields of the  generators of the Lorentz transformations 
$S^{st} $ ($S^{st}\,\omega_{stm}, (s,t)=(5,6,\cdots,14)$, Eqs.~(\ref{wholeaction},
\ref{faction}, \ref{so1+3})), with the vector index $m=(0,1,2,3)$ ,
 Eq.~(\ref{omegatildeomega}), Ref.%
~\cite{nd2017}.

In the {\it standard model} the scalar fields appear  
as the Higgs scalar and the Yukawa couplings by the assumption.  
In the  {\it spin-charge-family} theory both kinds of the gauge fields, 
$\sum_{s',t'} c^{s't'}\,\omega_{s't's}$, which are the gauge fields of $S^{st}$ with $(s',t') =(5,6,7,8)$,   
and $\sum_{a,b}\tilde{c}^{ab}\,\tilde{\omega}_{abs}$, which are the gauge fields of 
$\tilde{S}^{ab}$, with $(a,b)= (0,1,\cdots, 8)$, both with the scalar index  $s=(7,8)$,
manifesting properties of the Higss scalar (by carrying weak and hyper charges in the "fundamental 
representation"), 
define masses of quarks and leptons and of heavy bosons,  Eq.~(\ref{omegatildeomega}), 
Refs.~\cite{pikan2006,normaJMP2015,IARD2016}. \\ 
%
These scalar fields determine in  the {\it spin-charge-family} theory  masses of the two groups  
of four 
families~\cite{mdn2006,gmdn2008,gn2009,gn2013,gn2014, IARD2016,normaJMP2015}.
The lower group predicts the existence of the fourth family of quarks and leptons, coupled to the 
observed three families~\cite{mdn2006,gmdn2008,gn2014,gn2009,familiesNDproc}. 
Their from the symmetry of mass matrices predicted the $ 4 \times 4$
mixing matrix of quarks~\cite{gn2014} appear to be in better agreement with the 
experiments than if only three families are assumed~\cite{bereziani}. 

The lowest family of the upper four families offers 
the explanation for the existence of the dark matter~\cite{gn2009,nm2015}.

There are additional scalar fields in the {spin-charge-family} theory~\cite{n2014matterantimatter},
having the scalar space index $t\in (9,10,\dots,14)$. They carry colour charges in the 
"fundamental" representations, cause transitions of antileptons and 
antiquarks into quarks and back, enabling the decay of baryons. These scalar 
fields are  offering in the presence of the right handed neutrino condensate, Table~\ref{Table con.},
Ref.~\cite{n2014matterantimatter}, which breaks the ${\cal C}{\cal P}$ symmetry, 
the answer to the question about the matter-antimatter asymmetry in the universe%
~\cite{n2014matterantimatter}. 

Authors of this paper proved on the toy model of $d=(5+1)$ that breaking the symmetry in 
Kaluza-Klein theories can lead to massless
 fermions~\cite{NHD,ND012,familiesNDproc,
NHD2011}. 
The authors determine as well the discrete symmetries operators in observable dimensions 
$d=(3+1)$ for any $d$, Eqs.~(\ref{CPTN}), Ref.~\cite{nhds}. 

%
The breaking of the starting symmetry $SO(13+1)$ is in the {\it spin-charge-family} theory
triggered by the appearance of the condensate (Table~\ref{Table con.}) of the right handed 
neutrinos~\cite{n2014matterantimatter} and, like in the {\it standard model}, by the nonzero 
vacuum expectation values of the scalar fields with the space index $s=(7,8)$. 

In this paper it is demonstrated that the odd products of nilpotents and projectors, which are the
"egenfunctions" of the Cartan subalgebra of the Lorentz algebra in Clifford space, and which  
solves the Weyl equations for free massless fermions, 
fulfill  together with the corresponding Hermitian conjugated annihilation operators the
anti-commutation relations  as needed in the second quantized fermion fields~\cite{NH2005}. 
 No assumption of the Dirac kind about the creation and annihilation operators is needed.

The {\it spin-charge-family} theory  has many common points with other unifying 
theories~(\cite{Geor,FritzMin,PatiSal,GeorGlas,Cho,ChoFreu,KaluzaKlein,Witten,Duff,App,SapTin,%
Wetterich,zelenaknjiga,mil} and other references), and because of that and because of the 
fact that by starting  with the very simple action, Eq.~(\ref{wholeaction}), the theory is able 
to offer  explanations for so many observed phenomena, built into assumptions of the 
{\it standard model}(s)  of the elementary boson and fermion fields and also of  cosmology, 
and also in other unifying theories, it might be that it is the right next  step beyond the standard 
models.

The achievements of the {\it spin-charge-family} theory is discussed in 
more details in Subsect.~\ref{SCFT}. There also problems waiting to be solved are presented.

Let us present a very simple  starting action of the {\it spin-charge-family} theory of
N.S.M.B., in which  massless fermions in $d= (13+1)$-dimensional space  interact with 
massless bosons, that is only with gravity --- the vielbeins 
$f^{\alpha}{}_a$ (the gauge fields of moments $p_a$) and the two kinds of the spin connections 
($ \omega_{ab \alpha}$ and  $\tilde{\omega}_{ab \alpha}$, the gauge fields of the two kinds of the 
Clifford algebra objects $\gamma^a$ and $\tilde{\gamma}^a$, respectively). 
\begin{eqnarray}
{\cal A}\,  &=& \int \; d^dx \; E\;\frac{1}{2}\, (\bar{\psi} \, \gamma^a p_{0a} \psi) + h.c. +
\nonumber\\  
               & & \int \; d^dx \; E\; (\alpha \,R + \tilde{\alpha} \, \tilde{R})\,,
%
\label{wholeaction}
\end{eqnarray}
with $p_{0a } = f^{\alpha}{}_a p_{0\alpha} + \frac{1}{2E}\, \{ p_{\alpha},
E f^{\alpha}{}_a\}_- $, 
$ p_{0\alpha} =  p_{\alpha}  - \frac{1}{2}  S^{ab} \omega_{ab \alpha} - 
                    \frac{1}{2}  \tilde{S}^{ab}   \tilde{\omega}_{ab \alpha} $ and                     
$R =  \frac{1}{2} \, \{ f^{\alpha [ a} f^{\beta b ]} \;(\omega_{a b \alpha, \beta} 
- \omega_{c a \alpha}\,\omega^{c}{}_{b \beta}) \} + h.c. $,  
$\tilde{R}  =  \frac{1}{2} \, \{ f^{\alpha [ a} f^{\beta b ]} \;(\tilde{\omega}_{a b \alpha,\beta} - 
\tilde{\omega}_{c a \alpha} \,\tilde{\omega}^{c}{}_{b \beta})\} + h.c.$. 
Here~\footnote{$f^{\alpha}{}_{a}$ are inverted vielbeins to 
$e^{a}{}_{\alpha}$ with the properties $e^a{}_{\alpha} f^{\alpha}{\!}_b = \delta^a{\!}_b,\; 
e^a{\!}_{\alpha} f^{\beta}{\!}_a = \delta^{\beta}_{\alpha} $, $ E = \det(e^a{\!}_{\alpha}) $.
Latin indices  
$a,b,..,m,n,..,s,t,..$ denote a tangent space (a flat index),
while Greek indices $\alpha, \beta,..,\mu, \nu,.. \sigma,\tau, ..$ denote an Einstein 
index (a curved index). Letters  from the beginning of both the alphabets
indicate a general index ($a,b,c,..$   and $\alpha, \beta, \gamma,.. $ ), 
from the middle of both the alphabets   
the observed dimensions $0,1,2,3$ ($m,n,..$ and $\mu,\nu,..$), indexes from 
the bottom of the alphabets
indicate the compactified dimensions ($s,t,..$ and $\sigma,\tau,..$). 
We assume the signature $\eta^{ab} =
diag\{1,-1,-1,\cdots,-1\}$.} 
$f^{\alpha [a} f^{\beta b]}= f^{\alpha a} f^{\beta b} - f^{\alpha b} f^{\beta a}$.

The two kinds of the Clifford algebra objects, $\gamma^a$ and $\tilde{\gamma}^a$,  
Eq.~(\ref{twocliffordsab}, \ref{gammatildeA}), 
anticommute and determine the infinitesimal generators of the Lorentz transformations 
in the internal space of fermions --- $S^{ab}$ for $SO(13,1)$, arranging  states into 
representations (Table~\ref{Table so13+1.}), and $\tilde{S}^{ab}$ for $\widetilde{SO}(13,1)$,
arranging states into families (Table~\ref{Table III.}). 
%
\begin{eqnarray}
\label{twocliffordsab}
&&\{\gamma^a, \gamma^b\}_{+}= 2 \eta^{ab} = 
\{\tilde{\gamma}^a, \tilde{\gamma}^b\}_{+}\,,\nonumber\\
&&\{\gamma^a, \tilde{\gamma}^b\}_{+} = 0\,, \nonumber\\
&& S^{ab} = \,\frac{i}{4} (\gamma^a\, \gamma^b - \gamma^b\, \gamma^a)\,, \nonumber\\
&&\tilde{S}^{ab} = \,\frac{i}{4} 
(\tilde{\gamma}^a\, \tilde{\gamma}^b - \tilde{\gamma}^b\, \tilde{\gamma}^a)\,.
\end{eqnarray}
The generators $S^{ab}$  
are used in the {\it spin-charge-family} theory to determine spins and charges of spinors of any 
family, Table~\ref{Table so13+1.}, another kind, $\tilde{S}^{ab}$, 
determines the family quantum numbers, Table~\ref{Table III.}.

The scalar curvatures $R$ and $\tilde{R}$ determine dynamics of the gauge fields --- the spin 
connections and the vielbeins --- manifesting in $d=(3+1)$ as all the known vector gauge fields 
as well as the scalar fields~\cite{nd2017}, which offer the explanation for the appearance of 
the Higgs and the Yukawa couplings, of the ordinary  matter-antimatter asymmetry%
~\cite{n2014matterantimatter} and the dark matter~\cite{gn2009}, 
 provided that the symmetry breaks from the starting $SO(13,1)$ to 
$SO(3,1) \times SU(3) \times U(1)$.

In this paper we start to study the possibility that fermions are described
in Grassmann space, in order to better understand how far can the simple starting action, 
Eq.~(\ref{wholeaction}), 
of the {\it spin-charge-family} theory  agree with the at low energies observed properties 
of fermions and bosons.

We demonstrate in this paper that besides Clifford space also Grassmann space offers the 
description of the internal degrees of freedom of fermions in the second quantized procedure. 
In both cases there exist the creation and annihilation operators, which fulfill the 
anticommutation relations required for fermions, Eqs.~(\ref{ijthetaprod}, \ref{alphagammaprod0}).
But while the internal spins determined by the generators of the Lorentz group of the Clifford 
objects $S^{ab}$ and $\tilde{S}^{ab}$ --- we repeat here that in the {\it spin-charge-family} 
theory  $S^{ab}$ determine the spin degrees of freedom and $\tilde{S}^{ab}$ the family 
degrees of freedom --- are half integer, 
the internal spin determined by ${\cal {\bf S}}^{ab}$ (expressible with 
$S^{ab} + \tilde{S}^{ab}$) is integer. 

Correspondingly Clifford space offers according to the {\it spin-charge family} theory the 
description of spins, charges and families, all in the fundamental representations of the subgroups 
of  the Lorentz group $SO(d-1,1)$, while Grassmann space offers spins and charges in the 
adjoint representations of the subgroups of  the Lorentz group $SO(d-1,1)$ and no family 
degrees of freedom. Fermions with integer spins would lead to completely different nucleons, 
nuclei, atoms, molecules, matter than the so far observed ones.

Let us make a short introduction into the Grassmann space as well.

In Grassmann space the infinitesimal generators of the Lorentz transformations 
${\cal {\bf S}}^{ab}$ are expressible with anticommuting coordinates $\theta^a$ and their 
conjugate momenta $p^{\theta a} = i \frac{\partial}{\partial \theta_a}$%
~\cite{norma93},
\begin{eqnarray}
\{\theta^a, \theta^b\}_{+} &=& 0\,, \quad \{p^{\theta a}, p^{\theta b}\}_{+} = 0\, , \quad 
\{p^{\theta a}, \theta^b\}_{+} = i \,\eta^{a b}\, ,\nonumber\\ 
\label{thetarelcom}
{\cal {\bf S}}^{ab} &=& \theta^a p^{\theta b} -  \theta^b p^{\theta a}\,. 
\end{eqnarray}
Taking into account that  $\gamma^a$ and $\tilde{\gamma}^a$ are expressible in terms of 
$\theta^a$
and their conjugate momenta $p^{\theta a}$~\cite{norma93}
\begin{eqnarray}
\label{cliffthetareltheta}
\gamma^a= (\theta^a - i \, p^{\theta a})\,, \quad 
\tilde{\gamma}^a=i \, (\theta^a + i \, p^{\theta a})\,, 
\end{eqnarray}
one recognizes  
\begin{eqnarray}
\label{Lorentztheta}
{\cal {\bf S}}^{ab} &=& S^{ab} + \tilde{S}^{ab} \,,
\end{eqnarray}
from where one concludes, if taking into account Eq.~(\ref{wholeaction}), that in the Grassmann case
the covariant momenta $p_{0 \alpha}$ are
\begin{eqnarray}
\label{p0alphaGrass}
p_{0 \alpha} &=& p_{\alpha} - \frac{1}{2} {\cal {\bf S}}^{ab} \Omega_{ab \alpha} \,,
\end{eqnarray}
with $\Omega_{ab \alpha}$ as the only kind of the connection fields (instead of the two 
kinds in the Clifford case --- $\omega_{ab \alpha}$, which is the gauge fields of $S^{ab}$ and 
$\tilde{\omega}_{ab \alpha}$, which is the gauge fields of $\tilde{S}^{ab}$).

It follows  for ${\cal {\bf S}}^{ab}$
\begin{eqnarray}
\label{Lorentzthetacom}
\{{\cal {\bf S}}^{ab}, {\cal {\bf S}}^{cd}\}_{-} &=&   i \{{\cal {\bf S}}^{ad} \eta^{bc} + 
{\cal {\bf S}}^{bc} \eta^{ad} - {\cal {\bf S}}^{ac} \eta^{bd}- 
{\cal {\bf S}}^{bd} \eta^{ac}\}\,,\nonumber\\
{\cal {\bf S}}^{ab \dagger} &=& \eta^{aa} \eta^{bb} {\cal {\bf S}}^{ab}\,.
\end{eqnarray}
The same relations are true also if ${\cal {\bf S}}^{ab}$ is replaced with either  $S^{ab}$ or 
$\tilde{S}^{ab}$.
These infinitesimal generators of the Lorentz group ---  the two kinds of the Clifford operators 
and  the Grassmann operators --- operate as follows 
%
\begin{eqnarray}
\label{sabspinall}
&&\{S^{ab}, \, \gamma^e\}_{-} = - i \,(\eta^{ae} \,\gamma^b - 
\eta^{be}\,\gamma^a)\,, \nonumber\\ 
&&\{\tilde{S}^{ab}, \,\tilde{\gamma}^e\}_{-} = - i \,(\eta^{ae} \,\tilde{\gamma}^b - 
\eta^{be}\,\tilde{\gamma}^a)\,,\nonumber\\
&&\{S^{ab}, \, \tilde{S}^{cd}\}_{-}=0\,, \nonumber\\
&&\{{\cal {\bf S}}^{ab}, \, \theta^e\}_{-} = - i \,(\eta^{ae} \,\theta^b - \eta^{be}\,\theta^a)\,,
\nonumber\\
&&\{{\cal {\bf S}}^{ab}, \, p^{\theta e}\}_{-} = - i \,(\eta^{ae} \,p^{\theta b} - 
\eta^{be}\,p^{\theta a})\,,
\nonumber\\
&&\{{\cal {\bf M}}^{ab}, \, A^{d\dots e \dots g}\}_{-} = - i \,(\eta^{ae} \,A^{d\dots b \dots g} - 
\eta^{be}\,A^{d\dots a \dots g})\,, 
\end{eqnarray}
where $ {\cal {\bf M}}^{ab}$ are defined in the Clifford case by the sum of $L^{ab}$ plus 
either $S^{ab}$ (if $\gamma^a$'s are chosen to describe the basis, otherwise $\tilde{S}^{ab}$
replace $S^{ab}$), while in the Grassmann case $ {\cal {\bf M}}^{ab}$ is 
$L^{ab} + {\cal {\bf S}}^{ab}$ (which is, Eq.~(\ref{Lorentztheta}), $ {\cal {\bf M}}^{ab}$$ =
L^{ab} + S^{ab} + \tilde{S}^{ab}$).

In Sect.~\ref{observations} the actions and norms for free massless fermions, with 
the internal degrees of freedom described in Clifford and in Grassmann space in $d$-dimensional
spaces are presented. The discrete symmetry operators in  $d$-dimensional space  --- Clifford and 
Grassmann --- and their manifestation in $d=(3+1)$-dimensional space are presented in 
Subsect.~\ref{CPT} of Sect.~\ref{secondquantization}. 
While the action and the discrete symmetry operators 
in Clifford space are known from before~\cite{normaJMP2015,nhds}, the action
in Grassmann space as well as the discrete symmetry operators are here assumed by N.S.M.B..

The new way of second quantization of fermion fields in both spaces is discussed in 
Sect.~\ref{secondquantization}.
 We treat in both spaces only massless free particles. 
Sect.~\ref{conclusions} presents what we learn from this work.

This work is a part of the project of both authors, which includes  the {\it fermionization} procedure
of boson fields (or the {\it bosonization} procedure of fermion fields), discussed in 
Refs.~\cite{procnh2015,procnh2017,HolMasDSB} for any dimension $d$ (by the authors of
this contribution, while one of them, H.B.F.N.~\cite{holger},  has succeeded with another author
to do the {\it fermionization} for $d=(1+1)$), and which would hopefully also help to understand  
a little better the content and dynamics of our universe.

\subsection{Comments on the achievements of the {\it spin-charge-family} theory so far 
and the open questions to be solved}
\label{SCFT}

Let us illustrate  the achievements of the {\it spin-charge-family} theory, 
presented in  the introduction, by adding  some comments.

{\bf I.} $\;\;\;$ In the action, Eq.~(\ref{wholeaction}), fermions carry in $d=(13+1)$ two
kinds of spins --- no charges and interact with gravity only --- with the vielbeins $f^{\alpha}{}_{a}$
and the two kinds of the spin connection fields, the gauge fields of  $S^{ab}$ --- 
$\omega_{ab\alpha}$ --- and the gauge fields of $\tilde{S}^{ab}$ --- $\tilde{\omega}_{ab\alpha}$.

One can formally rewrite the  fermion part of the action so that it manifests in $d=(3+1)$  the
free massless fermion part (first line in Eq.~(\ref{faction})), the interaction of fermions with the 
vector gauge fields (the second line in Eq.~(\ref{faction})), the interaction of fermions with 
the scalar fields (the third line in Eq.~(\ref{faction})), and the rest.
\begin{eqnarray}
{\mathcal L}_f &=& \sum_{m} \bar{\psi}\gamma^{m} p_{m} \psi \nonumber \\
&- &  \sum_{A,i} \bar{\psi}\,\gamma^{m}   \tau^{Ai} A^{Ai}_{m}\, \psi + \nonumber\\
&+ &  \sum_{s=7,8}  \bar{\psi} \gamma^{s} p_{0s} \; \psi   \nonumber\\ 
&+ &  \sum_{t=5,6,9,\dots, 14}  \bar{\psi} \gamma^{t} p_{0t} \; \psi \,,
\label{faction}
\end{eqnarray}
with $\tau^{Ai}= \sum_{st} c_{st}{}^{Ai} S^{st}, (s,t)= (5,6,\cdots,13,14)$, which are 
generators of the subgroups of $SO(13,1)$, determining charges of fermions, Eq.~(\ref{so42}, 
\ref{so64}, \ref{YQY'Q'andtilde}), with $ A^{Ai}_{m}$, which are the corresponding superposition 
of $\omega_{st m}$~(\cite{n2014matterantimatter,normaJMP2015} and the references therein),
$p_{0s} =  p_{s}  - \frac{1}{2}  S^{s' s"} \omega_{s' s" s} - 
                    \frac{1}{2}  \tilde{S}^{ab}   \tilde{\omega}_{ab s}$  and 
$p_{0t} =  p_{t}  - \frac{1}{2}  S^{t' t"} \omega_{t' t" t} - 
                    \frac{1}{2}  \tilde{S}^{ab}   \tilde{\omega}_{ab t}$,                    
while $ m \in (0,1,2,3)$, $s \in (7,8),\, (s',s") \in (5,6,7,8)$, $(a,b)$ (appearing in
 $\tilde{S}^{ab}$) run within $ (0,1,2,3)$ and $ (5,6,7,8)$, $t \in (5,6,9,\dots,13,14)$, 
$(t',t") \in  (5,6,7,8)$ and $\in (9,10,\dots,14)$.  

{\bf I.i} $\;\;$
The spinor function $\psi$ represents all the family members, $2^{\frac{d}{2}-1}=64$ for
$d=13+1$, of all the $2^{\frac{7+1}{2}-1}=8$ families, including fermions and antifermions. 
Tables~\ref{Table so13+1.} and 
\ref{Table III.} represent the creation operators for the states of one family and the 
creation operators for the eight families, respectively. The rest of families are assumed to 
have very large masses as discussed and proved for a toy model in 
Ref.~\cite{NHD,NHD2011,ND012,familiesNDproc,nh2008}. The creation operators operate 
on a vacuum state, Eq.~(\ref{vac1}).

{\bf I. A.} $\;\;$ The Clifford object $\gamma^a$ are in the {\it spin-charge-family} theory used 
to determine from the point of view of $d=(3+1)$ spins and all the charges of fermions.

{\bf I. A.i.} $\;\;$ 
$d=(13 +1)$-dimensional space offers $2^{\frac{d}{2}-1}= 64$ members of $SO(13,1)$.
 In Table~\ref{Table so13+1.} the properties of quarks and leptons and antiquarks and
antileptons, forming $64$ members, are presented  from the points of view of subgroups of 
$SO(13,1)$ breaking first into $SO(7,1)\times SU(3) \times U(1)$, keeping connection
between handedness  and the two $SU(2)_{I,II}$ charges, and further to ---  
$SU(2)_{R}\times SU(2)_{L}$ $ \times SU(2)_{I}$ $\times SU(2)_{II} \times SU(3) \times U(1)$
 --- representing in $d=(3+1)$ the spin  and handedness, the weak charge $\tau^{13}$ of 
$SU(2)_{I}$, the second $\tau^{23}$ of $SU(2)_{II}$, the colour charge $\tau^{33}$ and 
$\tau^{38}$ of $SU(3)$ and $\tau^4$ of $U(1)$ {\it for quarks and leptons and for antiquarks 
and antileptons}.\\
Cartan subalgebra has $\frac{d}{2}= 7$ members,  the {\it standard model} assumes one 
commuting operator less. 

{\bf I. A.ii.} $\;\;$
Due to the additional commuting operator (the member of the Cartan subalgebra of $S^{ab}$) 
in  the  {\it spin-charge-family} theory, the
neutrinos become a regular members of quarks and leptons, with masses determined by the
interaction with the scalar fields as all the rest of family members~\cite{mdn2006,gmdn2008,gn2009,%
gn2013,gn2014, IARD2016,normaJMP2015} (in Eq.~(\ref{faction}) the interaction of fermions with 
the scalar fields is contained in the third line). This is the case also in $SO(10)$ theories~\cite{Geor,%
FritzMin,PatiSal,GeorGlas}. The difference in the  {\it spin-charge-family} theory is, that spin and 
handedness are correlated with charges, while in $SO(10)$ this is not the case (and must be 
correlated by "hand"). This fact is discussed in details in Ref.~\cite{nh2017}.

Let us point out that colour chargeless leptons and quarks of any of the three colours have completely 
the same $SO(7,1)$ part. Quarks and leptons distinguish only in the $SU(3)\times U(1)$ part.

{\bf I. B.} $\;\;$
The second Clifford object $\tilde{\gamma}^a$ offers the explanation for the existence of families.

{\bf I. B.i.} $\;\;$ There are twice four families of quarks and leptons in the  {\it spin-charge-family}
 theory~(\cite{IARD2016} and the references therein) after the appearance of the condensate of the
 two right handed neutrinos, presented in Table~\ref{Table con.}, Ref.~\cite{n2014matterantimatter}. 
Since we have not really shown yet how this dynamically happens (we did this so far only for the toy 
model~\cite{NHD,ND012,familiesNDproc,
NHD2011}), this remains as an open problem. 
All eight families obtain masses when the scalar gauge fields with the space index (7,8) --- third 
line in Eq.~(\ref{faction}) --- gain nonzero vacuum expectation values at the electroweak phase 
transition.
Table~\ref{Table III.}  
represents in the left column eight families of creation operators of $\hat{u}^{c1 \dagger}_{R}$ --- 
the first member in Table~\ref{Table so13+1.} --- and of 
chargeless $\hat{\nu}^{\dagger}_{R}$ --- the $25^{th}$ member  in Table~\ref{Table so13+1.}.
  ($S^{11\, 12}$, 
for example, transforms $\hat{u}^{ci \dagger}_{R}$ into $\hat{\nu}^{\dagger}_{R}$ and 
opposite). 

 \begin{table}
 \begin{center}
 \begin{tabular}{|r|c|c|c|c|c c c c c|}
 \hline
 &&&&&$\tilde{\tau}^{13}$&$\tilde{\tau}^{23}$&$\tilde{N}_{L}^{3}$&$\tilde{N}_{R}^{3}$&
 $\tilde{\tau}^{4}$\\
 \hline
 $I$&$\hat{u}^{c1 \dagger}_{R\,1}$&
   $ \stackrel{03}{(+i)}\,\stackrel{12}{[+]}|\stackrel{56}{[+]}\,\stackrel{78}{(+)} ||
   \stackrel{9 \;10}{(+)}\;\;\stackrel{11\;12}{(-)}\;\;\stackrel{13\;14}{(-)}$ & 
    $\hat{\nu}^{\dagger}_{R\,1}$&
   $ \stackrel{03}{(+i)}\,\stackrel{12}{[+]}|\stackrel{56}{[+]}\,\stackrel{78}{(+)} ||
   \stackrel{9 \;10}{(+)}\;\;\stackrel{11\;12}{[+]}\;\;\stackrel{13\;14}{[+]}$ 
  &$-\frac{1}{2}$&$0$&$-\frac{1}{2}$&$0$&$-\frac{1}{2}$ 
 \\
  $I$&$\hat{u}^{c1 \dagger}_{R\,2}$&
   $ \stackrel{03}{[+i]}\,\stackrel{12}{(+)}|\stackrel{56}{[+]}\,\stackrel{78}{(+)} ||
   \stackrel{9 \;10}{(+)}\;\;\stackrel{11\;12}{(-)}\;\;\stackrel{13\;14}{(-)}$ & 
   $\hat{\nu}^{\dagger}_{R\,2}$&
   $ \stackrel{03}{[+i]}\,\stackrel{12}{(+)}|\stackrel{56}{[+]}\,\stackrel{78}{(+)} ||
   \stackrel{9 \;10}{(+)}\;\;\stackrel{11\;12}{[+]}\;\;\stackrel{13\;14}{[+]}$ 
  &$-\frac{1}{2}$&$0$&$\frac{1}{2}$&$0$&$-\frac{1}{2}$
 \\
  $I$&$\hat{u}^{c1 \dagger}_{R\,3}$&
   $ \stackrel{03}{(+i)}\,\stackrel{12}{[+]}|\stackrel{56}{(+)}\,\stackrel{78}{[+]} ||
   \stackrel{9 \;10}{(+)}\;\;\stackrel{11\;12}{(-)}\;\;\stackrel{13\;14}{(-)}$ & 
    $\hat{\nu}^{\dagger}_{R\,3}$&
   $ \stackrel{03}{(+i)}\,\stackrel{12}{[+]}|\stackrel{56}{(+)}\,\stackrel{78}{[+]} ||
   \stackrel{9 \;10}{(+)}\;\;\stackrel{11\;12}{[+]}\;\;\stackrel{13\;14}{[+]}$ 
  &$\frac{1}{2}$&$0$&$-\frac{1}{2}$&$0$&$-\frac{1}{2}$
 \\
 $I$&$\hat{u}^{c1 \dagger}_{R\,4}$&
  $ \stackrel{03}{[+i]}\,\stackrel{12}{(+)}|\stackrel{56}{(+)}\,\stackrel{78}{[+]} ||
  \stackrel{9 \;10}{(+)}\;\;\stackrel{11\;12}{(-)}\;\;\stackrel{13\;14}{(-)}$ & 
   $\hat{\nu}^{\dagger}_{R\,4}$&
  $ \stackrel{03}{[+i]}\,\stackrel{12}{(+)}|\stackrel{56}{(+)}\,\stackrel{78}{[+]} ||
  \stackrel{9 \;10}{(+)}\;\;\stackrel{11\;12}{[+]}\;\;\stackrel{13\;14}{[+]}$ 
  &$\frac{1}{2}$&$0$&$\frac{1}{2}$&$0$&$-\frac{1}{2}$
  \\
  \hline
  $II$& $\hat{u}^{c1 \dagger}_{R\,5}$&
        $ \stackrel{03}{[+i]}\,\stackrel{12}{[+]}|\stackrel{56}{[+]}\,\stackrel{78}{[+]}||
        \stackrel{9 \;10}{(+)}\;\;\stackrel{11\;12}{(-)}\;\;\stackrel{13\;14}{(-)}$ & 
         $\hat{\nu}^{\dagger}_{R\,5}$&
        $ \stackrel{03}{[+i]}\,\stackrel{12}{[+]}|\stackrel{56}{[+]}\,\stackrel{78}{[+]}|| 
        \stackrel{9 \;10}{(+)}\;\;\stackrel{11\;12}{[+]}\;\;\stackrel{13\;14}{[+]}$ 
        &$0$&$-\frac{1}{2}$&$0$&$-\frac{1}{2}$&$-\frac{1}{2}$
 \\ 
  $II$& $\hat{u}^{c1 \dagger}_{R\,6}$&
      $ \stackrel{03}{(+i)}\,\stackrel{12}{(+)}|\stackrel{56}{[+]}\,\stackrel{78}{[+]}||
      \stackrel{9 \;10}{(+)}\;\;\stackrel{11\;12}{(-)}\;\;\stackrel{13\;14}{(-)}$ & 
    $\hat{\nu}^{\dagger}_{R\,6}$&
      $ \stackrel{03}{(+i)}\,\stackrel{12}{(+)}|\stackrel{56}{[+]}\,\stackrel{78}{[+]}|| 
      \stackrel{9 \;10}{(+)}\;\;\stackrel{11\;12}{[+]}\;\;\stackrel{13\;14}{[+]}$ 
      &$0$&$-\frac{1}{2}$&$0$&$\frac{1}{2}$&$-\frac{1}{2}$
 \\ 
 $II$&$\hat{u}^{c1 \dagger}_{R\,7}$&
 $ \stackrel{03}{[+i]}\,\stackrel{12}{[+]}|\stackrel{56}{(+)}\,\stackrel{78}{(+)}||
 \stackrel{9 \;10}{(+)}\;\;\stackrel{11\;12}{(-)}\;\;\stackrel{13\;14}{(-)}$ & 
      $ \hat{\nu}^{\dagger}_{R\,7}$&
      $ \stackrel{03}{[+i]}\,\stackrel{12}{[+]}|\stackrel{56}{(+)}\,\stackrel{78}{(+)}|| 
      \stackrel{9 \;10}{(+)}\;\;\stackrel{11\;12}{[+]}\;\;\stackrel{13\;14}{[+]}$ 
    &$0$&$\frac{1}{2}$&$0$&$-\frac{1}{2}$&$-\frac{1}{2}$
  \\
   $II$& $\hat{u}^{c1 \dagger}_{R\,8}$&
    $ \stackrel{03}{(+i)}\,\stackrel{12}{(+)}|\stackrel{56}{(+)}\,\stackrel{78}{(+)}||
    \stackrel{9 \;10}{(+)}\;\;\stackrel{11\;12}{(-)}\;\;\stackrel{13\;14}{(-)}$ & 
    $\hat{\nu}^{\dagger}_{R\,8}$&
    $ \stackrel{03}{(+i)}\,\stackrel{12}{(+)}|\stackrel{56}{(+)}\,\stackrel{78}{(+)}|| 
    \stackrel{9 \;10}{(+)}\;\;\stackrel{11\;12}{[+]}\;\;\stackrel{13\;14}{[+]}$ 
    &$0$&$\frac{1}{2}$&$0$&$\frac{1}{2}$&$-\frac{1}{3}$
 \\ 
 \hline 
 \end{tabular}
 \end{center}
\caption{\label{Table III.} 
Eight families of creation operators of $\hat{u}^{c1 \dagger}_{R}$ --- the right handed 
$u$-quark with spin $\frac{1}{2}$ and the colour charge $(\tau^{33}=1/2$, 
$\tau^{38}=1/(2\sqrt{3}))$, appearing in the first line of Table~\ref{Table so13+1.} --- 
and of  the colourless right handed neutrino $\hat{\nu}^{\dagger}_{R}$ ---  of spin 
$\frac{1}{2}$, appearing  in the $25^{th}$ line of Table~\ref{Table so13+1.} --- 
are presented in the  left and in the right column, respectively. Table is taken 
from~\cite{normaJMP2015}. 
Families belong to two groups of four families, one ($I$) is a doublet with respect to 
($\vec{\tilde{N}}_{L}$ and  $\vec{\tilde{\tau}}^{(1)}$) and  a singlet with respect to 
($\vec{\tilde{N}}_{R}$ and  $\vec{\tilde{\tau}}^{(2)}$), the other ($II$) is a singlet with respect to 
($\vec{\tilde{N}}_{L}$ and  $\vec{\tilde{\tau}}^{(1)}$) and  a doublet with with respect to 
($\vec{\tilde{N}}_{R}$ and  $\vec{\tilde{\tau}}^{(2)}$), Eq.~(\ref{so1+3}).
All the families follow from the starting one by the application of the operators 
($\tilde{N}^{\pm}_{R,L}$, $\tilde{\tau}^{(2,1)\pm}$), Eq.~(\ref{plusminus}).  The generators 
($N^{\pm}_{R,L} $, $\tau^{(2,1)\pm}$) (Eq.~(\ref{plusminus}))
transform $\hat{u}^{\dagger}_{1R}$ to all the members of one family of the same colour. 
The same generators transform equivalently the right handed   neutrino 
$\hat{\nu}^{\dagger}_{1R}$  to all the colourless members of the same family.
}
 \end{table}

{\bf I. B.ii.} $\;\;$
The eight-plets separate into two groups of four families: One group  contains  doublets with respect 
to $\vec{\tilde{N}}_{R}$ and  $\vec{\tilde{\tau}}^{2}$, these families are singlets with respect to 
$\vec{\tilde{N}}_{L}$ and  $\vec{\tilde{\tau}}^{1}$. Another group of families contains  doublets 
with respect to  $\vec{\tilde{N}}_{L}$ and  $\vec{\tilde{\tau}}^{1}$, these families are singlets 
with respect to  $\vec{\tilde{N}}_{R}$ and  $\vec{\tilde{\tau}}^{2}$. 
Mass matrices of both groups manifest correspondingly, when the scalar fields --- the gauge fields of
($\vec{\tilde{N}}_{R}$, $\vec{\tilde{\tau}}^{2}$, $U(1)$)  and  ($\vec{\tilde{N}}_{l}$,
 $\vec{\tilde{\tau}}^{1}$, $U(1)$) --- obtain nonzero vacuum expectation values. Correspondingly 
both groups manifest $SU(2)\times SU(2)\times U(1)$ symmetry, with the same $U(1)$ and 
two different $SU(2)_{(L,R)}\times SU(2)_{(I,II)}$ symmetries, Ref.~\cite{NA2018}.
 
To the lower four families the observed three families of quarks and leptons 
contribute~\cite{mdn2006,gmdn2007,gmdn2008,gn2013,gn2014,NH2017newdata}. By the 
{\it spin-charge-family} theory predicted $SU(2)\times SU(2)\times U(1)$  symmetry of mass 
matrices, which limits the number of free parameters of mass matrices, the properties of the 
fourth family could be predicted by fitting free parameters to the experimental data.  However,
the accuracy of the so far measured $3\times 3$ mixing (sub)matrices are even for quarks far 
from the required precision, which would enable prediction of masses of the fourth family 
members~\cite{gn2013,gn2014}. We predict for the assumed  masses of the fourth
family of quarks the corresponding matrix elements. Calculations show~\cite{gn2014} that the
 larger the masses of the fourth family --- up to $6$ TeV seems to be allowed by 
experiments~\cite{bereziani} --- the smaller the unwanted mixing elements which could cause 
the flavour-changing neutral currents and the better is agreement with the experimental data, 
which require, due to the observations in Refs.~\cite{bereziani,Zurab}, that there should be the 
fourth family due to the nonunitarity of the $3\times 3$ so far measured mixing matrix for 
quarks and that the $4 \times 4$ mixing matrix elements should have the properties:
$V_{u_1 d_4}> V_{u_1 d_3}$\,, \, 
$V_{u_2 d_4}<V_{u_1 d_4}$\,,  and\,
$V_{u_3 d_4}<V_{u_1 d_4}$.  Here  $u_{i}, d_{i}, i =1,2,3,4$ represent $u,c,t,u_4$ and 
$d,s,b,d_4$ quarks.

The lowest of the upper four families is, as evaluated in Refs.~\cite{gn2009,nm2015}, the candidate, 
which can explain (or at least can contribute to) the appearance of the dark matter in the universe. 
Comparing the results from following the fifth family members in the expanding universe with the 
astrophysical observations of dark matter and the direct measurements of the dark matter, the
predicted masses of the fifth family quarks would be $10^2$ TeV $< m_{q_5}\, c^2 < 4 \cdot 10^2$ 
TeV, and the scattering cross section $\sigma$ for the fifth family neutron at least $10^{-6} \times$ 
smaller than the cross section for the first family neutron. These values change if the fifth family 
neutron is not the only source of the dark matter. 

The fifth family would correspondingly
manifest completely different "nuclear force" than the members of the lower four 
families~\cite{gn2009}, 
leading to different atoms and molecules, if they would success to form a matter in the expanding 
universe.

{\bf II.} $\;\;$
The gauge fields --- the vielbeins, $f^{\alpha}{}_{a}$, and the two kinds of the spin connection 
fields, $\omega_{ab \alpha}$ and $\tilde{\omega}_{ab \alpha}$, of Eq.~(\ref{wholeaction}),
appearing in the  $2^{nd}, 3^{rd}$ and $4^{th}$ lines in Eq.~(\ref{faction})
--- manifest in $d=(3+1)$ as the vector gauge fields of  $\vec{\tau}^{3}$, Eq~.(\ref{so64}), 
$\tau^{4}$, Eq.~(\ref{so64}), $\vec{\tau}^{1}$, Eq.~(\ref{so42}), and $\vec{\tau}^{2}$, 
Eq.~(\ref{so42}), if the space index is $m=(0,1,2,3)$ ($2^{nd}$ line in Eq.~(\ref{faction})),
 as well as the scalar gauge fields, if the space index is $s \ge 5$ ($3^{rd}$ and $4^{th}$ line
 in Eq.~(\ref{faction})), of the same operators as in the vector 
gauge fields case, Ref.~\cite{nd2017}. 

Only if there are no fermion present, then both, $\omega_{ab \alpha}$ and 
$\tilde{\omega}_{ab \alpha}$, are uniquely expressed by vielbeins, 
Ref.~(\cite{normaJMP2015}, Eq.~(C9)).
\begin{eqnarray}
\label{omegatildeomega}
\omega_{ab \alpha} = \tilde{\omega}_{ab \alpha} &=& -\frac{1}{2E}\biggl\{
   e_{e\alpha}e_{b\gamma}\,\partial_\beta(Ef^{\gamma[e}f^\beta{}_{a]} )
   + e_{e\alpha}e_{a\gamma}\,\partial_\beta(Ef^{\gamma}{}_{[b}f^{\beta e]})
  \nonumber\\
                  & &  \qquad\qquad  {} - e_{e\alpha}e^e{}_\gamma\,
     \partial_\beta\bigl(Ef^\gamma{}_{[a}f^\beta{}_{b]} \bigr)
   \biggr\} \nonumber\\
                  &-& \frac{1}{d-2}  
   \biggl\{ e_{a\alpha} 
            \frac{1}{E} e^d{}_\gamma \partial_\beta
            \left(Ef^\gamma{}_{[d}f^\beta{}_{b]}\right)
\nonumber\\
                  & & \qquad {} - e_{b\alpha} 
            \frac{1}{E} e^d{}_\gamma \partial_\beta
            \left(Ef^\gamma{}_{[d}f^\beta{}_{a]}\right)
 \biggr\}\,. 
            \end{eqnarray}
%
   %

{\bf II. A.} $\;\;$ 
 It is proven in Ref.~\cite{nd2017} that  the vector  (as well as the scalar gauge fields) can indeed 
be expressed  with the spin connections (rather than with the vielbeins), \\
%
\begin{eqnarray}
 A^{Ai}_{m}&=&
\sum_{s,t} c^{Ai}{}_{st}\, \omega^{st}{}_{m},
\end{eqnarray}
 demonstrating the symmetry of space with  $(s,t) \ge5$,
making  the {\it spin-charge-family} theory  transparent and correspondingly "elegant", so that 
it is easier to recognize that the origin of charges of the observed fermions, vector gauge fields, 
Higgs's scalar and Yukawa couplings might really be in $(d-4)$ space.

In the presence of the condensate, Table~\ref{Table con.}, of the right handed neutrinos, 
all the vector gauge fields and the scalar gauge fields, which interact with the condensate, 
gain masses. Only the weak ($SU(2)_I$), the colour ($SU(3)$) and the hyper ($U(1)$, $Y=\tau^{4}
+ \tau^{23}$) gauge fields, which do not interact with the condensate, remain massless.

{\bf II.  A.i.} $\;\;$
The weak vector gauge fields $\vec{A}^{1}_{m} $ , the gauge field of $SU(2)_{I}$,
 and $\vec{A}^{2}_{m}$, the gauge fields of $SU(2)_{II}$, are the 
superposition of gauge fields 
$\omega_{s't' s}$ (Ref.~\cite{normaJMP2015}, Eqs.~(8,9,10)),
\begin{eqnarray}
\label{a1a2omegas}
 \vec{A}^{1}_{m} &=& (\omega_{58m} - \omega_{67m}, 
\omega_{57m} + \omega_{68m}, \omega_{56m} - \omega_{78m})\,,
 \nonumber\\
\vec{A}^{2}_{m} &=& (\omega_{58m} + \omega_{67m}, 
\omega_{57m} - \omega_{68m}, \omega_{56m} + \omega_{78m}) \,.
\end{eqnarray}
Taking into account Eq.~(\ref{so64}) one easily finds the colour vector gauge field expressed with
$\omega_{st m}$.
$\vec{A}^{2}_{m}$ get masses by interaction with the condensate.

In Ref.~\cite{nd2017}, Eqs.~(24-25),  the reader can find Lagrange density for the $R^{(d-4)}$ 
part of the gravity field $R$, Eq.(\ref{wholeaction}), expressed by the vector gauge fields 
$\vec{A}^{A}_{m}$. 

{\bf II. B.} $\;\;$ 
The scalar gauge fields are the superposition of either  $\omega_{s't's}$, with $(s',t',s)=$ 
$ (5,6,\cdots,14)$, Ref.~\cite{nd2017}, or $\tilde{\omega}_{abs}$, with $(a,b) =$ 
$(0,1,\cdots,8)$ and $(s)=$ $ (5,6,7,8)$, Refs.~\cite{n2014matterantimatter,JMP2013,%
normaJMP2015}, the fourth line in Eq.~(\ref{faction}).  

Both kinds of scalar fields with $s=(7,8)$  contribute to the masses of the two groups of 
four families.  Scalar fields   $\omega_{s't's}$, with $(s',t')=$ 
$ (5,6,\cdots,14)$, $s=  (9,10,\cdots,14)$ contribute to matter-antimatter asymmetry 
and to proton decay~\cite{n2014matterantimatter}.

{\bf II. B.i.} $\;\;$ In the {\it spin-charge-family} theory the scalar fields with the space
 index $s=(7,8)$  carry  with respect to this space index the weak charge and the
 hyper charge ($\mp \frac{1}{2}, \pm \frac{1}{2}$), 
respectively, independent of whether they are superposition of $\omega_{s' t' s}$ or of
$\tilde{\omega}_{a b s}$, $s=(7,8)$, Refs.~\cite{normaJMP2015,IARD2016,%
n2014matterantimatter}. 

There are twice two triplets, the superposition of $\tilde{\omega}_{ab s}$, 
Eqs.~(\ref{so1+3}, \ref{so42})  with $S^{ab}$ replaced by $\tilde{S}^{ab}$,
the gauge scalar fields of either the group $\widetilde{SU}(2)_{\widetilde{SO}(3,1)_L}
\times \widetilde{SU}(2)_{I}$ or of the group $\widetilde{SU}(2)_{\widetilde{SO}(3,1)_R} 
\times \widetilde{SU}(2)_{II}$, the first two triplets 
interacting with one group of four families, the second two triplets  interacting with another 
group of four families, both groups presented in Table~\ref{Table III.}.  There are also
three singlets, the gauge scalar fields of $=(Q$,$Q',Y'$), Eq.~(\ref{YQY'Q'andtilde}), which 
are the superposition of $\omega_{s't' s}$ and interact with members of all the eight 
families of  Table~\ref{Table III.}~\cite{JMP2013,normaJMP2015,IARD2016,%
n2014matterantimatter}.

 Let us use  a  common notation  $ A^{Ai}_{s}$ for all the scalar 
fields, independently of whether  they originate in $\tilde{\omega}_{abs}$ or $\omega_{abs}$,
 $s=(7,8)$, 
\begin{eqnarray}
\label{commonAi}
 A^{Ai}_{s} &\in& (\,A^{Q}_{s}\,,A^{Q'}_{s}\,, A^{Y'}_{s}\,, 
 \vec{\tilde{A}}^{\tilde{1}}_{s}\,, 
 \vec{\tilde{A}}^{\tilde{N}_{\tilde{L}}}_{s}\,, \vec{\tilde{A}}^{\tilde{2}}_{s}\,, 
 \vec{\tilde{A}}^{\tilde{N}_{\tilde{R}}}_{s}\,)\,,\nonumber\\
\tau^{Ai} &\supset& (Q,\,Q',\,Y', \,\vec{\tilde{\tau}}^{1},\, \vec{\tilde{N}}_{L},\,
\vec{\tilde{\tau}}^{2},\,\vec{\tilde{N}}_{R})\,.
\end{eqnarray}
Here $\tau^{Ai}$ represent the operators of the groups the gauge scalar fields of which
are $A^{Ai}_{s}$.

Let us rewrite the third line in Eq.~(\ref{faction}) as follows, 
Ref.~(\cite{normaJMP2015}, Eqs.~(18-19)).
\begin{eqnarray}
\label{eigentau1tau2}
 & &\sum_{s=(7,8), A,i}\, \bar{\psi} \,\gamma^s\, ( - \tau^{Ai} \,A^{Ai}_{s}\,)\,\psi =
\nonumber\\
 & &\sum_{A,i} - \bar{\psi}\,\{\,\stackrel{78}{(+)}\, \tau^{Ai} \,(A^{Ai}_{7} - i   
 \,A^{Ai}_{8})\, + \stackrel{78}{(-)}(\tau^{Ai} \,(A^{Ai}_{7} + i \,A^{Ai}_{8})\,\}\,\psi\,,
 \nonumber\\
 & &\stackrel{78}{(\pm)} = \frac{1}{2}\, (\gamma^{7} \pm \,i \, \gamma^{8}\,)\,,\quad
 A^{Ai}_{\scriptscriptstyle{\stackrel{78}{(\pm)}}}: = (A^{Ai}_7 \,\mp i\, A^{Ai}_8)\,,
\end{eqnarray}
with the summation over $A,i$ performed, since $A^{Ai}_s$ represent the scalar fields 
($A^{Q}_{s}$, $A^{Q'}_{s}$, $A^{Y'}_{s}$, 
$\tilde{A}^{\tilde{4}}_{s}$, 
$\vec{\tilde{A}}^{\tilde{1}}_{s}$, $\vec{\tilde{A}}^{\tilde{2}}_{s}$,
 $\vec{\tilde{A}}^{\tilde{N}_{R}}_{s}$ and  $\vec{\tilde{A}}^{\tilde{N}_{L}}_{s}$).
In the low energy regime the momentum $p_{s}$, $s=(7,8)$ can be neglected.

Taking into account that $\tau^{13} = \frac{1}{2} (S^{56}- S^{78})$,
$Y= (\tau^{23} + \tau^{4})$,  $\tau^{23} = \frac{1}{2} (S^{56}+ S^{78})$, while $\tau^4=
- \frac{1}{3} (S^{9\, 10} + S^{11\, 12} + S^{13\, 14})$, 
and $S^{ab} A_{c} = i (A^a \delta^b_c- A^b \delta^a_c)$, one finds
\begin{eqnarray}
\label{checktau13Y}
\tau^{13}\,(A^{Ai}_7 \,\mp i\, A^{Ai}_8)&=& \pm \,\frac{1}{2}\,(A^{Ai}_7 \,
\mp i\, A^{Ai}_8)\,,\nonumber\\
Y\,(A^{Ai}_7 \,\mp i\, A^{Ai}_8)&=& \mp \,\frac{1}{2}\,(A^{Ai}_7 \,\mp i\, A^{Ai}_8)\,,\nonumber\\
Q\,(A^{Ai}_7 \,\mp i\, A^{Ai}_8)&=& 0\,.
\end{eqnarray}
This are quantum numbers of the by the {\it standard model}  assumed Higgs.
These scalar gauge fields with the space index $(7,8)$, gaining nonzero vacuum expectation
values (by assumption as in the {\it standard model} so far), cause the electroweak break, 
breaking the weak and the 
hyper charge, explaining the appearance of in the  {\it standard model} assumed Higgs and the 
Yukawa couplings, predicting the existence of several scalars --- two triplets and three singlets, 
which couple to the lower four families, making them massive and giving masses to weak 
bosons.

These scalar fields manifest the $SU(2) \times SU(2) \times U(1)$ symmetry, which
reduces the number of free parameters in mass matrices of quarks and leptons, enabling
predictions of properties of the four families~\cite{gn2013,gn2014,NA2018}.

{\bf II. B.ii.} $\;\;$
The scalar fields  with the space index $s=(9,10,\cdots,14)$, presented in Table~\ref{Table bosons.}, 
carry  triplet or antitriplet colour charges and the "spinor" charge equal to twice the quark or antiquark 
"spinor" charge,  and the fractional hyper and electromagnetic charge.  

They carry in addition the quantum numbers of the adjoint representations originating in $S^{ab}$ or in
$\tilde{S}^{ab}$. (Although carrying the colour charge of  one of the triplet or antitriplet quantum
numbers, these fields can not be interpreted as superpartners of the quarks, since they do not 
have quantum numbers as required by, let say, the $N=1$ supersymmetry.
The hyper charges  and the electromagnetic charges are namely not those required by the 
supersymmetric partners to the family members.)

 \begin{table}
 \begin{tiny}
 \begin{center}
 \begin{tabular}{|c|c|c|c|c|c|c|c|c|c|c|c|c|c|}
 \hline
 ${\rm field}$&prop. & $\tau^4$&$\tau^{13}$&$\tau^{23}$&($\tau^{33},\tau^{38}$)&$Y$&
$Q$&$\tilde{\tau}^4$ 
 &$\tilde{\tau}^{13}$&$\tilde{\tau}^{23}$&$\tilde{N}_{L}^{3}$ &$\tilde{N}_{R}^{3}$  \\
 \hline
 $A^{1\spm}_{\scriptscriptstyle{\stackrel{9\,10}{(\cpm)}}}$& scalar&  $\cmp \frac{1}{3}$&
$\spm 1$&$0$ 
 & ($\cpm\frac{1}{2},$ $\cpm \frac{1}{2\sqrt{3}}$)& $\cmp \frac{1}{3}$&$\cmp \frac{1}{3}+
 \smp 1$&$0$
 &$0$&$0$&$0$&$0$\\ 
 $A^{13}_{\scriptscriptstyle{\stackrel{9\,10}{(\cpm)}}}$   & scalar&  $\cmp \frac{1}{3}$&$0$&
$0$ 
 & ($\cpm\frac{1}{2},$ $\cpm \frac{1}{2\sqrt{3}}$)& $\cmp \frac{1}{3}$&$\cmp \frac{1}{3}$&
$0$&$0$&$0$&$0$&$0$\\ 
 $A^{1\spm}_{\scriptscriptstyle{\stackrel{11\,12}{(\cpm)}}}$& scalar&  $\cmp \frac{1}{3}$&
$\smp 1$&$0$ 
  & ($\cmp\frac{1}{2},$ $\cpm \frac{1}{2\sqrt{3}}$)& $\cmp \frac{1}{3}$&$\cmp \frac{1}{3}+ 
\smp 1$&$0$
  &$0$&$0$&$0$&$0$\\ 
  $A^{13}_{\scriptscriptstyle{\stackrel{11\,12}{(\cpm)}}}$   & scalar&  $\cmp \frac{1}{3}$&$0$&
$0$ 
 & ($\cmp\frac{1}{2},$ $\cpm \frac{1}{2\sqrt{3}}$)& $\cmp \frac{1}{3}$&$\cmp \frac{1}{3}$&
$0$&$0$&$0$&$0$&$0$\\ 
 $A^{1\spm}_{\scriptscriptstyle{\stackrel{13\,14}{(\cpm)}}}$& scalar&  $\cmp \frac{1}{3}$&
$\smp 1$&$0$ 
   & ($0,$ $\cmp \frac{1}{\sqrt{3}}$)& $\cmp \frac{1}{3}$&$\cmp \frac{1}{3}+ \smp 1$&$0$
   &$0$&$0$&$0$&$0$\\ 
   $A^{13}_{\scriptscriptstyle{\stackrel{13\,14}{(\cpm)}}}$   & scalar&  $\cmp \frac{1}{3}$&$0$&
$0$ 
 & ($0,$ $\cmp \frac{1}{\sqrt{3}}$)& $\cmp \frac{1}{3}$&$\cmp \frac{1}{3}$&$0$&$0$&$0$&
$0$&$0$\\ 
 \hline
 $A^{2\spm}_{\scriptscriptstyle{\stackrel{9\,10}{(\cpm)}}}$& scalar&  $\cmp \frac{1}{3}$&$0$&
$\spm 1$ 
  & ($\cpm\frac{1}{2},$ $\cpm \frac{1}{2\sqrt{3}}$)& $\cmp \frac{1}{3}+ \smp 1$&$\cmp 
\frac{1}{3}+ \smp 1$&$0$
  &$0$&$0$&$0$&$0$\\  
  $A^{23}_{\scriptscriptstyle{\stackrel{9\,10}{(\cpm)}}}$   & scalar&  $\cmp \frac{1}{3}$&$0$&
$0$ 
 & ($\cpm\frac{1}{2},$ $\cpm \frac{1}{2\sqrt{3}}$)& $\cmp \frac{1}{3}$&$\cmp \frac{1}{3}$&
$0$&$0$&$0$&$0$&$0$\\ 
 $\cdots$&&&&&&&&&&&&\\
 \hline
 $\tilde{A}^{1 \spm}_{\scriptscriptstyle{\stackrel{9 10}{(\cpm)}}}$& scalar& $\cmp \frac{1}{3}$&
$0$&$0$ 
 & ($\cpm\frac{1}{2},$ $\cpm \frac{1}{2\sqrt{3}}$)& $\cmp \frac{1}{3}$&$\cmp \frac{1}{3}$&
$0$
 &$\spm 1$&$0$&$0$&$0$\\ 
 $\tilde{A}^{13}_{\scriptscriptstyle{\stackrel{9 10}{(\cpm)}}}$& scalar& $\cmp \frac{1}{3}$&$0$&
$0$ 
  & ($\cpm\frac{1}{2},$ $\cpm \frac{1}{2\sqrt{3}}$)& $\cmp \frac{1}{3}$&$\cmp \frac{1}{3}$&
$0$
 &$0$&$0$&$0$&$0$\\ 
  $\cdots$&&&&&&&&&&&&\\
 \hline
 $\tilde{A}^{2 \spm}_{\scriptscriptstyle{\stackrel{9 10}{(\cpm)}}}$& scalar& $\cmp \frac{1}{3}$&
$0$&$0$ 
  & ($\cpm\frac{1}{2},$ $\cpm \frac{1}{2\sqrt{3}}$)& $\cmp \frac{1}{3}$&$\cmp \frac{1}{3}$&
$0$
  &$0$&$\spm 1$&$0$&$0$\\ 
  $\tilde{A}^{23}_{\scriptscriptstyle{\stackrel{9 10}{(\cpm)}}}$& scalar& $\cmp \frac{1}{3}$&$0$&
$0$ 
   & ($\cpm\frac{1}{2},$ $\cpm \frac{1}{2\sqrt{3}}$)& $\cmp \frac{1}{3}$&$\cmp \frac{1}{3}$&
$0$
  &$0$&$0$&$0$&$0$\\ 
  $\cdots$&&&&&&&&&&&&\\
 \hline
 $\tilde{A}^{N_{L} \spm}_{\scriptscriptstyle{\stackrel{9 10}{(\cpm)}}}$& scalar& $\cmp 
\frac{1}{3}$&$0$&$0$ 
   & ($\cpm\frac{1}{2},$ $\cpm \frac{1}{2\sqrt{3}}$)& $\cmp \frac{1}{3}$&$\cmp \frac{1}{3}$&
$0$
   &$0$&$0$&$\spm 1$&$0$\\ 
   $\tilde{A}^{N_{L} 3}_{\scriptscriptstyle{\stackrel{9 10}{(\cpm)}}}$& scalar& $\cmp \frac{1}{3}$
&$0$&$0$ 
    & ($\cpm\frac{1}{2},$ $\cpm \frac{1}{2\sqrt{3}}$)& $\cmp \frac{1}{3}$&$\cmp \frac{1}{3}$&
$0$
   &$0$&$0$&$0$&$0$\\ 
   $\cdots$&&&&&&&&&&&&\\
 \hline
 $\tilde{A}^{N_{R} \spm}_{\scriptscriptstyle{\stackrel{9 10}{(\cpm)}}}$& scalar& $\cmp 
\frac{1}{3}$&$0$&$0$ 
    & ($\cpm\frac{1}{2},$ $\cpm \frac{1}{2\sqrt{3}}$)& $\cmp \frac{1}{3}$&$\cmp \frac{1}{3}$&
$0$
    &$0$&$0$&$0$&$\spm 1$\\ 
    $\tilde{A}^{N_{R} 3}_{\scriptscriptstyle{\stackrel{9 10}{(\cpm)}}}$& scalar& $\cmp 
\frac{1}{3}$&$0$&$0$ 
     & ($\cpm\frac{1}{2},$ $\cpm \frac{1}{2\sqrt{3}}$)& $\cmp \frac{1}{3}$&$\cmp 
\frac{1}{3}$&$0$
    &$0$&$0$&$0$&$0$\\ 
    $\cdots$&&&&&&&&&&&&\\
 \hline
 $A^{3 i}_{\scriptscriptstyle{\stackrel{9\,10}{(\cpm)}}}$& scalar&  $\cmp \frac{1}{3}$&$0$&$0$ 
  & ($\spm 1 + \cpm\frac{1}{2},$ $\cpm \frac{1}{2\sqrt{3}}$)& $\cmp \frac{1}{3}$&$\cmp 
\frac{1}{3}$&$0$
 &$0$&$0$&$0$&$0$\\ 
    $\cdots$&&&&&&&&&&&&\\
  \hline
 $A^{4}_{\scriptscriptstyle{\stackrel{9\,10}{(\cpm)}}}$& scalar&  $\cmp \frac{1}{3}$&$0$&$0$ 
  & ($ \cpm\frac{1}{2},$ $\cpm \frac{1}{2\sqrt{3}}$)& $\cmp \frac{1}{3}$&$\cmp \frac{1}{3}$&
$0$
 &$0$&$0$&$0$&$0$\\ 
    $\cdots$&&&&&&&&&&&&\\   
\hline 
%
 \end{tabular}
 \end{center}
 \end{tiny}
 \caption{\label{Table bosons.}%
 Quantum numbers of the scalar gauge fields carrying the space index $t =(9,10,\cdots,14)$, 
 appearing in  the fourth line of Eq.~(\ref{faction}), are presented. To the colour charge of all
 these scalar  fields the space degrees  of freedom --- the space index --- contribute one of 
the triplets or antitriplet values. 
These  scalars are  with respect to the two $SU(2)$ charges, ($\vec{\tau}^1$  and 
$\vec{\tau}^2$),  and the two   $\widetilde{SU}(2)$  charges, ($\vec{\tilde{\tau}}^1$  and 
$\vec{\tilde{\tau}}^2$),   triplets  (that is in the adjoint representations of the  corresponding 
groups), and they all carry  twice the  "spinor" number ($\tau^{4}$) of the quarks or antiquarks.  
The quantum numbers of the two  vector gauge fields, the colour and the  $U(1)_{II}$ ones, 
are added. These Table is taken from Ref.~\cite{n2014matterantimatter}, Table I.
%
We invite the reader to visit  Ref.~\cite{n2014matterantimatter} for more details.
 }
  \end{table}

Let us have a look what do the scalar fields, appearing in the fourth line of Eq.~(\ref{faction}) and in 
the seventh line of Table~\ref{Table bosons.}, do when applying on the left handed members of 
the Weyl representation presented on Table~\ref{Table so13+1.}, 
containing quarks and leptons and antiquarks and antileptons~\cite{pikan2003,pikan2006,nhds}.

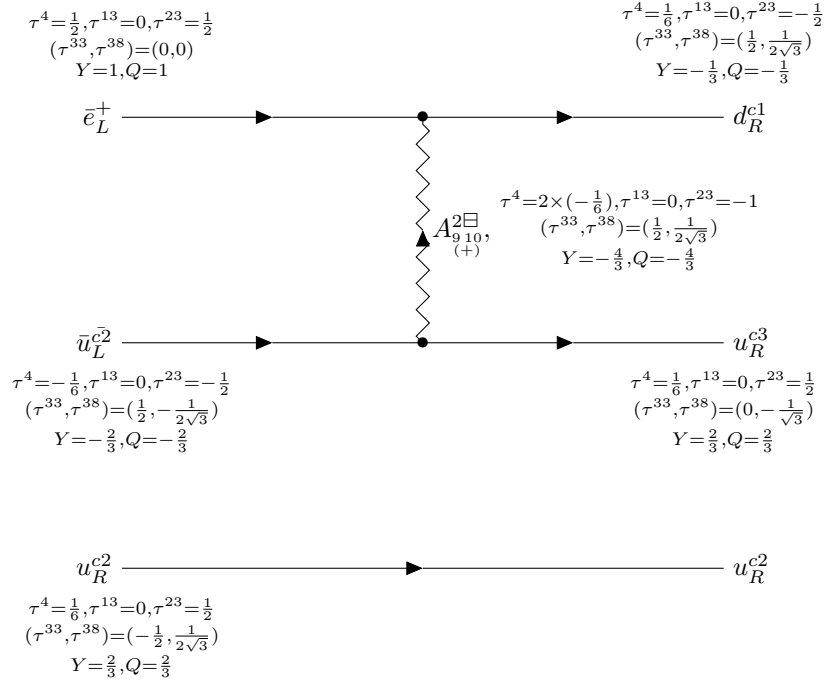
\begin{figure}
\begin{tikzpicture}[>=triangle 45]
\draw [->] (0,-3) node [anchor=east] {$u_R^{c2}$} 
node [anchor=north] {$\begin{smallmatrix} 
                                     { }\\ {}\\
                                     \tau^4=\frac{1}{6},
                                     \tau^{13}=0,
                                     \tau^{23}=\frac{1}{2}\\
                                     (\tau^{33},\tau^{38})=(-\frac{1}{2},\frac{1}{2\sqrt{3}})\\
                                      Y=\frac{2}{3}, Q=\frac{2}{3}
                                      \end{smallmatrix}$}  -- (4,-3);
\draw (4,-3) -- (8,-3) node [anchor=west] {$u_R^{c2}$};
\draw [->] (0,0) node [anchor=east] {$\bar{u}_L^{\bar{c2}}$}
node [anchor=north] {$\begin{smallmatrix} 
                                     {}\\ {}\\
                                     \tau^4=-\frac{1}{6},
                                     \tau^{13}=0,
                                     \tau^{23}=-\frac{1}{2}\\
                                     (\tau^{33},\tau^{38})=(\frac{1}{2},-\frac{1}{2\sqrt{3}})\\
                                      Y=-\frac{2}{3}, Q=-\frac{2}{3}
                                      \end{smallmatrix}$}  -- (2,0);
\draw (2,0) -- (4,0);
\draw [->] (4,0) -- (6,0);
\draw (6,0) -- (8,0) node [anchor=west] {$u_R^{c3}$}
node [anchor=north] {$\begin{smallmatrix} 
                                     {}\\ {}\\
                                     \tau^4=\frac{1}{6},
                                     \tau^{13}=0,
                                     \tau^{23}= \frac{1}{2}\\
                                     (\tau^{33},\tau^{38})=(0,-\frac{1}{\sqrt{3}})\\
                                      Y=\frac{2}{3}, Q=\frac{2}{3}
                                      \end{smallmatrix}$} ;
\draw [->] (0,3) node [anchor=east] {$\bar{e}_L^+$} 
node [anchor=south] {$\begin{smallmatrix} 
                                     \tau^4=\frac{1}{2},
                                     \tau^{13}=0,
                                     \tau^{23}=\frac{1}{2}\\
                                     (\tau^{33},\tau^{38})=(0,0)\\
                                      Y=1, Q=1\\
                                      { \ } \\ { \ }
                                      \end{smallmatrix}$} -- (2,3);
\draw (2,3) -- (4,3);
\draw [->] (4,3) -- (6,3);
\draw (6,3) -- (8,3) node [anchor=west] {$d_R^{c1}$} 
node [anchor=south] {$\begin{smallmatrix} 
                                     \tau^4=\frac{1}{6},
                                     \tau^{13}=0,
                                     \tau^{23}=-\frac{1}{2}\\
                                     (\tau^{33},\tau^{38})=(\frac{1}{2},\frac{1}{2\sqrt{3}})\\
                                      Y=-\frac{1}{3}, Q=-\frac{1}{3}\\
                                     { \ } \\ {\ }
                                      \end{smallmatrix}$};
\draw [->,snake]
(4,0) node {$\bullet$}-- (4,1.5) 
node [anchor=west]
{$A^{2\sminus}_{\scriptscriptstyle{\stackrel{9\,10}{(+)}}}, 
\begin{smallmatrix} 
                                     \tau^4=2\times(-\frac{1}{6}),
                                     \tau^{13}=0,
                                     \tau^{23}=-1\\
                                     (\tau^{33},\tau^{38})=(\frac{1}{2},\frac{1}{2\sqrt{3}})\\
                                      Y=-\frac{4}{3}, Q=-\frac{4}{3}
                                      \end{smallmatrix} $};
\draw [snake]
(4,1.5) -- (4,3) node {$\bullet$};
\end{tikzpicture}
\caption{\label{proton is born1.} The birth of a "right handed proton" out of a positron  
$\bar{e}^{+}_{L}$, antiquark $\bar{u}_L^{\bar{c2}}$ and quark (spectator) $u_{R}^{c2}$.  
The family quantum number can be any.} 
\end{figure}

 Fig.~\ref{proton is born1.} presents  the creation of proton due to the interaction of quarks
and leptons with these scalar fields. One can read on this 
Fig.~\ref{proton is born1.}  all the quantum numbers of a positron ($57^{th}$ line of 
Table~\ref{Table so13+1.}), an antiquark ($43^{rd}$ line of Table~\ref{Table so13+1.}), and 
a quark ($9^{th}$ line of Table~\ref{Table so13+1.}), as well as of the scalar field
$A^{2\sminus}_{\scriptscriptstyle{\stackrel{9\,10}{(+)}}}$, seventh line of 
Table~\ref{Table bosons.}, involved in the proton birth. 
The opposite transition at low energies would make the proton decay.

After the appearance of the condensate of the two right handed neutrinos, Table~\ref{Table con.}, 
the discrete symmetry ${\bf \mathbb{C}}_{N} {\cal P}_{N}$ is obviously broken.
In the expanding universe, fulfilling the Sakharov request for appropriate non-thermal equilibrium, 
the triplet scalars from Table~\ref{Table bosons.} have a chance to explain the matter-antimatter 
asymmetry in the universe~\cite{n2014matterantimatter}.

%

{\bf III.} $\;\;$
The {\it spin-charge-family} theory suggests two kinds of phase transitions --- two kinds of 
breaking symmetries: 
The appearance of the condensate and the nonzero vacuum expectation values of the 
scalar fields  with the space index $s= (7,8)$. 

{\bf III. A.} $\;\;$ Table~\ref{Table con.} represents the properties of the condensate of
the two right  handed neutrinos  $\hat{\nu}^{\dagger}_{R8}$ --- Table~\ref{Table III.} --- 
of spin up and spin down, breaking the discrete
 ${\bf \mathbb{C}}_{N} {\cal P}_{N}$ symmetry~Subsect.~\ref{CPT}, 
\cite{n2014matterantimatter,nhds}.

 \begin{table}
 \begin{center}
 \begin{tabular}{|c|c c c c c c c |c c c c c c c|}
 \hline
 state & $S^{03}$& $ S^{12}$ & $\tau^{13}$& $\tau^{23}$ &$\tau^{4}$& $Y$&$Q$&
$\tilde{\tau}^{13}$&
 $\tilde{\tau}^{23}$&$\tilde{\tau}^4$&$\tilde{Y} $& $\tilde{Q}$&$\tilde{N}_{L}^{3}$& 
$\tilde{N}_{R}^{3}$
 \\
 \hline
 ${\bf (|\nu_{1 R}^{VIII}>_{1}\,|\nu_{2 R}^{VIII}>_{2})}$
 & $0$& $0$& $0$& $1$& $-1$ & $0$& $0$& $0$ &$1$& $-1$& $0$& $0$& $0$& $1$\\ 
 \hline
 $ (|\nu_{1 R}^{VIII}>_{1}|e_{2 R}^{VIII}>_{2})$
 & $0$& $0$& $0$& $0$& $-1$ & $-1$& $-1$ & $0$ &$1$& $-1$& $0$& $0$& $0$& $1$\\ 
 $ (|e_{1 R}^{VIII}>_{1}|e_{2 R}^{VIII}>_{2})$
 & $0$& $0$& $0$& $-1$& $-1$ & $-2$& $-2$ & $0$ &$1$& $-1$& $0$& $0$& $0$& $1$\\ 
 \hline 
 \end{tabular}
 \end{center}
\caption{\label{Table con.} 
The condensate of the two right handed neutrinos $\nu_{R}$,  with the quantum numbers
of the $VIII^{th}$ 
family, coupled to spin zero and belonging to a triplet with respect to the generators 
$\tau^{2i}$, is presented, together with its two partners. 
The condensate carries $\vec{\tau}^{1}=0$, $\tau^{23}=1$, 
$\tau^{4}=-1$ and $Q=0=Y$. The triplet carries $\tilde{\tau}^4=-1$, $\tilde{\tau}^{23}=1$
 and $\tilde{N}_{R}^3 = 1$, $\tilde{N}_{L}^3 = 0$,  
$\tilde{Y}=0 $, $\tilde{Q}=0$. 
The family quantum numbers of quarks and leptons are presented in Table~\ref{Table III.}. 
}
 \end{table}
%

Due to the interaction with the condensate of Table~\ref{Table con.} the gauge vector 
fields of $\vec{\tau}^{2}$  and $\tau^{4}$ become massive. 
The colour vector gauge fields of $\vec{\tau}^{3}$, the weak vector gauge fields of 
$\vec{\tau}^{1}$ and the hyper vector gauge field of $Y$ do not interact with the 
condensate (the corresponding quantum numbers of the condensate are zero) and 
correspondingly remain massless,  the gravity in $d=(3+1)$, which is the gauge 
field of $S^{mn}$ and $p_m$, remains massless as well. 

Due to nonzero family quantum numbers of the condensate  the corresponding scalar 
gauge fields become massive.
The condensate gives masses to all the scalars from table~\ref{Table bosons.}, either 
because they couple to the condensate due to $\tau^{4}$ or $\tilde{\tau^{4}}$ or 
$\tau^{23}$ or $\tilde{\tau}^{23}$ quantum numbers. It gives masses also to all the 
scalar fields with $s \in (5,6,7,8)$, since they couple to the condensate due to the nonzero 
$\tau^{23}$. 
The scalar fields with the quantum numbers of the upper four families 
couple in addition through their family quantum numbers. 

{\bf III. B.} $\;\;$ The electroweak phase transition is caused by the nonzero vacuum 
expectation values of twice two triplets and three  singlet scalars, giving masses to the 
lower fourth families --- two of twice two triplets and three singlets --- and to the upper
four families --- another two triplets and the same three singlets.

{\bf IV.} $\;\;$
 Predictions of the {\it spin-charge-family} theory so far.

{\bf IV. A.} $\;\;$ The {\it spin-charge-family} theory predicts the fourth family to 
the observed three to be observed at the LHC~\cite{gmdn2008}. By predicting 
symmetry of mass matrices (in all orders of loop
corrections~\cite{NA2018}) the theory enables for accurate enough measured mixing 
matrices of the $3\times 3$ submatrices (the sensitivity of the fitting procedure on 
masses of the so far measured quarks and leptons is much smaller~\cite{gn2013,gn2014}),  
and due to other measured properties of quarks and leptons~\cite{bereziani}, to 
predict the properties of the $4\times 4$ mixing matrices and to explain correspondingly
the origin of Higgs and Yukawa couplings.  The $4 \times 4$ mixing matrix elements for quarks 
are predicted to have the properties:
$V_{u_1 d_4}> V_{u_1 d_3}$\,, \, 
$V_{u_2 d_4}<V_{u_1 d_4}$\,,  and\,
$V_{u_3 d_4}<V_{u_1 d_4}$, here  $u_{i}, d_{i}, i =1,2,3$ represent $u,c,t,u_4$ 
and $d,s,b,d_4$ quarks.

The theory explains~\cite{NH2017newdata} why the 
fourth family has not yet been observed, which is the main argument against the existence
of four families~\cite{AhmedAli,MatthiasNeubert} among experts in high energy physics.

{\bf IV. B.} $\;\;$ The theory predicts the existence of several scalar fields --- there are
two triplets and three singlets  which determine masses of the lower four 
families~\cite{normaJMP2015,JMP2013,IARD2016,n2012scalars} --- 
some of which will be observed in the  near future measurements.

{\bf IV. C.} $\;\;$ The theory predicts the second group of four families, the stable one of 
these four families contributing to the dark matter~\cite{gn2009}. The nuclear force among 
these baryons differs a lot from the so far observed nuclear force~\cite{gn2009,nm2015}.

{\bf IV. D.} $\;\;$ The masses of quarks and leptons are, according to these two groups
of four families, spread from $10^{-3}$ eV to  few TeV ---  at least $12$ orders of magnitude 
for the first four families --- and from $100$ TeV to $10^{13}$ TeV --- at least $11$ orders of 
magnitude for the second four families, offering the explanation  for the 
hierarchy problem. (The mass matrices of the two groups of mass matrices are very closed to
the democratic ones~\cite{gn2013,gn2014}). 

{\bf IV. E.} $\;\;$ The {\it spin-charge-family} theory predicts the masses of the  
dark matter baryons~\cite{gn2009}. 

{\bf IV. F.} $\;\;$ The {\it spin-charge-family} theory predicts the scalar fields which 
contribute to the matter-antimatter asymmetry in the universe~\cite{n2014matterantimatter}
and correspondingly also to  the proton decay.

{\bf V.} $\;\;$  The {\it spin-charge-family} theory has  (so far) several open problems, 
although it is also true that the more work 
is done, the more solutions of the open problems follow.

{\bf V. A.} $\;\;$  In the {\it spin-charge-family} the vector and scalar gauge fields 
originate in gravity as the two kinds of the spin connection fields and the vielbeins. 
 In the low energy region these vector and scalar gauge fields  can
be quantized in the usual way~\cite{nd2017}. Yet the
quantization of gravity remains as an open problem when the energies rise up to
$10^{16}$ GeV and above.

{\bf V. B.} $\;\;$ The dimension of space time --- $13+1$ --- remains as an open 
problem: Why $d= (13 +1)$, why not $\infty$? (Only $0$ and $\infty$ need no 
explanation.) How has the universe come to  $d=(13+1)$~\cite{NH2007Majorana}? 

{\bf V. C.} $\;\;$ Breaking the symmetry with the appearance of the 
condensate~\cite{n2014matterantimatter}, which lead to observable properties of fermion 
and boson fields, explaining all the assumptions of the {\it standard model}s, needs to be 
studied as a dynamical appearance of the condensate in the expanding universe. 

{\bf V. D.} $\;\;$ It should be demonstrated dynamically how do the scalar fields gain nonzero 
vacuum expectation values, leading to the effective fields as assumed for the Higgs. The 
demonstrations, made in Refs.~\cite{NHD,ND012,familiesNDproc,NHD2011}  for the toy 
model in $d=(5+1)$ must be done also for $d=(13+1)$.

{\bf V. F.} $\;\;$ The coupling constants of the gauge and scalar fields in the low energy 
regime should be evaluated when starting with the simple action of Eq.~(\ref{wholeaction})
in $d=(13+1)$, with only one (or already with two) coupling constants.

{\bf V. G.} $\;\;$ There are additional open problems which we already see and either solve, like the
one treated in this paper about the internal degrees of freedom of fermions in Clifford and 
Grassmann space and the new way of second quantization procedure, which explains 
the usual  way of second quantization, or they wait to be solved, like the  lepton number non 
conservation  in the {\it spin-charge-family} theory. And there are open problems which we do 
not see yet or which we could better understand if learning more from all the trials to understand 
the evolution of the universe and the creation of hadrons of all kinds in the literature.

\section{Fermions in Grassmann and in Clifford space}
\label{observations}

In the literature the Clifford algebra is frequently discussed as an useful tool to describe  
internal degrees of freedom of fermions~\cite{Hestenes84,Lounesto2001,MP1707.05695}. 
In the {\it spin-charge-family} theory Clifford space is used to describe all the internal degrees 
of fermions --- quarks and leptons with their families included~\cite{norma92,norma93,%
normaJMP2015}. 

In this paper we demonstrate that the Clifford algebra offers an  elegant and  transparent way 
to better understand fermions properties: In even dimensional spaces --- we make a choice 
of $d=2(2n+1)$, $n=3$ --- the creation operators of an odd Clifford character can be defined 
(they are superposition of odd numbers of the Clifford algebra objects ($\gamma^a$'s or 
$\tilde{\gamma}^a$'s, Eq.~(\ref{twocliffordsab})), each of them  is a product of $\frac{d}{2}$  
nilpotents and projectors, Eq.~(\ref{signature0},  \ref{signature})~\cite{nh02,nh03}, so that they
are the eigenvectors of twice all the $\frac{d}{2}$ members of the two kinds of the Cartan 
subalgebras of the Lorentz algebra --- $S^{ab}$ and $\tilde{S}^{ab}$ --- with  the half integer
eigenvalues, Eq.~(\ref{grapheigen}). \\
These creation operators, Eq.~(\ref{anycreation}), and their Hermitian conjugated partners  --- 
the annihilation operators, Eq.~(\ref{anyannihilation}) --- 
fulfill on the vacuum state, Eq.~(\ref{vac1}), the anti commutation relations required for fermions, 
Eq.~{\ref{alphagammaprod0}}. \\
The superposition of these creation operators solve for a particular momentum $p^a$ the 
equation of motions for free massless fermions, Eq.~(\ref{Weyl}), determining in $d=(3+1)$
spins, handedness, charges and family quantum numbers. Again they fulfill on the vacuum 
state, Eq.~(\ref{vac1}), together with their Hermitian conjugated annihilation operators, 
the anti commutation relations required for fermions, Eq.~(\ref{ijgammaprodordinary}).
Correspondingly the {\it creation and annihilation operators are indeed defined with the first 
quantized  fermion fields already}.

We demonstrate in this paper that there exist also in Grassmann space of anticommuting 
coordinates, Eq.~(\ref{thetarelcom}), the eigenvectors of the  Cartan commuting subalgebra of 
the Lorentz algebra ${\cal {\bf S}}^{ab}$, Eq.~(\ref{thetarelcom}, \ref{calsab}), the 
$\frac{d}{2}$ products of which form creation operators, Eq.~(\ref{start(2n+1)2theta}), 
and which fulfill together with their Hermitian conjugated partners the annihilation operators, 
Eq.~(\ref{grassher}), as well the anticommutation relations required for fermions, 
Eq.~(\ref{ijthetaprod}). 
However, the {\it eigenvalues of the Cartan subalgebra are in this case integer}. \\
Also in the Grassmann case the superposition of these creation operators solve for a particular 
momentum $p^a$ the equation of motions for free massless fermions, presented in 
Eq.~(\ref{Weylgrass}), determining in $d=(3+1)$ spins, handedness and charges.  There are no 
families in this case.

For both cases, Clifford and Grassmann, we present the proofs for the above statements and illustrate
the properties of fermions of both kinds on a few examples.

\subsection{ Actions and equation of motion in Clifford and in Grassmann space}
\label{action}

We define in $d=((d-1) +1)$-dimensional space states with integer spin --- in Grassmann space ---
and states with half  integer spin --- in Clifford space --- proving that  norms in both spaces can be
determined by the integral in  Grassmann space, Eqs.~(\ref{grassnorm}, \ref{grassgammanorm},
\ref{grasstildegammanorm}), since the Clifford algebra objects are expressible with the Grassmann
algebra objects, Eq.~(\ref{cliffthetareltheta})~\footnote{Observations in this paper might help also 
when fermionizing boson fields or bosonizing fermion fields~\cite{procnh2015}.}.
When reformulating the vacuum in the Clifford case, Eq.~(\ref{vac1}), half integer spinors 
presentation in Clifford space become more elegant, that is easier to recognize properties of fermions.

We present as well actions in both cases, Grassmann, Eq.~(\ref{actionWeylGrass}), and Clifford, 
Eq.~(\ref{Weyl}), leading to the equations of motion (in the Clifford case the Weyl equation is known
for a long time, in the Grassmann case it is present for the first time by N.S.M.B.).
We compare Euler-Lagrange equations in both cases to compare properties of
Grassmann "fermions" with the Clifford fermions.

\vspace{3mm}

{\bf a.}  {\it Fields with the integer spin in Grassmann space }

\vspace{2mm}

A point in $d$-dimensional Grassmann space of anticommuting coordinates $\theta^{a}$, 
$(a=0,1,2,3,5,\dots,d)$,
%
is determined by a vector $\{\theta^a \}= (\theta^0, \theta^1, \theta^2, \theta^3, 
\theta^5, \dots, \theta^d)$. A linear vector space over the coordinate Grassmann space has
correspondingly the dimension $2^d$, due to the 
fact that $(\theta^{a_i})^2 =0$ for any $a_i \in (0,1,2,3,5,\dots,d)$.

Correspondingly are fields in Grassmann space expressible in terms of the 
Grassmann algebra objects
\begin{eqnarray}
\label{grassmannboson}
{\bf B} &=& \sum_{k=0}^{d}\, a_{a_{1} a_{2} \dots a_{k}}\,
\theta^{a_1}\theta^{a_2} \dots \theta^{a_k} |\phi_{og}>\,,
\quad a_{i}\le a_{i+1}\,,
\end{eqnarray}
where $|\phi_{og}>$ is the vacuum state, here assumed to be  $|\phi_{og}> = |1\,>$, so that
$\frac{\partial}{\,\partial \theta^a} |\phi_{og}>=0$ for any $\theta^a$.
The {\it Kalb-Ramond} boson fields $a_{a_{1} a_{2} \dots a_{k}}$ are antisymmetric with 
respect to the permutation of indexes, since the Grassmann coordinates anticommute  
$\{\theta^a, \theta^b\}_{+} = 0$, Eq.~(\ref{thetarelcom}). 

The left derivative  $\frac{\partial}{\;\partial \theta_{a}}$  on vectors of the space of monomials
$ {\bf B}(\theta)$ is defined as follows  
\begin{eqnarray}
\label{grassvectorder}
\frac{\partial}{\partial \theta_{a}} \,{\bf B}(\theta)&=&
\frac{\partial {\bf B}(\theta)}{\partial \theta_{a}}\,,
\nonumber\\
\left \{\frac{\partial}{\partial \theta_{a}}, \frac{\partial}{\partial \theta_{b}}\right\}_{+}\, {\bf B}
 &=&
 0\,,\, \rm{for} \,\rm {all }\; {\bf B}\,.
\end{eqnarray}

The commutation relations are for $p^{\theta a} =i \frac{\partial}{\partial \theta_{a}}$
defined in Eq.~(\ref{thetarelcom}), where the metric tensor $\eta^{ab}$ ($= diag(1,-1,-1,\dots,-1)$)
lowers the indexes of a vector $\{\theta^a \}$: $\theta_a= \eta_{ab}\, \theta^b$
(the same metric tensor lowers the indexes of the ordinary vector $x^a$ of commuting 
coordinates). 

Defining~\footnote{In Ref.~\cite{norma93} the definition of $\theta^{a \dagger}$ was 
differently chosen. Correspondingly also the scalar product needed a (slightly) different weight function
in Eq.~(\ref{grassnorm}).}
\begin{eqnarray}
\label{grassher}
(\theta^a)^{\dagger} &=&  \frac{\partial}{\partial \theta_{a}} \eta^{aa}=
-i \,p^{\theta a} \eta^{aa}\, , 
\end{eqnarray}
it follows
\begin{eqnarray}
\label{grassp}
(\frac{\partial}{\partial \theta_{a}})^{\dagger} &=& \eta^{aa}\, \theta^a\,,\quad
(p^{\theta a})^{\dagger} = -i \eta^{aa} \theta^a\,.
\end{eqnarray}
Making a choice for the complex properties of $\theta^a$, and correspondingly 
of $\frac{\partial}{\partial \theta_{a}}$, as follows
\begin{eqnarray}
\label{complextheta}
\{\theta^a\}^* &=&  (\theta^0, \theta^1, - \theta^2, \theta^3, - \theta^5,
\theta^6,...,- \theta^{d-1}, \theta^d)\,, \nonumber\\
\{\frac{\partial}{\partial \theta_{a}}\}^* &=& (\frac{\partial}{\partial \theta_{0}},
\frac{\partial}{\partial \theta_{1}}, - \frac{\partial}{\partial \theta_{2}},
\frac{\partial}{\partial \theta_{3}}, - \frac{\partial}{\partial \theta_{5}}, 
\frac{\partial}{\partial \theta_{6}},..., - \frac{\partial}{\partial \theta_{d-1}}, 
\frac{\partial}{\partial \theta_{d}})\,, 
\end{eqnarray}
it follows for the two Clifford algebra objects $\gamma^a =(\theta^a + 
\frac{\partial}{\partial \theta_{a}})$, and $\tilde{\gamma}^a = 
i (\theta^a - \frac{\partial}{\partial \theta_{a}})$, 
Eq.~(\ref{cliffthetareltheta}), that $\gamma^a$ is real if $\theta^a$ is real,
and $\gamma^a$ is imaginary if $\theta^a$ is imaginary, while $\tilde{\gamma}^a$ is imaginary when 
$\theta^a$ is real and $\tilde{\gamma}^a$ is real  if $\theta^a$ is imaginary, just as it is required in
 Eq.~(\ref{complexgamatilde}). 

Applying the operator ${\cal {\bf S}}^{ab}$ of Eq.~(\ref{thetarelcom}) on the "state"
 $\frac{1}{\sqrt{2}} (\theta^a \pm \epsilon \theta^b)$,  it follows
%
\begin{eqnarray}
\label{calsab}
{\cal \bf{S}}^{ab}  (\theta^a \pm \epsilon \theta^b)&=& \pm (-i) \epsilon  
 (\theta^a \pm \epsilon \theta^b)\,,
\end{eqnarray}
$\epsilon = i $, if $\eta^{a a}=\eta^{b b}$ and $\epsilon = -1 $, if $\eta^{a a}\ne\eta^{b b}$.

We define here the commuting objects $\gamma^{a}_{G}$, which will be helpful when looking for  
the appropriate action for Grassmann fermions, Eq.~(\ref{actionWeylGrass}). These operators
will be needed also when looking for the definition of appropriate discrete symmetry operators in 
the Grassmann case. Following the definition of the discrete symmetry operators in the Clifford 
algebra case~\cite{nhds}  in $((d-1)+1)$ space-time and in $(3+1)$ space-time, the discrete 
symmetry operators  ($\cal{C}_{G}$, $\cal{T}_{G}$, $\cal{P}_{G}$) in 
($(d-1) +1$) 
and ($\cal{C}_{\cal{N G}}$, $\cal{T}_{\cal{N G}}$, $\cal{P}_{\cal{N G}}$) in $(3+1)$ 
will be defined, respectively.
\begin{eqnarray}
\label{grassgamma}
\gamma^{a}_{G} &=&(1- 2 \theta^a \eta^{aa} 
\frac{\partial}{\partial \theta_{a}}) 
 = - i \eta^{aa}\, \gamma^a \tilde{\gamma}^a\,, \quad 
\{\gamma^{a}_{G}, \gamma^{b}_{G} \}_- = 0\,.
\end{eqnarray}
Index ${}^{a}$ is not the Lorentz index in the usual sense. 
 $\gamma^{a}_{G} $ are commuting  operators 
for all $(a,b)$. They are real and Hermitian.
\begin{eqnarray}
\label{grassgammaprop}
\gamma^{a \dagger}_{G} &=& \gamma^{a}_{G}\,,
\;\;\;\; (\gamma^{a}_{G})^{*} = \gamma^{a}_{G}\,.
\end{eqnarray}
Correspondingly it follows: $\gamma^{a \dagger}_{G} \gamma^{a}_{G}= I$,
$\gamma^{a}_{G} \gamma^{a}_{G}= I$. $I$ represents the unit operator. 
 
By introducing~\cite{norma93} the generators of the  infinitesimal Lorentz transformations in 
 Grassmann space, as presented in Eq.~(\ref{thetarelcom}), and making use of the Cartan 
subalgebra of commuting operators, Eq.~(\ref{cartan}), the basic states in Grassmann space 
can be arranged into representations of the eigenstates of the Cartan subalgebra operators, 
Eq.~(\ref{calsab}), Ref.~\cite{norma93,nd2018}.   All these states have integer spin.
The starting state in $d$-dimensional space, for example, with the eigenvalues of the Cartan 
subalgebra equal to either $i$ or $1$ is: $(\theta^0 - 
\theta^3) (\theta^1 + i \theta^2) (\theta^5 + i \theta^6) \cdots (\theta^{d-1} + i \theta^d) 
|\phi_{og}>$, with $|\phi_{og}>=|1\,>$, Eq.~(\ref{calsab}). All the states of the representation,
which starts with this state, follow by the application of those $ {\cal {\bf S}}^{ab}$, which do
not belong to  the Cartan subalgebra of the Lorentz algebra. ${\cal {\bf S}}^{01}$, for example,
transforms  
this starting state into  $(\theta^0 \theta^3 +i \theta^1 i \theta^2) (\theta^5 + i \theta^6) 
\cdots (\theta^{d-1} +  i \theta^d) |\phi_{og}>$, while ${\cal {\bf S}}^{01} -i {\cal {\bf S}}^{02}$ 
transforms this state into  $(\theta^0 + \theta^3) (\theta^1 - i \theta^2) (\theta^5 + i \theta^6)
 \cdots (\theta^{d-1} + i \theta^d) |\phi_{og}>$.

\vspace{3mm}

{\bf b.} {\it Fields with the half integer spin in Clifford space }

\vspace{2mm}

Let us present as well the properties of the fermion fields with the half integer spin, expressed by the 
Clifford algebra objects~(\cite{norma92,norma93,normaJMP2015,IARD2016,nd2017,%
n2014matterantimatter,nh02} and the references therein) 
\begin{eqnarray}
\label{cliffordfermion}
{\bf F} &=& \sum_{k=0}^{d}\, a_{a_{1} a_{2} \dots a_{k}}\,
\gamma^{a_1}\gamma^{a_2} \dots \gamma^{a_k} |\psi_{oc}>\,,
\quad a_{i}\le a_{i+1}\,,
\end{eqnarray}
where $|\psi_{oc}>$ is the vacuum state. The {\it Kalb-Ramond} fields $ a_{a_{1} a_{2}
\dots a_{k}}$ are again in general  boson fields, which are antisymmetric with  respect to the 
permutation of indexes, since the Clifford objects have the anticommutation relations, 
Eq.~(\ref{twocliffordsab}),  $\{\gamma^a, \gamma^b\}_{+} = 2\eta^{a b}$.
The linear vector space over the Clifford coordinate space has, as in the Grassmann case, 
the dimension $2^d$, due to the 
fact that $(\gamma^{a_i})^2 =\eta^{a_i a_i}$ for any $a_i \in (0,1,2,3,5,\dots,d)$.

As written in Eq.~(\ref{cliffthetareltheta}), $\gamma^a$ are expressible in terms of the 
Grassmann coordinates and their 
conjugate momenta, as $\gamma^a= (\theta^a - i \, p^{\theta a})$,
and $\tilde{\gamma}^a=i \, (\theta^a + i \, p^{\theta a})$, with the anticommutation relation of 
Eq.~(\ref{twocliffordsab}), $\{\gamma, \gamma^b\}_{+}=2 \eta^{ab}$ 
$= \{\tilde{\gamma}^a, \tilde{\gamma}^b\}_{+}$, 
$ \{\gamma^a, \tilde{\gamma}^b\}_{+}=0$. 
Taking into account  Eqs.~(\ref{grassher}, \ref{grassp}, \ref{cliffthetareltheta})
one finds
\begin{eqnarray}
\label{cliffher}
(\gamma^a)^{\dagger} &=& \gamma^a \eta^{aa}\, ,\quad
(\tilde{\gamma}^a)^{\dagger} = \tilde{\gamma}^a \eta^{aa}\, ,\nonumber\\
\gamma^a \gamma^a &=& \eta^{aa}\,, \quad 
\gamma^a (\gamma^a)^{\dagger} =I\,,\quad
 \tilde{\gamma}^a  \tilde{\gamma}^a = \eta^{aa} \,,\quad
 \tilde{\gamma}^a  (\tilde{\gamma}^a)^{\dagger} =I\,,
\end{eqnarray}
where $I$ represents the unit operator.
Making a choice for the $\theta^a$ properties as presented in Eq.~(\ref{complextheta}), it 
follows for the Clifford objects
\begin{eqnarray}
\label{complexgamatilde}
\{\gamma^a\}^* &=&  (\gamma^0, \gamma^1, - \gamma^2, \gamma^3, - \gamma^5,
\gamma^6,...,- \gamma^{d-1}, \gamma^d)\,, \nonumber\\
\{\tilde{\gamma}^a\}^* &=&  (- \tilde{\gamma}^0, - \tilde{\gamma}^1, \tilde{\gamma}^2, 
- \tilde{\gamma}^3,  \tilde{\gamma}^5,- \tilde{\gamma}^6,..., \tilde{\gamma}^{d-1}, 
- \tilde{\gamma}^d)\,, 
\end{eqnarray}
Applying the operators $S^{ab}$ and $\tilde{S}^{ab}$, Eq.~(\ref{twocliffordsab}), on $\frac{1}{2}
(\gamma^a + \frac{\eta^{aa}}{ik} \gamma^b$) and on $\frac{1}{2}
(1 +  \frac{i}{k}  \gamma^a \gamma^b$) one obtains
\begin{eqnarray}
\label{signature0}
S^{ab} \frac{1}{2} (\gamma^a + \frac{\eta^{aa}}{ik} \gamma^b) &=& \frac{k}{2}  \,
\frac{1}{2} (\gamma^a + \frac{\eta^{aa}}{ik} \gamma^b)\,,\nonumber\\
\tilde{S}^{ab} \frac{1}{2} (\gamma^a + \frac{\eta^{aa}}{ik} \gamma^b) &=& \frac{k}{2}  \,
\frac{1}{2} (\gamma^a + \frac{\eta^{aa}}{ik} \gamma^b)\,,\nonumber\\
S^{ab} \frac{1}{2} (1 +  \frac{i}{k}  \gamma^a \gamma^b) &=&  \frac{k}{2}  \,
 \frac{1}{2} (1 +  \frac{i}{k}  \gamma^a \gamma^b)\,\nonumber\\
\tilde{S}^{ab} \frac{1}{2} (1 +  \frac{i}{k}  \gamma^a \gamma^b) &=& - \frac{k}{2}  \,
 \frac{1}{2} (1 +  \frac{i}{k}  \gamma^a \gamma^b)\,.
\end{eqnarray}
One could make a choice of  $\tilde{\gamma}^a$ instead of $\gamma^a$ and change 
correspondingly relations in Eq.~(\ref{signature0}).
 
All three choices for the linear vector space --- spanned over either the coordinate Grassmann space, 
or over the vector space of $\gamma^a$, or over the vector space of $\tilde{\gamma}^a$
 --- have the dimension $2^d$.

We can express Grassmann coordinates $\theta^a$ and momenta $p^{\theta a}= 
i \frac{\partial}{\partial \theta_{a}}$ in terms of
$\gamma^a$ and $\tilde{\gamma}^a$ as well~\footnote{In Ref.~\cite{MP2017} the author  
suggested in Eq.~(47) a choice of superposition of $\gamma^a$ and $\bar{\gamma}^a$, which  
resembles the choice of one of the authors (N.S.M.B.) in Ref.~\cite{norma93} and both authors in 
Ref.~\cite{nh02,nh03} and  in present article.} 
\begin{eqnarray}
\label{thetadertheta}
\theta^a &=&\frac{1}{2} (\gamma^a - i \tilde{\gamma}^a)\,,\nonumber\\
\frac{\,\partial}{\partial \theta_a} &=&\frac{1}{2}
 \, (\gamma^a + i \tilde{\gamma}^a)\,,
\end{eqnarray}
(and correspondingly with  $\gamma^a= \theta^{a} + \frac{\partial}{\partial \theta_a}$,  
$\tilde{\gamma}^a=i \,(\theta^{a} - \frac{\partial}{\partial \theta_a}$).
It then follows, taking into account  Eqs.~(\ref{gammabasis}, \ref{gammatildeA}), 
%
$\frac{\partial\;\,}{\partial \theta_b} \theta^a |1\,>=$
$ \eta^{a b} |1\,> $.
%

Correspondingly we can use either $\gamma^a$  or  $\tilde{\gamma}^a$
instead of $\theta^a$ to span the vector space. In this case we change the vacuum from
the one with the  property $\frac{\partial}{\partial \theta^a} |\phi_{og}>=0$ to  $|\psi_{oc}>$
with the property~\cite{norma93,JMP2013,normaJMP2015}
\begin{eqnarray}
\label{gammabasis}
<\psi_{oc}|\gamma^a |\psi_{oc}>&=& 0 \,,\quad \tilde{\gamma}^a |\psi_{oc}>=
i \gamma^a |\psi_{oc}> \,, \quad \tilde{\gamma}^a \gamma^b |\psi_{oc}>=- i \gamma^b 
\gamma^a  |\psi_{oc}>\,,\nonumber\\
\tilde{\gamma}^a \tilde{\gamma}^b |\psi_{oc}>|_{a\ne b} &=& - \gamma^a \gamma^b
 |\psi_{oc}>\,,\quad \tilde{\gamma}^a \tilde{\gamma}^b |\psi_{oc}>|_{a=b} = \eta^{ab} 
|\psi_{oc}>\,.
\end{eqnarray}
This is in agreement with the requirement
\begin{eqnarray}
\gamma^a \, {\bf F}(\gamma) \,|\psi_{oc} \,>: &=& (\, a_{0}\, \gamma^{a}\, + a_{a_1} 
\,\gamma^{a}\,\gamma^{a_1} + 
 a_{a_1 a_2}\,\gamma^{a}\, \gamma^{a_1} \gamma^{a_2} + \cdots + a_{a_1 \cdots a_d}
 \,\gamma^{a}\,\gamma^{a_1} 
 \cdots \gamma^{a_d}\,)\,|\psi_{oc} \,>\,, \nonumber\\
\tilde{\gamma}^a \, {\bf F}(\gamma) \,|\psi_{oc} \,>: &=& 
(\,i \,a_0 \gamma^a  -i\, a_{a_1} \gamma^{a_1}\, \gamma^a + i \,a_{a_1 a_2}
 \gamma^{a_1} \gamma^{a_2} \, \gamma^a 
+ \cdots + \nonumber\\
&&i\,(-1)^{d} \,a_{a_1 \cdots a_d} \gamma^{a_1} \cdots \gamma^{a_d}\, \gamma^a \,)\,
|\psi_{oc}\,>\,.
\label{tildecliffordB}
\end{eqnarray}
The basic states in Clifford space can be arranged in representations, in which any state is the 
eigenstate of the Cartan subalgebra operators of Eq.~(\ref{cartan}). The state, for example, 
in $d$-dimensional space with the eigenvalues  of either $S^{03}, S^{12},  S^{56}, \dots,  
S^{d-1\,d}$ or $\tilde{S}^{03}, \tilde{S}^{12}, \tilde{S}^{56}, \dots, \tilde{S}^{d-1\,d} $ 
equal to $\frac{1}{2} (i, 1,1,\dots,1)$ is $(\gamma^0 - \gamma^3) (\gamma^1 + i \gamma^2)
 (\gamma^5 + i \gamma^6) \cdots (\gamma^{d-1} + i \gamma^d)$, where the states are 
expressed in terms of $\gamma^a$. The states of one representation follow from the starting state 
 by the application of $S^{ab}$, which do not belong to the Cartan subalgebra operators, while 
$\tilde{S}^{ab}$, which operate on family quantum numbers, cause jumps from the starting 
family to the new one.

\subsubsection{Norms of vectors in Grassmann and Clifford space}
\label{norm}

Let us look for the norm of  vectors in Grassmann space, ${\bf B}=$ 
$\sum_{k=0}^{d} a_{a_{1} a_{2} \dots a_{k}}\,\theta^{a_1}\theta^{a_2} $$\dots$$
 \theta^{a_k}|\phi_{og}>$, and in Clifford space, ${\bf F}=$ $\sum_{k=0}^{d} a_{a_{1} a_{2}}$$
 \dots$$ a_{k}\,\gamma^{a_1}\gamma^{a_2} \dots \gamma^{a_k}|\psi_{oc}>$,  
where $|\phi_{og}>$ and $|\psi_{oc}>$ are the vacuum states in the Grassmann and Clifford case,
 respectively. In what follows we refer to Ref.~\cite{norma93}.

\vspace{3mm} 

{\bf a.} {\it  Norms of Grassmann vectors}

\vspace{2mm}

Let us define the integral over the Grassmann  space~\cite{norma93} of two functions of the 
Grassmann coordinates $<{\cal {\bf B}}|\theta> <{\cal {\bf C}}|\theta>$, $<{\cal {\bf B}}| \theta>= 
<\theta | {\cal {\bf B}}>^{\dagger}$, by requiring 
\begin{eqnarray}
\label{grassintegral}
\{ d\theta^a, \theta^b \}_{+} &=&0\,, \,\;\;  \int d\theta^a  =0\,,\,\;\; 
\int d\theta^a \theta^a =1\,,\;\;
\int d^d \theta \,\,\theta^0 \theta^1 \cdots \theta^d =1\,,
\nonumber\\
d^d \theta &=&d \theta^d \dots d\theta^0\,,\,\;\; 
\omega = \Pi^{d}_{k=0}(\frac{\partial}{\;\,\partial \theta_k} + \theta^{k})\,,
\end{eqnarray}
with $ \frac{\partial}{\;\,\partial \theta_a} \theta^c = \eta^{ac}$. We shall use the weight function 
$\omega= \Pi^{d}_{k=0}(\frac{\partial}{\;\,\partial \theta_k} + \theta^{k})$ to define the scalar
product  $<{\cal {\bf B}}|{\cal {\bf C}} >$ 
\begin{eqnarray}
\label{grassnorm}
<{\cal {\bf B}}|{\cal {\bf C}} > &=&  \int d^{d-1} x  d^d \theta^a\, \,\omega 
 <{\cal {\bf B}}|\theta>\, <\theta|{\cal {\bf C}}>  = \sum^{d}_{k=0} 
\int d^{d-1}x \, b^{*}_{b_{1} \dots b_{k}} c_{b_1 \dots b_{k}}\,,
\end{eqnarray}
where, according to Eq.~(\ref{grassher}), it follows: $<{\cal {\bf B}}|\theta>=$ $ 
\sum_{p=0}^{d} (-i)^p\, a^{*}_{a_1\dots a_p} p^{\theta a_p}\,\eta^{a_p a_p} 
\cdots  p^{\theta a_1}\, \eta^{a_1 a_1}$. The vacuum state is chosen to be 
$|\phi_{og}>= |1\,>$, as assumed in Eq.~(\ref{grassmannboson}).

The norm $<{\cal {\bf B}}|{\cal {\bf B}} >$ is correspondingly always nonnegative.
Let us notice that the choice of the Hermitian conjugated value of $\theta^a $ is 
$ \frac{\partial}{\partial \theta^a}$  ($(\theta^a)^{\dagger}=$
$\eta^{aa}\, \frac{\partial}{\partial \theta_a} $, Eq.~(\ref{grassher})) makes that
 we easily evaluate in any $d$ the scalar product 
$<\phi_{og}|(\frac{\partial}{\partial \theta^d} \frac{\partial}{\partial \theta^{d-1}}
 \frac{\partial}{\partial \theta^{d-2}} \cdots \frac{\partial}{\partial \theta^1}$
$ \frac{\partial}{\partial \theta^0})$ $(\theta^0 \theta^1 \cdots \theta^{d-2}
 \theta^{d-1} \theta^d) |\phi_{og}>=1$ for $ |\phi_{og}>=|1\,>$ (without  integration 
over coordinate space of $\theta^a$'s).

\vspace{3mm} 

{\bf b.} {\it  Norms of Clifford vectors}

\vspace{2mm}

To evaluate norms in the Clifford space for vectors of Eq.~(\ref{cliffordfermion}) we can use as well
Eqs.~(\ref{grassintegral}, \ref{grassnorm}), if expressing $\gamma^a$ (or equivalently 
$\tilde{\gamma}^a$) in terms of $\theta^a$ and $p^{\theta a}$: 
$<(\theta^a - i p^{\theta a}) |{\bf F}>$ (or equivalently
 $<(\theta^a + i p^{\theta a}) |{\bf \tilde{F}}>$), Eq.~(\ref{cliffthetareltheta}). In this case
$|\psi_{oc}>=|\phi_{og}>=|1\,>$. It follows
\begin{eqnarray}
\label{grassgammanorm}
<{\bf F}|{\bf G} > &=&  \int d^{d-1} x  d^d \theta^a\, \,\omega 
 < {\bf F}|\gamma>\, <\gamma|{\bf G}>  = \sum^{d}_{k=0} 
\int d^{d-1} x \, a^{*}_{a_{1} \dots a_{k}} b_{b_1 \dots b_{k}}\,.
\end{eqnarray}
\{Similarly one obtains, if  expressing  ${\bf \tilde{F}}=\sum_{k=0}^{d} a_{a_{1} a_{2} \dots a_{k}}\,
\tilde{\gamma}^{a_1} \tilde{\gamma}^{a_2} \dots \tilde{\gamma}^{a_k}|\psi_{oc}>$  
and ${\bf \tilde{G}}=\sum_{k=0}^{d} b_{b_{1} b_{2} \dots b_{k}}\,
\tilde{\gamma}^{b_1} \tilde{\gamma}^{b_2} \dots \tilde{\gamma}^{b_k}|\psi_{oc}>$ and 
taking $|\psi_{oc}>=|\phi_{og}> = |1\,>$, the scalar product 
\begin{eqnarray}
\label{grasstildegammanorm}
<{\bf \tilde{F}}|{\bf \tilde{G}} > &=&  \int d^{d-1} x  d^d \theta^a\, \,\omega 
 < {\bf \tilde{F}}|\tilde{\gamma}>\, <\tilde{\gamma}|{\bf \tilde{G}}>  = \sum^{d}_{k=0} 
\int d^{d-1}x \, a^{*}_{a_{1} \dots a_{k}} a_{b_1 \dots b_{k}}\,.\}
\end{eqnarray}
To simplify the evaluation we use instead~\cite{IARD2016,nh02} in the Clifford case the 
vacuum state $|\psi_{oc}>$,
Eq.~(\ref{vac1}), which is the product of projectors, Eq.~(\ref{signature}). It  takes care
of the orthogonality of states (like if we would evaluate the integration in Grassmann space).

\vspace{1mm}

Correspondingly we can write 
\begin{eqnarray}
\label{normKRfields}
 \int  d^d \theta^a\, \,\omega(a_{a_{1} a_{2} \dots a_{k}}\,
\gamma^{a_1}\gamma^{a_2} \dots \gamma^{a_k})^{\dagger} 
( a_{a_{1} a_{2} \dots a_{k}}\,
\gamma^{a_1}\gamma^{a_2} \dots \gamma^{a_k})
&=&a^*_{a_{1} a_{2} \dots a_{k}} \,a_{a_{1} a_{2} \dots a_{k}}\,. 
\end{eqnarray}
The norm of each scalar term in the sum of ${\bf F}$ is nonnegative. 

\vspace{3mm}

{\bf c.} We have learned:\\
In both spaces --- Grassmann and Clifford ---  norms of basic states 
can be defined so that  the states, which are eigenvectors of the Cartan subalgebra, are orthogonal 
and normalized using the same integral. 

Studying the second quantization procedure in 
Sect.~\ref{secondquantization} we learn that not all $2^d$ states can be represented as  
creation and annihilation operators, either in the Grassmann or in the Clifford case, since they must ---
in both cases --- fulfill the requirements for the second quantized operators, either for states with 
integer spins in Grassmann space or for states with a half integer spin in Clifford space. 

%
\subsubsection{Actions in Grassmann and Clifford space}
\label{action1}

We construct  an action for free massless fermion in which the internal degrees of freedom  is 
described: {\bf i.} in Grassmann  space, {\bf ii.} in Clifford  space. In the first case the internal 
degrees of freedom manifest integer spins, in the second case the half integer spin.

While the action in Clifford space is well known since long~\cite{Dirac}, the action in  
Grassmann space will be defined here (by N.S.M.B.). In both cases we present an action for 
free massless fermions in $((d-1)+1)$ space~\footnote{In  $d= (3+1) $ space masses of fermions
are in the {\it spin-charge-family} theory in the Clifford case caused by the interaction of fermions 
with scalar gauge fields with the space index $(7,8)$, that is the vielbeins and the spin connections  
of two kinds ---  the gauge scalar fields of $S^{ab}$ and of $\tilde{S}^{ab}$. We expect that 
masses of "fermions" appears also in the Grassmann case due to the interaction of fermions with 
scalar gauge fields with the space index $(7,8)$, but in this case due to the vielbeins and the spin 
connection of one kind only --- the gauge field of ${\cal {\bf S}}^{ab}$}. 
States in Grassmann space as well as states in Clifford space will be arranged to be the eigenstates 
of the Cartan subalgebra --- with respect to ${\cal {\bf S}}^{ab}$ in Grassmann space and with 
respect to $S^{ab}$ and $\tilde{S}^{ab}$ in Clifford space, Eq.~(\ref{cartan}), and orthogonal 
and normalized with respect to Eq.~(\ref{grassintegral})~\footnote{In the Clifford case the states 
can be orthogonalized also with respect to  Eq.~(\ref{vac1}).}.

In both spaces the requirement that states are obtained by the application of creation operators 
on the vacuum state --- in the Grassmann case  $\hat{b}^{\theta^k \dagger}_{i}$, 
Eq.~(\ref{general2}),  obeying together with the $\hat{b}^{\theta^k}_{i}$ the anti commutation
relations of Eq.~(\ref{ijthetaprod}) on the vacuum state $|\phi_{og}>= |1\,>$, and in the 
Clifford case $\hat{b}^{\alpha \dagger}_{i}$, Eq.(\ref{anycreation}), obeying together with the 
$\hat{b}^{\beta}_{i}$ the equivalent anti commutation relations of Eq.~(\ref{alphagammaprod0})
on the vacuum states $|\psi_{oc}>$, Eq.~(\ref{vac1}) --- reduces the number of 
states, in Clifford space more than in Grassmann space. But while in Clifford space all physically 
applicable states are reachable by either $S^{ab}$ (defining family members quantum numbers) 
or by  $\tilde{S}^{ab}$ (defining family quantum numbers), the states in Grassmann space, 
belonging to different representations with respect to the Lorentz  generators, seem not to be 
connected.

\vspace{3mm}

{\bf a.}  {\it Action in Clifford space }

\vspace{2mm}

In Clifford space the action for a free massless fermion must be Lorentz invariant
\begin{eqnarray}
{\cal A}\,  &=& \int \; d^dx \; \frac{1}{2}\, (\psi^{\dagger}\gamma^0 \, \gamma^a p_{a} \psi) +
 h.c.\,, 
\label{actionWeyl}
\end{eqnarray}
$p_{a} = i\, \frac{\partial}{\partial x^a}$, leading to the equations of motion 
\begin{eqnarray}
\label{Weyl}
\gamma^a p_{a}  |\psi>&= & 0\,, 
\end{eqnarray}
which fulfill also the Klein-Gordon equation
\begin{eqnarray}
\label{LtoKG}
\gamma^a p_{a} \gamma^b p_b |\psi>&= &   
p^a p_a |\psi>=0\,,\nonumber\\
\end{eqnarray}
for each of the basic states $ |\psi^{\alpha}_{i}>$. $\gamma^0$ appears in the 
action since we pay attention that
\begin{eqnarray}
\label{Linvariance}
S^{ab \dagger} \,\gamma^0 &= & \gamma^0\, S^{ab}\,,
\quad S^{\dagger}\gamma^0 = \gamma^0 S^{-1}\,,\nonumber\\
S&=&e^{- \frac{i}{2} \omega_{ab} (S^{ab} + L^{ab})}\,.
\end{eqnarray}

The Lagrange density,  Eq.~(\ref{actionWeyl}),
\begin{eqnarray}
{\cal L}_{C}\,  &=&  \frac{1}{2} \{\psi^{\dagger} \, \gamma^{0}\,\gamma^a\, 
\hat{p}_a \psi -  \hat{p}_a \psi^{\dagger}\, \gamma^{0}\,\gamma^a \,\psi \, \}\,,
\label{LDWeyl0}
\end{eqnarray}
leads to
\begin{eqnarray}
\frac{\partial {\cal L}_{C}}{\partial \psi^{\dagger}} -  
 \hat{p}_{a} \,\frac{\partial {\cal L}_{C}}{\partial \hat{p}_a \psi^{\dagger}}  &=& 
0 =\gamma^0 \gamma^a \,\hat{p}_a\,\psi\,, 
\nonumber\\
\frac{\partial {\cal L}_{C}}{\partial \psi} -  
 \hat{p}_{a} \,\frac{\partial {\cal L}_{C}}{\partial (\hat{p}_a \psi)}  &=& 0=
 - \hat{p}_a \,\psi^{\dagger} \gamma^0 \,\gamma^a\,.
\label{LDWeyl2}
\end{eqnarray}
%
All the states, belonging to different values of the Cartan subalgebra ---  they differ at least in one 
value of either the set of $S^{ab}$ or the set of $\tilde{S}^{ab}$, Eq.~(\ref{cartan})  --- are 
orthogonal according to the scalar product,  defined as the integral over the Grassmann coordinates, 
Eq.~(\ref{grassintegral}), for a chosen vacuum state $|1\,>$. Correspondingly the states generated 
by the creation operators, Eq.~(\ref{anycreation}), on the vacuum state, Eq.~(\ref{vac1}), are 
orthogonal as well. 

\vspace{3mm}

{\bf b.}  {\it Action in Grassmann space } 

\vspace{2mm}

We define here the action in Grassmann  space, for which we require --- similarly as in the Clifford 
case --- that the action for a free massless fermion is Lorentz invariant
\begin{eqnarray}
{\cal A}_{G}\,  &=&  \int \; d^dx \;d^d\theta\; \omega \, \{\phi^{\dagger} 
(1-2\theta^0 \frac{\partial}{\partial \theta^0}) \,\frac{1}{2}\,
\theta^a p_{a}
\phi \}  + h.c.\,.  
\label{actionWeylGrass}
\end{eqnarray}
%
We use the integral over $\theta^a$ coordinates with the weight function $\omega$ from 
Eq.~(\ref{grassintegral}, \ref{grassnorm}).  
Requiring the Lorentz invariance we add after $\phi^{\dagger}$ the operator $\gamma^0_{G}$
($\gamma^a_{G}$ $=(1-2\theta^a \frac{\partial}{\partial \theta^a})$), which takes care of the Lorentz 
invariance. Namely
\begin{eqnarray}
\label{Linvariancegrass}
{\cal {\bf S}}^{ab \dagger}\, (1-2\theta^0 \frac{\partial}{\partial \theta^0}) &= & 
 (1-2\theta^0 \frac{\partial}{\partial \theta^0})\,{\cal {\bf S}}^{ab}\,,\nonumber\\
{\cal {\bf S}}^{\dagger} \, (1-2\theta^0 \frac{\partial}{\partial \theta^0})&=& 
(1-2\theta^0 \frac{\partial}{\partial \theta^0})\, {\cal {\bf S}}^{-1}\,,\nonumber\\
{\cal {\bf S}} &=& e^{-\frac{i}{2} \omega_{ab} (L^{ab} + {\cal {\bf S}}^{ab})}\,,
\end{eqnarray}
while $\theta^a,  \frac{\partial}{\partial \theta_a}$ and $p^a$ transform as Lorentz vectors.
The equations of motion follow from the action, Eq.~(\ref{actionWeylGrass}),
%
\begin{eqnarray}
\label{Weylgrass}
\frac{1}{2}\,
\gamma^0_{G} \,(\theta^a - 
\frac{\partial}{\partial \theta^a})
\, p_{a} \,|\phi>\,&= & 0\,,\nonumber\\
\gamma^0_{G} = (1-2\theta^0 \frac{\partial}{\partial \theta^0})&&\,,
\end{eqnarray}
as well as the Klein-Gordon equation, $\gamma^0_{G} \,(\theta^a - 
\frac{\partial}{\partial \theta^a}) \, p_{a} \, \gamma^0_{G} \,(\theta^b - 
\frac{\partial}{\partial \theta^a}) \, p_{b} \,|\phi>=0$, leading to 
%
\begin{eqnarray}
\label{LtoKGgrass}
\{\theta^a p_{a}, \frac{\partial}{\partial \theta_b} \,p_{b}\}_{+} \,=
p^a p_a =0\,.
\end{eqnarray}

From the Lagrange density, presented in Eq.~(\ref{actionWeylGrass}),
using  Eqs.(\ref{grassher}, \ref{grassp}, \ref{thetadertheta}) ($ \gamma^{0}_{G}
= - i \eta^{aa} \gamma^a \tilde{\gamma}^{a}$, $(\theta^a - \frac{\partial}{\partial \theta_{a}})=
 -i \tilde{\gamma}^a $) it follows, up to the surface term, 
\begin{eqnarray}
{\cal L}_{G}\,  &=& -i \frac{1}{2} \phi^{\dagger} \, \gamma^0_{G}  \,
\tilde{ \gamma}^a\,( \hat{p}_a \phi)\, \nonumber\\
&=&  -i \frac{1}{4} \{ \phi^{\dagger} \, \gamma^0_{G}  \,
\tilde{ \gamma}^a\, \hat{p}_a \phi\, - \hat{p}_a  \phi^{\dagger}\,  \, \gamma^0_{G}  \,
\tilde{ \gamma}^a\, \phi\,\}.
\label{LDWeylGrass10}
\end{eqnarray}
One correspondingly finds
%
%
\begin{eqnarray}
\frac{\partial {\cal L}_{G}}{\partial \phi^{\dagger}} -  
 \hat{p}_{a} \,\frac{\partial {\cal L}_{G}}{\partial \hat{p}_a \phi^{\dagger}}  &=& 
0 =\frac{-i}{2} \gamma^0_{G}  \, \tilde{\gamma}^a\,\hat{p}_a\,\phi\}\,, 
\nonumber\\
\frac{\partial {\cal L}_{G}}{\partial \phi} -  
 \hat{p}_{a} \,\frac{\partial {\cal L}_{G}}{\partial (\hat{p}_a \phi)}  &=& 0=
 \frac{i}{2}\hat{p}_a \,\phi^{\dagger} \gamma^0_{G}  \, \tilde{\gamma}^a\,,
\label{LDWeylGrass2}
\end{eqnarray}
%
The solutions of these equations are presented in Eq.~(\ref{weylgenGrass}).

 We shall see that, if one identifies the creation  operators in both spaces with the 
products of odd numbers of either $\theta^a$  --- in the Grassmann case ---  or $\gamma^a$  
---in the Clifford case --- and the annihilation operators as the Hermitian conjugate operators of
 the creation operators, the creation and annihilation operators fulfill the  anticommutation relations, 
required for fermions. The internal parts of states are  then defined by the application of the creation 
operators on the vacuum state. 
 
But while the Clifford algebra defines states with the half integer "eigenvalues" of the Cartan 
subalgebra operators of the Lorentz algebra, the Grassmann algebra 
defines states with the integer "eigenvalues" of the Cartan subalgebra operators. 

\vspace{3mm}

{\bf c.} {\it We learn}:\\
In both spaces --- in Clifford and in Grassmann space --- there exists the action, which  leads
to the equations of motion and to the corresponding Klein-Gordon equation for free massless
fields.

\section{Second quantization of Grassmann and Clifford vectors}
\label{secondquantization}

It is proven in this section that solutions of the Weyl equations --- following from the Hermitian 
and Lorentz invariant actions for free massless fermions, using to describe their internal degrees 
of freedom either Clifford space, Eqs.~(\ref{actionWeyl}, \ref{Weyl}), or Grassmann space,
Eq.~(\ref{actionWeylGrass}, \ref{Weylgrass}),   --- can be represented as creation operators, 
operating on the appropriate vacuum state. The  corresponding Hermitian conjugated operators,
taken as their annihilation partners, fulfill  together with the creation operators, if both are of 
an odd either Clifford or Grassmann character, the anticommutation relations required for fermions. 

Correspondingly there is no need to assume the anticommutation relations as done 
in the Dirac theory~\cite{Dirac,Bethe,Itzykson}, since the creation and annihilation operators
by themselves fulfill the anticommutation  relations required for fermions either in Clifford or in 
Grassmann space. 

Creation operators in both spaces determine the Hilbert space of $n$ fermions for any
$n$ and have all the properties of the corresponding Slater determinants, if we recognize  that 
a product of two creation operators of two different moments in the ordinary space ($p_k$, 
$p_l$) ---  $\hat{b}^{\alpha \dagger}_{i \,p_k} \cdot \hat{b}^{\beta \dagger}_{j \, p_l} $, 
applying on the vacuum state $|\psi_{o}>$, are zero if  and only if  $i=j$, $\alpha=\beta$ 
and $p_k=p_l$.

Let us point out that fermions with the internal degrees of freedom described in Clifford space 
manifests  half integer spins, while "fermions" with the internal degrees of freedom described 
in Grassmann space demonstrate  integer spins.

We pay attention in this paper on $d=2(2n+1)$-dimensional spaces, arranging all the vectors 
to be "eigenvectors" of the Cartan subalgebra operators of $S^{ab}$ and $\tilde{S}^{ab}$ in 
the Clifford case and ${\cal {\bf S}}^{ab}$ in the Grassmann case, Eqs.~(\ref{cartan}, 
\ref{twocliffordsab}, \ref{thetarelcom}).

In $d$-dimensional spaces the linear vector space, spanned over either the Clifford coordinates 
or the Grassmann coordinates, has the dimension $2^d$. One can in both cases represent
the vector space as $2^d$ operators, which --- when applied on the vacuum state, --- 
create $2^d$ vectors. Half of these operators have an odd and half an even either Clifford 
(with respect to odd or even  products of $\gamma^a$'s) or Grassmann (with respect to 
odd or even products of $\theta^s$'s) character.

In the Clifford case there are in the group of an odd Clifford character  two groups of 
operators: each member of one group has its Hermitian conjugated partner in another group. 
One of the two groups can be therefore chosen to represent the creation operators, the 
other to represent the corresponding annihilation operators. Each of the two groups has 
$2^{d-2}$ members. 

Each of the two Clifford odd groups, one with $2^{d-2}$ creation the other with $2^{d-2}$ 
annihilation operators, further divides into $2^{\frac{d}{2}-1}$ subgroups with 
$2^{\frac{d}{2}-1}$  members. 
All the $2^{\frac{d}{2}-1}$ members of one particular subgroup  are  related by the
operators $S^{ab}$, while $\tilde{S}^{ab}$ transform each member of this subgroup of
particular family  into the same member  of one of  $2^{\frac{d}{2}-1}$ families.

In the group of the Clifford even operators there are again two groups, each with 
$2^{\frac{d}{2}-1}\cdot 2^{\frac{d}{2}-1}$ operators related by either $S^{ab}$ or by 
$\tilde{S}^{ab}$. 
Within each of the group there are $2^{\frac{d}{2}-1}$ subgroups  with $2^{\frac{d}{2}-1}$ 
members, related by the application of  $S^{ab}$, while $\tilde{S}^{ab}$ transform each 
member of a particular subgroup into the same member --- with respect to the operators 
$S^{ab}$ ---  of another subgroup with again $2^{\frac{d}{2}-1}$ members. 

These two groups are not related by the Hermitian conjugation as in the case of odd Clifford 
objects. 
In each of the two groups of an even Clifford character there are $2^{\frac{d}{2}-1}$ self 
adjoint operators. The rest of $2^{\frac{d}{2}-1}\cdot (2^{\frac{d}{2}-1}-1)$ Clifford even 
operators have the Hermitian conjugated partners within the same group. 

$\tilde{\gamma}^a \gamma^a$ transform $2^{\frac{d}{2}-1}$ self adjoint operators of one 
Clifford even group into $2^{\frac{d}{2}-1}$ self adjoint operators of another Clifford even 
group, while $\tilde{\gamma}^a \gamma^a$ transform the rest of this group --- that is   
$2^{\frac{d}{2}-1}\cdot (2^{\frac{d}{2}-1}-1)$ operators, having the Hermitian conjugated 
partners within the same subgroup --- into  $2^{\frac{d}{2}-1}\times (2^{\frac{d}{2}-1}-1)$ 
operators of another Clifford even group,  having again the Hermitian conjugated partners 
within the same subgroup.

Any odd Clifford member of the assumed (chosen to be) creation operators gives, when 
applied on one (only one) of the even self adjoint operators of only one of the two groups 
with $(2^{\frac{d}{2}-1})^2$ members, a nonzero contribution,  which is the same creation 
operator back. It gives nonzero contribution also on one (only one) of the rest 
$2^{\frac{d}{2}-1}\cdot (2^{\frac{d}{2}-1}-1)$ operators of the same group to which 
also the self adjoint operator belong, transforming it to one of creation operators, belonging to 
another family of the creation operators. 
On all the others Clifford even objects this creation operator gives zero.  

The annihilation operators manifest, when applied on the Clifford even objects, 
equivalent properties as creation operators. 

 Let $\hat{b}^{\alpha \dagger}_{i}$ be the creation operator of an odd Clifford character, 
$\alpha$ denoting the subgroup with a particular value of the Cartan subalgebra of 
$\tilde{S}^{ab}$ (family)  and with $i$ denoting a particular member of a family $\alpha$.
To all the members of particular $\alpha$ one and only one of the selfadjoint operators of 
an even Clifford character corresponds, which, when any of these members applies on it, 
gives the same creation operator back. 

 $(\hat{b}^{\alpha \dagger}_{i})^{\dagger}=$ $\hat{b}^{\alpha }_{i}$, denoting the 
corresponding annihilation operator of an odd Clifford character, gives zero when applied on 
the selfadjoint operators on which $\hat{b}^{\alpha  \dagger}_{i}$ gives nonzero contribution.

We choose the superposition of these selfadjoint operators to determine the vacuum state in 
the Clifford case, Eq.~(\ref{vac1}).
 
 All the members of the odd Clifford character, half of them creation operators and half of them
annihilation operators, fulfill the anticommutation relations, required for fermions. 
Correspondingly there are only $2^{\frac{d}{2}-1}\cdot 2^{\frac{d}{2}-1}$ creation operators,
determining $2^{\frac{d}{2}-1}$ families with $2^{\frac{d}{2}-1}$ family members each, 
which when applied on the superposition of selfadjoint operators of one group of Clifford even 
operators,  create fermion states. These creation operators determine $n$ fermions Hilbert 
space. 

In the Grassmann case there are two kinds of operators, $\theta^a$ and 
$\frac{\partial}{\partial \theta^a}$, Hermitian conjugated to each other, Eqs.~(\ref{grassher}, 
\ref{grassp}). If $\theta^a$ represent the creation operators, then
$\frac{\partial}{\partial \theta^a}$ are the corresponding annihilation operators. Not having 
the Hermitian conjugate partner the identity ($I$) can not belong either to creation or to 
annihilation operators. 

In $d=2(2n+1)$-dimensional Grassmann spaces there are $\frac{d!}{\frac{d}{2}! \frac{d}{2}!}$ 
creation operators and the same number of annihilation operators of an odd Grassmann character, 
Eq.~(\ref{nograssop}), chosen to be eigenstates of the Cartan subalgebra, Eq.~(\ref{cartan}), 
of ${\cal {\bf S}}^{ab}$. 
The application of the creation operators, which are products of $\frac{d}{2}$ $\theta^a$'s, 
on the identity ($I$)  gives them back, while the annihilation operators applied on $I$ give zero.

The $\frac{d!}{\frac{d}{2}! \frac{d}{2}!}$ creation operators split into two by the generators 
of the Lorentz transformations ${\cal {\bf S}}^{ab}$ unconnected groups, each with 
$\frac{1}{2}$ $\frac{d!}{\frac{d}{2}! \frac{d}{2}!}$ members.

Let $\hat{b}^{ \alpha  \dagger}_{i}$ be the creation operator of an odd Grassmann 
character with $\alpha =(1,2)$ denoting one of the two by  ${\cal {\bf S}}^{ab}$ unconnected 
subgroups and let $i$ denotes one of the  $\frac{1}{2}$ $\frac{d!}{\frac{d}{2}! \frac{d}{2}!}$ 
members related among themselves by ${\cal {\bf S}}^{ab}$.

We make a choice of the vacuum state in the Grassmann case to be  $|\psi_{o \alpha}>=
|\;\;1>$.

All members of two groups of $\frac{1}{2}$ $\frac{d!}{\frac{d}{2}! \frac{d}{2}!}$ number of 
creation operators of an odd Grassmann character,  and their Hermitian conjugated partners,
fulfill the anticommutation relations, required for fermions. 

The number of vectors in the Hilbert space of n-fermions depends for a chosen momentum 
$p_{k}^a$ on the number of the creation operators, creating  a particular fermion in the Clifford 
case or a particular "fermion" in the Grassmann case. 

There are for each $p_{k}^a$ in the Clifford case $2^{\frac{d}{2}-1}\cdot 2^{\frac{d}{2}-1}$ 
and in the Grassmann case $\frac{d!}{\frac{d}{2}! \frac{d}{2}!}$ creation operators 
$ \hat{{\bf b}}^{\alpha  \dagger}_{i\,p_k}$ of an odd character  --- either Clifford odd character, 
with $\alpha =(1,\cdots, 2^{\frac{d}{2}-1}), i =(1, \cdots, 2^{\frac{d}{2}-1}) $, or Grassmann 
odd character, with $\alpha =(1, 2), i=(1, \cdots, \frac{d!}{\frac{d}{2}! \frac{d}{2}!})$), creating 
the corresponding single particle states, when applied on the vacuum states $|\psi_{o}>$ --- in the
Clifford case is the vacuum state $|\psi_{oc}>$, the superposition of all selfadjoint operators, on 
which an odd $ \hat{{\bf b}}^{\alpha i \dagger}_{p_k}$ gives a nonzero contribution, and in the
Grassmann case the vacuum state is $|\psi_{og}>= |\;\;1>$.

Let the zero fermion state for any $p_{k}^a$ in either Clifford or Grassmann space, be 
written as $|\psi_{o}>$:
$=|0^{\alpha =1}_{i=1\, p_{1}}, 0^{\alpha =1}_{i=2 \,p_{1}},
 0^{\alpha= 1}_{i=3\, p_{1}},\dots,  0^{\alpha=1 }_{i_{\rm max}\, p_{1}}, \dots, 
 0^{\alpha= \alpha_{\rm max}}_{i_{\rm max}\, p_{1}}, \dots, 0^{\alpha =1}_{i=1\, p_{2}},
 0^{\alpha =1}_{i=2\, p_{2}},  0^{\alpha 1}_{i=3 \,p_{2}},\dots,  
0^{\alpha=1 }_{i_{\rm max}\, p_{2}}, \dots, 
 0^{\alpha= \alpha_{\rm max}}_{i_{\rm max}\, p_{k}},$ $\cdots, | \psi_{o}>$, 
with $| \psi_{o}> =(| \psi_{oc}>, |\;\;1>)$, in the Clifford  and the Grassmann case, respectively,
 and $\alpha_{max} =(2^{\frac{d}{2}-1}, 2)$ and $i_{max} =(2^{\frac{d}{2}-1}, \frac{1}{2}
\frac{d!}{\frac{d}{2}! \frac{d}{2}!})$, again  in the Clifford  and the Grassmann case, respectively.
Then the vector space with $n$ fermions in the Clifford case or $n$ "fermions" in the Grassmann
case, for any $n$ looks like
\begin{eqnarray}
\label{nfermions}
 \hat{{\bf b}}^{\alpha  \dagger}_{i\,p_k}\,|\psi_{o}>&=&|0^{\alpha =1}_{i=1\, p_{1}},
 0^{\alpha =1}_{i=2 \,p_{1}}, \dots, 
 0^{\alpha= \alpha_{\rm max}}_{i_{\rm max}\, p_{1}}, \dots, 1^{\alpha }_{i \, p_{k}}\,
\cdots,  | \psi_{o}>\nonumber\\
&&\cdots\nonumber\\
&&{\rm there\,\, are} \,\, \alpha_{\rm max}\cdot i_{\rm max} \,\, {\rm such\,\,  1-
 fermion \;states\, for \, each\, }  p_{k}\,,\nonumber\\
&&\hat{{\bf b}}^{\alpha  \dagger}_{i\,p_k}\, \hat{{\bf b}}^{\alpha  \dagger}_{j\,p_k}\,|\psi_{o}>
\,,\nonumber\\
&&\cdots
\nonumber\\
&&\cdots
\nonumber\\
&&\Pi_{\alpha=1, \alpha_{\rm max}}\, \Pi_{i=1, i_{\rm max}}\,
\hat{{\bf b}}^{\alpha  \dagger}_{i\,p_k}\,|\psi_{o}>\,,
\,,\nonumber\\
&&\cdots
\nonumber\\
&&\Pi_{\alpha=1, \alpha_{\rm max}}\, \Pi_{i=1, i_{\rm max}}\,
\hat{{\bf b}}^{\alpha  \dagger}_{i\,p_l}\,|\psi_{o}>\,,
\,,\nonumber\\
&&\cdots\nonumber\\
&&{\rm there\, are} \, 2^{\alpha_{\rm max}\cdot i_{\rm max}} \, {\rm Slater\, determinants\,
 of \; fermions\, for \, each\, }  p_{k}\,,\nonumber\\
&&\cdots
\end{eqnarray}
 $\alpha_{\rm max}=(2^{\frac{d}{2}-1}, 2)$ and $i_{\rm max}= (2^{\frac{d}{2}-1},
 \frac{1}{2}\, \frac{d!}{\frac{d}{2}! \frac{d}{2}!})$ in the Clifford and Grassmann case, 
respectively.

One sees that
\begin{eqnarray}
\label{SD}
&& \;\hat{{\bf b}}^{\alpha  \dagger}_{i\,p_k}\,  \hat{{\bf b}}^{\beta  \dagger}_{j\,p_l} \,
|0^{\alpha =1}_{i=1\, p_{1}}, 0^{\alpha =1}_{i=2 \,p_{1}},\dots,
 1^{\alpha'}_{i' p_{k'}},\dots, 0^{\alpha'''}_{i'''=1\, p_{k'''}},\dots, 
 1^{\alpha^{iv}}_{i^{iv} p_{k^{iv}}},\dots,
 1^{\beta' }_{j'\, p_{l'}}, \dots, | \psi_{o}>=\nonumber\\
&&- \hat{{\bf b}}^{\beta  \dagger}_{j\,p_l}\,  \hat{{\bf b}}^{\alpha  \dagger}_{i\,p_k} \,
|0^{\alpha =1}_{i=1\, p_{1}}, 0^{\alpha =1}_{i=2 \,p_{1}},\dots,
 1^{\alpha'}_{i' p_{k'}},\dots, 0^{\alpha'''}_{i'''=1\, p_{k'''}},\dots 
 1^{\alpha^{iv}}_{i^{iv} p_{k^{iv}}},\dots, 1^{\beta' }_{j'\, p_{l'}}, \dots, | \psi_{o}>\,,
\end{eqnarray}
and is zero only if any of the occupied states is the same as one (or both) of the two states 
determined by $\hat{{\bf b}}^{\alpha  \dagger}_{i\,p_k}$ or  
$\hat{{\bf b}}^{\beta  \dagger}_{j\,p_l}$ applied on 
$| \psi_{o}>$~\footnote{
Each single particle state caries its own internal space, described by a creation operator with  
a superposition of an odd number of $\gamma^a_{i}$'s, and its own coordinate space, 
described by $x^a_{i}$'s (or $p^{a}_{i}$). The creation operators of any two pairs of particles 
therefore anti-commute. 
Correspondingly the two states of two particles must distinguish in either internal space or in the 
coordinate space, as it follows from Eq.~(\ref{nfermionstate}).  The property of the creation 
operators $\hat{b}^{\alpha  \dagger}_{s p i} \hat{b}^{\alpha' \dagger}_{s' p' j}$ 
applying on the n-particle state $ |1^{\alpha}_{s p 1}, 1^{\alpha'}_{s' p' 2}, 
1^{\alpha"}_{s" p" 3}, \dots, 0^{\alpha"'}_{s"' p_i \,i}, \dots, 0^{\alpha^{iv}}_{s^{iv} p_j\, j},
 \dots,>$,presented in Eq.~(\ref{nfermionstate}),  
can be as well described by  (superposition of) Slater determinants of single particle states.
Let us add that the vacuum state, having the sum of the spins of both kinds of operators, $S^{ab}$ 
and $\tilde{S}^{ab}$, equal to zero and therefore neutral, remains neutral also when filled  with 
fermions of all the spins, $S^{ab}$ and $\tilde{S}^{ab}$. }.

One fermion states are either in Clifford or in Grassmann space already second quantized, since
in both cases they fulfill the anticommutation relations required for fermions, 
Eqs.~(\ref{grassordinary}, \ref{cliffordinary}).


All together there are $2^{2^{d -2}}$ Slater determinants for a chosen $p_k$ in the Clifford case
and $2^{ \frac{d!}{\frac{d}{2}! \frac{d}{2}!}}$ Slater determinants for a chosen $p_k$ in the 
Grassmann case,
$p_{k}$ has a continuously changing value,  $p^0 =(0,\infty)$, $- \infty \le p^{l} \le \infty$,
$l=(1, 2, 3, 5, \cdot,d)$. 

It can be concluded that there are only second quantized states, since the anticommuting
creation and annihilation operators, creating a Clifford fermion  or Grassmann "fermion" states, 
determine all the properties of the n-particle Hilbert space for any n. 

We shall as well recognize that no Dirac sea is needed either in the Clifford or in the Grassmann 
case, since the same Lorentz representation includes in both cases fermions and antifermions.

We discuss in the subsections the second quantization procedure in both spaces, Clifford and 
Grassmann when dimension of the space-time is larger than four. We demonstrate that if the 
dynamics manifests only in $d=(3+1)$, that is when momentum is different from zero only in 
$d=(3+1)$, $p^a=(p^0,p^1,p^2,p^3,0,0,\cdots,0)$  --- what happens at low energies after the
break of Lorentz symmetries in $d\ge 5$ --- spins in $d\ge 5$ manifest as charges in $d=(3+1)$.

While the Clifford case offers the explanation for all the properties of observable fermions, 
the Grassmann case, having difficulties in describing energy within the usual second quantized 
procedure, as long as the Lorentz invariance in internal space is unbroken, leads to unobserved 
"fermions" with integer spins.

Let us point out that states in  Grassmann space as well as states in Clifford space are organized 
to be --- within each of the two spaces --- orthogonal and normalized with respect to Eq.~(\ref{grassintegral}, 
\ref{grassnorm}, \ref{grassgammanorm}). 
All the states in each of spaces are chosen to be eigenstates of the Cartan subalgebra --- with 
respect to ${\cal {\bf S}}^{ab}$ in Grassmann space, Eqs.~(\ref{thetarelcom}, \ref{Lorentztheta},
\ref{cartan}), and with respect to $S^{ab}$ and $\tilde{S}^{ab}$, Eq.~(\ref{twocliffordsab}), in 
Clifford space, Eq.~(\ref{cartan}).

We pay attention in this paper almost only to spaces with $d=2(2n+1)$%
~\footnote{The main reason that we treat here mostly $d=2(2n+1)$ spaces is that one Weyl 
representation, expressed by the product of the Clifford algebra objects, manifests in $d=(1+3)$ 
all the observed properties of quarks and leptons, if $d\ge 2(2n+1), n=3$, and that the breaks of 
the starting symmetry down to $d=(3+1)$ can lead to massless fermions~\cite{NHD,ND012}. 
}. 

\subsection{Second quantization in Grassmann space }
\label{secondGrassmann}

There are $2^d$ states in Grassmann space, orthogonal to each other with respect to 
Eqs.~(\ref{grassintegral}, \ref{grassnorm}). To any coordinate there exists the conjugate 
momentum. We pay attention
in what follows mostly  to  spaces with $d=2(2n+1)$.
In $d=2(2n+1)$ spaces there are $\frac{d!}{\frac{d}{2}! \frac{d}{2}!}$ states, divided into 
two separated groups of states, Eq.~(\ref{nograssop}).  All states of one group are reachable 
from a starting state by the application of ${\cal {\bf S}}^{ab}$. The states, which contribute in
the second quantization procedure and manifest anticommutation relations required for fermions, 
are Grassmann odd products of eigenstates of the 
Cartan subalgebra.   Any Grassmann odd state can be written as a creation operator, operating 
on the vacuum state, while the Hermitian conjugated creation operator is the
corresponding annihilation operator. Creation and annihilation operators  of an odd Grassmann 
character fulfill the commutation relations of Eq.~(\ref{btheta1}, \ref{ijthetaprod}).
Let us see how it works.
 
If $\hat{b}^{\theta \dagger}_i$ is a creation operator, which creates a state in the Grassmann
space when operating on a vacuum state
$|\psi_{og}>$ and $\hat{b}^{\theta}_i=(\hat{b}^{\theta \dagger}_i)^{\dagger}$ is the 
corresponding  annihilation operator, then for a set of creation operators 
$\hat{b}^{\theta \dagger}_i$ and the corresponding annihilation operators $\hat{b}^{\theta}_i$   
it must be 
\begin{eqnarray}
\label{btheta0}
 \hat{b}^{\theta}_{i}|\phi_{og}>&=& 0\,, \nonumber\\
\hat{b}^{\theta \dagger}_{i}|\phi_{og}>&\ne& 0\,.
\end{eqnarray}
We first pay attention on only the internal degrees of freedom of the Grassmann "fermions": 
the spin. 
 
Choosing $\hat{b}^{\theta \dagger}_{a} =\theta^a$, then it follows that 
$(\hat{b}^{\,\theta \dagger}_{a})^{\dagger} =$$ \frac{\partial}{\partial \theta^a}$,
Eqs.~(\ref{grassher}, \ref{grassp}). 
One correspondingly finds 
\begin{eqnarray}
\hat{b}^{\theta \dagger}_{a} = \theta^a \,,  &&  
\hat{b}^{\theta}_{a} = \frac{\partial}{\;\partial \theta^a}\,,\nonumber\\
\{ \hat{b}^{\theta}_a, \hat{b}^{\theta \dagger}_{b} \}_{+}  |\phi_{og}>
&=&\delta_{a b} |\phi_{og}>
\,,\nonumber\\
\{\hat{b}^{\theta}_a, \hat{b}^{\theta}_{b}\}_{+} |\phi_{og}>
&=&0\,,\nonumber\\
\{\hat{b}^{\theta\dagger}_a,\hat{b}^{\theta \dagger}_{b}\}_{+} |\phi_{og}>
&=&0\,,\nonumber\\
 \hat{b}^{\dagger \theta}_{a}\;|\phi_{og}>&=& \theta^a|\phi_{og}>\,, \nonumber\\
 \hat{b}^{\theta}_{a} \;|\phi_{og}>&=& 0\,.
\label{btheta1}
\end{eqnarray}
The vacuum state $|\phi_{og}>$ is in this case chosen to be $|1\,>$. 

The number operator $\hat{N}^{\theta}_a = \hat{b}^{\theta \dagger}_{a} \hat{b}^{\theta}_{a}$
has the property, due to the second line in Eq.~(\ref{btheta1}), that $(\hat{N}^{\theta}_a)^2
=$ $\hat{N}^{\theta}_a$, with  the eigenvalue $0$ or $1$.

 The identity $I $ ($I^{\dagger} = I$) can not be taken as a creation operator, since its 
annihilation partner does not fulfill Eq.~(\ref{btheta0}).

We can use the products of superposition of $\theta^a$'s as creation operators and
correspondingly products of superposition of $\frac{\partial}{\; \partial \theta_a}$'s as 
annihilation operators, provided that they fulfill the requirements for the creation and 
annihilation operators, Eq.~(\ref{ijthetaprod}), with the vacuum state $|\phi_{og}> =|1\,>$. 
In general they would not. Only an odd number of $\theta^{a}$ in any product would have 
the required anticommutation properties.

To construct creation operators it is convenient to take products of such superposition of vectors 
$\theta^a$ and $\theta^b$ 
that each factor is the "eigenstate" of one of the Cartan subalgebra members of the Lorentz 
algebra~(\ref{cartan}).  Let us start with the creation operator, which is a products of $\frac{d}{2}$
"eigenstates" of the Cartan subalgebra $S^{ab}$: $\hat{b}^{\theta \dagger}_{a_i b_i}=
 \frac{1}{\sqrt{2}} (\theta^{a_i} \pm \epsilon \theta^{b_i})$. Then the corresponding annihilation 
operator has as well $\frac{d}{2}$ factors: 
$\hat{b}^{\theta}_{a_i b_i}= $ $\frac{1}{\sqrt{2}} (\frac{\partial}{\; \partial \theta^{a_i}} 
 \pm \epsilon^{*} \frac{\partial}{\; \partial \theta_{b_i}})$, $\epsilon = i $, if 
$\eta^{a_i a_i}=\eta^{b_i b_i}$ and $\epsilon = -1 $, if $\eta^{a_i a_i}\ne\eta^{b_i b_i}$. 

 Let us in $d=2(2n+1)$, $n$ is a positive integer, start with the state 
\begin{eqnarray}
|\phi^{1}_{1}> &=&(\frac{1}{\sqrt{2}})^{\frac{d}{2}} \,
(\theta^0 - \theta^3) (\theta^1 + i \theta^2) (\theta^5 + i \theta^6) \cdots (\theta^{d-1} +
 i \theta^{d}) \hat{b}^{\theta 1}_{1} \, |1\,>\,,\nonumber\\
&=& \hat{b}^{\theta 1\dagger}_{1} \, |1\,>\,, \quad {\rm with} \nonumber\\
\hat{b}^{\theta 1\dagger}_{1}:&=& (\frac{1}{\sqrt{2}})^{\frac{d}{2}} \,
(\theta^0 - \theta^3) (\theta^1 + i \theta^2) (\theta^5 + i \theta^6) \cdots (\theta^{d-1} +
 i \theta^{d}) \,.
\label{start(2n+1)2theta}
\end{eqnarray}
 Recognizing that 
${\cal {\bf S}}^{ab} (\theta^a\mp \theta^b) \,|1\,>= \pm \,i \,(\theta^a\mp \theta^b)\, |1\,>$,
 for $(a=0$ and  $b\ne a)$, ${\cal {\bf S}}^{ab} (\theta^a\pm i \theta^b) \,|1\,>= 
\pm\,1\, (\theta^a\pm i \theta^b)\,|1\,>$, for $(a,  b) \ne 0)$ and 
${\cal {\bf S}}^{ab} (\theta^a \theta^b) \, |1\,>= 0 \,|1\,>$, it follows that the 
eigenvalues of the Cartan subalgebra operators, Eq.~(\ref{cartan}), are 
($+i,+1, +1, \dots +1$).

The rest of states, belonging to the same Lorentz representation, follow from the starting 
state by the application of the operators ${\cal {\bf S}}^{cf}$, which do not belong to
the Cartan subalgebra operators.

One can find creation and annihilation operators for $d=4n$ in App.~\ref{4n}. 

\vspace{3mm} 

$\;\;$ {\bf i.} 
We proposed in Eq.~(\ref{start(2n+1)2theta}) the starting creation operator 
$\hat{b}^{\theta 1\dagger}_{1}$, the upper index indicates one of the two groups, the lower 
index indicates the starting state. By taking into account Eqs.~(\ref{grassher},~\ref{grassp}) 
 the starting creation operator and its annihilation partner  are  for $d=2(2n+1)$ equal to
\begin{eqnarray}
\hat{b}^{\theta 1 \dagger}_{1} &=& (\frac{1}{\sqrt{2}})^{\frac{d}{2}} \,(\theta^0 - \theta^3) 
(\theta^1 + i \theta^2) (\theta^5 + i \theta^6) \cdots (\theta^{d-1} + i \theta^{d})\,,
 \nonumber\\
\hat{b}^{\theta 1}_{1} &=& (\frac{1}{\sqrt{2}})^{\frac{d}{2}}\,
 (\frac{\partial}{\;\partial \theta^{d-1}} - i \frac{\partial}{\;\partial \theta^{d}})
\cdots (\frac{\partial}{\;\partial \theta^{0}}
-\frac{\partial}{\;\partial \theta^3})\,,\nonumber\\
&& {\rm for} \; d=2(2n+1)\,.
\label{graphicstheta2(2n+1)1}
\end{eqnarray}
The rest of creation operators belonging to this group (group $1$) in $d=2(2n+1)$ follow
by the application of all the operators ${\cal {\bf S}}^{ef}$, which do not belong to the Cartan 
subalgebra operators. The corresponding annihilation operators are the Hermitian 
conjugated values of a particular creation operator.  For $d=2(2n+1)$ one finds by the application 
of ${\cal {\bf S}}^{01}$ another creation operator and the corresponding annihilation operator 
as follows 
\begin{eqnarray}
\hat{b}^{\theta 1 \dagger}_{2} &=& (\frac{1}{\sqrt{2}})^{\frac{d}{2}-1} \,
(\theta^0  \theta^3 +i \theta^1 \theta^2)  (\theta^5 + i \theta^6) \cdots
 (\theta^{d-1} + i \theta^{d})\,,\nonumber\\
\hat{b}^{\theta 1}_{2} &=&   (\frac{1}{\sqrt{2}})^{\frac{d}{2}-1}
 \, (\frac{\partial}{\;\partial \theta^{d-1}} - i \frac{\partial}{\;\partial \theta^{d}})
\cdots (\frac{\partial}{\;\partial \theta^{3}} \,\frac{\partial}{\;\partial \theta^0}
-i \frac{\partial}{\;\partial \theta^{2}} \,\frac{\partial}{\;\partial \theta^1})\,, \nonumber\\
{\rm in} \; {\rm general}: \nonumber\\
\hat{b}^{\theta 1 \dagger}_{i} &\propto& {\cal {\bf S}}^{ab} \cdots {\cal {\bf S}}^{ef}
\hat{b}^{\theta 1 \dagger}_{1}\,,\nonumber\\
\hat{b}^{\theta 1}_{i} &=& (\hat{b}^{\theta 1 \dagger}_{i})^{\dagger}\,.
\label{graphicstheta2(2n+1)2}
\end{eqnarray}
It was taken into account in the above equation that any ${\cal {\bf S}}^{ac}$ ($a \ne c $),
which does not belong to the Cartan subalgebra, Eq.(\ref{cartan}), transforms $(\frac{1}{\sqrt{2}})^{2}
 (\theta^a + i \theta^b) (\theta^c + i \theta^d) $ ($a \ne c$ and $a \ne d $, $ b \ne c$ 
and $b \ne d$, $\eta^{aa}= \eta^{bb}$) into $\,\frac{1}{\sqrt{2}} 
(\theta^a  \theta^b + \theta^c \theta^d)$. The states are normalized and the simplest
phases are assumed.
One evaluates that ${\cal {\bf S}}^{ab} (\theta^a \pm \epsilon \theta^b)= \mp i $ 
$\frac{\eta^{aa}}{\epsilon} (\theta^a \pm \epsilon \theta^b)$, $\epsilon = 1$ for $\eta^{aa}=1$ 
and $\epsilon = i $ for $\eta^{aa}= -1$,  while  either ${\cal {\bf S}}^{ab}$ or ${\cal {\bf S}}^{cd}$,
applied on $ (\theta^a \theta^b \pm \epsilon \theta^c \theta^d)$, gives zero. The vacuum state 
is in all these cases $|1\,>$.

Although all the states, generated by  creation operators, which  include one  
$(I \pm \epsilon \theta^a  \theta^b)$ 
or several $ (I \pm \epsilon \theta^{a_1}  \theta^{b_1})\cdots ( I \pm \epsilon \theta^{a_k} 
\theta^{a_k})$,  are eigenstates of the Cartan subalgebra operators and are orthogonal with respect
 to the scalar product, Eq.(\ref{grassnorm}), to all the other states, their 
 Hermitian conjugate values include  $I^{\dagger}$, which,  when applying on the vacuum state 
$|\phi_{og}>=|1\,>$,  does not give zero. Correspondingly such creation operators do not have 
appropriate annihilation partners, which would  fulfill Eqs.~(\ref{btheta0}, \ref{btheta1}).

However, creation operators which are products of several  $\theta$'s,  
$\theta^{a_1}\cdots \theta^{a_n}$,   
let say $n$ with $n=2,4\dots \frac{d}{2}-1$ (always of an even number of $\theta$'s, 
since ${\cal {\bf S}}^{ab}$ is a Grassmann 
even operator, Eq.~(\ref{thetarelcom})),  
and are "eigenstates" of the Cartan subalgebra operators (Eq.~(\ref{cartan}), namely 
${\cal {\bf S}}^{ab} \theta^a \theta^b |1\,> =0$), fulfill the requirements of 
Eq.~(\ref{ijthetaprod}), provided that the rest of expression has an odd number of 
factors  --- ($\frac{d}{2} - n$).
\begin{eqnarray}
\{ \hat{b}^{\theta k}_i, \hat{b}^{\theta l \dagger}_{j} \}_{+} 
 |\phi_{og}> &=& \delta_{i j}\; \delta_{k l}\;|\phi_{og}>\,,\nonumber\\
\{ \hat{b}^{\theta k}_i, \hat{b}^{\theta l}_{j} \}_{+}  |\phi_{og}>
&=& 0\; |\phi_{og}> \,,\nonumber\\
\{\hat{b}^{\theta k \dagger}_i,\hat{b}^{\theta l \dagger}_{j}\}_{+} |\phi_{og}>
&=&0\; |\phi_{og}> \,,\nonumber\\
\hat{b}^{\theta k \dagger}_{i} |\phi_{og}>& =& \, |\phi^{k}_{i}>\,, \nonumber\\
\hat{b}^{\theta k}_{j} |\phi_{og}>& =&0\,  |\phi_{og}> \,, \nonumber\\
(k,l) &=&(1,2)\,.
\label{ijthetaprod}
\end{eqnarray}
Since there is another group of states, not reachable from the starting state by 
${\cal {\bf S}}^{ab} $ from the above starting state, presented in 
Eq.~(\ref{startgrass02}),
we denote creation operator with $\hat{b}^{\theta k \dagger}_{i}$ and the annihilation
operator by $\hat{b}^{\theta k}_{i}$, to generalize the notation.

It is not difficult to see that states included into one representation, which started with  
$\hat{b}^{\theta 1 \dagger}_{i} \,|1\,> $  as presented in Eq.~(\ref{graphicstheta2(2n+1)1}) 
for $d=(2n+1)2$ 
 {\it have the properties, required  by} Eq.~(\ref{ijthetaprod}) for $k=1$:

$\;\;$ {\bf i.a.} In any  $d$-dimensional space 
the product  $\frac{\partial}{\;\partial \theta^{a_1}} \cdots \frac{\partial}{\;\partial \theta^{a_k}}$, 
with all  different $a_i$ (if all or some of them are equal, then this is trivially true since 
$(\frac{\partial}{\partial \theta^{a}})^2 =0$), if 
applied on the vacuum $|1\,>$, is equal to zero. Correspondingly the second equation and 
the fifth equation of Eq.~(\ref{ijthetaprod}) are fulfilled.

$\;\;$ {\bf i.b.} In any  $d$-dimensional space the product of different $\theta^a$s --- $\theta^{a_1} 
\theta^{a_2}  \cdots \theta^{a_k}$ with all different $\theta^a$'s ($a_i \ne a_j$ for all $a_i$ and 
$a_j$) --- applied on the vacuum $|1\,>$, is different from zero. 
Since all the $\theta$'s, appearing in Eqs.~(\ref{graphicstheta2(2n+1)1},  
\ref{graphicstheta2(2n+1)2}), are different, forming  orthogonal and normalized states, the fourth 
equation of Eq.~(\ref{ijthetaprod}) is fulfilled.

$\;\;$ {\bf i.c.} The third equation of Eq.~(\ref{ijthetaprod}) is fulfilled provided that there is an odd 
number of $\theta^s$ in the expression for a creation operator. Then, when in the anticommutation 
relation different $\theta^a$'s  appear (like in the case of $d=6$ $\{\theta^{0} \theta^{3} 
\theta^{5}, \theta^{1} \theta^{2} \theta^{6}\}_{+}$), such a contribution gives zero. When  
 two or several equal $\theta$'s appear in the anticommutation relation, the contribution is zero
(since $(\theta^a)^2=0$).

$\;\;$ {\bf i.d.} Also for the first equation in Eq.~(\ref{ijthetaprod}) it is not difficult to show that it 
is fulfilled only for a particular creation operator and its Hermitian conjugate: Let us show this for 
$d=1+3$ and the creation operator $\frac{1}{\sqrt{2}} (\theta^{0} - \theta^{3})\, 
\theta^1 \theta^2 $ and  its Hermitian conjugate (annihilation) operator: 
$\frac{1}{\sqrt{2}}\, \{\frac{\partial}{\;\partial \theta^{2}}\; 
\frac{\partial}{\;\partial \theta^{1}}\, (\frac{\partial}{\;\partial \theta^{0}} - 
\frac{\partial}{\;\partial \theta^{3}}), \frac{1}{\sqrt{2}} (\theta^{0} - \theta^{3})\, 
\theta^1 \theta^2\}_{+} $. Applying $(\frac{\partial}{\;\partial \theta^{0}} - 
\frac{\partial}{\;\partial \theta^{3}})$  on $ (\theta^{0} - \theta^{3})  $ gives two, while 
$ \frac{\partial}{\;\partial \theta^{2}}\; \frac{\partial}{\;\partial \theta^{1}}$ applied on 
$\theta^1 \theta^2$ gives one. 

$\;\;$ {\bf i.e.} If we define the number operator $N^{\theta k}_{i}$ as follows
\begin{eqnarray}
\label{thetanop}
\hat{N}^{\theta k}_{i}&=& \hat{b}^{\theta k \dagger}_{i} \hat{b}^{\theta k}_{i}\,, 
\end{eqnarray}
it follows, taking into account the third equation of Eq.~(\ref{ijthetaprod}), that
 $(\hat{N}^{\theta k}_{i})^2= \hat{b}^{\theta k \dagger}_{i} \hat{b}^{\theta k}_{i}
\hat{b}^{\theta k \dagger}_{i} \hat{b}^{\theta k}_{i}= \hat{N}^{\theta k}_{i}$,
requiring that the eigenvalue of this operator on the state 
$\hat{b}^{\theta k \dagger}_{i}|\phi^{k}_{i}>$ is $0$ or $1$. 

\vspace{3mm}

$\;\;$ {\bf ii.}
There is one additional group of creation and annihilation operators in $d=2(2n+1)$, which follow from the
starting state 
\begin{eqnarray}
|\phi^{2}_{1}> &=& \hat{b}^{\theta 2 \dagger}_{01} |1\,>\,, \nonumber\\
\hat{b}^{\theta 2 \dagger}_{01}: &=& (\frac{1}{\sqrt{2}})^{\frac{1}{2}} (\theta^0+\theta^3) 
(\theta^1+i\theta^2) (\theta^5+i\theta^6)
 \cdots  (\theta^{d-3}+i\theta^{d-2}) (\theta^{d-1}+i \theta^d)\,,\nonumber\\
&&{\rm for}\; d=2(2n+1)\,.
\label{startgrass02}
\end{eqnarray}
This state can not be obtained from the  previous group of states, presented in
Eqs.~(\ref{graphicstheta2(2n+1)1}, \ref{graphicstheta2(2n+1)2})
by the application of 
${\cal {\bf S}}^{ef}$, since  each ${\cal {\bf S}}^{ef}$ changes an even number of factors, never
an odd one. Correspondingly the above starting state forms a new group of states in $d=2(2n+1)$. 
 All the other states of this new group of states 
follow from the starting one by the application of ${\cal {\bf S}}^{ef}$. The corresponding creation and 
annihilation operators are
\begin{eqnarray}
\label{creatorgrass2(2n+1)2}
\hat{b}^{\theta 2 \dagger}_{1} &=& (\frac{1}{\sqrt{2}})^{\frac{d}{2}} \,(\theta^0 + \theta^3) 
(\theta^1 + i \theta^2) (\theta^5 + i \theta^6) \cdots (\theta^{d-1} + i \theta^{d})\,,
 \nonumber\\
\hat{b}^{\theta 2}_{1} &=& (\frac{1}{\sqrt{2}})^{\frac{d}{2}}\,
 (\frac{\partial}{\;\partial \theta^{d-1}} - i \frac{\partial}{\;\partial \theta^{d}})
\cdots (\frac{\partial}{\;\partial \theta^{0}}
+ \frac{\partial}{\;\partial \theta^3})\,,\nonumber\\
&& {\rm for}\;  d=2(2n+1)\,.
\end{eqnarray}
As in the  case of the first group all the rest of creation operators are  obtainable from the starting 
one by the application of ${\cal {\bf S}}^{ac}$, and the
annihilation operators by the Hermitian conjugation of the creation operators.
\begin{eqnarray}
\label{general2}
\hat{b}^{\theta 2 \dagger}_{i} &\propto& {\cal {\bf S}}^{ab} \cdots {\cal {\bf S}}^{ef}
\hat{b}^{\theta 2 \dagger}_{1}\,,\nonumber\\
\hat{b}^{\theta 2}_{i} &=& (\hat{b}^{\theta 2 \dagger}_{i})^{\dagger}\,.
\end{eqnarray}
Also all  these creation and annihilation operators fulfill the requirements for the creation and 
annihilation operators, presented in Eq.~(\ref{ijthetaprod}),  due to the same reasons as 
in the first case.

It is true also in this case, as stated below Eq.~(\ref{thetanop}), that $\hat{N}^{\theta k}_{i}$
applied on the state $|\phi^{k}_i>$ gives $0$ or $1$, due to the fact that 
$(\hat{N}^{\theta k}_{i})^2=$ $\hat{N}^{\theta k}_{i}$. Thus the basic states, determined by
the application of creation operators of Eqs.(\ref{graphicstheta2(2n+1)2}, \ref{general2}) on 
the vacuum state $|\,1>$ have the properties required for fermions.

\vspace{3mm}

Let us now count the number of states of the odd Grassmann character in $d=2(2n+1)$.

There are  in ($d=2$) two creation ($(\theta^0 \mp \theta^1 $, for $\eta^{ab}  = diag(1,-1)$) 
and correspondingly two annihilation operators $(\frac{\partial}{\;\partial \theta^{0}} \mp 
\frac{\partial}{\;\partial \theta^{1}}$), 
each belonging to its own group with respect to the Lorentz transformation operators, 
both fulfilling Eq.~(\ref{ijthetaprod}). 

It is not difficult to see that the number of all  creation operators of an odd Grassmann character 
in $d=2(2n+1)$-dimensional space is equal to $\frac{d!}{\frac{d}{2}! \frac{d}{2}!}$.

We namely ask: In how many ways can one put on $\frac{d}{2}$  places $d$ different $\theta^a$'s.
And the answer is --- the  central binomial coefficient for $x^{\frac{d}{2}}\, 1^{\frac{d}{2}}$ --- with
all $x$ different. This is just $\frac{d!}{\frac{d}{2}! \frac{d}{2}!}$. But we have counted all the 
states with an odd Grassmann character, while we know that these states belong to two different 
groups of representations with respect to the Lorentz group.

Correspondingly one concludes: {\it There are two groups of states in} $d=2(2n+1)$ {\it with an 
odd Grassmann character}, {\it each of these two groups has} 
%
\begin{eqnarray}
\label{nograssop}
\frac{1}{2}\;\frac{d!}{\frac{d}{2}! \frac{d}{2}!}\,
\end{eqnarray}  
{\it members}.

In $d=2$ we have two groups with one state, which have an odd Grassmann character, in $d=6$
we have two groups of $10$ states, in $d=10$ we have two groups of $126$ states with an odd 
Grassmann characters. And so on.

Correspondingly we have  in  $d=2(2n+1)$-dimensional spaces two groups of creation operators 
with $\frac{1}{2}\; \frac{d!}{\frac{d}{2}! \frac{d}{2}!} $ members each, creating states with an 
odd Grassmann character and the same number of annihilation operators. Creation and annihilation 
operators fulfill anticommutation relations presented in Eq.~(\ref{ijthetaprod}).

The rest of creation operators [and the corresponding annihilation operators] with the opposite 
Grassmann character than the ones studied so far --- like  $\theta^0 \theta^1$ 
[$\frac{\partial}{\partial \theta^1}\frac{\partial}{\partial \theta^0}$] in $d=(1+1)$
 $(\theta^0 \mp \theta^3) (\theta^1\pm i \theta^2)$ 
[$ (\frac{\partial}{\partial \theta^1}\mp i \frac{\partial}{\partial \theta^2}) 
(\frac{\partial}{\partial \theta^0} \mp \frac{\partial}{\partial \theta^3}$],   $\theta^0 \theta^3
 \theta^1  \theta^2$ [$\frac{\partial}{\partial \theta^2} \,\frac{\partial}{\partial \theta^1}\,
\frac{\partial}{\partial \theta^3}\,\frac{\partial}{\partial \theta^0}$] in $d=(3+1)$, do not fulfill the
anticommutation relations required for fermions in Eq.~(\ref{ijthetaprod}), with  
$\hat{b}^{\theta 1}_{i}$ and $\hat{b}^{\theta 1 \dagger}_{i}$ replaced by  
$\hat{b}^{\theta k}_{i}$ and $\hat{b}^{\theta k \dagger}_{i}$, $k=(1,2)$ and 
correspondingly with $\{\hat{b}^{\theta k}_{i}$, $\hat{b}^{\theta l \dagger}_{j}\}|\phi_{og}>
= \delta_{k l}\,\delta_{i j}|\phi_{og}>$, $(k, l)=(1,2), (i , j)$ running from 
$(1,\dots,\frac{1}{2}\; \frac{d!}{\frac{d}{2}! \frac{d}{2}!} )$. 

All the states $|\phi^{k}_{i}>$, $k=(1,2)$, generated by the creation operators, 
Eqs.~(\ref{ijthetaprod}, \ref{general2}), 
on the vacuum state $|\phi_{og}>(=|1\,>)$ are the eigenstates of the Cartan subalgebra
operators and are orthogonal and normalized with respect to the norm of Eq.~(\ref{grassintegral})
\begin{eqnarray}
\label{statesgrass}
<\phi^{k}_{i}|\phi^{k'}_{j}>&=& \delta_{i j}\, \delta^{k k'}\,,\nonumber\\
(k,k') =(1,2)\,,\; (i,j) &=& (1,2,\dots, \frac{1}{2} \frac{d!}{\frac{d}{2}! \frac{d}{2}!})\,.
\end{eqnarray}
All these basic states describing the internal degrees of freedom can be used to solve 
Eq.~(\ref{Weylgrass}) for free massless "fermions", with the part in ordinary space 
proportional to $e^{-i p^a x_a}$.  The eigenstates of Eq.~(\ref{Weylgrass}) are
superposition of the basic states $|\phi^{k}_{i}>$ with coefficients depending on momentum
$p^a, a=(1,\dots,d)$
\begin{eqnarray}
\hat{b}^{\theta k \dagger}_{s p}&=&\sum_{i} c^{k}{}_{s p i} \, 
\hat{b}^{\theta k \dagger}_{i}\,,\nonumber\\
|\phi^{k}_{s p}> &=&\hat{b}^{\theta k \dagger}_{s p}\,|\phi_{og}>\,,\nonumber\\
|\phi^{k}_{s p}> &=& \sum_{i} c^{k}{}_{s p i}\,|\phi^{k}_{i}>\,, 
\label{ptheta}
\end{eqnarray}
$s$ represents different solutions of the equations of motion, and, since they are orthogonalized, 
they fulfill  the relation $<\phi^{k}_{s p}|\phi^{k'}_{s' p'}>=
\delta_{k k'}\, \delta_{s s'}\,\delta^{p p'}$, where we assumed the discretization of momenta.

The corresponding creation operators, creating the basic states describing free massless 
"fermions" --- $ \hat{b}^{\theta k\dagger}_{s p}$ --- are superposition of creation operators 
$\hat{b}^{\theta k\dagger}_i$, $ \hat{b}^{\theta k\dagger}_{s p}=\sum_{i} c^{k}{}_{s p i}
\hat{b}^{\theta k\dagger}_i $  and fulfill together with the corresponding annihilation operators 
$\hat{b}^{\theta k}_{s p}= ( \hat{b}^{\theta k\dagger}_{s p})^{\dagger}$ the relations
\begin{eqnarray}
\{ \hat{b}^{\theta k}_{s p}, \hat{b}^{\theta k' \dagger}_{s' p'}\}_{+} |\phi_{og}> 
&=&\delta^{k}_{k'}\,\delta_{s s'}\delta_{p p'}\;|\phi_{og}>
\,,\nonumber\\
\{ \hat{b}^{\theta k}_{s p}, \hat{b}^{\theta k'}_{s' p'}\}_{+}  |\phi_{og}>
&=&0 \;|\phi_{og}>\,,\nonumber\\
\{\hat{b}^{\theta k \dagger}_{s p}, \hat{b}^{\theta k' \dagger}_{s' p'}\}_{+}
|\phi_{og}> &=&0\; |\phi_{og}>\,,\nonumber\\
\hat{b}^{\theta k}_{s p} |\phi_{og}>& =& 0 \;|\phi_{og}>\,,\nonumber\\
\hat{b}^{\theta k \dagger}_{s p} |\phi_{og}>& =& |\phi^{k}_{s p}>\,,\nonumber\\
|\phi_{og}>&=& |1\,>\,.
\label{ijthetaprodordinary}
\end{eqnarray}
Again indexes $k=(1,2)$ in ($\hat{b}^{\theta 1}_{s p}$, $\hat{b}^{\theta 1 \dagger}_{s p}$) 
($\hat{b}^{\theta 2}_i$, $\hat{b}^{\theta \dagger 2}_i$) denote creation and annihilation 
operators of one of the two groups of states describing the internal space of "fermions", 
reachable by ${\cal \bf{S}^{ab}} $ and  $\hat{b}^{\theta k \dagger }_{s p}$ creates the 
state for a particular momentum in ordinary space $p^a$, solving Eq.~(\ref{Weylgrass}).

The number operator for  a "fermion" state $|\phi^{k}_{p}>$ is now
%
\begin{eqnarray}
\label{theta}
 \hat{N}^{\theta k}_{s p} &=& \hat{b}^{\theta k\dagger}_{s p} \hat{b}^{\theta k}_{s p}\,,
\nonumber\\
(\hat{N}^{\theta k}_{s p})^2 &=& \hat{N}^{\theta k}_{s p}\,,
\end{eqnarray}
with the eigenvalues $0$ or $1$, since the states of a chosen discretized $p^a$ are orthogonal.
Correspondingly each state can be occupied or empty. 
 If $|1^{\theta k}_{s_{1}p_{1}},1^{\theta k}_{s_{2} p_{2}},1^{\theta k}_{s_{3}p_{3}},\dots,  
0^{\theta k}_{s_{k}p_{k}}, \dots, 0^{\theta k}_{s_{l}p_{l}}, \dots,>$ is a $n$ particle 
state of "fermions" (and "antifermions"), 
 where $1$ denotes the occupied 
state and $0$ the unoccupied state, then it follows, for example, due to the 
third line in Eq.~(\ref{ijthetaprodordinary}), that  
\begin{eqnarray}
\label{nfermionthetastate}
&&\hat{b}^{\theta k \dagger}_{s_{i} p_{i}} 
    \hat{b}^{\theta k \dagger}_{s_{j} p_{j}}\,
|1^{\theta k}_{s_{1} p_{1}}, 1^{\theta k}_{s_{2} p_{2}},
 1^{\theta k}_{s_{3} p_{3}},\dots, 
 0^{\theta k}_{s_{i} p_{i}}, \dots, 
 0^{\theta k}_{s_{j} p_{j}}, \dots,> =\nonumber\\
&& - \hat{b}^{\theta k \dagger}_{s_{j} p_{j}} 
       \hat{b}^{\theta k \dagger}_{s_{i}  p_{i}}
|1^{\theta k}_{s_{1} p_{1}}, 
 1^{\theta k}_{s_{2} p_{2}}, 1^{\theta k}_{s_{3} p_{3}},\dots,
 0^{\theta k}_{s_{i} p_{i}}, \dots, 
 0^{\theta k}_{s_{j} p_{j}}, \dots,>\,.
\end{eqnarray}
Any $n$ "fermion" state is therefore a product of   $n$ creation operators 
$\,\hat{b}^{\theta k \dagger}_{i\, p_{k}}$ as presented in Eq.~(\ref{nfermions}).    

The number operator for "fermions" in the n-particle state of Eq.~(\ref{nfermionthetastate}) is 
correspondingly
\begin{eqnarray}
\label{thetanopgen}
\hat{N}^{\theta}&=& \sum_{k,s_{i}p_i} \hat{N}^{\theta k}_{s_{i}p_i}
\end{eqnarray}
When coefficients $c^{k}{}_{s p i}$   depend also on coordinates $x^a$ (for free "fermions"  
${\bf c}^{k}{}_{s p i} (x) =$ $c^{k}{}_{s p i}\cdot e^{-i p_a x^a} $), it follows 
\begin{eqnarray}
\label{grasscreationx}
{\bf \hat{b}}^{\theta k \dagger}_{s} (x^0,\vec{x}) &=& \sum_{i} \int\,
 \frac{d^{d-1} p}{(2\pi)^{d-1}}\,
{\bf c}^{k}{}_{s p i} (x)\, \hat{b}^{\theta k \dagger }_{i}\,.\nonumber
\end{eqnarray}
\begin{eqnarray}
\label{grassordinary}
\{ {\bf \hat{b}}^{\theta k}_{s} (x^0,\vec{x}),\, {\bf \hat{b}}^{\theta k' \dagger}_{s'} 
(x^0,\vec{y})\}_{+}   |\psi_{oc}>
&=& \delta^{k k'}\, \delta^{s s'}\delta^{d-1} (\vec{x}-\vec{y}) \,|\psi_{oc}>
\,,\nonumber\\
\{ {\bf \hat{b}}^{\theta k \dagger}_{s} (x^0,\vec{x}),\, {\bf \hat{b}}^{\theta k' \dagger}_{s'} 
(x^0,\vec{y})\}_{+}   |\psi_{oc}>&=&0\,,\quad
\{ {\bf \hat{b}}^{\theta k }_{s} (x^0,\vec{x}),\, {\bf \hat{b}}^{\theta k' \dagger}_{s'} 
(x^0,\vec{y})\}_{+} |\psi_{oc}>=0\,.
\end{eqnarray}

It is discussed in the subsection~\ref{CPT} how do discrete symmetry operators in the 
Grassmann case take care of "fermion" and "antifermion" states. 

 Let us now take into account  Eq.~(\ref{LDWeylGrass10}) with ${\cal L}_{G}=\frac{1}{4}\{
\hat{\phi}^{\dagger} \gamma^0_{G} \tilde{\gamma}^a (\hat{p}_{a}\phi) - (\hat{p}_{a} 
\phi^{\dagger}) \gamma^0_{G} \tilde{\gamma}^a \phi\} $. The Euler-Lagrange equations lead to
$-i \frac{1}{2} \gamma^0_{G} \tilde{\gamma}^a \hat{p}_{a}\phi=0$ and 
$i \frac{1}{2}\hat{p}_{a}\phi^{\dagger} \gamma^0_{G} \tilde{\gamma}^a =0$.

Let us find the Hamilton function for a second quantized field:  $  {\bf \hat{\phi}}(x^0,\vec{y})$, 
generated by one of the creation operators ${\bf \hat{b}}^{\theta  \dagger}_{s}$ on the
vacuum state $|\phi_{og}>$,
\begin{eqnarray}
\label{GrassH0}
\Pi_{{\bf \hat{\phi}}}&=& \frac{\partial {\cal L}_{G}}{\partial (\hat{p}_0 {\bf \hat{\phi}})}= 
\frac{1}{4}\, {\bf \hat{\phi}}^{\dagger} \gamma^0_{G} \tilde{\gamma}^0\,,\quad 
\Pi_{{\bf \hat{\phi}}^{\dagger}}= 
\frac{\partial {\cal L}_{G}}{\partial (\hat{p}_0 {\bf \hat{\phi}}^{\dagger})}=
-\frac{1}{4}\, \gamma^0_{G} \tilde{\gamma}^0  {\bf \hat{\phi}}\,,
\nonumber\\
{\cal H}_{G} &=&\Pi_{\phi} \, (\hat{p}_0 {\bf \hat{\phi}}) + (\hat{p}_0 {\bf \hat{\phi}}^{\dagger})
\Pi_{{\bf \hat{\phi}}^{\dagger}} \, - {\cal L}_{G}\,, \nonumber\\
&=& \frac{i}{4 }\, {\big [} {\bf \hat{\phi}}^{\dagger} \gamma^0_{G} 
\tilde{\gamma}^i (\hat{p}_{i} {\bf \hat{\phi}}) - 
(\hat{p}_{i} {\bf \hat{\psi}}^{\dagger}) \gamma^0_{G} \tilde{\gamma}^i {\bf \hat{\phi}} \, {\big ]}\,,
\nonumber\\
H_{G}&=& \int d^{d-1} x \,{\cal H}_{G}\, .
\end{eqnarray}
A vector ${\bf \hat{\phi}} $ depends on $k=(I,II)$ and on spins (which in $d=(3+1)$ 
 manifest as in spin and charges).

Hamilton function is obviously an odd Grassmann object and {\it does not define the energy of 
the system}. However, if assuming the relation: 
$\frac{i}{2}\gamma^0 p_0 \,{\bf \hat{\phi}}^{k} (x^0,\vec{x})=\{ {\bf \hat{\phi}}^{k} 
(x^0,\vec{x}) ,\, H_{G}\}_{-} $, 
 one still ends up with the equations of motion, Eq.~(\ref{LDWeylGrass10}). One namely obtains  
\begin{eqnarray}
\label{HGrassphieq}
\gamma^0 \hat{p_0} {\bf \hat{\phi}}^k (t,\vec{x}) &=& 
{\bf {\Big\{}} \,{\bf \hat{\phi}}^k (t,\vec{x})\,,
\, H_{G} {\bf {\Big\}}}_{-} = - \gamma^0_{G} \tilde{\gamma}^i \hat{p}_{i}
 {\bf \hat{\phi}}^k (t,\vec{x})\,,
\end{eqnarray}
what might help to find the procedure to define the energy for the interacting 
"Grassmann fermions".
One must at this point either give up with the Grassmann "fermions" with the integer spins or 
find a consistent unconventional way to define the energy, like the one suggested in 
Eq.~(\ref{HGrassphieq}).

\subsection{Second quantization in Clifford space} 
\label{secondClifford}

In Grassmann space the requirement that products of "eigenstates" of the Cartan subalgebra 
operators form the creation and annihilation operators, obeying the relations of 
Eq.~(\ref{ijthetaprod}), 
reduces the number of creation operators and correspondingly the number of states from 
$2^d$ (allowed for "eigenstates" of the Cartan subalgebra operators)  to two isolated groups 
of $\frac{1}{2}\frac{d!}{\frac{d}{2}! \frac{d}{2}!}$ creation operators. 
(There are no generators of the Lorentz transformations ${\cal \bf{S}}^{ab}$  that would 
connect both groups of states and correspondingly there are no families.) 

Let us study what happens, when, let say, $\gamma^{a}$'s are used to create the basis and 
correspondingly also to create the creation and annihilation operators. Here we briefly follow 
Ref.~\cite{NH2005}.

Let us point out that  $\gamma^{a}$ is expressible with $\theta^a$ and its derivative 
($\gamma^a= (\theta^a + \frac{\partial}{\partial \theta_{a}})$), Eq.~(\ref{cliffthetareltheta}), 
and that we again require that creation (annihilation) operators create (annihilate) states, which are 
"eigenstates" (Eq.~(\ref{grapheigen})) of the Cartan subalgebra operators, Eq.~(\ref{cartan}). 
Then the application of $\tilde{\gamma}^a$ on any Clifford algebra object $A(\gamma^a)$,
(determined by $\gamma^a$'s), can be evaluated as follows, Eq.~(\ref{gammabasis},
 \ref{tildecliffordB}),
\begin{eqnarray}
(\tilde{\gamma}^a A = i (-)^{(A)} A \gamma^a)|\psi_{oc}>\,,
\label{gammatildeA}
\end{eqnarray}
where  $(-)^{(A)} = -1$,  if $A$ is an odd Clifford algebra object and $(-)^{(A)} = 1$, 
 if $A$  is an even Clifford algebra object, while $|\psi_{oc}>$ is the vacuum state, 
replacing  the vacuum state in the Grassmann case $|\psi_{og}>= |1\,>$ with the one 
of Eq.~(\ref{vac1}), in accordance with the relation of Eqs.~({\ref{cliffthetareltheta},
\ref{grassnorm}, \ref{grassintegral}}), Refs.~\cite{NH2005,nh2018}.
We could as well make a choice of $\tilde{\gamma}^{a}=
 i (\theta^a - \frac{\partial}{\partial \theta_{a}})$ instead of $\gamma^a$'s to create the basic 
states, exchanging correspondingly the role of $\gamma^{a}$ and $\tilde{\gamma}^{a}$~%
\footnote{In the case that we 
would  choose $\tilde{\gamma}^a$'s instead of $\gamma^a$'s, Eq.(\ref{cliffthetareltheta}), the
role of $\tilde{\gamma}^a$ and $\gamma^a$ should be then correspondingly exchanged in 
Eq.~(\ref{gammatildeA}).}).
%

Making a choice of the Cartan subalgebra "eigenstates" of $S^{ab}$, Eq.~(\ref{signature0}), 
one defines nilpotents $\stackrel{ab}{(k)}$  and projectors $\stackrel{ab}{[k]}$
\begin{eqnarray}
\stackrel{ab}{(k)}:&=& \frac{1}{2}(\gamma^a + \frac{\eta^{aa}}{ik} \gamma^b)\,,
\quad \stackrel{ab}{(k)}{}^2 =0\,, \nonumber\\
\stackrel{ab}{[k]}:&=& \frac{1}{2}(1+ \frac{i}{k} \gamma^a \gamma^b)\,,
\quad \stackrel{ab}{[k]}{}^2 =\stackrel{ab}{[k]}\,,  
\label{signature}
\end{eqnarray}
where $k^2 = \eta^{aa} \eta^{bb}$.   
Recognizing that the Hermitian conjugate values of 
$\stackrel{ab}{(k)}$ and $\stackrel{ab}{[k]}$ are
\begin{eqnarray}
\stackrel{ab}{(k)}^{\dagger}=\eta^{aa}\stackrel{ab}{(-k)},\quad
\stackrel{ab}{[k]}^{\dagger}= \stackrel{ab}{[k]}\,,
\label{graphher}
\end{eqnarray}
while the corresponding "eigenvalues" of $S^{ab}$ and  $\tilde{S}^{ab}$ on nilpotents and 
projectors, Eq.~(\ref{signature0}), are
\begin{eqnarray}
S^{ab}\stackrel{ab}{(k)}&=&\frac{k}{2} \stackrel{ab}{(k)}\,,\quad 
S^{ab}\stackrel{ab}{[k]}=\;\;\frac{k}{2} \stackrel{ab}{[k]}\,, \nonumber\\
\tilde{S}^{ab} \stackrel{ab}{(k)}&=& \frac{k}{2}\stackrel{ab}{(k)}\,,\quad
\tilde{S}^{ab} \stackrel{ab}{[k]}  = -\frac{k}{2} \stackrel{ab}{[k]}\,, 
\label{grapheigen}
\end{eqnarray}
we find for $d=2(2n+1)$ that from the starting state made as a product of  an odd number
of only nilpotents 
\begin{eqnarray}
|\psi^{1}_{1}>&=&{\hat b}^{1\dagger}_{1} |\psi_{oc}>\,,\nonumber\\
{\hat b}^{1 \dagger}_{1}:&=& \stackrel{03}{(+i)} \stackrel{12}{(+)} \stackrel{35}{(+)}\cdots 
\stackrel{d-3\;d-2}{(+)}\;\;\stackrel{d-1\;d}{(+)}\,,\nonumber\\
{\hat b}^{1}_{1}\;\;\;&=& ({\hat b}^{1 \dagger}_{1})^{\dagger}= \stackrel{d-1\;d}{(-)} \;\;
\stackrel{d-3\;d-2}{(-)}\cdots  \stackrel{35}{(-)}  \stackrel{12}{(-)}  \stackrel{01}{(-i)}\,,
\label{start(2n+1)2cliff}
\end{eqnarray}
having correspondingly an odd Clifford character, %
all other states of the same Lorentz representation, there are $2^{\frac{d}{2}-1}$ members,
follow by the application of $S^{cd}$ (which do not belong to the Cartan subalgebra) on the starting 
state%
~\footnote{The smallest number of all the 
generators $S^{ac}$, which do not belong to the Cartan subalgebra, Eq.~(\ref{cartan}), needed 
to create from the 
starting state all the other members, is  $2^{\frac{d}{2}-1}-1$. This is true for both even dimensional 
spaces -- $2(2n+1)$ and $4n$.},  
Eq.~(\ref{cartan}),   ($S^{cd}$ $|\psi^{1}_{1}>= $  $|\psi^{1}_{i}>$). 
\begin{eqnarray}
\hat{b}^{1\dagger}_i &\propto & S^{ab} \dots S^{ef} \hat{b}^{1\dagger}_1\,,
\quad |\psi^{1}_i> = S^{ab} ..S^{ef} |\psi^{1}_{1}>\,, \nonumber\\
\hat{b}^{1}_i&\propto & \hat{b}^{1}_1 S^{ef} \dots S^{ab}\,,
\label{b1i}
\end{eqnarray}
with $S^{ab\dagger} = \eta^{aa} \eta^{bb} S^{ab}$. We make a choice of the proportionality 
factors so that the corresponding states $|\psi^{1}_{i}>= \hat{b}^{1\dagger}_i |\psi_{oc}>$  
 are normalized~\cite{NH2005,nh2018}. 

The operators $\tilde{S}^{cd}$, which belong  to the Cartan subalgebra  of $\tilde{S}^{ab}$,
Eq.~(\ref{cartan}), generate "eigenstates" of the Cartan subalgebra operators
 ($\tilde{S}^{03}$, $\tilde{S}^{12}$, $\tilde{S}^{56}$, $\cdots, \tilde{S}^{d-1\, d}$), with the 
eigenvalues which determine the "family" quantum numbers. There are $2^{\frac{d}{2}-1}$
families. 
From the  starting new  member with a different "family" quantum number the whole Lorentz 
representation of family members with this "family" quantum number follows by the application 
of $S^{ef}$: \, 
$S^{ab} \cdots S^{ef}$ $\tilde{S}^{cd} |\psi^{1}_{1}>=$ $|\psi^{\alpha}_{i}>$. All states of 
one Lorentz representation of any particular "family" quantum number have an odd Clifford character,
since neither $S^{cd}$ nor  $\tilde{S}^{cd}$ --- both of an even Clifford character --- can change the 
odd character of the starting state. 

Any vector $|\psi^{\alpha}_{i}>$ follows from the starting vector, Eqs.~(\ref{start(2n+1)2cliff}), 
by the application of either $\tilde{S}^{ef}$, which change the family quantum number, 
or $S^{gh}$, which change the family member quantum number of a particular family 
or with the corresponding product of $S^{ef}$ and $\tilde{S}^{ef}$
\begin{eqnarray}
|\psi^{\alpha}_i>&\propto & \tilde{S}^{ab} \cdots \tilde{S}^{ef}|\psi^{1}_{i}>  
\propto \tilde{S}^{ab} \cdots \tilde{S}^{ef}{S}^{mn}\cdots {S}^{pr} |\psi^{1}_{1}>\,.
\label{anyfun}
\end{eqnarray}
Again, $\alpha$ denotes "family" quantum numbers, $i $ denotes family member quantum number.
Correspondingly we define $\hat{b}^{\alpha \dagger}_{i}$ (up to a constant) to be 
\begin{eqnarray}
\hat{b}^{\alpha \dagger}_i&\propto & \tilde{S}^{ab} \cdots \tilde{S}^{ef}{S}^{mn}\cdots {S}^{pr}
\hat{b}^{1\dagger}_{1}\nonumber\\
&\propto & {S}^{mn}\cdots {S}^{pr} \hat{b}^{1\dagger}_{1} {S}^{ab} \cdots {S}^{ef}\,.
\label{anycreation}
\end{eqnarray}
This last expression follows due to the property of the Clifford object $\tilde{\gamma}^{a}$ and 
correspondingly of $\tilde{S}^{ab}$, presented in Eqs.~(\ref{gammatildeA}, \ref{gammatildeAapp}).

We accordingly have for an annihilation operator $\hat{b}^{\alpha }_i 
(= $ $(\hat{b}^{\alpha \dagger}_i)^{\dagger}$)
\begin{eqnarray}
\hat{b}^{\alpha }_i = (\hat{b}^{\alpha  \dagger}_i)^{\dagger} &\propto & {S}^{ef} \cdots 
{S}^{ab}\hat{b}^{1}_{1} {S}^{pr}\cdots {S}^{mn}\,.
\label{anyannihilation}
\end{eqnarray}
The proportionality factor ought to be chosen so that the corresponding states 
$|\psi^{\alpha}_{i}> = \hat{b}^{\alpha \dagger}_i |\psi_{oc}> $ are normalized when  
 the vacuum state $|\psi_{oc}> $  is normalized, $<\psi_{oc}|\psi_{oc}>=1$, while
all the states belonging to the physically acceptable states, like 
$ \stackrel{03}{[+i]} \stackrel{12}{[+]} \stackrel{56}{[-]}  \stackrel{78}{[-]}\cdots 
\stackrel{d-3\;d-2}{(+)}\;\;\stackrel{d-1\;d}{(+)}|\psi_{oc}>$, 
must not give
zero for either $d=2(2n+1)$ or for $d=4n$. We also want that states, obtained by the 
application of ether $S^{cd} $ or $\tilde{S}^{cd}$ or both, are orthogonal. To make a choice 
of the vacuum it is needed to know the relations of Eq.~(\ref{graphbinoms}). It must be
\begin{eqnarray}
<\psi_{oc}|\cdots \stackrel{ab}{(k)}^{\dagger} \cdots| \cdots  \stackrel{ab}{(k')}
\cdots |\psi_{oc} >
 &=& \delta_{k k'}\,,  \nonumber\\
<\psi_{oc}|\cdots \stackrel{ab}{[k]}^{\dagger}\cdots| \cdots \stackrel{ab}{[k']}\cdots
|\psi_{oc}> &=& \delta_{k k'}\,, \nonumber\\
<\psi_{oc}|\cdots \stackrel{ab}{[k]}^{\dagger} \cdots| \cdots  \stackrel{ab}{(k')}
\cdots |\psi_{oc} >&=&0\,.
\label{scalgamma0}
\end{eqnarray}
We must choose the vacuum state in a way that fulfills the above requirements as well as the
 requirements
$ \hat{b}^{\beta \dagger}_i |\psi_{oc}>\ne 0$ and  $\hat{b}^{\beta}_i |\psi_{oc}>=0$ 
for all members $i$ of any family $\beta$. Since any $\tilde{S}^{eg}$ changes $\stackrel{ef}{(+)}
\, \stackrel{gh}{(+)}$ into $\stackrel{ef}{[+]}\, \stackrel{gh}{[+]}$ and 
$\stackrel{ab}{[+]}{}^{\dagger}=$$\stackrel{ab}{[+]}$, while $\stackrel{ab}{(+)}{}^{\dagger} \,
\stackrel{ab}{(+)}= \stackrel{ab}{[-]}$, the vacuum state  $|\psi_{oc}> $ must be
\begin{eqnarray}
|\psi_{oc}>&=& \stackrel{03}{[-i]} \stackrel{12}{[-]} \stackrel{56}{[-]}\cdots
\stackrel{d-1\;d}{[-]} + \stackrel{03}{[+i]} \stackrel{12}{[+]} \stackrel{56}{[-]} \cdots
\stackrel{d-1\;d}{[-]} + \stackrel{03}{[+i]} \stackrel{12}{[-]} \stackrel{56}{[+]}\cdots
\stackrel{d-1\;d}{[-]} + \cdots |0>\,, \quad \nonumber\\
&&{\rm for}\; d=2(2n+1)\,,
\label{vac1}
\end{eqnarray}
$n$ is a positive integer. There are $2^{\frac{d}{2}-1}$ summands, since we can start with the 
vacuum state $ \stackrel{03}{[-i]} \stackrel{12}{[-]}
 \stackrel{56}{[-]}\cdots \stackrel{d-1\;d}{[-]}|0>$, which fulfills the requirement for 
$ \hat{b}^{1 \dagger}_1 |\psi_{oc}>\ne0$ and $\hat{b}^{1}_1 |\psi_{oc}>=0$, and then we 
must step by step replace all 
possible pairs of $\stackrel{ab}{[-]}  \cdots \stackrel{ef}{[-]} $ in the starting 
part $\stackrel{03}{[-i]} \stackrel{12}{[-]} \stackrel{35}{[-]}\cdots
\stackrel{d-1\;d}{[-]}$ 
into $\stackrel{ab}{[+]}  \cdots \stackrel{ef}{[+]} $ and include new terms into the vacuum state 
so that the last $(2n+1)$ summands have for $d=2(2n+1)$ case, $n$ is a positive integer,  only 
one factor $[-]$ and all the rest $[+]$,  each $[-]$ at different position~\footnote{The choice of 
Eq.~(\ref{vac1})  for the vacuum state is not unique. If one would multiply any of summands
by a number $\beta_{\alpha}$, where $\alpha$ represents the $\alpha$-th family, and then 
multiply each of $2^{\frac{d}{2}-1}$ members of creation operators belonging to this family
$\hat{b}^{\alpha \dagger}_{i}$ by $\sqrt{\beta_{\alpha}}$ and the corresponding annihilation 
operator $\hat{b}^{\alpha}_{i}$  by $\sqrt{\beta^{*}_{\alpha}}$, $\beta^*_{\alpha}$ is 
the complex conjugated value of $\beta_{\alpha}$, it would still be true that $\hat{b}^{\alpha}_{i}$ $\hat{b}^{\beta \dagger}_{j}$
$= \delta^{\alpha \beta} \delta_{ij} $ times the corresponding summand of the vacuum back.}.

This vacuum has all the spins, either with respect to $S^{ab}$ or with respect to $\tilde{S}^{ab}$,
equal to zero.


The vacuum state has then the normalization factor $1/\sqrt{2^{d/2-1}}$,\\
while there is   
\begin{eqnarray}
\label{nocliffop}
2^{\frac{d}{2} -1} \, 2^{\frac{d}{2} -1}  
\end{eqnarray}  
number of creation operators, defining the orthonormalized states when applying on the vacuum 
state of  Eqs.~(\ref{vac1})  and the same number of annihilation operators, which are Hermitian 
conjugated to creation operators. Again, operators $\tilde{S}^{ab} $ connect members of 
different families, operators $S^{ab} $ generate all the members of one family.

Paying attention on only internal degrees of freedom, that is on the spin, the creation and annihilation 
operators must fulfill the relations
\begin{eqnarray}
\{ \hat{b}^{\alpha }_i, \hat{b}^{\alpha{'} \dagger}_{j} \}_{+} 
 |\psi_{oc}> &=&\delta^{\alpha  \alpha{'}} \, \delta_{i j}|\psi_{oc}>\,,\nonumber\\
\{ \hat{b}^{\alpha}_i, \hat{b}^{\alpha{'}}_{j} \}_{+}  |\psi_{oc}>
&=& 0\; |\psi_{oc}> \,,\nonumber\\
\{\hat{b}^{\alpha  \dagger}_i,\hat{b}^{\alpha{'} \dagger}_{j}\}_{+} |\psi_{oc}>
&=&0 \;|\psi_{oc}> \,,\nonumber\\
\hat{b}^{\alpha}_{i} |\psi_{oc}>& =& 0\; |\psi_{oc}>\,, \nonumber\\
\hat{b}^{\alpha \dagger}_{i} |\psi_{oc}>& =&  |\psi^{\alpha }_{i}>\,,
\label{alphagammaprod0}
\end{eqnarray}
with ($i,j$) determining family members quantum numbers and ($\alpha , \alpha{'}$) denoting
"family" quantum numbers.

Only Clifford odd objects fulfill the relations of Eq.~(\ref{alphagammaprod0}), since the odd 
Clifford objects anti-commute (like: $\{(\gamma^0 - \gamma^3), (\gamma^1 + 
i\gamma^2)\}_{+} =0$), while the Clifford even objects commute (like: $\{(1-\gamma^0  \gamma^3),
  (1-i \gamma^1 \gamma^2)\}_{+}= 2\, (1-\gamma^0  \gamma^3) (1-i \gamma^1 \gamma^2)$).

The reader can find the detailed proofs for the above statements, for either $d=2(2n+1)$ or $d=4n$, 
in Refs.~\cite{NH2005,nh2018}. 

Let us again, like in the Grassmann case, Eq.~(\ref{ijthetaprodordinary}), look for the creation 
(and their annihilation operators) which, when applied on the vacuum state, Eq.~(\ref{vac1}), solve
the equation of motion, Eq.~(\ref{Weyl}). The solution for each momentum $p^a_{k},\, 
a=(1,\dots,d$), for discretized  values of momenta, is a superposition of  
$ \hat{b}^{\alpha \dagger}_i$,
\begin{eqnarray}
\hat{b}^{\alpha \dagger}_{s \, p_k} &=& \sum_{i} c^{\alpha}{}_{s i} (p_k)\,
\hat{b}^{\alpha \dagger}_{i}\,, 
\label{pgamma}
\end{eqnarray}
applied on the vacuum state, Eq.~(\ref{vac1}). 
Since $ \hat{b}^{\alpha \dagger}_{i}$ and $ \hat{b}^{\alpha}_{j}$ fulfill 
the relations of Eq.~(\ref{alphagammaprod0})  and, if the states for two different momenta
are orthogonalized, it follows 
\begin{eqnarray}
\{ \hat{b}^{\alpha}_{s\, p_k},\, \hat{b}^{\alpha{'} \dagger}_{s' \, p_l} \}_{+}  |\psi_{oc}>
&=& \delta^{\alpha \alpha{'}}\,\delta_{s s'}\, \delta_{p_k\, p_{l}} \,|\psi_{oc}>
\,,\nonumber\\
\{ \hat{b}^{\alpha}_{s\, p_k},\, \hat{b}^{\alpha{'}}_{s'\, p_l} \}_{+}  |\psi_{oc}>
&=&0 \;|\psi_{oc}>\,,\nonumber\\
\{ \hat{b}^{\alpha \dagger}_{s \,p_k}, \hat{b}^{\alpha{'} \dagger}_{s' \,p_l} \}_{+} |\psi_{oc}>
&=&0\; |\psi_{oc}>\,,\nonumber\\
\hat{b}^{\alpha}_{s\, p_{k}} |\psi_{oc}>& =& 0\; |\psi_{oc}>\,, \nonumber\\
\hat{b}^{\beta \dagger}_{s\, p_{k}} |\psi_{oc}>& =& |\psi^{\alpha}_{s\, p_{k}}>\,,
\label{ijgammaprodordinary}
\end{eqnarray}
with the vacuum state $|\psi_{oc}>$ defined in Eq.~(\ref{vac1}), with $s$ denoting the corresponding
solution of equations of motion and for a discretized momentum space. 

 The number operator of a particular solution $s$, a particular momentum $p_{k}$ and a particular 
"family" $\alpha$, 
\begin{eqnarray}
\label{gammanop}
\hat{N}^{\alpha}_{s p_{k}}&=& \hat{b}^{\alpha \dagger}_{s p_{k}} 
\hat{b}^{\alpha}_{s p_{k}}\,,\quad
(\hat{N}^{\alpha}_{s p_{k}})^2 =\hat{N}^{\alpha}_{s p_{k}}\,,
\end{eqnarray}
has the eigenvalues $1$ or $0$. 

The number of fermions in the $n$-particle state of Eq.~(\ref{nfermionstate})
 is correspondingly
\begin{eqnarray}
\label{gammanopall}
\hat{N} &=& \sum_{\alpha, s, p_k } \hat{N}^{\alpha}_{s p_{k}}\,.
\end{eqnarray}
%

For a $n$ fermion  and  antifermion state, Eqs.~(\ref{nfermions}, \ref{SD}) in  the introduction to 
Sect.~\ref{secondquantization},  
$|1^{\alpha=1}_{s=1\, p_1}, 1^{\alpha=1}_{s=2\, p_1},1^{\alpha=1}_{s=3 \,p_1}, \dots, 
0^{\alpha}_{s\,p_i }, \dots, 0^{\alpha}_{s^{iv} p_j }, \dots, >$ it follows, for example, 
due to the third line in Eq.~(\ref{ijthetaprodordinary}), that  
\begin{eqnarray}
\label{nfermionstate}
&&\hat{b}^{\alpha{'}  \dagger}_{s' p_{i}} \,\hat{b}^{\alpha^{"}  \dagger}_{s^{"} p_{j}} 
|1^{\alpha=1}_{s=1\, p_1}, 1^{\alpha=1}_{s=2\, p_1},1^{\alpha=1}_{s=3 \,p_1}, \dots, 
0^{\alpha}_{s\,p_i }, \dots, 0^{\alpha}_{s^{iv} p_j }, \dots, >=
\nonumber\\
 - && \hat{b}^{\alpha^{"}  \dagger}_{s^{"} p_{j}} 
\hat{b}^{\alpha{'} \dagger}_{s' p_i} 
|1^{\alpha=1}_{s=1\, p_1}, 1^{\alpha=1}_{s=2\, p_1},1^{\alpha=1}_{s=3 \,p_1}, \dots, 
0^{\alpha}_{s\,p_i }, \dots, 0^{\alpha}_{s^{iv} p_j }, \dots, >\,,
\end{eqnarray}
 where $1$ denotes the occupied state and $0$ the unoccupied state, and 
$|1^{\alpha=1}_{s=1\, p_1}> = \hat{b}^{\alpha=1  \dagger}_{s=1 p_{1}}\,|\psi{oc>} $.

Eq.~(\ref{nfermionstate}, \ref{nfermions}) demonstrates properties of Slater determinants. 
One fermion state is obviously second quantized by construction.

Two states with $n_1$ and $n_2$ fermions each, defined by $\hat{A}^{a \dagger}$ as $n_1$ 
products of  $\hat{b}^{\alpha \dagger}_{s p_i}$ (which distinguish among themselves in at least 
one of the properties $(\alpha, s, p_i$)) and by $\hat{A}^{b \dagger}$ as $n_2$ products of
$\hat{b}^{\beta \dagger}_{s' p_j}$ (which distinguish among themselves in at least one of the 
properties $(\alpha', s', p_j$)),  applying on $|\psi_{oc}>$,  must distinguish in either 
internal space or in the coordinate space, as it follows from Eq.~(\ref{nfermionstate}), that the 
product of $\hat{A}^{a \dagger}$ and $\hat{A}^{b \dagger}$  applying on $|\psi_{oc}>$
would give a state with ($n_1+n_2$) fermions.  

Let us add that the vacuum state, having the sum of the spins of both kinds of operators, $S^{ab}$ 
and $\tilde{S}^{ab}$, equal to zero and therefore neutral, remains neutral also when filled  with 
fermions of all the spins, $S^{ab}$ and $\tilde{S}^{ab}$. 

When coefficients $c^{\alpha}{}_{s i} (p_k)$  depend also on coordinates $x^a$ (for free fermions  
${\bf c}^{\alpha}{}_{s i} (p_{k}, x) =$ $c^{\alpha}{}_{s i} (p_k)\cdot e^{-i p_a x^a} $), it follows 
\begin{eqnarray}
\label{cliffcreationx}
{\bf \hat{b}}^{\alpha \dagger}_{s}(x^0,\vec{x}) &=& \sum_{i} \int\, \frac{d^{d-1} p}{(2\pi)^{d-1}}\,
{\bf c}^{\alpha}{}_{s i} (p_{k}, x)\, \hat{b}^{\alpha \dagger }_{i}\,.\nonumber
\end{eqnarray}
\begin{eqnarray}
\label{cliffordinary}
\{ {\bf \hat{b}}^{\alpha}_{s}(x^0,\vec{x}),\, {\bf \hat{b}}^{\alpha' \dagger}_{s'}(x^0,\vec{y})\}_{+}
  |\psi_{oc}>
&=& \delta^{\alpha \alpha{'}}\,\delta_{s s'}\, \delta^{d-1} (\vec{x}-\vec{y}) \,|\psi_{oc}>
\,,\nonumber\\
\{ {\bf \hat{b}}^{\alpha \dagger}_{s}(x^0,\vec{x}),\, {\bf \hat{b}}^{\alpha' \dagger}_{s'}(x^0,\vec{y})\}_{+}
  |\psi_{oc}>&=&0\,,\quad
\{ {\bf \hat{b}}^{\alpha}_{s}(x^0,\vec{x}),\, {\bf \hat{b}}^{\alpha' }_{s'}(x^0,\vec{y})\}_{+}
  |\psi_{oc}>=0\,.
\end{eqnarray}

 Let us now take into account  Eq.~(\ref{actionWeyl}) with ${\cal L}_{C}=\frac{1}{2}\{
\hat{\psi}^{\dagger} \gamma^0  \gamma^a (\hat{p}_{a} \psi) - (\hat{p}_{a} 
\psi^{\dagger}) \gamma^0 \gamma^a \psi\} $. The Euler-Lagrange equations lead to
$ \gamma^0 \gamma^a \hat{p}_{a}\psi=0$ and 
$- \hat{p}_{a}\phi^{\dagger} \gamma^0 \gamma^a =0$.

Let us look for the Hamilton function for fermions determined by one of the creation 
operators, like  $  {\bf \hat{\psi}}^{\alpha}_{s} (x^0,\vec{x})$ 
$={\bf \hat{b}}^{\alpha \dagger}_{s} (x^0,\vec{x})] |\psi_{oc}>$, which is already the 
second quantized state.. 
 
For a vector $ {\bf \hat{\psi}}$ and ${\bf \hat{\psi}}^{\dagger} $ it therefore follows  
\begin{eqnarray}
\label{HCliffpsigen}
\Pi_{{\bf \hat{\psi}}}&=& \frac{\partial {\cal L}_{C}}{\partial (\hat{p}_0 {\bf \hat{\psi}})}= 
\frac{1}{2}\, {\bf \hat{\psi}}^{\dagger}\,,\quad 
\Pi_{{\bf \hat{\psi}}^{\dagger}}= 
\frac{\partial {\cal L}_{C}}{\partial (\hat{p}_0 {\bf \hat{\psi}}^{\dagger})}=
\frac{1}{2}\, {\bf \hat{\psi}}\,,
\nonumber\\
{\cal H}_{C} &=&\Pi_{\psi} \, (\hat{p}_0 {\bf \hat{\psi}}) +  (\hat{p}_0 {\bf \hat{\psi}}^{\dagger})
 \Pi_{{\bf \hat{\psi}}^{\dagger}} \, 
- {\cal L}_{C}\,, \nonumber\\
&=& -\frac{1}{2}\, {\big [} {\bf \hat{\psi}}^{\dagger} \,\gamma^0 \,\gamma^i\,
 (\hat{p}_{i} {\bf \hat{\psi}})
- (\hat{p}_{i} {\bf \hat{\psi}}^{\dagger}) \gamma^0 \gamma^i {\bf \hat{\psi}} \, {\big ]}\,,\nonumber\\
H_{C}&=& \int d^{d-1} x \,{\cal H}_{C}\, ,
\end{eqnarray}
Correspondingly one finds for a component  ${\bf \hat{\psi}}^{\alpha}_{s} (x^0,\vec{x}) 
$~\cite{Bethe},  $\vec{x}  $ is a vector in $(d-1)$-dimensional coordinate space, 
\begin{eqnarray}
\label{HCliffpsieq}
\hat{p_0} {\bf \hat{\psi}}^{\alpha}_{s} \,(x^0,\vec{x}) &=& {\bf {\Big\{}} \,
{\bf \hat{\psi}}^{\alpha}_{s}\, (x^0,\vec{x})\,,\, H_{C} {\bf {\Big\}}}_{-} \nonumber\\
&=&  {\bf {\Big\{}} \,{\bf \hat{\psi}}^{\alpha}_{s} (x^0,\vec{x})\,,\,\int\, d^{d-1}x'\,
 \sum_{\alpha',s'}\, 
 {\bf \hat{\psi}}^{\alpha' \dagger}_{s'}\, (x^0,\vec{x'})\, \gamma^0 \gamma^i\, (\hat{p'}_{i} 
{\bf \hat{\psi}}^{\alpha'}_{s'} (x^0,\vec{x'})) \,{\bf {\Big\}}}_{-}
 \nonumber\\
&=& \int\, d^{d-1} x'\, \sum_{\alpha' s'}\,  {\bf {\Big\{}} \,
{\bf \hat{\psi}}^{\alpha}_{s} (x^0,\vec{x})\,,
 {\bf \hat{\psi}_{s'}}^{\alpha' \dagger} (x^0,\vec{x'})\,{\bf {\Big\}}}_{+} \gamma^0 \gamma^i 
\,(\hat{p'}_{i} {\bf \hat{\psi}^{\alpha'}_{s'}} (x^0,\vec{x'}))\nonumber\\
&=&- \gamma^0 \gamma^i \,(\hat{p}_{i} {\bf \hat{\psi}^{\alpha}_{s}} (x^0,\vec{x}))\,.
\end{eqnarray}
(We took into account that $\gamma^0 \gamma^{i} $ transforms 
 ${\bf \hat{\psi}}^{\alpha'}_{s'} \,(x^0,\vec{x'})$ into  $\sum_{s''} c^{\alpha'}{}_{s' s''} \,
{\bf \hat{\psi}}^{\alpha'}_{s''}(x^0,\vec{x'})$, which anticommute with
${\bf \hat{\psi}}^{\alpha}_{s}(x^0,\vec{x})$ (Eq.~(\ref{cliffordinary})), we also
assumed that states, obtained when operators operate on a vacuum state,
do not contribute to the surface term. Integrating per partes and dropping the surface term  
simplifies $H_{C}$ into  $- \int\, \sum d^{d-1} x' \,
{\bf \hat{\psi}}^{\alpha' \dagger}_{s'} (x^0,\vec{x'})\, \gamma^0 \gamma^i \,
 (\hat{p'}_{i}{\bf \hat{\psi}}^{\alpha'}_{s'} (x^0,\vec{x'}))$.)
The obtained equations of motion agree with the ones from Eqs.~(\ref{LDWeyl0}, \ref{LDWeyl2}).
Correspondingly the energy of the n-fermion state is $E = \sum_{\alpha s} \hat{N}^{\alpha}_{s} 
$, for free massless spinor this is equal to the zero component of the momentum of a state 
${\bf \hat{b}}^{\alpha \dagger}_{s} |\psi_{oc}>$, solving the Weyl equation Eqs.~(\ref{Weyl},
\ref{LDWeyl2}).  
The current is correspondingly 
${\bf \hat{j}}^a= {\bf \hat{\psi}_{s}}^{\alpha\dagger} \gamma^0 \gamma^a 
{\bf \hat{\psi}^{\alpha}_{s}}\,$.
%

The observed fermions --- quarks and leptons --- manifest their properties obviously in $d=(3+1)$. 
The internal space in $d=(3+1)$ can therefore be used to describe the spin 
and handedness of massless fermions,  in the {\it spin-charge-family} theory also families, while 
the internal space in $d\ge5$ can be used to describe charges of fermions, contributing  in the 
{\it spin-charge-family} theory as well to families.

One family representation --- connecting all the members by $S^{ab}$ --- contains in  
$d=2(2n+1), n=3$,  from the point of view of $d=(3+1)$  quarks and leptons and antiquarks 
and antileptons. Correspondingly there is no need for the negative energy "Dirac sea", 
all the quarks and leptons and antiquarks and antileptons appear as products of creation operators 
on the vacuum state$|\psi_{oc}>$, Eq.~(\ref{vac1}).

We discuss below discrete symmetry operators for both cases, the Clifford one and the Grassmann 
one, in $d$ and in observable dimension $d=(3+1)$. In Subsect.~\ref{examples} we present a 
few examples.

\subsection{ Discrete symmetries in Grassmann space and in Clifford space in $d$ and in 
$d=(3+1)$ part of the space}
\label{CPT}

\vspace{2mm}

We have treated so far free massless fermions in Grassmann and in Clifford space. The fermion "nature"
of states are in both spaces demonstrated by the fact that the corresponding creation and 
annihilation operators  fulfill the anticommutation relations of Eq.~(\ref{ijthetaprodordinary}) in Grassmann 
case and of  Eq.~(\ref{ijgammaprodordinary}) in Clifford space.
Fermions --- in both spaces --- are in general in superposition of eigenstates of the Cartan subalgebra 
operators of ${\cal {\bf S}}^{ab}$ in the Grassmann case,  in the Clifford case they are in superposition of the 
Cartan subalgebra operators of $S^{ab}$ as well as of $\tilde{S}^{ab}$.

We distinguish in $d$-dimensional space two kinds of discrete symmetry ${\cal C}, {\cal P}$ 
and ${\cal T}$  operators with respect to the internal space in which the fermion properties are described.

In the Clifford case we have~\cite{nhds}
\begin{eqnarray}
\label{calCPTH}
{\cal C}_{{\cal H}}&=& \prod_{\gamma^a \in \Im} \gamma^a \,\, K\,,\nonumber\\
{\cal T}_{{\cal H}}&=& \gamma^0 \prod_{\gamma^a \in \Re} \gamma^a \,\, K\, I_{x^0}\,,\nonumber\\
{\cal P}^{(d-1)}_{{\cal H}} &=& \gamma^0\,I_{\vec{x}}\,,\nonumber\\
I_{x} x^a &=&- x^a\,, \quad I_{x^0} x^a = (-x^0,\vec{x})\,, \quad I_{\vec{x}} \vec{x} = -\vec{x}\,, \nonumber\\
I_{\vec{x}_{3}} x^a &=& (x^0, -x^1,-x^2,-x^3,x^5, x^6,\dots, x^d)\,.
\end{eqnarray}
The product $\prod \, \gamma^a$ is meant in the ascending order in $\gamma^a$, $K$ 
stands for complex conjugation.

In the Grassmann case we correspondingly define
\begin{eqnarray}
\label{calCPTG}
{\cal C}_{G}&=& \prod_{\gamma^a_{ G} \in \Im \gamma^a} \, \gamma^a_{ G}\, K\,,\nonumber\\
{\cal T}_{G}&=& \gamma^0_{G} \prod_{\gamma^a_{G} \in \Re
 \gamma^a} \,\gamma^a_{ G}\, K \, I_{x^0}\,,\nonumber\\
{\cal P}^{(d-1)}_{G} &=& \gamma^0_{G} \,I_{\vec{x}}\,,
\end{eqnarray}
$\gamma^a_{G}$  is defined in Eq.~(\ref{grassgamma})  as 
%
$\gamma^{a}_{G} = 
(1- 2 \theta^a \eta^{aa} \frac{\partial}{\partial \theta_{a}})$, 
%
while $I_{x} x^a =- x^a\,, 
I_{x^0} x^a = (-x^0,\vec{x})\,,I_{\vec{x}} \vec{x} = -\vec{x}\,,
I_{\vec{x}_{3}} x^a = (x^0, -x^1,-x^2,-x^3,x^5, x^6,\dots, x^d)$, like in the Clifford case.
Let be noticed, that since $\gamma^a_{G}$ ($= - i \eta^{aa}\, \gamma^a \tilde{\gamma}^a$) is 
always real as we see in Eq.~(\ref{thetadertheta})~\footnote{If we choose a real $\theta^a$,
then $\gamma^a$ is real and $\tilde{\gamma}^{a}$ imaginary, if $\theta^a$ is imaginary,  
then $\gamma^a$ is imaginary and $\tilde{\gamma}^{a}$ real, as is demonstrated in 
Eq.~(\ref{thetadertheta}).}
Since $ \gamma^a$ is either real or imaginary (Eq.~(\ref{grassgamma})), we use in 
Eq.~(\ref{calCPTG}) $\gamma^a$  to make 
a choice of appropriate $\gamma^a_{G}$. 
In what follows we shall use the notation as in Eq.~(\ref{calCPTG}).


Let us define in the Clifford case and in the Grassmann case the operator "emptying"~\footnote{ 
The operator "emptying" empties the "Dirac sea" of negative energies~\cite{nhds}, although in the 
{\it spin-charge-family} theory is no need for the "Dirac sea" of negative energies, as we discussed 
already in the introduction of Sect.~\ref{secondquantization}, for either Clifford or Grassmann 
fermions. The operation of "emptying$_{NH}$" after the charge conjugation 
${\cal C }_{{\cal H}}$ in the Clifford case,  
which transforms the state put on the top of the Clifford "Dirac sea" into the 
corresponding negative energy state, namely creates the anti-particle state to the starting particle
state, each  anti-particle state, put on the top of the "Dirac sea",   
solving the Weyl equation in the Clifford case, Eq.~(\ref{Weyl}). }.
The operation $"{\rm emptying}_{NH}$"  after the charge conjugation 
${\cal C }_{{\cal H}}$ in the Clifford case~\cite{nhds,JMP2013,normaJMP2015} (arxiv:1312.1541) 
and $"{\rm emptying}_{NG}$" after the charge conjugation ${\cal C }_{G}$ in the Grassmann case,
namely transforms the positive energy fermions into positive energy antifermions in both cases, 
solving Eq.~(\ref{Weyl}) in the Clifford case, and Eq.~(\ref{Weylgrass})
 in the Grassmann case.
\begin{eqnarray}
\label{empt}
"{\rm emptying}_{NH}"&=& \prod_{\Re \gamma^a}\, \gamma^a \,K\, \quad {\rm in} \, \;
{\rm Clifford}\, {\rm space}\,, 
\nonumber\\
"{\rm emptying}_{NG}"&=& \prod_{\Re \gamma^a}\, \gamma^a_{G} \,K\, \quad {\rm in}\;\, 
{\rm Grassmann} \,  {\rm space}\,. 
\end{eqnarray}
%
Then the anti-particle  state creation operator  
to the corresponding  particle state creation operator can be obtained by the application of 
%
\begin{eqnarray}
\label{makingantip}
{\bf \mathbb{C}}_{\cal H,G} &=& "{\rm emptying}_{NH,NG}"\,\cdot\, {\cal C}_{{\cal H,G}}  \,
\end{eqnarray}
 ${\bf \mathbb{C}}_{\cal H} $ and ${\bf \mathbb{C}}_{G}$, 
with indexes ${}_{{{\bf \cal H}}}$ and ${}_{NH}$ denoting the Clifford case
and  with ${}_{{{\bf \cal G}}}$ and  ${}_{NG}$ denoting the Grassman case, on the creation 
operator for a particle state, or opposite. Let us remind the reader that in the {\it spin-charge-family} 
theory, using  the Clifford algebra,  the family members of each family include fermions and 
antifermions --- quarks and leptons and antiquarks and antileptons. This is the case also for 
Grassmann fermions and antifermions, but in this casethere are  instead of families two by 
${\cal {\bf S}}^{ab}$ unconnected representations.
%
%
%
%

Ref.~\cite{nhds} proposes in the Clifford case the following discrete symmetry operators,
manifesting dynamics in $d=(3+1)$
\begin{eqnarray}
\label{CPTN}
{\cal C}_{{\cal N}}  &= & \prod_{\Im \gamma^m, m=0}^{3} \gamma^m\,\, \,\Gamma^{(3+1)} \,
K \,I_{x^6,x^8,\dots,x^{d}}  \,,\nonumber\\
{\cal T}_{{\cal N}}  &= & \prod_{\Re \gamma^m, m=1}^{3} \gamma^m \,\,\,\Gamma^{(3+1)}\,K \,
I_{x^0}\,I_{x^5,x^7,\dots,x^{d-1}}\,,\nonumber\\
{\cal P}^{(d-1)}_{{\cal N}}  &= & \gamma^0\,\Gamma^{(3+1)}\, \Gamma^{(d)}\, I_{\vec{x}_{3}}
\,, \nonumber\\
{\bf \mathbb{C}}_{{\cal N}} &=& \gamma^0 \gamma^5 \gamma^7 \cdots \gamma^{d-1} 
I_{\vec{x}_{3}} \,I_{x^6,x^8,\dots,x^{d}}\, \nonumber\\
{\cal C}_{{\cal N}} {\cal P}^{(d-1)}_{{\cal N}}  &= &  \gamma^0\,\gamma^{2}\, I_{\vec{x}_{3}}
K \,I_{x^6,x^8,\dots,x^{d}} \,, \nonumber\\
{\bf \mathbb{C}}_{{\cal N}} {\cal P}^{(d-1)}_{{\cal N}}  &= &  \gamma^0\,\gamma^{5} \,
\cdots  \gamma^{d-1 }\,  I_{\vec{x}_{3}} \,I_{x^6,x^8,\dots,x^{d}} \,\,. 
\end{eqnarray}

In the Grassmann case we use the Grassmann even, Hermitian and real operators
$\gamma^{a}_{G}$, Eq.~(\ref{grassgamma}), to determine discrete symmetries in
$((d-1)+1)$ space (as presented in Eq.~(\ref{calCPTG})) and in $d=(3+1)$  space. 
%
%
In 
 $(3+1)$  space we proceed --- in analogy with the operators in the Clifford case~\cite{nhds} --- 
as follows 
\begin{eqnarray}
\label{calCPTNG}
{\cal C}_{NG}&=& \prod_{\gamma^m_{ G} \in \Re  \gamma^m} \, \gamma^m_{ G}\, K \,
 I_{x^6 x^8...x^d}\,,\nonumber\\
{\cal T}_{NG}&=& \gamma^0_{G} \, \prod_{\gamma^m_{G} \in \Im \gamma^m}
 \gamma^m_{G}\,  K \, I_{x^0} I_{x^5 x^7...x^{d-1}}\,,\nonumber\\
{\cal P}^{(d-1)}_{NG} &=& \gamma^0_{G} \, \prod_{s=5}^{d}\, \gamma^s_{ G} I_{\vec{x}}\,,
\nonumber\\
{\bf \mathbb{C}}_{NG} &=&\prod_{\gamma^s_{ G} \in \Re  \gamma^s} \,\gamma^s_{ G}\,, 
 I_{x^6 x^8...x^d}\,,\nonumber\\
{\cal C}_{NG} {\cal P}^{(d-1)}_{NG} &=&\gamma^0_{G}\, \gamma^2_{G} \, K 
\, I_{\vec{x}_{3}}\, I_{x^6 x^8...x^d}\,,
\nonumber\\
{\bf \mathbb{C}}_{NG} {\cal P}^{(d-1)}_{NG} &=& \gamma^0_{G} \, \prod_{
\gamma^s_{ G} \in \Im  \gamma^s, s=5}^{d}\, \gamma^s_{ G}\, I_{\vec{x}_{3}}\,
 I_{x^6 x^8...x^d}\,,
\nonumber\\
{\bf \mathbb{C}}_{NG} {\cal T}_{NG} {\cal P}^{(d-1)}_{NG} &=& \prod_{\gamma^s_{ G} \in 
\Im  \gamma^a} \,\gamma^a_{ G}\,I_x K\,.
\end{eqnarray}
\subsection{Examples of massless fermion and antifermion states  in Clifford and in Grassmann space}
\label{examples}

Let us illustrate solutions for free fermion states, represented by the creation operators applied
on the vacuum states for the Clifford and the Grassmann case in $((d-1)+1)$-dimensional space, 
representing indeed the contribution of a one fermion second quantized state in the  Fock space 
of any number of fermions. 
We analyze states in both cases from the point of view $d=(3+1)$-dimensional space, with  the 
momentum in ordinary space $p^a= (p^0, p^1, p^2, p^3, 0, \cdots, 0)$, so that the charges 
"seen" in $d=(3+1)$ are determined by the generators of the Lorentz transformations in the 
internal space --- $S^{st}, (s,t)= (5,6,7,\cdots, d)$ in the Clifford case and ${\cal {\bf S}}^{st},
 (s,t)= (5,6,7,\cdots, d)$ in the Grassmann case. In the Clifford case we discuss one family in 
details (let be reminded that the generators $S^{ab}$ connect all the members  belonging to one 
family, while $\tilde{S}^{ab}$ transform a particular member of one family into the same member 
of another family), commenting also on the appearance of families (all the families are reachable 
by $\tilde{S}^{ab}$) and present them briefly. 
In the Grassmann case different representations can not be reached by the generators of the 
Lorentz representations ${\cal {\bf S}}^{ab}$.  
The discrete symmetry operators are  in the Clifford case presented in Eq.~(\ref{CPTN}), and  
in the Grassmann case in Eq.~(\ref{calCPTNG}).

We start with examples in  $d=(5+1)$-dimensional space, with charges determined by $S^{st}, 
(s,t)= (5,6)$  in the Clifford case and ${\cal {\bf S}}^{st}, (s,t)=(5,6)$  in the Grassmann case.  

The dimension $(13+1)$, used in the {\it spin-charge-family} theory to describe quarks and lepton 
as well the gauge fields  and scalar fields, offers to free fermions 
at low energies additional charges, what explains observable properties of quarks and leptons.
We present the creation operators creating all the states of one family members in Clifford space. 
The family members creation operators are reachable by $S^{ab}$. All the rest of the families 
are reachable by $\tilde{S}^{ab}$. 

In Ref.~\cite{NH2005,NHD,ND012,familiesNDproc} the $(d=5+1)$-dimensional space is studied 
as a toy model to manifest that the break of symmetry from the higher dimensional space to the 
$(3+1)$-dimensional space {\it can} lead to massless fermions. Fermions were described in Clifford
space. We briefly follow these references, and Refs.~\cite{nhds,TDN},  adding  new observations. 

The first study of Grassmann case can be found in Ref.~\cite{nd2018}.

\subsubsection{{\bf Clifford fermions and antifermions}}
\label{Cliffordfermions}

Let us start with the examples in the Clifford case. To make discussions transparent let us
first treat the $d=(5+1)$ case. The $d=(13+1)$ case is not so easy to present in particular
when also families are treated.

\vspace{2mm}

$\;\;$ {\bf   Clifford case in $d=(5+1)$:}

\vspace{2mm}

In Table~\ref{cliff basis5+1.} the basic creation operators $\hat{b}^{\alpha \dagger}_{i=(ch, s)}$ 
and their annihilation partners $\hat{b}^{\alpha}_{i=(ch, s)}$ in $d=(5+1)$ are presented for all 
four ($2^{\frac{d}{2}-1}$) families $\alpha =(I,II,III,IV)$. Index $i$ is split into $ch$ and $s$ to 
point out that $S^{56}$ represents the charge from the point of view of $d=(3+1)$, having two 
values, $+\frac{1}{2}$ and $-\frac{1}{2}$. The vacuum state, Eq.~(\ref{vac1}) is the 
superposition of $\stackrel{03}{[-i]}\,\stackrel{12}{[-]}| \stackrel{56}{[-]}$, $\stackrel{03}{[+i]}\,
\stackrel{12}{[+]}| \stackrel{56}{[-]}$, $\stackrel{03}{[+i]}\,\stackrel{12}{[-]}| \stackrel{56}{[+]}$,
and  $\stackrel{03}{[-i]}\,\stackrel{12}{[+]}| \stackrel{56}{[+]}$, needed that the first, second,
third and fourth family creation operators, respectively, applying on the vacuum state, give nonzero value.
 \begin{table}
\begin{tiny}
 \begin{center}
 \begin{tabular}{|r|r|r|r|r|r|r|r|r|r|r|r|}
 \hline
${\rm family}\, \alpha$&$i=(ch,s)$&$\hat{b}^{\alpha \dagger}_{ch, s}$&$\hat{b}^{\alpha}_{ch, s}$&
$S^{03}$&$ S^{1 2}$&$S^{5 6}$&$\Gamma^{3+1}$ &
$\tilde{S}^{03}$&$\tilde{S}^{1 2}$& $\tilde{S}^{5 6}$\\
\hline
$I$&$(\frac{1}{2},\frac{1}{2})$&$
\stackrel{03}{(+i)}\,\stackrel{12}{(+)}| \stackrel{56}{(+)}$&
$
{\scriptstyle (-)} \stackrel{56}{(-)}|{\scriptstyle (-)} 
\stackrel{12}{(-)} \stackrel{03}{(-i)}$&$\frac{i}{2}$&$\frac{1}{2}$&$\frac{1}{2}$&$1$
&$\frac{i}{2}$&$\frac{1}{2}$&$\frac{1}{2}$\\
$I$ &$(\frac{1}{2},-\frac{1}{2})$&$
\stackrel{03}{[-i]}\,\stackrel{12}{[-]}|\stackrel{56}{(+)}$&$
 {\scriptstyle (-)} \stackrel{56}{(-)}| \stackrel{12}{[-]} \stackrel{03}{[-i]}$&
$-\frac{i}{2}$&$-\frac{1}{2}$&$\frac{1}{2}$&$1$
&$\frac{i}{2}$&$\frac{1}{2}$&$\frac{1}{2}$\\
$I$ &$(-\frac{1}{2},\frac{1}{2})$&$
\stackrel{03}{[-i]}\,\stackrel{12}{(+)}|\stackrel{56}{[-]}$&$
 \stackrel{56}{[-]}|{\scriptstyle (-)} \stackrel{12}{(-)} \stackrel{03}{[-i]}$&
$-\frac{i}{2}$&$ \frac{1}{2}$&$-\frac{1}{2}$&$-1$
&$\frac{i}{2}$&$\frac{1}{2}$&$\frac{1}{2}$\\
$I$ &$(-\frac{1}{2},-\frac{1}{2})$&$
\stackrel{03}{(+i)}\,\stackrel{12}{[-]}|\stackrel{56}{[-]}$&$
\stackrel{56}{[-]}| \stackrel{12}{[-]} \stackrel{03}{(-i)}$&
$\frac{i}{2}$&$- \frac{1}{2}$&$-\frac{1}{2}$&$-1$
&$\frac{i}{2}$&$\frac{1}{2}$&$\frac{1}{2}$\\
\hline 
$II$&$(\frac{1}{2},\frac{1}{2})$&$
\stackrel{03}{[+i]}\,\stackrel{12}{[+]}| \stackrel{56}{(+)}$&
$
{\scriptstyle (-)} \stackrel{56}{(-)}|\stackrel{12}{[+]} \stackrel{03}{[+i]}$&$\frac{i}{2}$&
$\frac{1}{2}$&$\frac{1}{2}$&$1$&$-\frac{i}{2}$&$-\frac{1}{2}$&$\frac{1}{2}$\\
$II$ &$(\frac{1}{2},-\frac{1}{2})$&$
\stackrel{03}{(-i)}\,\stackrel{12}{(-)}|\stackrel{56}{(+)}$&$
 {\scriptstyle (-)} \stackrel{56}{(-)}| {\scriptstyle (-)} \stackrel{12}{(+)}
 \stackrel{03}{(+i)}$&
$-\frac{i}{2}$&$-\frac{1}{2}$&$\frac{1}{2}$&$1$
&$-\frac{i}{2}$&$-\frac{1}{2}$&$\frac{1}{2}$\\
$II$ &$(-\frac{1}{2},\frac{1}{2})$&$
\stackrel{03}{(-i)}\,\stackrel{12}{[+]}|\stackrel{56}{[-]}$&$
 \stackrel{56}{[-]}| \stackrel{12}{[+]} \stackrel{03}{(+i)}$&
$-\frac{i}{2}$&$ \frac{1}{2}$&$-\frac{1}{2}$&$-1$
&$-\frac{i}{2}$&$-\frac{1}{2}$&$\frac{1}{2}$\\
$II$ &$(-\frac{1}{2},-\frac{1}{2})$&$
\stackrel{03}{[+i]}\, \stackrel{12}{(-)}|\stackrel{56}{[-]}$&$
\stackrel{56}{[-]}|{\scriptstyle (-)}  \stackrel{12}{(+)} \stackrel{03}{[+i]}$&
$\frac{i}{2}$&$- \frac{1}{2}$&$-\frac{1}{2}$&$-1$
&$-\frac{i}{2}$&$-\frac{1}{2}$&$\frac{1}{2}$\\ 
%
%
 \hline
$III$&$(\frac{1}{2},\frac{1}{2})$&$
\stackrel{03}{[+i]}\,\stackrel{12}{(+)}| \stackrel{56}{[+]}$&
$
 \stackrel{56}{[+]}|{\scriptstyle (-)}\stackrel{12}{(-)} \stackrel{03}{[+i]}$&$\frac{i}{2}$&
$\frac{1}{2}$&$\frac{1}{2}$&$1$&$-\frac{i}{2}$&$\frac{1}{2}$&$-\frac{1}{2}$\\
$III$ &$(\frac{1}{2},-\frac{1}{2})$&$
\stackrel{03}{(-i)}\,\stackrel{12}{[-]}|\stackrel{56}{[+]}$&$
 \stackrel{56}{[+]}|  \stackrel{12}{[-]} \stackrel{03}{(+i)}$&
$-\frac{i}{2}$&$-\frac{1}{2}$&$\frac{1}{2}$&$1$
&$-\frac{i}{2}$&$\frac{1}{2}$&$-\frac{1}{2}$\\
$III$ &$(-\frac{1}{2},\frac{1}{2})$&$
\stackrel{03}{(-i)}\,\stackrel{12}{(+)}|\stackrel{56}{(-)}$&$
 {\scriptstyle (-)}\stackrel{56}{(+)}| {\scriptstyle (-)} \stackrel{12}{(-)} \stackrel{03}{(+i)}$&
$-\frac{i}{2}$&$ \frac{1}{2}$&$-\frac{1}{2}$&$-1$
&$-\frac{i}{2}$&$\frac{1}{2}$&$-\frac{1}{2}$\\
$III$ &$(-\frac{1}{2},-\frac{1}{2})$&$
\stackrel{03}{[+i]} \stackrel{12}{[-]}|\stackrel{56}{(-)}$&$
{\scriptstyle (-)} \stackrel{56}{(+)}|  \stackrel{12}{[-]} \stackrel{03}{[+i]}$&
$\frac{i}{2}$&$- \frac{1}{2}$&$-\frac{1}{2}$&$-1$
&$-\frac{i}{2}$&$\frac{1}{2}$&$-\frac{1}{2}$\\
\hline
$IV$&$(\frac{1}{2},\frac{1}{2})$&$
\stackrel{03}{(+i)}\,\stackrel{12}{[+]}| \stackrel{56}{[+]}$&
$
 \stackrel{56}{[+]}|\stackrel{12}{[+]} \stackrel{03}{(-i)}$&$\frac{i}{2}$&
$\frac{1}{2}$&$\frac{1}{2}$&$1$&$\frac{i}{2}$&$-\frac{1}{2}$&$-\frac{1}{2}$\\
$IV$ &$(\frac{1}{2},-\frac{1}{2})$&$
\stackrel{03}{[-i]}\,\stackrel{12}{(-)}|\stackrel{56}{[+]}$&$
 \stackrel{56}{[+]}| {\scriptstyle (-)} \stackrel{12}{(+)}
 \stackrel{03}{[-]}$&
$-\frac{i}{2}$&$-\frac{1}{2}$&$\frac{1}{2}$&$1$
&$\frac{i}{2}$&$-\frac{1}{2}$&$-\frac{1}{2}$\\
$IV$ &$(-\frac{1}{2},\frac{1}{2})$&$
\stackrel{03}{[-i]}\,\stackrel{12}{[+]}|\stackrel{56}{(-)}$&$
{\scriptstyle (-)} \stackrel{56}{(+)}| \stackrel{12}{[+]} \stackrel{03}{[-i]}$&
$-\frac{i}{2}$&$ \frac{1}{2}$&$-\frac{1}{2}$&$-1$
&$\frac{i}{2}$&$-\frac{1}{2}$&$-\frac{1}{2}$\\
$IV$ &$(-\frac{1}{2},-\frac{1}{2})$&$
\stackrel{03}{(+i)}\, \stackrel{12}{(-)}|\stackrel{56}{(-)}$&$
{\scriptstyle (-)} \stackrel{56}{(+)}|{\scriptstyle (-)}  \stackrel{12}{(+)} \stackrel{03}{(-i)}$&
$\frac{i}{2}$&$- \frac{1}{2}$&$-\frac{1}{2}$&$-1$
&$\frac{i}{2}$&$-\frac{1}{2}$&$-\frac{1}{2}$\\ 
\hline 
 \end{tabular}
 \end{center}
\end{tiny}
 \caption{\label{cliff basis5+1.}  The basic creation operators --- $\hat{b}^{\alpha\dagger}_{ch, s}$,  $ch$ 
and $s$ replace the index $i$ --- and
their annihilation partners --- $\hat{b}^{\alpha}_{ch, s}$ --- are presented for the $d= (5+1)$-dimensional
case. The basic creation operators are the products of nilpotents and projectors, which are the 
"eigenstates" of the Cartan subalgebra generators, ($S^{03}$, $S^{12}$, $S^{56}$),  
($\tilde{S}^{03}$, $\tilde{S}^{12}$, $\tilde{S}^{56}$), 
presented in Eq.~(\ref{cartan}). Operators $\hat{b}^{\dagger}_{ch, s}$ and $\hat{b}_{ch, s}$
fulfill the commutaion relations of Eq.~(\ref{alphagammaprod0}).
}
 \end{table}

There are superposition of the basic creation operators --- $\hat{b}^{\alpha\dagger}_{i=(ch, s)}$ ---  
which solve, applied on the vacuum state, the Weyl equation Eq.~({\ref{Weyl}}).
Let us make the choice of $p^a =(p^0,p^1,p^2,p^3,0,\cdots,0)$ to see how the spin in $d=(5,6)$
manifest charges in $d=(3+1)$.
\begin{eqnarray}
\label{Cliff basis3+1from5+1Weyl}
p^a &=&
(p^0,p^1,p^2,p^3,0,\cdots,0)\,,\nonumber\\
\hat{b}^{\alpha}_{ch,{\rm sol}} (p) |\psi_{oc}> &=& \sum_{s} 
c^{\alpha\, i=(ch,s)}{}_{i=(ch,sol)} (p) \,
\hat{b}^{\alpha \dagger}_{ch,s} e^{-i p_a x^a} |\psi_{oc}>\,,
\end{eqnarray}
%
where index ${}_{(ch,sol)}$, represents charges and different solutions, respectively, of the Weyl  
equation for massless free fermions.

We present in Eq.~(\ref{weylgen0})  the creation operators, the superposition of the first family 
members, presented in Table~\ref{cliff basis5+1.}, which solve the Weyl equation, 
Eq.~(\ref{Weyl}),  for $p^{a}=(p^0, p^1,p^2,p^3,0,0)$. 
The corresponding annihilation operators follow by the Hermitian conjugation of the creation
operators. 

There are two fermion solutions with the charge $\frac{1}{2}$ and two antifermion solutions with 
the charge $- \frac{1}{2}$, both having the positive energy. The first two creation operators are
related  by the time reversal operator (${\cal T}_{{\cal N}}$ 
$=\gamma^1\,\gamma^3 \,K\,I_{x^0}\, 
I_{x^5, x^7, \cdots, x^{d-1}}$), while the second two follow from the first two  by the application 
of ${\bf \mathbb{C}}_{{\cal N}} {\cal P}^{(d-1)}_{{\cal N}}$  $=  \gamma^0\,\gamma^{5} \,
\cdots  \gamma^{d-1 }\,  I_{\vec{x}_{3}} \,I_{x^6,x^8,\dots,x^{d}} $, both are presented in
Eq.~(\ref{CPTN}).

\begin{tiny}
\begin{eqnarray}
\label{weylgen0}
&& {\rm Creation \; operators\; for\;the \; fermion} \; {\rm states}\; {\rm in \; 
Clifford \; space\; for \; d=(5+1)}
\, \nonumber\\
p^0&=&|p^0|\,,\nonumber\\
\hat{b}^{I_{1}  \dagger}_{\frac{1}{2},\frac{1}{2}}\,(\vec{p})  &=&\beta\, \left( \stackrel{03}{(+i)}\,
\stackrel{12}{(+)}| \stackrel{56}{(+)} + \frac{p^1 +i p^2}{ |p^0| + |p^3|} 
\stackrel{03}{[-i]}\,\stackrel{12}{[-]}|\stackrel{56}{(+)}\right)\,
e^{-i(|p^0| x^0 - \vec{p}\cdot\vec{x})}\,,\nonumber\\
\hat{b}^{I_2 \dagger}_{\,\frac{1}{2},-\frac{1}{2}}\,(\vec{p})&=& \beta^*\, \left(\stackrel{03}{[-i]}\,
\stackrel{12}{[-]}|\stackrel{56}{(+)} - \frac{p^1 -i p^2}{ |p^0| + |p^3|}\,
 \stackrel{03}{(+i)}\,\stackrel{12}{(+)}|\stackrel{56}{(+)}\right)\,
e^{-i(|p^0| x^0 + \vec{p}\cdot\vec{x})}\,,\nonumber\\
&&  {\rm Creation \; operators\; for\;the \; antifermion} \; {\rm states}\; {\rm in \; 
Clifford \; space\; for\; d=(5+1)}\nonumber\\
p^0&=&|p^0|\,,\nonumber\\
\hat{b}^{I_3 \dagger}_{-\frac{1}{2},\,\frac{1}{2}}\,(\vec{p})  &=& - \beta \, 
\left( \stackrel{03}{[-i]}\,\stackrel{12}{(+)}| \stackrel{56}{[-]} + \frac{p^1 +i p^2}{ |p^0| + |p^3|}
\stackrel{03}{(+i)}\,\stackrel{12}{[-]}|\stackrel{56}{[-]}\right)\,
e^{-i(|p^0| x^0 + \vec{p}\cdot\vec{x})}\,,\nonumber\\
\hat{b}^{I_4 \dagger}_{-\frac{1}{2},-\frac{1}{2}}\,(\vec{p})&=& - \beta^*\,\left(\stackrel{03}{(+i)}
\,\stackrel{12}{[-]}| \stackrel{56}{[-]} - \frac{p^1 -i p^2}{ |p^0| + |p^3|} 
\stackrel{03}{[-i]}\,\stackrel{12}{(+)}|\stackrel{56}{[-]}\right)\,
e^{-i(|p^0| x^0 - \vec{p}\cdot\vec{x})}\,,
%
\end{eqnarray}
\end{tiny}

Index ${}_{i}$ counts the solutions, while $\beta^* \beta= \frac{|p^0| + |p^3|}{2|p^0|} $ takes 
care that the corresponding states are normalized. All the states are correspondingly orthogonalized.
The coefficients $c^{\alpha\, i=(ch,s)}{}_{i=(ch,sol)} (p)$ can be read from the solutions.
The solutions  have the definite handedness and orientation of the spin
 with respect to the momentum: 
$\hat{b}^{I_1 \dagger}_{\,\frac{1}{2},\,\frac{1}{2}}$ defines the state with $\Gamma^{(3+1)}
= 1$  and the spin and momentum both up,  $\hat{b}^{I_2 \dagger}_{\,\frac{1}{2},\,-\frac{1}{2}}$
defines the state with $\Gamma^{(3+1)}= 1$  and with spin and momentum both down,
$\hat{b}^{I_3 \dagger}_{\,-\frac{1}{2},\,\frac{1}{2}}$ defines the state with $\Gamma^{(3+1)}
= - 1$  and the spin up and the  momentum down, $\hat{b}^{I_4 \dagger}_{\,-\frac{1}{2},\,
-\frac{1}{2}}$ defines the state with $\Gamma^{(3+1)}= - 1$, the spin down and the  momentum 
up.

The same indexes --- $c^{\alpha\, i=(ch,s)}{}_{i=(ch,sol)} (p) $ --- solve the Weyl equation also for 
the rest three families presented in Table~\ref{cliff basis5+1.}.

%
The phases of creation operators are in agreement  with the application of discrete symmetry
operators ${\bf \mathbb{C}}_{{ \cal N}} \cdot {\cal P}_{{ \cal N}}$,
and ${\cal T}_{{\cal N}}$.

Let us point out that the scalar fields, interacting with fermions (in the {\it spin-charge-family}
theory~[\cite{n2014matterantimatter,IARD2016} and the references cited therein] the scalar fields 
origin in the spin connection fields --- $\omega_{abc}$, the gauge fields of $S^{ab}$, and 
$\tilde{\omega}_{abc}$, the gauge fields of $\tilde{S}^{ab}$, appearing in Eq.~(\ref{wholeaction}) --- 
with the space indexes $c \ge 5$) 
can make massless fermions massive~\cite{NHD,ND012,nh2008,NHD2011,TDN}. In this case, the creation operators 
and correspondingly the annihilation operators,  start to be superposition of basic operators of 
different charges $ch$ as well :
$(\hat{b}^{\alpha \dagger }_{\rm{sol'}} (p)$ $= \sum_{ch,sol} c^{\alpha, ch, sol}{}_{ch,sol'} (p) \,
\hat{b}^{\alpha \dagger}_{ch,sol} e^{-i p_a x^a} )\,|\psi_{oc}>$.
In this case the solutions of the corresponding equations of motion, presented in Eq.~(\ref{weylgen0}) 
for massless states, become
superposition  of different charges and different families.

For $p^{m}=(0,0,0), m=(1,2,3)$ and one massive family only~\cite{TDN}  the creation operators for 
the basic states (usually used in text books~\cite{Bethe,Itzykson} for massive states) 
are presented at Table~\ref{cliff basis5+1m.}. 
 \begin{table}
 \begin{center}
 \begin{tabular}{|c|c|r|}
 \hline
${\rm family}\, \alpha$&
$\hat{b}^{\alpha \dagger}_{{\rm s,m}}$&
$\tilde{S}^{1 2}$%
\\
$1$&
$\;\;\;\;\frac{1}{\sqrt{2}}\; (\stackrel{03}{(+i)} \stackrel{12}{(+)} \stackrel{56}{(+)} + 
\frac{m\;}{m_{+}} \,\stackrel{03}{[-i]} \stackrel{12}{(+)} \stackrel{56}{[-]})$&
$\frac{1}{2}$\\
$2$&$\,\;\;\;\frac{1}{\sqrt{2}}\; (\stackrel{03}{[-i]} \stackrel{12}{[-]} \stackrel{56}{(+)}\; + 
\frac{m\;}{m_{+}} \,\stackrel{03}{(+i)} \stackrel{12}{[-]} \stackrel{56}{[-]})$&
$-\frac{1}{2}$\\
\hline 
 \end{tabular}
 \end{center}
 \caption{\label{cliff basis5+1m.}  The basic creation operators --- $\hat{b}^{\alpha\dagger}_{s,m}$ --- 
for massive states, the first 
 with spin up and the second 
with spin down, are 
presented. $\hat{b}^{\alpha\dagger}_{s,m}\, e^{-i m x^0}$, $s=\pm\frac{1}{2}$, solve 
the equations of motion $\{p_{0} + \gamma^0 (\stackrel{56}{(+)} \,m_{+}  + 
\stackrel{56}{(-)}\, m_{-} )\}\,\hat{b}^{\alpha\dagger}_{s,m}\, e^{-i m x^0} =0$, 
for the two  positive energy states, ($1$,$2$), 
(one with spin up and the other with spin down).
$m^2= m_{+} m_{-}$, $m_{+} = -m_{-}$, $(p_{0})^2 = m^2$, $p_{0 a} =    - \frac{1}{2} 
 S^{cd} \omega_{cd a}$ is assumed to be real~\cite{TDN}.
}
 \end{table}
The creation operators, presented in Table~\ref{cliff basis5+1m.}, define orthonormal states when 
applied on the vacuum state and fulfill, together with the annihilation operators, the 
anticommutation relations presented in Eq.~(\ref{ijgammaprodordinary}).

\vspace{2mm}

$\;\;$ {\bf   Clifford case in $d=(13+1)$:}

\vspace{2mm}

There are $2^{\frac{14}{2}-1}=64$  creation operators for family members of one family, all 
reachable from the starting one by $S^{ab}$. They are presented in Table~\ref{Table so13+1.},
analyzed so that the internal degrees of freedom manifest in $d=(3+1)$ quantum numbers of the 
observed quarks and leptons. 
Applied on the vacuum state $|\psi_{oc}>$ they form
in the {\it spin-charge-family} theory $64$ basic states {\it for quarks and leptons and anti-quarks and 
anti-leptons} for each family. In the {\it spin-charge-family} theory  there are  two times four families 
--- $2^{\frac{8}{2}-1}$ --- getting  masses after the two triplet scalar fields, the superposition of  
$\tilde{\omega}_{abc}$, $(a,b)=(0,1,\cdots,8)$  and three singlet scalar fields, the superposition 
of  $\omega_{abc}$, $(a,b) = (5,6)$ or $(7,8)$ or $(9,\cdots14)$, while $c=(5,6,7,8)$ for all 
these scalar fields, get nonzero vacuum expectation values at low 
energies~\cite{normaJMP2015,IARD2016,n2014matterantimatter,n2012scalars,JMP2013}. 

Table~\ref{Table III.} represents the creation operators creating  $8$  families 
of  $\hat{u}^{c1 \dagger}_{R}$ and of  $\hat{\nu}^{ \dagger}_{R}$. All the family 
members of each of these families follow by the application of $S^{ab}$.

 All the rest of families not included in these eight families get in the {spin-charge-family} theory masses 
by the interaction with the condensate~\cite{normaJMP2015,IARD2016,n2014matterantimatter,%
n2012scalars,JMP2013}. 

To the lower four families  the three so far observed families of quarks and leptons belong.

\bottomcaption{\label{Table so13+1.}%
\tiny{
The left handed ($\Gamma^{(13,1)} = -1$),
multiplet of  creation operators of spinors --- the members of the fundamental representation of the 
$SO(13,1)$ group, manifesting the subgroup $SO(7,1)$
 of the colour charged quarks and anti-quarks and the colourless
leptons and anti-leptons --- is presented in the massless basis using the technique presented in
App.~\ref{technique}. 
It represent the left handed  ($\Gamma^{(3+1)}=-1$, App.~\ref{technique}) weak ($SU(2)_{I}$)
charged  ($\tau^{13}=\pm \frac{1}{2}$, 
($\vec{\tau}^{1}= \frac{1}{2} (S^{58}- S^{67}, S^{57}+ S^{68}, S^{56}- S^{78})$)
and $SU(2)_{II}$ chargeless ($\tau^{23}=0$, 
$\vec{\tau}^{2}= \frac{1}{2} (S^{58}+ S^{67}, S^{57}- S^{68}, S^{56}+ S^{78})$)
quarks and leptons and the right handed  ($\Gamma^{(3+1)}=1$), 
 weak  ($SU(2)_{I}$) chargeless and $SU(2)_{II}$
charged ($\tau^{23}=\pm \frac{1}{2}$) quarks and leptons, both with the spin $ S^{12}$  up and
down ($\pm \frac{1}{2}$, respectively). 
The creation operators of quarks distinguish from those of leptons only in the 
$SU(3) \times U(1)$ part: Quarks are triplets of three colours  ( $= (\tau^{33}, \tau^{38})$ 
$ = [(\frac{1}{2},\frac{1}{2\sqrt{3}}),
(-\frac{1}{2},\frac{1}{2\sqrt{3}}), (0,-\frac{1}{\sqrt{3}}) $], 
($\vec{\tau}^{3}= \frac{1}{2}(S^{9\,12}- S^{10\,11},S^{9\,11}+ S^{10\,12},S^{9\,10}-
S^{11\,12},$ $S^{9\,14}- S^{10\,13},S^{9\,13}+ S^{10\,14},S^{11\,14}- S^{12\,13},$
$S^{11\,13}+ S^{12\,14},\frac{1}{\sqrt{3}}(S^{9\,10}+ S^{11\,12} - 2S^{13\,14})$),
carrying  the "fermion charge" ($\tau^{4}=\frac{1}{6}$, 
$=-\frac{1}{3}(S^{9\,10}+ S^{11\,12}+ S^{13\,14})$. 
The colourless leptons carry the "fermion charge" ($\tau^{4}=-\frac{1}{2}$).
The same multiplet of creation operators represents also the left handed weak ($SU(2)_{I}$) 
chargeless and $SU(2)_{II}$ charged anti-quarks and anti-leptons and the right handed weak 
($SU(2)_{I}$) charged and $SU(2)_{II}$ chargeless anti-quarks and anti-leptons.
Anti-quarks distinguish from anti-leptons again only in the $SU(3) \times U(1)$ part: Anti-quarks are
anti-triplets, 
 carrying  the "fermion charge" ($\tau^{4}=-\frac{1}{6}$).
The anti-colourless anti-leptons carry the "fermion charge" ($\tau^{4}=\frac{1}{2}$).
 $Y=(\tau^{23} + \tau^{4})$ is the hyper charge, the electromagnetic charge
is $Q=(\tau^{13} + Y$).
The creation operators of opposite charges (anti-particle creation operators) are reachable  from 
the particle ones besides by $S^{ab}$  also by the application of the discrete symmetry operator
${\bf \mathbb{C}}_{{\cal N}}$ ${\cal P}_{{\cal N}}$, presented in Refs.~\cite{nhds,TDN}.
%
The reader can find this  Weyl representation also in
Refs.~\cite{n2014matterantimatter,pikan2003,pikan2006,normaJMP2015} and in the references
therein. }
}
\tablehead{\hline
i&$$&$^a\hat{b}^{\dagger}_i $&$\Gamma^{(3+1)}$&$ S^{12}$&
$\tau^{13}$&$\tau^{23}$&$\tau^{33}$&$\tau^{38}$&$\tau^{4}$&$Y$&$Q$\\
\hline
&& ${\rm (Anti)octet},\,\Gamma^{(7+1)} = (-1)\,1\,, \,\Gamma^{(6)} = (1)\,-1$&&&&&&&&& \\
&& ${\rm of \;(anti) quarks \;and \;(anti)leptons}$&&&&&&&&&\\
\hline\hline}
\tabletail{\hline \multicolumn{12}{r}{\emph{Continued on next page}}\\}
\tablelasttail{\hline}
\begin{center}
\tiny{
\begin{supertabular}{|r|c||c||c|c||c|c||c|c|c||r|r|}
1&$ \hat{u}_{R}^{c1\dagger}$&$ \stackrel{03}{(+i)}\,\stackrel{12}{[+]}|
\stackrel{56}{[+]}\,\stackrel{78}{(+)}
||\stackrel{9 \;10}{(+)}\;\;\stackrel{11\;12}{[-]}\;\;\stackrel{13\;14}{[-]} $ &1&$\frac{1}{2}$&0&
$\frac{1}{2}$&$\frac{1}{2}$&$\frac{1}{2\,\sqrt{3}}$&$\frac{1}{6}$&$\frac{2}{3}$&$\frac{2}{3}$\\
\hline
2&$\hat{u}_{R}^{c1 \dagger}$&$\stackrel{03}{[-i]}\,\stackrel{12}{(-)}|\stackrel{56}{[+]}\,\stackrel{78}{(+)}
||\stackrel{9 \;10}{(+)}\;\;\stackrel{11\;12}{[-]}\;\;\stackrel{13\;14}{[-]}$&1&$-\frac{1}{2}$&0&
$\frac{1}{2}$&$\frac{1}{2}$&$\frac{1}{2\,\sqrt{3}}$&$\frac{1}{6}$&$\frac{2}{3}$&$\frac{2}{3}$\\
\hline
3&$\hat{d}_{R}^{c1 \dagger}$&$\stackrel{03}{(+i)}\,\stackrel{12}{[+]}|\stackrel{56}{(-)}\,\stackrel{78}{[-]}
||\stackrel{9 \;10}{(+)}\;\;\stackrel{11\;12}{[-]}\;\;\stackrel{13\;14}{[-]}$&1&$\frac{1}{2}$&0&
$-\frac{1}{2}$&$\frac{1}{2}$&$\frac{1}{2\,\sqrt{3}}$&$\frac{1}{6}$&$-\frac{1}{3}$&$-\frac{1}{3}$\\
\hline
4&$\hat{d}_{R}^{c1 \dagger} $&$\stackrel{03}{[-i]}\,\stackrel{12}{(-)}|
\stackrel{56}{(-)}\,\stackrel{78}{[-]}
||\stackrel{9 \;10}{(+)}\;\;\stackrel{11\;12}{[-]}\;\;\stackrel{13\;14}{[-]} $&1&$-\frac{1}{2}$&0&
$-\frac{1}{2}$&$\frac{1}{2}$&$\frac{1}{2\,\sqrt{3}}$&$\frac{1}{6}$&$-\frac{1}{3}$&$-\frac{1}{3}$\\
\hline
5&$\hat{d}_{L}^{c1\dagger}$&$\stackrel{03}{[-i]}\,\stackrel{12}{[+]}|\stackrel{56}{(-)}\,\stackrel{78}{(+)}
||\stackrel{9 \;10}{(+)}\;\;\stackrel{11\;12}{[-]}\;\;\stackrel{13\;14}{[-]}$&-1&$\frac{1}{2}$&
$-\frac{1}{2}$&0&$\frac{1}{2}$&$\frac{1}{2\,\sqrt{3}}$&$\frac{1}{6}$&$\frac{1}{6}$&$-\frac{1}{3}$\\
\hline
6&$\hat{d}_{L}^{c1 \dagger} $&$  \stackrel{03}{(+i)}\,\stackrel{12}{(-)}|\stackrel{56}{(-)}\,\stackrel{78}{(+)}
||\stackrel{9 \;10}{(+)}\;\;\stackrel{11\;12}{[-]}\;\;\stackrel{13\;14}{[-]} $&-1&$-\frac{1}{2}$&
$-\frac{1}{2}$&0&$\frac{1}{2}$&$\frac{1}{2\,\sqrt{3}}$&$\frac{1}{6}$&$\frac{1}{6}$&$-\frac{1}{3}$\\
\hline
7&$ \hat{u}_{L}^{c1\dagger}$&$  \stackrel{03}{[-i]}\,\stackrel{12}{[+]}|\stackrel{56}{[+]}\,\stackrel{78}{[-]}
||\stackrel{9 \;10}{(+)}\;\;\stackrel{11\;12}{[-]}\;\;\stackrel{13\;14}{[-]}$ &-1&$\frac{1}{2}$&
$\frac{1}{2}$&0 &$\frac{1}{2}$&$\frac{1}{2\,\sqrt{3}}$&$\frac{1}{6}$&$\frac{1}{6}$&$\frac{2}{3}$\\
\hline
8&$\hat{u}_{L}^{c1 \dagger}$&$\stackrel{03}{(+i)}\,\stackrel{12}{(-)}|\stackrel{56}{[+]}\,\stackrel{78}{[-]}
||\stackrel{9 \;10}{(+)}\;\;\stackrel{11\;12}{[-]}\;\;\stackrel{13\;14}{[-]}$&-1&$-\frac{1}{2}$&
$\frac{1}{2}$&0&$\frac{1}{2}$&$\frac{1}{2\,\sqrt{3}}$&$\frac{1}{6}$&$\frac{1}{6}$&$\frac{2}{3}$\\
\hline\hline
\shrinkheight{0.2\textheight}
9&$\hat{u}_{R}^{c2 \dagger}$&$ \stackrel{03}{(+i)}\,\stackrel{12}{[+]}|
\stackrel{56}{[+]}\,\stackrel{78}{(+)}
||\stackrel{9 \;10}{[-]}\;\;\stackrel{11\;12}{(+)}\;\;\stackrel{13\;14}{[-]} $ &1&$\frac{1}{2}$&0&
$\frac{1}{2}$&$-\frac{1}{2}$&$\frac{1}{2\,\sqrt{3}}$&$\frac{1}{6}$&$\frac{2}{3}$&$\frac{2}{3}$\\
\hline
10&$\hat{u}_{R}^{c2 \dagger}$&$\stackrel{03}{[-i]}\,\stackrel{12}{(-)}|\stackrel{56}{[+]}\,\stackrel{78}{(+)}
||\stackrel{9 \;10}{[-]}\;\;\stackrel{11\;12}{(+)}\;\;\stackrel{13\;14}{[-]}$&1&$-\frac{1}{2}$&0&
$\frac{1}{2}$&$-\frac{1}{2}$&$\frac{1}{2\,\sqrt{3}}$&$\frac{1}{6}$&$\frac{2}{3}$&$\frac{2}{3}$\\
\hline
11&$\hat{d}_{R}^{c2 \dagger}$&$\stackrel{03}{(+i)}\,\stackrel{12}{[+]}|\stackrel{56}{(-)}\,\stackrel{78}{[-]}
||\stackrel{9 \;10}{[-]}\;\;\stackrel{11\;12}{(+)}\;\;\stackrel{13\;14}{[-]}$
&1&$\frac{1}{2}$&0&
$-\frac{1}{2}$&$ - \frac{1}{2}$&$\frac{1}{2\,\sqrt{3}}$&$\frac{1}{6}$&$-\frac{1}{3}$&$-\frac{1}{3}$\\
\hline
12&$ \hat{d}_{R}^{c2 \dagger} $&$\stackrel{03}{[-i]}\,\stackrel{12}{(-)}|
\stackrel{56}{(-)}\,\stackrel{78}{[-]}
||\stackrel{9 \;10}{[-]}\;\;\stackrel{11\;12}{(+)}\;\;\stackrel{13\;14}{[-]} $
&1&$-\frac{1}{2}$&0&
$-\frac{1}{2}$&$-\frac{1}{2}$&$\frac{1}{2\,\sqrt{3}}$&$\frac{1}{6}$&$-\frac{1}{3}$&$-\frac{1}{3}$\\
\hline
13&$\hat{d}_{L}^{c2 \dagger}$&$\stackrel{03}{[-i]}\,\stackrel{12}{[+]}|\stackrel{56}{(-)}\,\stackrel{78}{(+)}
||\stackrel{9 \;10}{[-]}\;\;\stackrel{11\;12}{(+)}\;\;\stackrel{13\;14}{[-]}$
&-1&$\frac{1}{2}$&
$-\frac{1}{2}$&0&$-\frac{1}{2}$&$\frac{1}{2\,\sqrt{3}}$&$\frac{1}{6}$&$\frac{1}{6}$&$-\frac{1}{3}$\\
\hline
14&$\hat{d}_{L}^{c2 \dagger} $&$  \stackrel{03}{(+i)}\,\stackrel{12}{(-)}|\stackrel{56}{(-)}\,\stackrel{78}{(+)}
||\stackrel{9 \;10}{[-]}\;\;\stackrel{11\;12}{(+)}\;\;\stackrel{13\;14}{[-]} $&-1&$-\frac{1}{2}$&
$-\frac{1}{2}$&0&$-\frac{1}{2}$&$\frac{1}{2\,\sqrt{3}}$&$\frac{1}{6}$&$\frac{1}{6}$&$-\frac{1}{3}$\\
\hline
15&$ \hat{u}_{L}^{c2 \dagger}$&$  \stackrel{03}{[-i]}\,\stackrel{12}{[+]}|\stackrel{56}{[+]}\,\stackrel{78}{[-]}
||\stackrel{9 \;10}{[-]}\;\;\stackrel{11\;12}{(+)}\;\;\stackrel{13\;14}{[-]}$ &-1&$\frac{1}{2}$&
$\frac{1}{2}$&0 &$-\frac{1}{2}$&$\frac{1}{2\,\sqrt{3}}$&$\frac{1}{6}$&$\frac{1}{6}$&$\frac{2}{3}$\\
\hline
16&$\hat{u}_{L}^{c2 \dagger}$&$\stackrel{03}{(+i)}\,\stackrel{12}{(-)}|\stackrel{56}{[+]}\,\stackrel{78}{[-]}
||\stackrel{9 \;10}{[-]}\;\;\stackrel{11\;12}{(+)}\;\;\stackrel{13\;14}{[-]}$&-1&$-\frac{1}{2}$&
$\frac{1}{2}$&0&$-\frac{1}{2}$&$\frac{1}{2\,\sqrt{3}}$&$\frac{1}{6}$&$\frac{1}{6}$&$\frac{2}{3}$\\
\hline\hline
17&$ \hat{u}_{R}^{c3 \dagger}$&$ \stackrel{03}{(+i)}\,\stackrel{12}{[+]}|
\stackrel{56}{[+]}\,\stackrel{78}{(+)}
||\stackrel{9 \;10}{[-]}\;\;\stackrel{11\;12}{[-]}\;\;\stackrel{13\;14}{(+)} $ &1&$\frac{1}{2}$&0&
$\frac{1}{2}$&$0$&$-\frac{1}{\sqrt{3}}$&$\frac{1}{6}$&$\frac{2}{3}$&$\frac{2}{3}$\\
\hline
18&$\hat{u}_{R}^{c3 \dagger}$&$\stackrel{03}{[-i]}\,\stackrel{12}{(-)}|\stackrel{56}{[+]}\,\stackrel{78}{(+)}
||\stackrel{9 \;10}{[-]}\;\;\stackrel{11\;12}{[-]}\;\;\stackrel{13\;14}{(+)}$&1&$-\frac{1}{2}$&0&
$\frac{1}{2}$&$0$&$-\frac{1}{\sqrt{3}}$&$\frac{1}{6}$&$\frac{2}{3}$&$\frac{2}{3}$\\
\hline
19&$\hat{d}_{R}^{c3 \dagger}$&$\stackrel{03}{(+i)}\,\stackrel{12}{[+]}|\stackrel{56}{(-)}\,\stackrel{78}{[-]}
||\stackrel{9 \;10}{[-]}\;\;\stackrel{11\;12}{[-]}\;\;\stackrel{13\;14}{(+)}$&1&$\frac{1}{2}$&0&
$-\frac{1}{2}$&$0$&$-\frac{1}{\sqrt{3}}$&$\frac{1}{6}$&$-\frac{1}{3}$&$-\frac{1}{3}$\\
\hline
20&$\hat{d}_{R}^{c3 \dagger} $&$\stackrel{03}{[-i]}\,\stackrel{12}{(-)}|
\stackrel{56}{(-)}\,\stackrel{78}{[-]}
||\stackrel{9 \;10}{[-]}\;\;\stackrel{11\;12}{[-]}\;\;\stackrel{13\;14}{(+)} $&1&$-\frac{1}{2}$&0&
$-\frac{1}{2}$&$0$&$-\frac{1}{\sqrt{3}}$&$\frac{1}{6}$&$-\frac{1}{3}$&$-\frac{1}{3}$\\
\hline
21&$\hat{d}_{L}^{c3 \dagger}$&$\stackrel{03}{[-i]}\,\stackrel{12}{[+]}|\stackrel{56}{(-)}\,\stackrel{78}{(+)}
||\stackrel{9 \;10}{[-]}\;\;\stackrel{11\;12}{[-]}\;\;\stackrel{13\;14}{(+)}$&-1&$\frac{1}{2}$&
$-\frac{1}{2}$&0&$0$&$-\frac{1}{\sqrt{3}}$&$\frac{1}{6}$&$\frac{1}{6}$&$-\frac{1}{3}$\\
\hline
22&$\hat{d}_{L}^{c3 \dagger} $&$  \stackrel{03}{(+i)}\,\stackrel{12}{(-)}|\stackrel{56}{(-)}\,\stackrel{78}{(+)}
||\stackrel{9 \;10}{[-]}\;\;\stackrel{11\;12}{[-]}\;\;\stackrel{13\;14}{(+)} $&-1&$-\frac{1}{2}$&
$-\frac{1}{2}$&0&$0$&$-\frac{1}{\sqrt{3}}$&$\frac{1}{6}$&$\frac{1}{6}$&$-\frac{1}{3}$\\
\hline
23&$ \hat{u}_{L}^{c3 \dagger}$&$  \stackrel{03}{[-i]}\,\stackrel{12}{[+]}|\stackrel{56}{[+]}\,\stackrel{78}{[-]}
||\stackrel{9 \;10}{[-]}\;\;\stackrel{11\;12}{[-]}\;\;\stackrel{13\;14}{(+)}$ &-1&$\frac{1}{2}$&
$\frac{1}{2}$&0 &$0$&$-\frac{1}{\sqrt{3}}$&$\frac{1}{6}$&$\frac{1}{6}$&$\frac{2}{3}$\\
\hline
24&$\hat{u}_{L}^{c3 \dagger}$&$\stackrel{03}{(+i)}\,\stackrel{12}{(-)}|\stackrel{56}{[+]}\,\stackrel{78}{[-]}
||\stackrel{9 \;10}{[-]}\;\;\stackrel{11\;12}{[-]}\;\;\stackrel{13\;14}{(+)}$&-1&$-\frac{1}{2}$&
$\frac{1}{2}$&0&$0$&$-\frac{1}{\sqrt{3}}$&$\frac{1}{6}$&$\frac{1}{6}$&$\frac{2}{3}$\\
\hline\hline
25&$ \hat{\nu}^{ \dagger}_{R}$&$ \stackrel{03}{(+i)}\,\stackrel{12}{[+]}|
\stackrel{56}{[+]}\,\stackrel{78}{(+)}
||\stackrel{9 \;10}{(+)}\;\;\stackrel{11\;12}{(+)}\;\;\stackrel{13\;14}{(+)} $ &1&$\frac{1}{2}$&0&
$\frac{1}{2}$&$0$&$0$&$-\frac{1}{2}$&$0$&$0$\\
\hline
26&$\hat{\nu}^{ \dagger}_{R}$&$\stackrel{03}{[-i]}\,\stackrel{12}{(-)}|\stackrel{56}{[+]}\,\stackrel{78}{(+)}
||\stackrel{9 \;10}{(+)}\;\;\stackrel{11\;12}{(+)}\;\;\stackrel{13\;14}{(+)}$&1&$-\frac{1}{2}$&0&
$\frac{1}{2}$ &$0$&$0$&$-\frac{1}{2}$&$0$&$0$\\
\hline
27&$\hat{e}^{ \dagger}_{R}$&$\stackrel{03}{(+i)}\,\stackrel{12}{[+]}|\stackrel{56}{(-)}\,\stackrel{78}{[-]}
||\stackrel{9 \;10}{(+)}\;\;\stackrel{11\;12}{(+)}\;\;\stackrel{13\;14}{(+)}$&1&$\frac{1}{2}$&0&
$-\frac{1}{2}$&$0$&$0$&$-\frac{1}{2}$&$-1$&$-1$\\
\hline
28&$\hat{e}^{ \dagger}_{R} $&$\stackrel{03}{[-i]}\,\stackrel{12}{(-)}|
\stackrel{56}{(-)}\,\stackrel{78}{[-]}
||\stackrel{9 \;10}{(+)}\;\;\stackrel{11\;12}{(+)}\;\;\stackrel{13\;14}{(+)} $&1&$-\frac{1}{2}$&0&
$-\frac{1}{2}$&$0$&$0$&$-\frac{1}{2}$&$-1$&$-1$\\
\hline
29&$\hat{e}^{ \dagger}_{L}$&$\stackrel{03}{[-i]}\,\stackrel{12}{[+]}|\stackrel{56}{(-)}\,\stackrel{78}{(+)}
||\stackrel{9 \;10}{(+)}\;\;\stackrel{11\;12}{(+)}\;\;\stackrel{13\;14}{(+)}$&-1&$\frac{1}{2}$&
$-\frac{1}{2}$&0&$0$&$0$&$-\frac{1}{2}$&$-\frac{1}{2}$&$-1$\\
\hline
30&$\hat{e}^{ \dagger}_{L} $&$  \stackrel{03}{(+i)}\,\stackrel{12}{(-)}|\stackrel{56}{(-)}\,\stackrel{78}{(+)}
||\stackrel{9 \;10}{(+)}\;\;\stackrel{11\;12}{(+)}\;\;\stackrel{13\;14}{(+)} $&-1&$-\frac{1}{2}$&
$-\frac{1}{2}$&0&$0$&$0$&$-\frac{1}{2}$&$-\frac{1}{2}$&$-1$\\
\hline
31&$ \hat{\nu}^{ \dagger}_{L}$&$  \stackrel{03}{[-i]}\,\stackrel{12}{[+]}|\stackrel{56}{[+]}\,\stackrel{78}{[-]}
||\stackrel{9 \;10}{(+)}\;\;\stackrel{11\;12}{(+)}\;\;\stackrel{13\;14}{(+)}$ &-1&$\frac{1}{2}$&
$\frac{1}{2}$&0 &$0$&$0$&$-\frac{1}{2}$&$-\frac{1}{2}$&$0$\\
\hline
32&$\hat{\nu}^{ \dagger}_{L}$&$\stackrel{03}{(+i)}\,\stackrel{12}{(-)}|\stackrel{56}{[+]}\,\stackrel{78}{[-]}
||\stackrel{9 \;10}{(+)}\;\;\stackrel{11\;12}{(+)}\;\;\stackrel{13\;14}{(+)}$&-1&$-\frac{1}{2}$&
$\frac{1}{2}$&0&$0$&$0$&$-\frac{1}{2}$&$-\frac{1}{2}$&$0$\\
\hline\hline
33&$\hat{\bar{d}}_{L}^{\bar{c1} \dagger}$&$ \stackrel{03}{[-i]}\,\stackrel{12}{[+]}|
\stackrel{56}{[+]}\,\stackrel{78}{(+)}
||\stackrel{9 \;10}{[-]}\;\;\stackrel{11\;12}{(+)}\;\;\stackrel{13\;14}{(+)} $ &-1&$\frac{1}{2}$&0&
$\frac{1}{2}$&$-\frac{1}{2}$&$-\frac{1}{2\,\sqrt{3}}$&$-\frac{1}{6}$&$\frac{1}{3}$&$\frac{1}{3}$\\
\hline
34&$\hat{\bar{d}}_{L}^{\bar{c1} \dagger}$&$\stackrel{03}{(+i)}\,\stackrel{12}{(-)}|\stackrel{56}{[+]}\,\stackrel{78}{(+)}
||\stackrel{9 \;10}{[-]}\;\;\stackrel{11\;12}{(+)}\;\;\stackrel{13\;14}{(+)}$&-1&$-\frac{1}{2}$&0&
$\frac{1}{2}$&$-\frac{1}{2}$&$-\frac{1}{2\,\sqrt{3}}$&$-\frac{1}{6}$&$\frac{1}{3}$&$\frac{1}{3}$\\
\hline
35&$\bar{u}_{L}^{\bar{c1} \dagger}$&$  \stackrel{03}{[-i]}\,\stackrel{12}{[+]}|\stackrel{56}{(-)}\,\stackrel{78}{[-]}
||\stackrel{9 \;10}{[-]}\;\;\stackrel{11\;12}{(+)}\;\;\stackrel{13\;14}{(+)}$&-1&$\frac{1}{2}$&0&
$-\frac{1}{2}$&$-\frac{1}{2}$&$-\frac{1}{2\,\sqrt{3}}$&$-\frac{1}{6}$&$-\frac{2}{3}$&$-\frac{2}{3}$\\
\hline
36&$ \bar{u}_{L}^{\bar{c1} \dagger} $&$  \stackrel{03}{(+i)}\,\stackrel{12}{(-)}|
\stackrel{56}{(-)}\,\stackrel{78}{[-]}
||\stackrel{9 \;10}{[-]}\;\;\stackrel{11\;12}{(+)}\;\;\stackrel{13\;14}{(+)} $&-1&$-\frac{1}{2}$&0&
$-\frac{1}{2}$&$-\frac{1}{2}$&$-\frac{1}{2\,\sqrt{3}}$&$-\frac{1}{6}$&$-\frac{2}{3}$&$-\frac{2}{3}$\\
\hline
37&$\hat{\bar{d}}_{R}^{\bar{c1} \dagger}$&$\stackrel{03}{(+i)}\,\stackrel{12}{[+]}|\stackrel{56}{[+]}\,\stackrel{78}{[-]}
||\stackrel{9 \;10}{[-]}\;\;\stackrel{11\;12}{(+)}\;\;\stackrel{13\;14}{(+)}$&1&$\frac{1}{2}$&
$\frac{1}{2}$&0&$-\frac{1}{2}$&$-\frac{1}{2\,\sqrt{3}}$&$-\frac{1}{6}$&$-\frac{1}{6}$&$\frac{1}{3}$\\
\hline
38&$\hat{\bar{d}}_{R}^{\bar{c1} \dagger} $&$  \stackrel{03}{[-i]}\,\stackrel{12}{(-)}|\stackrel{56}{[+]}\,\stackrel{78}{[-]}
||\stackrel{9 \;10}{[-]}\;\;\stackrel{11\;12}{(+)}\;\;\stackrel{13\;14}{(+)} $&1&$-\frac{1}{2}$&
$\frac{1}{2}$&0&$-\frac{1}{2}$&$-\frac{1}{2\,\sqrt{3}}$&$-\frac{1}{6}$&$-\frac{1}{6}$&$\frac{1}{3}$\\
\hline
39&$\hat{\bar{u}}_{R}^{\bar{c1} \dagger}$&$\stackrel{03}{(+i)}\,\stackrel{12}{[+]}|\stackrel{56}{(-)}\,\stackrel{78}{(+)}
||\stackrel{9 \;10}{[-]}\;\;\stackrel{11\;12}{(+)}\;\;\stackrel{13\;14}{(+)}$ &1&$\frac{1}{2}$&
$-\frac{1}{2}$&0 &$-\frac{1}{2}$&$-\frac{1}{2\,\sqrt{3}}$&$-\frac{1}{6}$&$-\frac{1}{6}$&$-\frac{2}{3}$\\
\hline
40&$\hat{\bar{u}}_{R}^{\bar{c1} \dagger}$&$\stackrel{03}{[-i]}\,\stackrel{12}{(-)}|\stackrel{56}{(-)}\,\stackrel{78}{(+)}
||\stackrel{9 \;10}{[-]}\;\;\stackrel{11\;12}{(+)}\;\;\stackrel{13\;14}{(+)}$
&1&$-\frac{1}{2}$&
$-\frac{1}{2}$&0&$-\frac{1}{2}$&$-\frac{1}{2\,\sqrt{3}}$&$-\frac{1}{6}$&$-\frac{1}{6}$&$-\frac{2}{3}$\\
\hline\hline
41&$ \hat{\bar{d}}_{L}^{\bar{c2} \dagger}$&$ \stackrel{03}{[-i]}\,\stackrel{12}{[+]}|
\stackrel{56}{[+]}\,\stackrel{78}{(+)}
||\stackrel{9 \;10}{(+)}\;\;\stackrel{11\;12}{[-]}\;\;\stackrel{13\;14}{(+)} $
&-1&$\frac{1}{2}$&0&
$\frac{1}{2}$&$\frac{1}{2}$&$-\frac{1}{2\,\sqrt{3}}$&$-\frac{1}{6}$&$\frac{1}{3}$&$\frac{1}{3}$\\
\hline
42&$\hat{\bar{d}}_{L}^{\bar{c2} \dagger}$&$\stackrel{03}{(+i)}\,\stackrel{12}{(-)}|\stackrel{56}{[+]}\,\stackrel{78}{(+)}
||\stackrel{9 \;10}{(+)}\;\;\stackrel{11\;12}{[-]}\;\;\stackrel{13\;14}{(+)}$
&-1&$-\frac{1}{2}$&0&
$\frac{1}{2}$&$\frac{1}{2}$&$-\frac{1}{2\,\sqrt{3}}$&$-\frac{1}{6}$&$\frac{1}{3}$&$\frac{1}{3}$\\
\hline
43&$\hat{\bar{u}}_{L}^{\bar{c2} \dagger}$&$  \stackrel{03}{[-i]}\,\stackrel{12}{[+]}|\stackrel{56}{(-)}\,\stackrel{78}{[-]}
||\stackrel{9 \;10}{(+)}\;\;\stackrel{11\;12}{[-]}\;\;\stackrel{13\;14}{(+)}$
&-1&$\frac{1}{2}$&0&
$-\frac{1}{2}$&$\frac{1}{2}$&$-\frac{1}{2\,\sqrt{3}}$&$-\frac{1}{6}$&$-\frac{2}{3}$&$-\frac{2}{3}$\\
\hline
44&$ \hat{\bar{u}}_{L}^{\bar{c2} \dagger} $&$  \stackrel{03}{(+i)}\,\stackrel{12}{(-)}|
\stackrel{56}{(-)}\,\stackrel{78}{[-]}
||\stackrel{9 \;10}{(+)}\;\;\stackrel{11\;12}{[-]}\;\;\stackrel{13\;14}{(+)} $
&-1&$-\frac{1}{2}$&0&
$-\frac{1}{2}$&$\frac{1}{2}$&$-\frac{1}{2\,\sqrt{3}}$&$-\frac{1}{6}$&$-\frac{2}{3}$&$-\frac{2}{3}$\\
\hline
45&$\hat{\bar{d}}_{R}^{\bar{c2} \dagger}$&$\stackrel{03}{(+i)}\,\stackrel{12}{[+]}|\stackrel{56}{[+]}\,\stackrel{78}{[-]}
||\stackrel{9 \;10}{(+)}\;\;\stackrel{11\;12}{[-]}\;\;\stackrel{13\;14}{(+)}$
&1&$\frac{1}{2}$&
$\frac{1}{2}$&0&$\frac{1}{2}$&$-\frac{1}{2\,\sqrt{3}}$&$-\frac{1}{6}$&$-\frac{1}{6}$&$\frac{1}{3}$\\
\hline
46&$\hat{\bar{d}}_{R}^{\bar{c2} \dagger} $&$  \stackrel{03}{[-i]}\,\stackrel{12}{(-)}|\stackrel{56}{[+]}\,\stackrel{78}{[-]}
||\stackrel{9 \;10}{(+)}\;\;\stackrel{11\;12}{[-]}\;\;\stackrel{13\;14}{(+)} $
&1&$-\frac{1}{2}$&
$\frac{1}{2}$&0&$\frac{1}{2}$&$-\frac{1}{2\,\sqrt{3}}$&$-\frac{1}{6}$&$-\frac{1}{6}$&$\frac{1}{3}$\\
\hline
47&$ \hat{\bar{u}}_{R}^{\bar{c2} \dagger}$&$\stackrel{03}{(+i)}\,\stackrel{12}{[+]}|\stackrel{56}{(-)}\,\stackrel{78}{(+)}
||\stackrel{9 \;10}{(+)}\;\;\stackrel{11\;12}{[-]}\;\;\stackrel{13\;14}{(+)}$
 &1&$\frac{1}{2}$&
$-\frac{1}{2}$&0 &$\frac{1}{2}$&$-\frac{1}{2\,\sqrt{3}}$&$-\frac{1}{6}$&$-\frac{1}{6}$&$-\frac{2}{3}$\\
\hline
48&$\hat{\bar{u}}_{R}^{\bar{c2} \dagger}$&$\stackrel{03}{[-i]}\,\stackrel{12}{(-)}|\stackrel{56}{(-)}\,\stackrel{78}{(+)}
||\stackrel{9 \;10}{(+)}\;\;\stackrel{11\;12}{[-]}\;\;\stackrel{13\;14}{(+)}$
&1&$-\frac{1}{2}$&
$-\frac{1}{2}$&0&$\frac{1}{2}$&$-\frac{1}{2\,\sqrt{3}}$&$-\frac{1}{6}$&$-\frac{1}{6}$&$-\frac{2}{3}$\\
\hline\hline
49&$ \hat{\bar{d}}_{L}^{\bar{c3} \dagger}$&$ \stackrel{03}{[-i]}\,\stackrel{12}{[+]}|
\stackrel{56}{[+]}\,\stackrel{78}{(+)}
||\stackrel{9 \;10}{(+)}\;\;\stackrel{11\;12}{(+)}\;\;\stackrel{13\;14}{[-]} $ &-1&$\frac{1}{2}$&0&
$\frac{1}{2}$&$0$&$\frac{1}{\sqrt{3}}$&$-\frac{1}{6}$&$\frac{1}{3}$&$\frac{1}{3}$\\
\hline
50&$\hat{\bar{d}}_{L}^{\bar{c3} \dagger}$&$\stackrel{03}{(+i)}\,\stackrel{12}{(-)}|\stackrel{56}{[+]}\,\stackrel{78}{(+)}
||\stackrel{9 \;10}{(+)}\;\;\stackrel{11\;12}{(+)}\;\;\stackrel{13\;14}{[-]} $&-1&$-\frac{1}{2}$&0&
$\frac{1}{2}$&$0$&$\frac{1}{\sqrt{3}}$&$-\frac{1}{6}$&$\frac{1}{3}$&$\frac{1}{3}$\\
\hline
51&$\hat{\bar{u}}_{L}^{\bar{c3} \dagger}$&$  \stackrel{03}{[-i]}\,\stackrel{12}{[+]}|\stackrel{56}{(-)}\,\stackrel{78}{[-]}
||\stackrel{9 \;10}{(+)}\;\;\stackrel{11\;12}{(+)}\;\;\stackrel{13\;14}{[-]} $&-1&$\frac{1}{2}$&0&
$-\frac{1}{2}$&$0$&$\frac{1}{\sqrt{3}}$&$-\frac{1}{6}$&$-\frac{2}{3}$&$-\frac{2}{3}$\\
\hline
52&$ \hat{\bar{u}}_{L}^{\bar{c3} \dagger} $&$  \stackrel{03}{(+i)}\,\stackrel{12}{(-)}|
\stackrel{56}{(-)}\,\stackrel{78}{[-]}
||\stackrel{9 \;10}{(+)}\;\;\stackrel{11\;12}{(+)}\;\;\stackrel{13\;14}{[-]}  $&-1&$-\frac{1}{2}$&0&
$-\frac{1}{2}$&$0$&$\frac{1}{\sqrt{3}}$&$-\frac{1}{6}$&$-\frac{2}{3}$&$-\frac{2}{3}$\\
\hline
53&$\hat{\bar{d}}_{R}^{\bar{c3} \dagger}$&$\stackrel{03}{(+i)}\,\stackrel{12}{[+]}|\stackrel{56}{[+]}\,\stackrel{78}{[-]}
||\stackrel{9 \;10}{(+)}\;\;\stackrel{11\;12}{(+)}\;\;\stackrel{13\;14}{[-]} $&1&$\frac{1}{2}$&
$\frac{1}{2}$&0&$0$&$\frac{1}{\sqrt{3}}$&$-\frac{1}{6}$&$-\frac{1}{6}$&$\frac{1}{3}$\\
\hline
54&$\hat{\bar{d}}_{R}^{\bar{c3} \dagger} $&$  \stackrel{03}{[-i]}\,\stackrel{12}{(-)}|\stackrel{56}{[+]}\,\stackrel{78}{[-]}
||\stackrel{9 \;10}{(+)}\;\;\stackrel{11\;12}{(+)}\;\;\stackrel{13\;14}{[-]} $&1&$-\frac{1}{2}$&
$\frac{1}{2}$&0&$0$&$\frac{1}{\sqrt{3}}$&$-\frac{1}{6}$&$-\frac{1}{6}$&$\frac{1}{3}$\\
\hline
55&$ \hat{\bar{u}}_{R}^{\bar{c3} \dagger}$&$\stackrel{03}{(+i)}\,\stackrel{12}{[+]}|\stackrel{56}{(-)}\,\stackrel{78}{(+)}
||\stackrel{9 \;10}{(+)}\;\;\stackrel{11\;12}{(+)}\;\;\stackrel{13\;14}{[-]} $ &1&$\frac{1}{2}$&
$-\frac{1}{2}$&0 &$0$&$\frac{1}{\sqrt{3}}$&$-\frac{1}{6}$&$-\frac{1}{6}$&$-\frac{2}{3}$\\
\hline
56&$\hat{\bar{u}}_{R}^{\bar{c3} \dagger}$&$\stackrel{03}{[-i]}\,\stackrel{12}{(-)}|\stackrel{56}{(-)}\,\stackrel{78}{(+)}
||\stackrel{9 \;10}{(+)}\;\;\stackrel{11\;12}{(+)}\;\;\stackrel{13\;14}{[-]} $&1&$-\frac{1}{2}$&
$-\frac{1}{2}$&0&$0$&$\frac{1}{\sqrt{3}}$&$-\frac{1}{6}$&$-\frac{1}{6}$&$-\frac{2}{3}$\\
\hline\hline
57&$ \hat{\bar{e}}^{ \dagger}_{L}$&$ \stackrel{03}{[-i]}\,\stackrel{12}{[+]}|
\stackrel{56}{[+]}\,\stackrel{78}{(+)}
||\stackrel{9 \;10}{[-]}\;\;\stackrel{11\;12}{[-]}\;\;\stackrel{13\;14}{[-]} $ &-1&$\frac{1}{2}$&0&
$\frac{1}{2}$&$0$&$0$&$\frac{1}{2}$&$1$&$1$\\
\hline
58&$\hat{\bar{e}}^{ \dagger}_{L}$&$\stackrel{03}{(+i)}\,\stackrel{12}{(-)}|\stackrel{56}{[+]}\,\stackrel{78}{(+)}
||\stackrel{9 \;10}{[-]}\;\;\stackrel{11\;12}{[-]}\;\;\stackrel{13\;14}{[-]}$&-1&$-\frac{1}{2}$&0&
$\frac{1}{2}$ &$0$&$0$&$\frac{1}{2}$&$1$&$1$\\
\hline
59&$\hat{ \bar{\nu}}^{ \dagger}_{L}$&$  \stackrel{03}{[-i]}\,\stackrel{12}{[+]}|\stackrel{56}{(-)}\,\stackrel{78}{[-]}
||\stackrel{9 \;10}{[-]}\;\;\stackrel{11\;12}{[-]}\;\;\stackrel{13\;14}{[-]}$&-1&$\frac{1}{2}$&0&
$-\frac{1}{2}$&$0$&$0$&$\frac{1}{2}$&$0$&$0$\\
\hline
60&$\hat{ \bar{\nu}}^{ \dagger}_{L} $&$  \stackrel{03}{(+i)}\,\stackrel{12}{(-)}|
\stackrel{56}{(-)}\,\stackrel{78}{[-]}
||\stackrel{9 \;10}{[-]}\;\;\stackrel{11\;12}{[-]}\;\;\stackrel{13\;14}{[-]} $&-1&$-\frac{1}{2}$&0&
$-\frac{1}{2}$&$0$&$0$&$\frac{1}{2}$&$0$&$0$\\
\hline
61&$\hat{ \bar{\nu}}^{ \dagger}_{R}$&$\stackrel{03}{(+i)}\,\stackrel{12}{[+]}|\stackrel{56}{(-)}\,\stackrel{78}{(+)}
||\stackrel{9 \;10}{[-]}\;\;\stackrel{11\;12}{[-]}\;\;\stackrel{13\;14}{[-]}$&1&$\frac{1}{2}$&
$-\frac{1}{2}$&0&$0$&$0$&$\frac{1}{2}$&$\frac{1}{2}$&$0$\\
\hline
62&$\hat{\bar{\nu}}^{ \dagger}_{R} $&$  \stackrel{03}{[-i]}\,\stackrel{12}{(-)}|\stackrel{56}{(-)}\,\stackrel{78}{(+)}
||\stackrel{9 \;10}{[-]}\;\;\stackrel{11\;12}{[-]}\;\;\stackrel{13\;14}{[-]} $&1&$-\frac{1}{2}$&
$-\frac{1}{2}$&0&$0$&$0$&$\frac{1}{2}$&$\frac{1}{2}$&$0$\\
\hline
63&$ \hat{\bar{e}}^{ \dagger}_{R}$&$\stackrel{03}{(+i)}\,\stackrel{12}{[+]}|\stackrel{56}{[+]}\,\stackrel{78}{[-]}
||\stackrel{9 \;10}{[-]}\;\;\stackrel{11\;12}{[-]}\;\;\stackrel{13\;14}{[-]}$ &1&$\frac{1}{2}$&
$\frac{1}{2}$&0 &$0$&$0$&$\frac{1}{2}$&$\frac{1}{2}$&$1$\\
\hline
64&$\hat{\bar{e}}^{ \dagger}_{R}$&$\stackrel{03}{[-i]}\,\stackrel{12}{(-)}|\stackrel{56}{[+]}\,\stackrel{78}{[-]}
||\stackrel{9 \;10}{[-]}\;\;\stackrel{11\;12}{[-]}\;\;\stackrel{13\;14}{[-]}$&1&$-\frac{1}{2}$&
$\frac{1}{2}$&0&$0$&$0$&$\frac{1}{2}$&$\frac{1}{2}$&$1$\\
\hline
\end{supertabular}
}
\end{center}
%
 Table~\ref{Table so13+1.} represents in the {\it spin-charge-family} theory the basic creation 
operators for observed {\it quarks and leptons and anti-quarks and anti-leptons} for a particular family. 
Hermitian conjugation of the  creation operators of Table~\ref{Table so13+1.} generates 
 the corresponding annihilation operators, fulfilling together with the creation operators anticommutation 
relations for fermions of Eq.~(\ref{alphagammaprod0}).

In observable dimension $d=(3+1)$ the $d=(13+1)$ case differs from $d=(5+1)$ case, 
Table~\ref{cliff basis5+1m.}, in a much reacher offer of charges.
The kinematics of the fermion states in $d=(13 +1)$, Table~\ref{Table so13+1.}, in 
$d=(3+1)$ is, however, very similar to the one of  Table~
\ref{weylgen0}.

The coefficients of the superposition of the basic creation operators --- $\hat{b}^{\alpha \dagger}_{i}$ ---  
which solve, applied on the vacuum state, the Weyl equation, Eq.~({\ref{Weyl}}), for 
 the choice of $p^a =(p^0,p^1,p^2,p^3,0,\cdots,0)$,  can be taken  from Eq.~(\ref{weylgen0}).
For the positive energy solution of spin $\frac{1}{2}$ one only has to 
replace $\stackrel{03}{(+i)} \stackrel{12}{(+)}\stackrel{56}{(+)}$  by $\hat{u}_{R,1/2}^{c1\dagger}$
with spin $\frac{1}{2}$ and $\stackrel{03}{[-i]} \stackrel{12}{[-]}\stackrel{56}{(+)}$ by 
$\hat{u}_{R, -1/2}^{c1\dagger}$ with spin $-\frac{1}{2}$. The coefficients, $\beta$ and 
$\frac{p^1+i p^2}{|p^0| + |p^3|}$, remain the one of the case with $d=(5+1)$. 

The operator ${\cal T}_{{\cal N}}= \gamma^1\, \gamma^3\, K\, I_{x^0}\, 
I_{x^5, x^7,\cdots,x^{d-1}}$ transforms this superposition of creation operators  
$\beta\, (\hat{u}_{R,1/2}^{c1\dagger} + \frac{p^1+i p^2}{|p^0| + |p^3|}
\hat{u}_{R, -1/2}^{c1\dagger}) \cdot e^{-i (p^0 x^0 - \vec{p} \cdot \vec{x})}$  into 
$\beta^*\, (\hat{u}_{R, - 1/2}^{c1\dagger} - \frac{p^1- i p^2}{|p^0| + |p^3|}
\hat{u}_{R, 1/2}^{c1\dagger}) \cdot  e^{-i (p^0 x^0 + \vec{p} \cdot \vec{x})}$.

The operator  ${\bf \mathbb{C}}_{{\cal N}} {\cal P}^{(d-1)}_{{\cal N}}=\gamma^0\, \gamma^5
\,\gamma^7\cdots \gamma^{d-1}\,I_{\vec{x}_{3}}\cdots I_{x^6, x^8,\cdots,x^{d}} $ transforms
the positive energy solution creation operator  for $u$ quark 
$\beta\, (\hat{u}_{R, 1/2}^{c1\dagger} + \frac{p^1+i p^2}{|p^0|
 + |p^3|} \,\hat{u}_{R, -1/2}^{c1\dagger}) \cdot e^{-i (p^0 x^0 - \vec{p} \cdot \vec{x})}$ 
into the positive energy  solution of anti-$u$ quark
 $-\beta\, (\hat{\bar{u}}_{L, 1/2}^{\bar{c1}\dagger}  + \frac{p^1+i p^2}{|p^0|+ |p^3|} 
\,\hat{\bar{u}}_{L,- 1/2}^{\bar{c1}\dagger}) \cdot e^{-i (p^0 x^0 + \vec{p} \cdot \vec{x})}$.
   
One can  proceed in the same way also for the $\hat{u}_{L}^{c1\dagger}$, 
$\hat{d}_{R}^{c1\dagger}$, and all the other quarks  ${}^{ci}$, as well as for  leptons.

Spins in higher dimensional space manifest charges in 
$d=(3+1)$, 
Table~\ref{Table so13+1.}, provided that the angular momentum in ordinary space at higher 
dimensions do not contribute, which is supposed to be  the case at low energies. 
All the creation operators of any family and any family member, or the orthogonal superposition 
of them, together with their Hermitian conjugate annihilation operators fulfill the anticommutation
 relations of Eqs.~(\ref{alphagammaprod0}, \ref{pgamma}, \ref{ijgammaprodordinary}).

The commuting operators of $S^{ab}$, Eq.~(\ref{cartan}), determine in  $d=(3+1)$
the handedness  ($\Gamma^{(3+1))}= -4i \cdot S^{03} S^{12})$), the spin ($S^{12}$), 
the  third component of the weak $SU(2)$ charge ($\tau^{13}$), the  third component of the 
second $SU(2)$ charge ($\tau^{23}$), the two components of the $SU(3)$ colour charge 
($\tau^{33}, \tau^{38}$) and the "fermion charge" ($\tau^4$, originating in $U(1)$ from 
$SO(6)$, which includes $SU(3) \times U(1)$).  
The hypercharge $Y$,  which is in the {\it standard model} "guessed" from the experimental data, 
is in the {\it spin-charge-family} theory equal to  ($\tau^{4} + \tau^{23}$), while
electromagnetic charge $Q$ is, like in the {\it standard model}, equal to  ($Y + \tau^{13}$). 

One representation of creation operators with $2^{\frac{d}{2}-1}$ members includes all the 
left and the right handed coloured quarks and colourless leptons and left and right handed 
(anti coloured) antiquarks and (anti colourless)  antileptons. The right handed 
neutrinos and the left handed antineutrinos, like all the other members of one Lorentz 
representation, carry the additional hypercharge (the additional superposition of 
$\tau^{4}$ and $\tau^{23}$) and are correspondingly not chargeless like in the 
{\it standard model}.

The sum of the charges, the sum of the spins and the sum of the handedness ---properties 
defined  with respect to $d=(3+1)$ --- over all the members of one representation are 
equal to  zero in any $d$, as it is the case of $d=(5+1)$.
However, in the $d=(13+1)$ case this is true even within quarks and leptons separately 
and within antiquarks and antileptons separately.  Let be repeated that this is so since 
the right handed neutrinos and the left handed antineutrinos are the regular members of 
one representation, as it is true for quarks and charged leptons. This can be checked 
in Table~\ref{Table so13+1.}. Exclusion of the right handed neutrinos and left handed 
antineutrinos makes nonzero the sum of $(\Gamma^{(3+1))}$, $\tau^{23}$ and
$\tau^4$ over  the spinor part separately and correspondingly also over the antispinor part. 
The whole representation has even in this case sums over all the quantum numbers 
of spins and charges equal to zero.

\subsubsection{{\bf Grassmann "fermions" and "antifermions"}}
\label{Grassmannfermions}

Let us represent creation and annihilation operators in Grassmann space, like as we did in the 
Clifford case. 

In the Grassmann case the representations ind $d=(13+1)$ space start to be very large 
and correspondingly almost uncontrollable,  Eq.~(\ref{nograssop}).
We learn in the Clifford case that at the low energy regime, when we treat 
the equations of motion for free massless fermions with nonzero momentum
only in $d=(3+1)$, the higher dimensional space contributes charges, which are reacher 
the larger is  space, but  kinematics in $d=(3+1)$ are in all such cases the same. We treat 
therefore  only the $d=(5+1)$ case. 

In Table~\ref{Table grassdecuplet.} the basic creation  operators for $d=(5+1)$ case, with
Grassmann space used to describe internal degrees of freedom of "fermions" and "antifermions", 
are presented.  "Fermions" carry in Grassmann space integer spins and correspondingly integer charges.

There are  two independent decuplets (unconnected by ${\cal {\bf S}}^{ab}$).
 \begin{table}
\begin{tiny}
 \begin{center}
 \begin{tabular}{|c|r|r|r|r|r|}
 \hline
$I$&$i$ &$\rm{decuplet\; of\; creation \;operators}\;\; 
\hat{b}^{\theta k \dagger}_{i}$&${\cal {\bf S}}^{03}$&
${\cal {\bf S}}^{1 2}$&
${\cal {\bf S}}^{5 6}$\\
 \hline 
& $1$  & $ (\theta^{0} - \theta^{3}) (\theta^{1} + i \theta^{2})
 (\theta^{5} + i \theta^{6})$ &$ i$&$ 1$&$1$\\
\hline
&$2$  & $ (\theta^{0} \theta^{3} + i \theta^{1} \theta^{2}) 
 (\theta^{5} + i \theta^{6})$ & $ 0$ & $0 $ &$1$\\
\hline
&$3$  & $ (\theta^{0} +  \theta^{3}) (\theta^{1} - i \theta^{2})
  (\theta^{5} + i \theta^{6})$ &$-i $&$-1$&$1$\\
\hline
&$4$  & $ (\theta^{0} -  \theta^{3}) (\theta^{1} - i \theta^{2})
  (\theta^{5} - i \theta^{6})$ &$ i $&$-1$&$-1$\\
\hline
&$5$  & $ (\theta^{0} \theta^{3} - i \theta^{1} \theta^{2}) 
 (\theta^{5} - i \theta^{6})$ & $ 0 $& $0 $&$-1$\\
\hline
&$6$  & $ (\theta^{0} +  \theta^{3}) (\theta^{1} + i \theta^{2})
  (\theta^{5} - i \theta^{6})$ &$-i $&$ 1$&$-1$\\
\hline
& $7$  & $ (\theta^{0} - \theta^{3}) (\theta^{1} \theta^{2} +
 \theta^{5} \theta^{6})$ &$ i$&$ 0$&$0$\\
\hline
& $8$  & $ (\theta^{0} + \theta^{3}) (\theta^{1} \theta^{2} -
 \theta^{5} \theta^{6})$ &$- i$&$ 0$&$0$\\
\hline
& $9$  & $ (\theta^{0}  \theta^{3} +i \theta^{5} \theta^{6}) (\theta^{1}+i \theta^{2}) 
$ &$ 0$&$ 1$&$0$\\
\hline
& $10$  & $ (\theta^{0}  \theta^{3} - i \theta^{5} \theta^{6}) (\theta^{1}-i \theta^{2}) 
$ &$ 0$&$- 1$&$0$\\
\hline\hline 
$II$&$i$ &$\rm{decuplet\; of\; creation \;operators}\;\;
 \hat{b}^{\theta k \dagger}_{i}$&${\cal {\bf S}}^{03}$&${\cal {\bf S}}^{1 2}$&
${\cal {\bf S}}^{5 6}$\\ 
 \hline 
& $1$  & $ (\theta^{0} + \theta^{3}) (\theta^{1} + i \theta^{2})
 (\theta^{5} + i \theta^{6})$ &$- i$&$ 1$&$1$\\
\hline
&$2$  & $ (\theta^{0} \theta^{3} - i \theta^{1} \theta^{2}) 
 (\theta^{5} + i \theta^{6})$ & $ 0$&$ 0 $&$1$\\
\hline
&$3$  & $ (\theta^{0} -  \theta^{3}) (\theta^{1} - i \theta^{2})
  (\theta^{5} + i \theta^{6})$ &$ i $&$-1$&$1$\\
\hline
&$4$  & $ (\theta^{0} +  \theta^{3}) (\theta^{1} - i \theta^{2})
  (\theta^{5} - i \theta^{6})$ &$- i $&$-1$&$-1$\\
\hline
&$5$  & $ (\theta^{0} \theta^{3} + i \theta^{1} \theta^{2}) 
 (\theta^{5} - i \theta^{6})$ & $ 0$& $0 $&$-1$\\
\hline
&$6$  & $ (\theta^{0} -  \theta^{3}) (\theta^{1} + i \theta^{2})
  (\theta^{5} - i \theta^{6})$ &$ i $&$ 1$&$-1$\\
\hline
& $7$  & $ (\theta^{0} + \theta^{3}) (\theta^{1} \theta^{2} +
 \theta^{5} \theta^{6})$ &$- i$&$ 0$&$0$\\
\hline
& $8$  & $ (\theta^{0} - \theta^{3}) (\theta^{1} \theta^{2} -
 \theta^{5} \theta^{6})$ &$ i$&$ 0$&$0$\\
\hline
& $9$  & $ (\theta^{0} \theta^{3} - i \theta^{5} \theta^{6}) (\theta^{1}+i \theta^{2}) 
$ &$ 0$&$ 1$&$0$\\
\hline
& $10$  & $ (\theta^{0}  \theta^{3} + i \theta^{5} \theta^{6}) (\theta^{1}-i \theta^{2}) 
$ &$ 0$&$- 1$&$0$\\
\hline\hline 
 \end{tabular}
 \end{center}
\end{tiny}
 \caption{\label{Table grassdecuplet.} Two decuplets of the basic creation operators  
$\hat{b}^{\theta k \dagger}_{i}$, $k=(I,II), i=(1,\dots,10)$,  of the orthogonal group 
$SO(5,1)$ in Grassmann space are presented. 
The creation operators form "eigenstates" of
the Cartan subalgebra, Eq.~(\ref{cartan}), (${\cal {\bf S}}^{0 3}, {\cal {\bf S}}^{1 2}$, 
${\cal {\bf S}}^{5 6}$ for $SO(5,1)$) with integer spins and charges, defining 
"fermions" and "antifermions". 
The creation operators within each decuplet are
reachable from any member by (a product of) ${\cal {\bf S}}^{ab}$'s (which do not belong to 
the Cartan subalgebra). 
Creation operators $\hat{b}^{\theta k \dagger}_{i}$ and their Hermitian conjugated annihilation 
operators $\hat{b}^{\theta k }_{i}$ fulfill the anticommutation relations for fermions, 
Eq.~(\ref{ijthetaprodordinary}).
The product of the discrete symmetry operators ${\bf \mathbb{C}}_{NG}$ 
and ${\cal P}^{(d-1)}_{NG}$,  Eq.~(\ref{calCPTNG}), 
(${\bf \mathbb{C}}_{NG} {\cal P}^{(d-1)}_{NG}$ $= \gamma^0_{G}  \gamma^5_{G}$
$I_{\vec{x}_{3}}  I_{x^6}$ in $d=(5+1)$) transforms, for example, $\hat{b}^{\theta I \dagger}_{1}$ 
into  $\hat{b}^{\theta I \dagger}_{6}$, $\hat{b}^{\theta I \dagger }_{2}$ $\hat{b}^{\theta I \dagger }_{5}$ 
and $\hat{b}^{\theta I \dagger}_{3}$ into  $\hat{b}^{\theta I \dagger}_{4}$, 
transforming "fermions" with the charge $1$ into "antifermions" with the charge $-1$. 
}
 \end{table}

Both decuplets~\cite{nd2018} of creation operators are of an odd Grassmann character,
representing the second quantized $n =1$ "fermion" states, Eq.~(\ref{ijthetaprod}) of the 
$n$ (any $n$) "fermion" states.
There are, from the point of view of $d=(3+1)$ space, two triplets, one doublet and two singlets 
in each of the two decouplets. 

In Subsect.~\ref{CPT} the discrete symmetry operators in Grassmann space are discussed, 
with the discrete symmetry operators for the case that  "fermions" manifest kinematics only in
 $d=(3+1)$-dimensional space, while the higher dimensions contribute charges, included.

Let us notice that the Grassmann even operator
 ${\bf \mathbb{C}}_{NG} {\cal P}^{(d-1)}_{NG}$,
Eq.~(\ref{calCPTNG}), transforms $\hat{b}^{\theta I \dagger}_{1}$ with 
$p^a= (|p^0|,0,0, |p^3|,0,0)$ 
into the anti-particle state $\hat{b}^{\theta I \dagger}_{6}$,  with  the positive energy 
$|p^0|$ and with $- |p^3|$, for example. Correspondingly  
${\bf \mathbb{C}}_{NG} {\cal P}^{(d-1)}_{NG}$, 
Eq.~(\ref{calCPTNG}), transforms the particle state $\hat{b}^{\theta I \dagger}_{3}$ 
with the positive energy  into the anti-particle state $\hat{b}^{\theta I \dagger}_{4}$ with
the positive energy. All these states belong to the same representation, the same decuplet.

In Eq.~(\ref{weylgenGrass}) the superposition of the creation operators of the two triplets of the 
first decuplet of creation operators --- 
($\hat{b}^{\theta I \dagger}_{1}$, 
$\hat{b}^{\theta I \dagger}_{2}$, $\hat{b}^{\theta I \dagger}_{3}$) ---
which solves  Eq.~(\ref{Weylgrass}) for free massless "fermions" in Grassmann space, 
with the space function $e^{-i p_a x^a}$, $p^a= (p^0,p^1,p^2,p^3,0,0)$, 
Eq.~(\ref{grassordinary}), is presented. Two indexes --- ($ch, s$) --- replace the index $i$,
$ch$ represents the charge ${\cal {\bf S}}^{56}$ and  $s$ the spin ${\cal {\bf S}}^{12}$.

\begin{tiny}
\begin{eqnarray}
\label{weylgenGrass}
&& {\rm Creation \;operators\;for \;"fermion"\; states} \; {\rm in \; 
Grassmann \; space\; for d=(5+1)} \nonumber\\
p^0&=&|p^0|\,,\nonumber\\
\hat{b}^{\theta 1 \dagger}_{1,\,1}\,(\vec{p})  &=&\beta \, 
\{(\frac{1}{\sqrt{2}})^3\, (\theta^0 - \theta^3)  (\theta^1 +i \theta^2)  - 
\frac{2(|p^0| - |p^3|)}{p^1- i p^2} \,(\frac{1}{\sqrt{2}})^2
 (\theta^0  \theta^3 +i \theta^1 \theta^2) \nonumber\\
&& - (\frac{p^1 +i p^2}{ |p^0| + |p^3|})^2 \,(\frac{1}{\sqrt{2}})^3\,
 (\theta^0 + \theta^3)  (\theta^1 - i \theta^2)\}\, (\theta^5 +i \theta^6)
e^{-i(|p^0| x^0 - \vec{p}\cdot\vec{x})}\,,\nonumber\\
\hat{b}^{\theta 2 \dagger}_{1,\,-1}\,(\vec{p})  &=&\beta^*\,
\{(\frac{1}{\sqrt{2}})^3  (\theta^0 + \theta^3)  (\theta^1 - i \theta^2)  - 
\frac{2(|p^0| - |p^3|)}{p^1+ i p^2}\,(\frac{1}{\sqrt{2}})^2 
(\theta^0  \theta^3 +i \theta^1 \theta^2) \nonumber\\
&& - (\frac{p^1 - i p^2}{ |p^0| + |p^3|})^2 \, (\frac{1}{\sqrt{2}})^3
 (\theta^0 - \theta^3)  (\theta^1 + i \theta^2)\}\, (\theta^5 +i \theta^6)
\,
e^{-i (|p^0| x^0 + \vec{p}\cdot\vec{x})}\,,\nonumber\\
&& {\rm Creation \;operators\;for\; "anti-fermion" \; states} \; {\rm in \; 
Grassmann \; space\; for d=(5+1)}\nonumber\\
p^0&=&|p^0|\,,\nonumber\\
\hat{b}^{\theta 3 \dagger}_{- 1,\,1}\,(\vec{p})  &=& \beta \, 
\{(\frac{1}{\sqrt{2}})^3  (\theta^0 + \theta^3)  (\theta^1 + i \theta^2)  - 
\frac{2(|p^0| - |p^3|)}{p^1- i p^2}\,(\frac{1}{\sqrt{2}})^2 
(\theta^0  \theta^3 - i \theta^1 \theta^2) \nonumber\\
&& - (\frac{p^1 +i p^2}{ |p^0| + |p^3|})^2\, (\frac{1}{\sqrt{2}})^3
 (\theta^0 - \theta^3)  (\theta^1 - i \theta^2)\} \,(\theta^5 - i \theta^6)
\,
e^{-i (|p^0| x^0 +\vec{p}\cdot\vec{x})}\,,\nonumber\\
\hat{b}^{\theta 4 \dagger}_{- 1,\,- 1}\,(\vec{p})  &=& \beta^*\, 
\{(\frac{1}{\sqrt{2}})^3 (\theta^0 - \theta^3)  (\theta^1 - i \theta^2)  - 
\frac{2(|p^0| - |p^3|)}{p^1+ i p^2}\,(\frac{1}{\sqrt{2}})^2 
(\theta^0  \theta^3 - i \theta^1 \theta^2) \nonumber\\
&& - (\frac{p^1 -i p^2}{ |p^0| + |p^3|})^2 (\frac{1}{\sqrt{2}})^3
 (\theta^0 + \theta^3)  (\theta^1 + i \theta^2)\}\, (\theta^5 - i \theta^6) \,
e^{-i (|p^0| x^0 -\vec{p}\cdot\vec{x})}\,,
\end{eqnarray}
\end{tiny}

Here $\beta^* \beta= \frac{(|p^0| + |p^3|)^2}{2 (3(p^0)^2 - (p^3)^2)} $. All the 
corresponding states are  orthonormal. 

The corresponding annihilation operators follow from the creation ones by taking into account
Eq.~(\ref{grassher}).
Let us write down, as an example, the annihilation operator partner to the creation operator 
$\hat{b}^{\theta 1 \dagger}_{1,\,1}\,(\vec{p})  $  from Eq.~(\ref{weylgenGrass}). 
Taking into account  Eq.~(\ref{grassher}) (saying that  $\theta^{a \dagger} =\eta^{aa}
\frac{\partial}{\partial \theta_a} = \frac{\partial}{\partial \theta^a}$), it follows
$ \hat{b}^{\theta 1}_{1,\,1}\,(\vec{p})  =
(\frac{1}{\sqrt{2}})^3 \beta^* (\partial_{\theta^5} - i
\partial_{\theta^6}) \Bigl\{ (\partial_{\theta^1} - i \partial_{\theta^2}) 
(\partial_{\theta^0} - \partial_{\theta^3})
 - \frac{2(|p|^0 - |p^3|)}{p^1+ i p^2} \,\sqrt{2}
   (\partial_{\theta^3} \partial_{\theta^0} - i \partial_{\theta^2}
\partial_{\theta^1}) - (\frac{p^1 - i p^2}{ |p^0| + |p^3|})^2 \,
(\partial_{\theta^1} + i \partial_{\theta^2}) (\partial_{\theta^0} +
\partial_{\theta^3})
\Bigr\}  e^{i(|p^0| x^0 - \vec{p}\cdot\vec{x})}$.

The creation and annihilation operators fulfill the 
anti-commutation relations of Eq.~(\ref{ijthetaprodordinary}).

Creation operators $\,\hat{b}^{\theta k \dagger}_{ch,\, s} (\vec{p}) \, e^{-i (p_m x^m)},\, 
m=(0,\cdots,3)$, while $p^5=0=p^6$, generate states, which solve the equation of motion
$(\theta^a - \frac{\partial}{\partial \theta_{a}})\, p_a\, 
\phi^{\theta k}_{ch,\,s} (x^0,\vec{x}) =0$, Eq.~(\ref{Weylgrass}), 
\footnote{The equation $(\theta^a - \frac{\partial}{\partial \theta_{a}}) \,p_a\,
 \phi (\theta, x)=0$
can be rewritten into $- i \tilde{\gamma}^a \,p_a\, \phi=0$, from where the equation
 $\{\tilde{\Gamma}^{(3+1)}\, \hat{p}^{0} = 2 (\tilde{S}^{23}\, \hat{p}^{1} +
\tilde{S}^{31} \,\hat{p}^{2} + \tilde{S}^{12}\,\hat{p}^{3} ) \} \, \phi (\theta, x)$ follows,
leading to the same solutions as presented in Eq.~(\ref{weylgenGrass}). 
Similar relation appears also in the Clifford case.}.

Let be noticed that the second creation operator $ \hat{b}^{\theta 2 \dagger}_{1,- 1}$   follows 
from the first one --- $ \hat{b}^{\theta 1 \dagger}_{1,1}$ ---   by the application of the operator 
${\cal T}_{{\cal NG}}= \gamma^{1}_{G} \, \gamma^{3}_{G}\, K\, I_{x^0} \,
I_{x^5, x^7, \cdots, x^{d-1}}$, Eq.~(\ref{calCPTNG}).
 
When applying on the first two creation operators of  positive charge 
($ \hat{b}^{\theta 1 \dagger}_{1,1}$,  $ \hat{b}^{\theta 2 \dagger}_{1,- 1}$),  
defining the "fermion" states of positive energy, the operator 
$\mathbb{C}_{{ \cal NG}} \cdot {\cal P}^{(d-1)}_{{ \cal NG}} (= \gamma^{0}_{G} \gamma^{5}_{G} 
\gamma^{7}_{G} \cdots \gamma^{d-1}_{G}\,  I_{\vec{x}_3}\,  I_{x^6, x^8, \cdots,x^d})$, 
the third and the fourth creation operators follow, defining the "antifermion" states of negative charge
and positive energy ($ \hat{b}^{\theta  \dagger}_{-1,1}$,  $\hat{b}^{\theta 4 \dagger}_{-1,-1}$).

Solutions of the equation of motion of the second decouplet, and correspondingly the creation 
and annihilation operators, can be obtained in equivalent way.

{\it We learned that states transform 
under the application of the discrete symmetry operators} 
(defined in the Clifford case in Eq.~(\ref{calCPTH}) and Eq.~(17) in Ref.~\cite{nhds}, or Eq.~(10) in 
Ref.~\cite{TDN}, and in the Grassmann case in Eqs.~(\ref{calCPTG}, \ref{calCPTNG}))  
{\it equivalently in the Clifford and in the Grassmann case.}

\subsection{What do we learn from the second quantization procedure in Grassmann 
and in Clifford space?} 

\label{Whatdowelearn}

We proved that in both spaces, in Clifford space and in Grassmann space, the corresponding
creation operators and their Hermitian conjugated annihilation operators of an odd character 
fulfill anticommutation relations as required for fermions, Eqs~(\ref{ijgammaprodordinary}, 
\ref{ijthetaprodordinary}), if operating on an appropriate vacuum state, representing in both spaces
a $n=1$ fermion space out of $n$, any $n$, fermion Hilbert space.  

{\it No postulated creation operators are needed as in ordinary second quantization procedure.}
  
In Clifford space the creation operators are products of odd numbers of $\gamma^a$'s,
arranged into nilpotents and projectors, Eq.~(\ref{signature}), which are the "eigenstates" of the 
Cartan subalgebras of $S^{ab}$, Eq.~(\ref{grapheigen}), generating spins and charges,  
and of $\tilde{S}^{ab}$, generating families,  Eqs.~(\ref{twocliffordsab}, \ref{cliffthetareltheta}). 
In Grassmann space they are products of $\theta^a$, 
arranged in "eigenstates" of the Cartan subalgebra of ${\cal {\bf S}}^{ab}$, 
Eq.~(\ref{Lorentztheta}, \ref{graphicstheta2(2n+1)1})).

While in the Grassmann case the vacuum state is simple, $|\phi_{og}>= |1\,>$, in the 
Clifford case the vacuum state is a sum of $2^{\frac{d}{2}-1}$ products of  projectors, 
Eq.~(\ref{vac1}). 

In $2(2n+1)$-dimensional spaces there are in the Clifford case in one representation 
$2^{\frac{d}{2}-1}$ creation operators. The whole representation is reachable from the (any) 
starting operator by products of  $S^{ab}$, while products of $\tilde{S}^{ab}$ transform 
each of these creation operators into the creation operator of the same family member,
but belonging to another family, Eq.~(\ref{anycreation}). There are correspondingly 
$2^{\frac{d}{2}-1}$ $\cdot$ $2^{\frac{d}{2}-1}$ creation operators, and correspondingly
the same number of states,  reachable by products of $S^{ab}$'s or $\tilde{S}^{ab}$'s or of
 both, $S^{ab}$'s  and $\tilde{S}^{ab}$'s.  Each state follows  by the corresponding creation 
operator on the vacuum state and it is annihilated by its 
Hermitian conjugated operator, Eq.(\ref{graphher}).

In $2(2n+1)$-dimensional spaces there are in the Grassmann case two decoupled representations, 
each with $\frac{1}{2}\,\frac{d!}{\frac{d}{2}! \frac{d}{2}!}$ creation operators, and 
correspondingly with the same number of states. Each state can be obtained by the 
corresponding creation operator operating on the vacuum state 
and any state is annihilated by the corresponding Hermitian conjugated creation operator.  
While all of $2^{\frac{d}{2}-1}\cdot 2^{\frac{d}{2}-1}$ states in Clifford space are 
reachable from any of states by  even Clifford objects, either by products of $S^{ab}$'s or 
by products of $\tilde{S}^{ab}$'s or by products of both, in Grassmann space the two groups 
of representations are decoupled --- no products of ${\cal {\bf S}}^{ab}$'s transform states of 
one group into states of another group.

The creation (annihilation) operators --- which are superposition of the 
creation (annihilation) operators defining the eigenstates of the Cartan subalgebra in the 
internal space, fulfilling the relations of Eqs.~(\ref{ijthetaprodordinary}, \ref{ijgammaprodordinary}),
respectively --- 
form the eigenstates of the equations of motion for free massless "fermions" with integer 
spins and no families in the Grassmann case, Eqs.~(\ref{Weylgrass}, \ref{ptheta}),  and for 
free massless fermions with half integer spins and families in the Clifford case, Eqs.~(\ref{Weyl}, 
\ref{pgamma}).

The number operators  have in both cases the eigenvalues $0$ or $1$, Eqs.~%
(\ref{thetanop}, \ref{gammanop}).

One can as well define in both cases the Hamilton functions, which lead to
the equations of motion in the Grassmann case, Eqs.~(\ref{GrassH0}, \ref{HGrassphieq}), 
and in the Clifford case, Eqs.~(\ref{HCliffpsigen}, \ref{HCliffpsieq}). 
But while in the Clifford case the procedure to find the Hamilton function is the usual one, 
the Grassmann case is not. It remains therefore to understand better the  Hamilton 
function in the Grassmann case.

Comparing solutions for free massless states in a toy model with $d=(5+1)$ from the point 
of view of $d=(3+1)$ (assuming that $p^a= (p^0, p^1, p^2, p^3, 0, \cdots, 0)$)
for the Clifford case and for the Grassmann case, one observes  several similarities. 
The main differences are: {\bf i.} that spins and charges  are in the Clifford case half integer while in
the Grassmann case are integer, {\bf ii.} that Clifford space offers the existence of families, while 
Grassmann space does not, and {\bf iii.} that the requirement that the action is Lorentz invariant lead in
Clifford space to well defined Hamilton function, while in the Grassmann case this point needs 
further study.

We can conclude:
{\bf a.} The --- odd part of the --- Clifford algebra presentation of the internal degrees of freedom 
of fermions offers the $n=1$  second quantized fermion  part of the $n$ second quantized Hilbert 
space,  offering the fermion creation and annihilation operators, fulfilling the required relations,  
explaining therefore the assumption of Dirac about introducing creation and annihilation  operators in 
the second quantized fields. \\
{\bf b.} The {\it spin-charge-family} theory of N.S.M.B., assuming $d\ge (13+1)$-dimensional space
and the Clifford algebra to explain internal degrees of freedom of  fermions, enables to justify the 
assumption of the usual second quantized procedure. The group theory alone, without connecting 
the internal degrees of freedom with the {\it Clifford objects for explaining spins, charges, and families}, 
can not do that.\\
{\bf c.} Table~\ref{Table so13+1.}  demonstrates that any family contains all the fermions and 
antifermions, what in the {\it spin-charge-family} theory means all the quarks and the antiquarks 
and leptons and anileptons, left and right handed. No Dirac sea of negative energy states is needed 
to explain the existence of antifermions. Correspondingly the vacuum state is simple, of an even 
Clifford character, with the sum of all the quantum numbers over the family members equal to zero.\\
{\bf d.} The sum of all the quantum numbers within one family  representation,
but also separately within fermions and separately within antifermions within the same 
representation, is zero.  Also the sum over family quantum numbers is zero.\\
{\bf e.} In the Clifford case the operator $ {\bf \mathbb{C}}_{{\cal N}} {\cal P}^{(d-1)}_{{\cal N}}$, 
Eq.~(\ref{CPTN}), transforms the fermion state into the anti-fermion state. \\
In the Grassmann case it is the operator ${\bf \mathbb{C}}_{NG}  
{\cal P}^{(d-1)}_{NG}$, which transforms the Grassmann "fermion" into the "antifermion".

\section{Conclusions}
\label{conclusions}

We have learned in the present study that both Clifford and  Grassmann space offer the $n=1$ 
fermion second quantized part of vector space, with creation and annihilation operators --- defined as an 
odd products of either Clifford or Grassmann eigenstates of the corresponding Cartan subalgebra
operators in even dimensional  space, Eq.~(\ref{cartan}) --- fulfilling the desired anticommutation 
relations for fermions, Eqs. ~(\ref{ijthetaprodordinary}, \ref{ijgammaprodordinary}). The 
corresponding number operators have the eigenvalues $0$ or $1$ in both cases. Correspondingly the 
second quantization postulate is not needed.

The $1$-fermion second quantized vector space has for a chosen momentum $p^a_k$ in the Clifford 
case $2^{\frac{d}{2}-1}\cdot 2^{\frac{d}{2}-1}$ members --- that is  $2^{\frac{d}{2}-1}$ families,  
each family having $2^{\frac{d}{2}-1}$ members, --- and in the Grassmann case  
$\frac{d!}{\frac{d}{2}! \frac{d}{2}!}$ members --- separated into two by ${\cal {\bf S^{ab}}}$ 
uncoupled representations.  

In both spaces the members of one representation include fermions and 
antifermions and correspondingly there is no need for the Dirac sea of negative energies filled by 
fermions.

In both cases the creation and annihilation operators of different momentum and the same internal 
part represent different creation operators.

The $n$ (any $n$) second quantized vector space of fermions (or "fermions") follows in both cases
as products of $n$ creation operators defining  one fermion states on the corresponding vacuum 
state, $|\psi_{oc}>$, Eq.~(\ref{vac1}), in the Clifford case ($|\psi_{og}>=|\;\;1>$ in the 
Grassmann case), if the creation operators distinguish at least either in one of the quantum 
numbers of the Cartan subalgebra or in momentu $p^a_k$. 

But while in the Clifford case states carry spin and charges from the point of view of $d=(3+1)$ 
in the fundamental representations of the Lorentz group carrying therefore half integer spins, states 
in the Grassmann case are in adjoint representations of the Lorentz group, carrying therefore 
integer spins.

We present in this paper as well the action (Eq.~(\ref{actionWeylGrass},  \ref{Linvariancegrass})), 
describing free massless "fermions" with the internal degrees
of freedom describable in Grassmann space. %
The action leads to the equations of motion (Eq.~(\ref{Weylgrass})), analogous to the Weyl 
equation in Clifford space  (Eq.~(\ref{Weyl})), fulfilling as well the Klein-Gordon equation  
(Eq.~(\ref{LtoKGgrass})).   We also present the discrete symmetry operators in the Grassmann
case.

Since the Clifford objects $\gamma^a$ and $\tilde{\gamma}^a$ are expressible with the Grassmann 
coordinates $\theta^a$ and their conjugate moments $\frac{\partial}{\partial \theta^a}$ ---
$\gamma^a=(\theta^a + \frac{\,\partial}{\partial \theta_a})$, 
$\tilde{\gamma}^a= i (\theta^a - \frac{\,\partial}{\partial \theta_a}) $, Eq.~(\ref{cliffthetareltheta})
--- either basic 
states in Grassmann space, Eq.~(\ref{grassmannboson}), or basic states in Clifford space, 
Eq.~(\ref{start(2n+1)2cliff}), can be normalized with the same integral, Eq.~(\ref{grassintegral},
\ref{grassnorm}, \ref{grassgammanorm}).

To understand better the difference in the description of the fermion internal degrees of freedom  
either with Clifford or with Grassmann  space, let us replace in the starting action of the 
{\it spin-charge-family} theory, Eq.~(\ref{wholeaction}), using the Clifford algebra to describe 
fermion degrees of freedom, the covariant momentum $p_{0a}= f^{\alpha}{}_a$  
$p_{0\alpha}$,  $p_{0\alpha}=  p_{\alpha}  -  \frac{1}{2}  S^{ab} \omega_{ab \alpha} - 
\frac{1}{2}  \tilde{S}^{ab}   \tilde{\omega}_{ab \alpha}$, 
with $p_{0\alpha} =  p_{\alpha}  - \frac{1}{2}\, {\cal {\bf S}}^{ab} \Omega_{ab \alpha}$, where 
${\cal {\bf S}}^{ab} = S^{ab}  + \tilde{S}^{ab}$, Eq.~(\ref{Lorentztheta}), and 
$\Omega_{ab \alpha}$ are  the spin connection gauge fields of $ {\cal {\bf S}}^{ab}$ (which are 
the generators of the Lorentz transformations in Grassmann space), while 
$f^{\alpha}{}_{a}\,p_{0\alpha}$ replaces the ordinary  momentum when massless objects start to 
interact with the gravitational field through the vielbeins and the spin connections. 
Let us add that it follows, if varying the action with respect to either $\omega_{ab \alpha}$ or 
$\tilde{\omega}_{ab \alpha}$ when no fermions are present, that both spin connections 
are uniquely determined by the vielbeins~(\cite{normaJMP2015,IARD2016,nd2017} and references 
therein) and correspondingly in this particular case $\Omega_{ab \alpha} =\omega_{ab \alpha}= $ 
$\tilde{\omega}_{ab \alpha}$.

The present study was stimulated by one of the author in order to better understand whether and to
which extend the {\it spin-charge-family} theory does offer the next step to both {\it standard models} 
--- the one of the fermion and boson fields and the cosmological one. Correspondingly we present in
Subsect.~\ref{SCFT} of the introductory Sect.~\ref{introduction}, the so far achievements 
of the  {\it spin-charge-family} theory as well as the open problems of this theory, both suggested 
by the referees.

In shortly,  the {\it spin-charge-family} theory (using Clifford objects to describe the 
internal space of fermions)  offers, while starting with the simple action 
in $d\ge (13 +1)$ with fermions interacting with gravity only (the vielbeins and the two kinds 
of the spin connection fields, the gauge fields of moments and the generators of the Lorentz 
transformations $S^{ab}$ and $\tilde{S}^{ab}$, respectively),
Eq.~(\ref{wholeaction}), the explanation for all the assumptions 
of the {\it standard model} --- for quarks and leptons, antiquarks and antileptons, for fermion families, 
for the vector gauge fields, for the scalar Higgs and Yukawa couplings --- explaining also the 
phenomena like the existence of the 
dark matter~\cite{gn2009}, of the matter-antimatter asymmetry~\cite{n2014matterantimatter}, 
offering correspondingly the next step beyond both standard models --- cosmological one and the one 
of the elementary fields, Sect.~\ref{SCFT}.  This theory predicts the fourth family to the observed 
three, Sect.~\ref{SCFT}, and the new scalar fields, some of those which explains the properties of
the observed Higgs and Yukawa couplings, Sect.~\ref{SCFT}, 
and which will be observed at the LHC and other experiments in the future. This theory predicts also 
the existence of the stable fifth family, manifesting the dark matter and with the "new nuclear"
force among the hadrons of these much heavier families, Sect.~\ref{SCFT}. 

To these achievements the present study adds the recognition that the creation operators for 
one fermion states are in Clifford space already second quantized, and that the creation 
operators for any  $n$ fermion second quantized vectors are  products of one fermion creation 
operators, operating on the empty vacuum state. The {\it spin-charge-family} theory namely
describes all the internal degrees of freedom of fermions in Clifford space --- spins and charges. 

There is in this theory no need for the existence of the negative energy states filled with fermions.

The most severe among the open problems of the {\it spin-charge-family} theory is the quantization of 
gravity gauge fields, although the  {\it spin-charge-family} theory is explaining the phenomena in 
the low energy regime  where all the vector and scalar gauge fields can be quantized in the known 
procedure. There are also other open problems, some of them needing only time to be solved, 
presented in Sect.~\ref{SCFT}.

The second quantization of "fermions" with the internal degrees of freedom described in Grassmann
space might help to understand better the properties of scalars and vectors in the 
{\it spin-charge-family} theory.

\appendix 
\section{Creation and annihilation operators in Grassmann and Clifford space for $d=4n$}
\label{4n}
We discuss in Subsect.~\ref{secondquantization} mainly cases with $d=2(2n+1)$, since 
if assuming no conserved charges in the fundamental theory with fermions, which carry only spins 
and interact with only the gravity --- as the {\it spin-charge-family} theory assumes --- the dimensions 
$4n$, $n$ is positive integer, as well as all odd dimensions, are excluded under the requirement of 
mass protection~\cite{NH2007Majorana}.

Let us nevertheless add in this appendix comments on the second quantization procedure  in 
$d=4n$ spaces. 

\vspace{3mm}

$\;\;${\bf i.} {\bf Grassmann space}

In Eq.~(\ref{start(2n+1)2theta}) we define in Grassmann space a possible starting creation operator 
for $d=2(2n+1)$ spaces. In $d=4n$ we correspondingly start with the state 
\begin{eqnarray}
|\phi^{ 1}_{1}> &=& b^{\theta 1 \dagger}_{1}  |1\,>\,,\nonumber\\
b^{\theta 1 \dagger}_{1} &=&(\frac{1}{\sqrt{2}})^{\frac{d}{2}-1} \,
(\theta^0 - \theta^3) (\theta^1 + i \theta^2) (\theta^5 + i \theta^6) \cdots (\theta^{d-3} +
 i \theta^{d-2})  \theta^{d-1} \theta^d\,,
\label{start4ntheta}
\end{eqnarray}
generated by the creation operator $b^{\theta 1 \dagger}_{1}$, which is, as it ought to be --- like in the 
$d=2(2n+1)$ case --- of an odd Grassmann character to fulfill the anticommutation relations for 
fermions, Eq.~(\ref{ijthetaprodordinary}). Again the rest of states, belonging to the same Lorentz
representation, follow from 
the starting state by the application of the operators ${\cal {\bf S}}^{cf}$, which do not belong to
the Cartan subalgebra operators. Their annihilation partners follow by Hermitian conjugation. 

One finds therefore for the (chosen) starting creation and the corresponding annihilation operator 
\begin{eqnarray}
\hat{b}^{\theta 1 \dagger}_{1} &=& (\frac{1}{\sqrt{2}})^{\frac{d}{2}-1} \,(\theta^0 - \theta^3) 
(\theta^1 + i \theta^2) (\theta^5 + i \theta^6) \cdots (\theta^{d-3} + i \theta^{d-2}) 
\theta^{d-1} \theta^{d}\,,
 \nonumber\\
\hat{b}^{\theta 1}_{1} &=& (\frac{1}{\sqrt{2}})^{\frac{d}{2}-1}\,\frac{\partial}{\;\partial \theta^{d}}
\frac{\partial}{\;\partial \theta^{d-1}} (\frac{\partial}{\;\partial \theta^{d-3}} -
 i \frac{\partial}{\;\partial \theta^{d-2}})\cdots (\frac{\partial}{\;\partial \theta^{0}}
-\frac{\partial}{\;\partial \theta^3})\,,\nonumber\\
&&  d=4n\,.
\label{graphicstheta4n1}
\end{eqnarray}

The application of ${\cal {\bf S}}^{01}$, for example, generates 
\begin{eqnarray}
\hat{b}^{\theta 1 \dagger}_{2} &=& (\frac{1}{\sqrt{2}})^{\frac{d}{2}-2} \,
(\theta^0  \theta^3 +i \theta^1 \theta^2)  (\theta^5 + i \theta^6) \cdots
 (\theta^{d-3} + i \theta^{d-2})\; \theta^{d-1}  \theta^{d}\,,\nonumber\\
\hat{b}^{\theta 1}_{2} &=&   (\frac{1}{\sqrt{2}})^{\frac{d}{2}-2}
 \,\frac{\partial}{\;\partial \theta^{d}} \frac{\partial}{\;\partial \theta^{d-1}}\;
(\frac{\partial}{\;\partial \theta^{d-3}} - i \frac{\partial}{\;\partial \theta^{d-2}})
\cdots (\frac{\partial}{\;\partial \theta^{3}} \,\frac{\partial}{\;\partial \theta^0}
-i \frac{\partial}{\;\partial \theta^{2}} \,\frac{\partial}{\;\partial \theta^1})\,. 
\label{graphicstheta4n2}
\end{eqnarray}
%
There is the additional group of creation and annihilation operators in $d=4n$, which follows from the
starting creation operator $\hat{b}^{\theta 2 \dagger}_{1}$
\begin{eqnarray}
\hat{b}^{\theta 2 \dagger}_{1} &=& (\frac{1}{\sqrt{2}})^{\frac{d}{2}-1} (\theta^0+\theta^3) 
(\theta^1+i\theta^2) (\theta^5+i\theta^6)
 \cdots  (\theta^{d-3}+i\theta^{d-2})\;\theta^{d-1}\theta^d\,,\nonumber\\
\hat{b}^{\theta 2}_{1} &=& (\hat{b}^{\theta 2 \dagger}_{1})^{\dagger}= 
(\frac{1}{\sqrt{2}})^{\frac{d}{2}-1}\,\frac{\partial}{\;\partial \theta^{d}}
\frac{\partial}{\;\partial \theta^{d-1}} (\frac{\partial}{\;\partial \theta^{d-3}} -
 i \frac{\partial}{\;\partial \theta^{d-2}})\cdots (\frac{\partial}{\;\partial \theta^{0}}
+\frac{\partial}{\;\partial \theta^3})\,,\nonumber\\
{\rm for}\; d=4n\,.
\label{startgrass024n}
\end{eqnarray}
All the rest of creation operators follow from the starting creation operator of each of the two 
groups by the (left) application of products of $ {\cal {\bf S}}^{ab}$ 
\begin{eqnarray}
\label{general24n}
\hat{b}^{\theta k \dagger}_{i} &\propto& {\cal {\bf S}}^{ab} \cdots {\cal {\bf S}}^{ef}
\hat{b}^{\theta k \dagger}_{1}\,,\nonumber\\
\hat{b}^{\theta k}_{i} &=& (\hat{b}^{\theta 2 \dagger}_{i})^{\dagger}\,,\quad k=1,2\,.
\end{eqnarray}

{\it Only creation and annihilation operators with an odd Grassmann character, fulfill, applied 
on the vacuum state $|1\,>$, the anticommutation relations required for fermions}, 
Eq.~(\ref{ijthetaprod}).

\vspace{3mm}

$\;\;${\bf i.} {\bf Clifford space}

In Eq.~(\ref{start(2n+1)2cliff}) we define in Clifford space a possible starting creation operator 
for $d=2(2n+1)$ spaces. In $d=4n$ we correspondingly start with the state  with an odd number
 of nilpotents and with one projector
\begin{eqnarray}
|\psi^{1}_{1}>&=&{\hat b}^{1\dagger}_{1} |\psi_{oc}>\,,\nonumber\\
{\hat b}^{1 \dagger}_{1}:&=& \stackrel{03}{(+i)} \stackrel{12}{(+)} \stackrel{35}{(+)}\cdots 
\stackrel{d-3\;d-2}{(+)}\;\;\stackrel{d-1\;d}{[+]}\,,\nonumber\\
{\hat b}^{1}_{1}&=& ({\hat b}^{1 \dagger}_{1})^{\dagger}= \stackrel{d-1\;d}{[+]} \;\;
\stackrel{d-3\;d-2}{(-)}\cdots  \stackrel{35}{(-)}  \stackrel{12}{(-)}  \stackrel{01}{(-i)}
\label{start4ncliff}
\end{eqnarray}
All the other creation operators, creating all the members of the representation of this particular family,
are obtainable by the application of products of $S^{ab}$ on this creation operator from the left hand 
side. There are $2^{\frac{d}{2}-1}$ members of each family. All the other families follows from
the starting one by the application of products of $\tilde{S}^{ab}$. There are  $2^{\frac{d}{2}-1}$
families with  $2^{\frac{d}{2}-1}$ members each.

A general creation operator in $d=4n$ follows by the application of $S^{ab}$ and $\tilde{S}^{ab}$ on the 
starting  creation operator of Eq.~(\ref{start4ncliff})  and the corresponding annihilation operator is its 
Hermitian conjugated value.

Correspondingly we define $\hat{b}^{\alpha \dagger}_{i}$ (up to a constant) to be 
\begin{eqnarray}
\hat{b}^{\alpha \dagger}_i&\propto & \tilde{S}^{ab} \cdots \tilde{S}^{ef}{S}^{mn}\cdots {S}^{pr}
\hat{b}^{1\dagger}_{1}\nonumber\\
&\propto & {S}^{mn}\cdots {S}^{pr} \hat{b}^{1\dagger}_{1} {S}^{ab} \cdots {S}^{ef}\,,
\nonumber\\
\hat{b}^{\alpha}_i = (\hat{b}^{\alpha \dagger}_i)^{\dagger} &\propto & {S}^{ef} \cdots 
{S}^{ab}\hat{b}^{1}_{1} {S}^{pr}\cdots {S}^{mn}\,, \nonumber\\
&& d=4n\,.
\label{general4n}
\end{eqnarray}
These creation and annihilation operators --- again of an odd Clifford character in $4n$ ---
fulfill the anticommutation relations of Eq.~(\ref{ijgammaprodordinary}), if applied on the 
vacuum state of Eq.~(\ref{vac1}), 
\begin{eqnarray}
|\psi_{oc}>&=& \stackrel{03}{[-i]} \stackrel{12}{[-]} \stackrel{35}{[-]}\cdots
\stackrel{d-3\;d-2}{[-]}\stackrel{d-1\;d}{[+]} + \stackrel{03}{[+i]} \stackrel{12}{[+]} \stackrel{56}{[-]}
\cdots \stackrel{d-3\;d-2}{[-]} \;\,\stackrel{d-1\;d}{[+]} + \cdots |0>\,, \quad  \nonumber\\
&& d=4n,
\label{vac14n}
\end{eqnarray}
$n$ is a positive integer. There are $2^{\frac{d}{2}-1}$ summands, since we step by step replace all 
possible pairs of $\stackrel{ab}{[-]}  \cdots \stackrel{ef}{[-]} $ in the starting 
part $\stackrel{03}{[-i]} 
\stackrel{12}{[-]} \stackrel{35}{[-]}\cdots \stackrel{d-3\;d-2}{[-]}\stackrel{d-1\;d}{[+]}$ into 
$\stackrel{ab}{[+]}  \cdots \stackrel{ef}{[+]} $ and include new terms into the vacuum state 
so that the last $2n+1$ summand has for  $d=4n$ 
also the factor $\stackrel{d-1\;d}{[+]}$ in the starting term  $\stackrel{03}{[-i]} 
\stackrel{12}{[-]} \stackrel{35}{[-]}\cdots \stackrel{d-3\;d-2}{[-]}\stackrel{d-1\;d}{[+]}$ changed into 
 $\stackrel{d-1\;d}{[-]}$. The vacuum state has then the normalization factor $1/\sqrt{
 2^{d/2-1}}$.\\

\section{Lorentz algebra and representations in Grassmann and Clifford space}
\label{Lorentz}
The Lorentz transformations of vector components $\theta^a$, $\gamma^a$, or 
$ \tilde{\gamma}^a$ ---  usable for the description of the internal degrees of freedom of fermion 
fields obeying in the second quantization the anticommutation relations for fermions --- and of 
vector components  $x^{a}$, which are 
real (ordinary) commuting coordinates,   
$ \theta'^{a} = \Lambda^{a}{}_b\, \theta^b$, $\,\; \gamma'^{a} = \Lambda^{a}{}_b\,
 \gamma^b$, $\,\; \tilde{\gamma}'^{a} = \Lambda^{a}{}_b\, \tilde{\gamma}^b\;\,$ and 
$ x^a=\Lambda^{a}{}_b \,x^b$,
leave forms $a_{a_{1} a_{2} ... a_{i}}$ $\theta^{a_1} \theta^{a_2} \dots \theta^{a_i}$, 
$\;\,a_{a_{1} a_{2} ... a_{i}}$ $\gamma^{a_1} \gamma^{a_2} \dots \gamma^{a_i}$,
$\;\,a_{a_{1} a_{2} ... a_{i}}$ $\tilde{\gamma}^{a_1} \tilde{\gamma}^{a_2} \dots
\tilde{\gamma}^{a_i}$ and 
$\;\,b_{a_{1} a_{2} \dots a_{i}} \,x^{a_{1}} x^{a_{2}} \dots x^{a_{i}}$, $\;\,i=(1,\dots,d),\,\;$
invariant. 

While $b_{a_{1} a_{2} \dots a_{i}}$ ($=\eta_{a_{1} b_{1}} \eta_{a_{2} b_{2}} 
\dots \eta_{a_{i} b_{i}} \,b^{b_{1} b_{2} \dots b_{i}} $) is a symmetric tensor field,  
$a_{a_{1} a_{2} \dots a_{i}}$ ($=  \eta_{a_{1} b_{1}} \eta_{a_{2} b_{2}} 
\dots \eta_{a_{i} b_{i}}\, a^{b_{1} b_{2} \dots b_{i}}$) are antisymmetric  
{\it Kalb-Ramond} fields.

The requirements:  $x^{'a}\,x^{'b}\eta_{ab} = $ $x^{c}\,
x^{d}\eta_{cd}$,  $\theta'^a  \theta'^b \varepsilon_{ab} = $ 
$\theta^c  \theta^d \varepsilon_{cd}$,  $\gamma'^a  \gamma'^b \varepsilon_{ab} = $ 
$\gamma^c  \gamma^d \varepsilon_{cd}$ and $\tilde{\gamma}'^a  \tilde{\gamma}'^b 
\varepsilon_{ab} = $ $\tilde{\gamma}^c  \tilde{\gamma}^d \varepsilon_{cd}$
 lead to $\Lambda^a{}_b \,\Lambda^c{}_d \,\eta_{ac}= \eta_{bd}$.
Here 
  $\eta^{ab}$ (in our case $\eta^{ab}= diag(1,-1,-1,\dots,-1)$) is  the metric tensor lowering
the indexes of vectors ($\{x^{a}\}$ $= \eta^{ab} x_b$, $\{\theta^a \}$ $= 
\eta^{ab}\, \theta_b$, $\{\gamma^a \}$ $=\eta^{ab}\, \gamma_b$ and 
$\{\tilde{\gamma}^a \}$ $=\eta^{ab}\, \tilde{\gamma}_b$) and 
$\varepsilon_{ab}$ is the antisymmetric tensor.
 An infinitesimal Lorentz
transformation for the case with $\det \Lambda=1, \Lambda^0{}_0\ge 0$ can be written as 
$\Lambda^a{}_b = \delta^{a}_{b} + \omega^a{}_{b}$, where $\omega^a{}_{b} + 
\omega_{b}{}^{a}=0$.

In Eqs.~(\ref{cliffthetareltheta}, \ref{sabspinall})
the commutation relations among the above objects are presented.
\subsection{Lorentz properties of basic vectors}
\label{Lorentzbasicvectors}

What follows is taken from Ref.~\cite{norma93} and Ref.~\cite{normaJMP2015}, Appendix B.

Let us first repeat some properties of the anticommuting Grassmann and Clifford coordinates,
taking into account Eqs.~(\ref{thetarelcom},\ref{cliffthetareltheta}).
An infinitesimal Lorentz transformation of the proper ortochronous Lorentz group is then
\begin{eqnarray}
\label{inflorentz}
\delta \theta^c &=& - \frac{i}{2} \omega_{ab} {\cal {\bf S}}^{ab} \theta^c=
 \omega^{c}{}_{a}  \theta^a\,,\nonumber\\
\delta \gamma^c &=& - \frac{i}{2} \omega_{ab} S^{ab} \gamma^c= 
\omega^{c}{}_{a}  \gamma^a\,,\nonumber\\
\delta \tilde{\gamma}^c &=& - \frac{i}{2} \omega_{ab} \tilde{S}^{ab}  
\tilde{\gamma}^c= \omega^{c}{}_{a}   \tilde{\gamma}^a\,,\nonumber\\
\delta x^c &=& - \frac{i}{2} \omega_{ab} L^{ab} x^c= \omega^{c}{}_{a} x^a \,,
\end{eqnarray}
where $\omega_{ab}$ are parameters of a transformation and $\gamma^a$ and  
$\tilde{\gamma}^a$ are expressible  by $\theta^a$ and $\frac{\partial}{\partial \theta_a}$ in
Eqs.~(\ref{thetarelcom},\ref{cliffthetareltheta}). 

Let us write the operator of finite Lorentz transformations as follows
\begin{eqnarray}
\label{finitelorentz0}
{\cal {\bf S }} &=& e^{- \frac{i}{2} \omega_{ab} ({\cal {\bf S}}^{ab}+L^{ab})} \,,
\end{eqnarray}
${\cal {\bf S}}^{ab}$ have to be replaced by $S^{ab}$ and $\tilde{S}^{ab}$ in the Clifford case.
We see that the Grassmann  $\theta^a$ and the ordinary $x^a$  coordinates and  the Clifford
objects  $\gamma^a$ and $\tilde{\gamma}^a$ transform as vectors 
\begin{eqnarray}
\label{finitelorentz1}
\theta'^c &=& e^{- \frac{i}{2} \omega_{ab} ({\cal {\bf S}}^{ab}+L^{ab})}\, \theta^c \, 
e^{ \frac{i}{2}\, \omega_{ab} ({\cal {\bf S}}^{ab}+L^{ab})} \nonumber\\
&=& \theta^c - \frac{i}{2} \omega_{ab} \{{\cal {\bf S}}^{ab},\theta^c\}_{-} + \dots = 
\theta^c + \omega^{c}{}_{a} \theta^a + \dots = \Lambda^{c}{}_{a} \theta^a \,,\nonumber\\
x'^c &=&
\Lambda^{c}{}_{a} x^a \,,\quad \gamma'^{c}=
\Lambda^{c}{}_{a} \gamma^a \,,\quad \tilde{\gamma}'^{c}=
\Lambda^{c}{}_{a} \tilde{\gamma}^a \,.
\end{eqnarray}
Correspondingly one finds that  compositions like $\gamma^a p_{a}$ and 
$\tilde{\gamma}^a p_{a}$,  here $p_{a}$ are $p^{{\small x}}_{a}$ 
($=i\frac{\partial}{\partial x^{a}}$),  
 transform as scalars (remaining  invariants), while $ S^{ab}\, \omega_{abc}$ and 
$\tilde{S}^{ab}\, \tilde{\omega}_{abc}$ transform as vectors. 
%
Objects like $R=  \frac{1}{2} \, f^{\alpha [ a} f^{\beta b ]} \;(\omega_{a b \alpha, \beta} 
- \omega_{c a \alpha}\,\omega^{c}{}_{b \beta})$ and   
$\tilde{R} = \frac{1}{2}\,   f^{\alpha [ a} f^{\beta b ]} \;(\tilde{\omega}_{a b \alpha,\beta} - 
\tilde{\omega}_{c a \alpha} \tilde{\omega}^{c}{}_{b \beta})$ from Eq.~(\ref{wholeaction})
transform with respect to the Lorentz transformations as scalars.

Making a choice of the Cartan subalgebra set of the algebra ${\cal {\bf S}}^{ab}$, $ S^{ab}$ 
and $\tilde{S}^{ab}$, Eqs.~(\ref{twocliffordsab}, \ref{Lorentztheta}, \ref{Lorentzthetacom}),
\begin{eqnarray}
{\cal {\bf S}}^{03}, {\cal {\bf S}}^{12}, {\cal {\bf S}}^{56}, \cdots, {\cal {\bf S}}^{d-1 \;d}\,, \nonumber\\
S^{03}, S^{12}, S^{56}, \cdots, S^{d-1 \;d}\,, \nonumber\\
\tilde{S}^{03}, \tilde{S}^{12}, \tilde{S}^{56}, \cdots,  \tilde{S}^{d-1\; d}\,,
\label{cartan}
\end{eqnarray}
one can arrange the basic vectors so that they are eigenstates of the Cartan subalgebra, belonging
to representations of ${\cal {\bf S}}^{ab}$, or of $ S^{ab}$ and $\tilde{S}^{ab}$, with $ab$ 
from Eq~(\ref{cartan}).

\section{Technique to generate spinor representations in terms of Clifford algebra objects}
\label{technique}

Here we briefly repeat the main points of the technique for generating spinor representations from 
Clifford algebra objects, following Ref.~\cite{norma93,nh02}. We advise the reader to look for details and 
proofs in these references. No requirements for the  second quantization is taken  
into account.

We assume the objects $\gamma^a$, Eq.~(\ref{cliffthetareltheta}), which fulfill the Clifford algebra
relations of Eq.~(\ref{twocliffordsab}),  
%
%
$\{\gamma^a, \gamma^b \}_+ = I \cdot 2\eta^{ab}\,, {\rm for} \; a,b  \in 
\{0,1,2,3,5,\cdots,d \}\,$,
%
for any $d$, even or odd.  $I$ is the unit element in the Clifford algebra, while
 $\{\gamma^a, \gamma^b \}_{\pm} = \gamma^a \gamma^b \pm \gamma^b \gamma^a $.

The ``Hermiticity'' property for $\gamma^a$'s and $\tilde{\gamma}^a$'s, Eq.~(\ref{cliffher}), follows 
from  Eq.~(\ref{grassher}), 
%
$\gamma^{a\dagger} = \eta^{aa} \gamma^a$ $\tilde{\gamma}^{a\dagger} = \eta^{aa} 
\tilde{\gamma}^a $\,,
%
 leading to $\gamma^{a \dagger} \gamma^a=I$, $\tilde{\gamma}^{a \dagger} 
\tilde{\gamma}^a=I$. 

The Clifford algebra objects 
%
$S^{ab}$  
 close the Lie algebra of the Lorentz group 
$ \{S^{ab},S^{cd}\}_- = i (\eta^{ad} S^{bc} + \eta^{bc} S^{ad} - \eta^{ac} S^{bd} - 
\eta^{bd} S^{ac})$, Eq.~(\ref{Lorentzthetacom}).
One finds from Eq.(\ref{cliffher}) that $(S^{ab})^{\dagger} = \eta^{aa} \eta^{bb}S^{ab}$ and
 that $\{S^{ab}, S^{ac} \}_+= \frac{1}{2} \eta^{aa} \eta^{bc}$.

Recognizing 
that two Clifford algebra objects ($S^{ab}, S^{cd}$) with all indexes different 
commute, we  select (out of many possibilities) the Cartan subalgebra set of the algebra 
of the Lorentz group  of Eq.~(\ref{cartan}) 
%

Let us present the operators of subgroups of the $SO(13+1)$ group
\begin{eqnarray}
\label{so1+3}
\vec{N}_{\pm}(= \vec{N}_{(L,R)}): &=& \,\frac{1}{2} (S^{23}\pm i S^{01},S^{31}\pm i S^{02}, 
S^{12}\pm i S^{03} )\,,
\end{eqnarray}
 \begin{eqnarray}
 \label{so42}
 \vec{\tau}^{1}:=\frac{1}{2} (S^{58}-  S^{67}, \,S^{57} + S^{68}, \,S^{56}-  S^{78} )\,,\;\;
 \vec{\tau}^{2}:= \frac{1}{2} (S^{58}+  S^{67}, \,S^{57} - S^{68}, \,S^{56}+  S^{78} )&&\,,
 \end{eqnarray}
  \begin{eqnarray}
 \label{so64}
 \vec{\tau}^{3}: = &&\frac{1}{2} \,\{  S^{9\;12} - S^{10\;11} \,,
  S^{9\;11} + S^{10\;12} ,\, S^{9\;10} - S^{11\;12} ,\nonumber\\
 && S^{9\;14} -  S^{10\;13} ,\,  S^{9\;13} + S^{10\;14} \,,
  S^{11\;14} -  S^{12\;13}\,,\nonumber\\
 && S^{11\;13} +  S^{12\;14} ,\, 
 \frac{1}{\sqrt{3}} ( S^{9\;10} + S^{11\;12} - 
 2 S^{13\;14})\}\,,\nonumber\\
 \tau^{4}: = &&-\frac{1}{3}(S^{9\;10} + S^{11\;12} + S^{13\;14})\,.
 \end{eqnarray}
 \begin{eqnarray}
  \label{YQY'Q'andtilde}
  Y:= \tau^{4} + \tau^{23}\,,\;\; Y':= -\tau^{4}\tan^2\vartheta_2 + \tau^{23}\,,\;\;
  Q: =  \tau^{13} + Y\,,\;\; Q':= -Y \tan^2\vartheta_1 + \tau^{13}&&\,.
  \end{eqnarray}
The equivalent expressions for the group $\widetilde{SO}(13,1)$ follows from the above one, if 
replacing $S^{ab}$ by $\tilde{S}^{ab}$.

To make the technique simple, we introduce the graphic representation, \cite{nh02},
 Eq.~(\ref{signature}),
 \begin{eqnarray}
\stackrel{ab}{(k)}:&=& 
\frac{1}{2}(\gamma^a + \frac{\eta^{aa}}{ik} \gamma^b)\, , \nonumber\\
\stackrel{ab}{[k]}: &=&
\frac{1}{2}(1+ \frac{i}{k} \gamma^a \gamma^b)\,, \nonumber
\end{eqnarray}
where $k^2 = \eta^{aa} \eta^{bb}$.
One can easily check by taking into account the Clifford algebra relation (Eqs.~(\ref{cliffthetareltheta}, 
\ref{grassher})) and the
definition of $S^{ab}$ (Eq.~(\ref{twocliffordsab}))
that if one multiplies from the left hand side by $S^{ab}$ the Clifford algebra objects $\stackrel{ab}{(k)}$
and $\stackrel{ab}{[k]}$,
it follows that, Eq.~(\ref{grapheigen}),
%
$S^{ab}\stackrel{ab}{(k)}=\frac{1}{2}k \stackrel{ab}{(k)}$,
$S^{ab}\stackrel{ab}{[k]}=\frac{1}{2}k \stackrel{ab}{[k]}$.
%
This means that
$\stackrel{ab}{(k)}$ and $\stackrel{ab}{[k]}$ acting from the left hand side on the
vacuum state $|\psi_{oc}\rangle$), Eqs.~(\ref{vac1}, \ref{vac14n}) for $d=2(2n+1)$ and 
$d=4n$ respectively, are eigenvectors of $S^{ab}$.

We further find 
\begin{eqnarray}
\gamma^a \stackrel{ab}{(k)}&=&\eta^{aa}\stackrel{ab}{[-k]},\nonumber\\
\gamma^b \stackrel{ab}{(k)}&=& -ik \stackrel{ab}{[-k]}, \nonumber\\
\gamma^a \stackrel{ab}{[k]}&=& \stackrel{ab}{(-k)},\nonumber\\
\gamma^b \stackrel{ab}{[k]}&=& -ik \eta^{aa} \stackrel{ab}{(-k)}\,.
\label{graphgammaaction}
\end{eqnarray}
It follows that
$S^{ac}\stackrel{ab}{(k)}\stackrel{cd}{(k)} = -\frac{i}{2} \eta^{aa} \eta^{cc} 
\stackrel{ab}{[-k]}\stackrel{cd}{[-k]}$, 
$S^{ac}\stackrel{ab}{[k]}\stackrel{cd}{[k]} = \frac{i}{2} \stackrel{ab}{(-k)}\stackrel{cd}{(-k)}$, 
$S^{ac}\stackrel{ab}{(k)}\stackrel{cd}{[k]} = -\frac{i}{2} \eta^{aa}  
\stackrel{ab}{[-k]}\stackrel{cd}{(-k)}$, 
$S^{ac}\stackrel{ab}{[k]}\stackrel{cd}{(k)} = \frac{i}{2} \eta^{cc}  
\stackrel{ab}{(-k)}\stackrel{cd}{[-k]}$.

It is useful to deduce the following relations

\begin{eqnarray}
\stackrel{ab}{(k)}\stackrel{ab}{(k)}& =& 0\,, \quad \quad \stackrel{ab}{(k)}\stackrel{ab}{(-k)}
= \eta^{aa}  \stackrel{ab}{[k]}\,, \quad \stackrel{ab}{(-k)}\stackrel{ab}{(k)}=
\eta^{aa}   \stackrel{ab}{[-k]}\,,\quad
\stackrel{ab}{(-k)} \stackrel{ab}{(-k)} = 0\,, \nonumber\\
\stackrel{ab}{[k]}\stackrel{ab}{[k]}& =& \stackrel{ab}{[k]}\,, \quad \quad
\stackrel{ab}{[k]}\stackrel{ab}{[-k]}= 0\,, \;\;
\quad \quad  \quad \stackrel{ab}{[-k]}\stackrel{ab}{[k]}=0\,,
 \;\;\quad \quad \quad \quad \stackrel{ab}{[-k]}\stackrel{ab}{[-k]} = \stackrel{ab}{[-k]}\,,
 \nonumber\\
\stackrel{ab}{(k)}\stackrel{ab}{[k]}& =& 0\,,\quad \quad \quad \stackrel{ab}{[k]}\stackrel{ab}{(k)}
=  \stackrel{ab}{(k)}\,, \quad \quad \quad \stackrel{ab}{(-k)}\stackrel{ab}{[k]}=
 \stackrel{ab}{(-k)}\,,\quad \quad \quad 
\stackrel{ab}{(-k)}\stackrel{ab}{[-k]} = 0\,,
\nonumber\\
\stackrel{ab}{(k)}\stackrel{ab}{[-k]}& =&  \stackrel{ab}{(k)}\,,
\quad \quad \stackrel{ab}{[k]}\stackrel{ab}{(-k)} =0\,,  \quad \quad 
\quad \stackrel{ab}{[-k]}\stackrel{ab}{(k)}= 0\,, \quad \quad \quad \quad
\stackrel{ab}{[-k]}\stackrel{ab}{(-k)} = \stackrel{ab}{(-k)}\,.
\label{graphbinoms}
\end{eqnarray}
We recognize in  the first equation of the first row and the first equation of the second row
the demonstration of the nilpotent and the projector character of the Clifford algebra objects $\stackrel{ab}{(k)}$ and 
$\stackrel{ab}{[k]}$, respectively. 

{\em Whenever the Clifford algebra objects apply from the left hand side,
they always transform } $\stackrel{ab}{(k)}$ {\em to} $\stackrel{ab}{[-k]}$, {\em never to} $\stackrel{ab}{[k]}$,
{\em and similarly } $\stackrel{ab}{[k]}$ {\em to} $\stackrel{ab}{(-k)}$, {\em never to} $\stackrel{ab}{(k)}$.

We define in Eq.~(\ref{vac1}, \ref{vac14n}) the vacuum state $|\psi_{oc}>$ so that one finds
\begin{eqnarray}
< \;\stackrel{ab}{(k)}^{\dagger}
 \stackrel{ab}{(k)}\; > = 1\,, \nonumber\\
< \;\stackrel{ab}{[k]}^{\dagger}
 \stackrel{ab}{[k]}\; > = 1\,.
\label{graphherscal}
\end{eqnarray}

Taking the above equations into account it is easy to find a Weyl spinor irreducible representation
for $d$-dimensional space, with $d$ even or odd. (We advise the reader to see 
Ref.~\cite{norma93,nh02} in particular for $d$ odd.) 

For $d$ even,  we simply set the starting state as a product of $d/2$, let us say, only nilpotents 
$\stackrel{ab}{(k)}$ for $d=2(2n+1)$, Eq.~(\ref{start(2n+1)2cliff}), or nilpotents and one 
projector, Eq.~(\ref{start4ncliff}), for $d=4n$, one for each $S^{ab}$ of the Cartan subalgebra  elements 
(Eq.~(\ref{cartan})),  applying it on the vacuum state, Eqs.~(\ref{vac1}, \ref{vac14n}). 
Then the generators $S^{ab}$, which do not belong to the Cartan subalgebra, applied to the 
starting state from the left hand side, generate all the members of one Weyl spinor.  
\begin{eqnarray}
\stackrel{0d}{(k_{0d})} \stackrel{12}{(k_{12})} \stackrel{35}{(k_{35})}\cdots 
\stackrel{d-1\;d-2}{(k_{d-1\;d-2})}
|\psi_{oc}>\,, \nonumber\\
\stackrel{0d}{[-k_{0d}]} \stackrel{12}{[-k_{12}]} \stackrel{35}{(k_{35})}\cdots
 \stackrel{d-1\;d-2}{(k_{d-1\;d-2})}
|\psi_{oc}>\,, \nonumber\\
\stackrel{0d}{[-k_{0d}]} \stackrel{12}{(k_{12})} \stackrel{35}{[-k_{35}]}\cdots 
\stackrel{d-1\;d-2}{(k_{d-1\;d-2})}
|\psi_{oc}>\,, \nonumber\\
\vdots \nonumber\\
\stackrel{od}{(k_{0d})} \stackrel{12}{[-k_{12}]} \stackrel{35}{[-k_{35}]}\cdots 
\stackrel{d-1\;d-2}{[-k_{d-1\;d-2}]}
|\psi_{oc}>\,, \nonumber\\
{\rm for} \, d=2(2n+1)\,,\quad n = {\rm positive}\,\, {\rm integer}\,.
\label{graphicd}
\end{eqnarray}
\begin{eqnarray}
\stackrel{0d}{(k_{0d})} \stackrel{12}{(k_{12})} \stackrel{35}{(k_{35})}\cdots 
\stackrel{d-1\;d-2}{[k_{d-1\;d-2}]}
|\psi_{oc}>\,, \nonumber\\
\stackrel{0d}{[-k_{0d}]} \stackrel{12}{[-k_{12}]} \stackrel{35}{(k_{35})}\cdots
 \stackrel{d-1\;d-2}{[k_{d-1\;d-2}]}
|\psi_{oc}>\,, \nonumber\\
\stackrel{0d}{[-k_{0d}]} \stackrel{12}{(k_{12})} \stackrel{35}{[-k_{35}]}\cdots 
\stackrel{d-1\;d-2}{[k_{d-1\;d-2}]}
|\psi_{oc}>\,, \nonumber\\
\vdots \nonumber\\
\stackrel{od}{(k_{0d})} \stackrel{12}{[-k_{12}]} \stackrel{35}{[-k_{35}]}\cdots 
\stackrel{d-1\;d-2}{[k_{d-1\;d-2}]}
|\psi_{oc}>\,, \nonumber\\
{\rm for} \, d=4n\,,\quad n = {\rm positive}\,\, {\rm integer}\,.
\label{graphicd4n}
\end{eqnarray}

\subsection{Technique to generate ''families'' of spinor representations in terms of Clifford algebra 
objects}
\label{families}

We found in this paper that for $d$ even there are $2^{d/2-1}$ ''family members'' and 
$2^{d/2-1}$ ''families'' 
of spinors, which can be second quantized. (The reader is advised to see also Refs.~%
\cite{norma93,pikan2003,nh02,nh03,pikan2006,normaJMP2015}.) We shall here
pay attention on only even $d$.

One Weyl representation forms a left ideal with respect to the multiplication with the Clifford 
algebra objects. 
We  proved in Refs.~(\cite{normaJMP2015,nh03}, and the references therein) 
that there is the application of the Clifford algebra object from the right hand side, which 
generates ''families'' of spinors.

Right multiplication with the Clifford algebra objects namely transforms the state with
the quantum numbers of one "family member" belonging to one
''family'' into the state of the same "family member" (into the same state with respect to 
the generators $S^{ab}$ when the multiplication from the left hand side is performed) 
of another ''family''.

We defined in Ref.\cite{norma93,nh03} the Clifford algebra objects
$\tilde{\gamma}^a$'s  as operations which operate formally from the left hand side (as 
$\gamma^a$'s do) on any Clifford algebra object A  as follows, Eq.~(\ref{gammatildeA}),
\begin{eqnarray}
\tilde{\gamma}^a A = i (-)^{(A)} A \gamma^a\,,
\label{gammatildeAapp}
\end{eqnarray}
with $(-)^{(A)} = -1$,  if $A$ is an odd Clifford algebra object and 
$(-)^{(A)} = 1$,  if $A$  is an even Clifford algebra object.

Then it follows, in accordance  with Eq.~(\ref{cliffthetareltheta}),
that $\tilde{\gamma}^a$ obey the same Clifford algebra relation as
 $\gamma^a$.
\begin{eqnarray}
(\tilde{\gamma}^a \tilde{\gamma}^b + \tilde{\gamma}^b \tilde{\gamma}^a) A =
 -i i ((-)^{(A)})^2 A (\gamma^a \gamma^b + \gamma^b \gamma^a) =I\cdot 2.\eta^{ab} A
\label{gammatildeAa}
\end{eqnarray}
and that $\tilde{\gamma}^a$ and $\gamma^a$ anticommute
\begin{eqnarray}
(\tilde{\gamma}^a \gamma^b + \gamma^b \tilde{\gamma}^a) A =  i(-)^{(A)}( -\gamma^b
 A \gamma^a  + \gamma^b A \gamma^a)  = 0\,.
\label{gammatildeACL}
\end{eqnarray}
We may write
\begin{eqnarray}
\{ \tilde{\gamma}^a, \gamma^b \}_+ &=& 0, \quad {\rm while} \quad
\{\tilde{\gamma}^a, \tilde{\gamma}^b \}_+ =I\cdot 2 \eta^{ab}\,.
\label{gammatildegammaapp}
\end{eqnarray}

One accordingly finds
\begin{eqnarray}
\tilde{\gamma}^a \stackrel{ab}{(k)}: &=& -i\stackrel{ab}{(k)} \gamma^a =
 - i\eta^{aa}\stackrel{ab}{[k]}\,,\nonumber\\
\tilde{\gamma}^b \stackrel{ab}{(k)}: &=& -i\stackrel{ab}{(k)} \gamma^b = - k
\stackrel{ab}{[k]}\,, \nonumber\\
\tilde{\gamma}^a \stackrel{ab}{[k]}: &=& \;\; i \stackrel{ab}{[k]} \gamma^a = 
\;\;i\stackrel{ab}{(k)}\,,\nonumber\\
\tilde{\gamma}^b \stackrel{ab}{[k]}: &=& \;\; i \stackrel{ab}{[k]} \gamma^b =
 - k \eta^{aa} \stackrel{ab}{(k)}\,.
\label{gammatilde}
\end{eqnarray}

If we define, Eq.~(\ref{twocliffordsab}),
\begin{eqnarray}
\tilde{S}^{ab} = \frac{i}{4}\; [\tilde{\gamma}^a,\tilde{\gamma}^b] = 
\frac{i}{4}\;\{\tilde{\gamma}^a,\tilde{\gamma}^b\}_{-} =
\frac{1}{4} (\tilde{\gamma}^a
\tilde{\gamma}^b - \tilde{\gamma}^b\tilde{\gamma}^a)\,,
\label{tildesab}
\end{eqnarray}
it follows
\begin{eqnarray}
\tilde{S}^{ab} A = A \frac{1}{4} (\gamma^b \gamma^a - \gamma^a \gamma^b)\,,
\label{tildesab1}
\end{eqnarray}
manifesting accordingly that $\tilde{S}^{ab}$ fulfill the Lorentz algebra relation as $S^{ab}$ do. 
Taking into account Eq.~(\ref{gammatildeA}), we further find
\begin{eqnarray}
\{\tilde{S}^{ab}, S^{ab}\}_- =0\,,\quad
\{\tilde{S}^{ab}, \gamma^c \}_-=0\,,\quad
\{S^{ab}, \tilde{\gamma}^c \}_-=0\,.
\label{sabtildesab}
\end{eqnarray}
One also finds 
\begin{eqnarray}
\{\tilde{S}^{ab}, \Gamma \}_- &=&0\,,\quad \{ \tilde{\gamma}^a, \Gamma \}_- = 0\,,
\quad \{\tilde{S}^{ab}, \tilde{\Gamma} \}_- = 0\,, 
\quad {\rm for \;\; d\;\; even}\,,\nonumber\\
\Gamma^{(d)} :&=&(i)^{d/2}\; \;\;\;\;\;\prod_a \quad (\sqrt{\eta^{aa}} \gamma^a)\,, 
\quad {\rm if } \quad d = 2n\,,
\nonumber\\
\tilde{\Gamma}^{(d)} :&=&(i)^{d/2}\; \;\;\;\;\;\prod_a \quad (\sqrt{\eta^{aa}} \tilde{\gamma}^a)\,, 
\quad {\rm if } \quad d = 2n\,,  
\label{sabtildeGAMMA}
\end{eqnarray}
where handedness $\Gamma$ ($\{\Gamma, S^{ab}\}_- =0$) 
 is a Casimir of the Lorentz group, 
which means that in $d$ even transformation of one ''family'' into another with either $\tilde{S}^{ab}$ 
or $\tilde{\gamma}^a$ leaves handedness $\Gamma$ unchanged.

We advise the reader to read \cite{norma93} where the two kinds of 
Clifford algebra objects follow as two different superpositions of a Grassmann coordinate and its 
conjugate momentum.

Below some useful relations~\cite{pikan2003,pikan2006} are presented 
\begin{eqnarray}
\label{plusminus}
N^{\pm}_{+}         &=& N^{1}_{+} \pm i \,N^{2}_{+} = 
 - \stackrel{03}{(\mp i)} \stackrel{12}{(\pm )}\,, \quad N^{\pm}_{-}= N^{1}_{-} \pm i\,N^{2}_{-} = 
  \stackrel{03}{(\pm i)} \stackrel{12}{(\pm )}\,,\nonumber\\
\tilde{N}^{\pm}_{+} &=& - \stackrel{03}{\tilde{(\mp i)}} \stackrel{12}{\tilde{(\pm )}}\,, \quad 
\tilde{N}^{\pm}_{-}= 
  \stackrel{03} {\tilde{(\pm i)}} \stackrel{12} {\tilde{(\pm )}}\,,\nonumber\\ 
\tau^{1\pm}         &=& (\mp)\, \stackrel{56}{(\pm )} \stackrel{78}{(\mp )} \,, \quad   
\tau^{2\mp}=            (\mp)\, \stackrel{56}{(\mp )} \stackrel{78}{(\mp )} \,,\nonumber\\ 
\tilde{\tau}^{1\pm} &=& (\mp)\, \stackrel{56}{\tilde{(\pm )}} \stackrel{78}{\tilde{(\mp )}}\,,\quad   
\tilde{\tau}^{2\mp}= (\mp)\, \stackrel{56}{\tilde{(\mp )}} \stackrel{78}{\tilde{(\mp )}}\,.
\end{eqnarray}
%
%
\begin{eqnarray}
\tilde{S}^{ab} \stackrel{ab}{(k)}&=& \;\;\frac{k}{2}\stackrel{ab}{(k)}\,,\nonumber\\
\tilde{S}^{ab} \stackrel{ab}{[k]}&=& -\frac{k}{2} \stackrel{ab}{[k]}\,, \nonumber\\
\tilde{S}^{ac}\stackrel{ab}{(k)}\stackrel{cd}{(k)}& = &\frac{i}{2} \eta^{aa} \eta^{cc} 
\stackrel{ab}{[k]}\stackrel{cd}{[k]}\,, \nonumber\\
\tilde{S}^{ac}\stackrel{ab}{[k]}\stackrel{cd}{[k]}& = & -\frac{i}{2}  
\stackrel{ab}{(k)}\stackrel{cd}{(k)}\,, \nonumber\\
\tilde{S}^{ac}\stackrel{ab}{(k)}\stackrel{cd}{[k]}& = & -\frac{i}{2} \eta^{aa}  
\stackrel{ab}{[k]}\stackrel{cd}{(k)}\,, \nonumber\\
\tilde{S}^{ac}\stackrel{ab}{[k]}\stackrel{cd}{(k)}& = &\frac{i}{2} \eta^{cc}  
\stackrel{ab}{(k)}\stackrel{cd}{[k]}\,.
\label{tildesac}
\end{eqnarray}

We transform the state of one ''family'' to the state of another ''family'' by 
the application of 
$\tilde{S}^{ac}$ (formally from the left hand side) on a state of the first
''family'' for a chosen 
$a,c$. To transform all the states of one ''family'' into states
of another ''family'', we apply  
 $\tilde{S}^{ac}$ to each state of the starting ''family''.
It is, of course, sufficient to apply 
$\tilde{S}^{ac}$ to only one state of a ''family'' and then use generators of the 
Lorentz group ($S^{ab}$) 
to generate all the states of one Dirac spinor $d$-dimensional space.

One must notice that nilpotents $\stackrel{ab}{(k)}$ and projectors $\stackrel{ab}{[k]}$ are 
"eigenvectors"not only of the Cartan subalgebra $S^{ab}$ but also of $\tilde{S}^{ab}$. Accordingly 
only $\tilde{S}^{ac}$, which do not carry the Cartan subalgebra indices, cause the transition from 
one ''family'' to another ''family''.

\section*{Acknowledgement}
The author N.S.M.B. thanks Department of Physics, FMF, University of Ljubljana, Society of 
Mathematicians, Physicists and Astronomers of Slovenia,  for supporting the research on the 
{\it spin-charge-family} theory, the author H.B.N. thanks the Niels Bohr Institute for
being allowed to staying as emeritus, both authors thank DMFA and  Matja\v z Breskvar of Beyond 
Semiconductor for donations, in particular for the annual workshops entitled 
"What comes beyond the standard models" at Bled.


\begin{thebibliography}{99}
\bibitem{norma92}  N. Manko\v c Bor\v stnik, "Spin connection as a 
              superpartner of a vielbein", {\it Phys. Lett.} {\bf B 292} (1992)  25-29.
\bibitem{norma93} N. Manko\v c Bor\v stnik, "Spinor and vector representations in four dimensional Grassmann
              space", {\it J. of Math. Phys.} {\bf 34} (1993) 3731-3745.
\bibitem{IARD2016} N.S. Manko\v c Bor\v stnik, "Spin-charge-family theory is offering next step
             in understanding elementary particles and fields and correspondingly universe", 
             Proceedings to the Conference on Cosmology, Gravitational Waves and Particles, 
             IARD conferences, Ljubljana, 6-9 June 2016, The $10^{th}$ Biennial Conference 
             on Classical and Quantum Relativistic Dynamics of Particles and Fields,  
                          J. Phys.: Conf. Ser. 845 012017 
    [arXiv:1409.4981, arXiv:1607.01618v2].
\bibitem{n2014matterantimatter}  N.S. Manko\v c Bor\v stnik, "Matter-antimatter asymmetry in the 
{\it spin-charge-family} theory", {\it Phys. Rev.} {\bf D 91} (2015) 065004  [arXiv:1409.7791]. 
\bibitem{nd2017} N.S. Manko\v c Bor\v stnik, D. Lukman, "Vector and scalar gauge fields with
              respect to $d=(3+1)$ in Kaluza-Klein theories and in the {\it spin-charge-family theory}",
              {\it Eur. Phys. J. C} {\bf 77} (2017) 231.
\bibitem{n2012scalars} N.S. Manko\v c Bor\v stnik, "The {\it spin-charge-family} theory explains  
              why  the scalar Higgs carries the weak charge $\pm \frac{1}{2}$ and the hyper charge 
              $ \mp \frac{1}{2}$",
              Proceedings to 
              the $17^{th}$ Workshop "What comes beyond the standard models", Bled, 20-28 of July, 2014, 
              Ed. N.S. Manko\v c Bor\v stnik, H.B. Nielsen, D. Lukman, DMFA  Zalo\v zni\v stvo, 
              Ljubljana December 2014, p.163-82 [ arXiv:1502.06786v1] [arXiv:1409.4981].  
\bibitem{JMP2013} N.S. Manko\v c Bor\v stnik N S, "The spin-charge-family theory is explaining the
 origin of families, of the Higgs and the Yukawa couplings", {\it J. of Modern Phys.} {\bf 4} (2013) 823
[arXiv:1312.1542].
\bibitem{nh2017} N.S. Manko\v c Bor\v stnik, H.B.F. Nielsen, "The spin-charge-family theory 
             offers understanding of the triangle anomalies cancellation in the standard model",
             {\it Fortschrite der Physik, Progress of Physics} (2017) 1700046.
\bibitem{normaJMP2015} N.S. Manko\v c Bor\v stnik, "The explanation for the origin of the 
Higgs  scalar and for the Yukawa couplings by the {\it spin-charge-family} theory", 
{\it J.of Mod. Physics} {\bf 6} (2015) 2244-2274, http://dx.org./10.4236/jmp.2015.615230
              [arXiv:1409.4981].
\bibitem{nh2018}  N.S. Manko\v c Bor\v stnik and H.B.  Nielsen, "Why nature made a choice of 
Clifford and not  Grassmann coordinates",   Proceedings  to  the $20^{th}$ Workshop "What comes 
beyond the standard models", Bled, 9-17 of July, 2017, Ed. N.S. Manko\v c Bor\v stnik, H.B. Nielsen,
D. Lukman, DMFA  Zalo\v zni\v stvo, Ljubljana, December 2017, p. 89-120 
[arXiv:1802.05554v1v2].
\bibitem{Geor} H. Georgi, in {\it Particles and Fields} (edited by C. E. Carlson), A.I.P., 1975; Google Scholar.\\
\bibitem{FritzMin} H. Fritzsch and P. Minkowski, {\it Ann. Phys.} {\bf 93} (1975) 193.\\
\bibitem{PatiSal} J. Pati and A. Salam, {\it Phys.Rev.} {\bf D 8} (1973) 1240.\\
\bibitem{GeorGlas} H. Georgy and S.L. Glashow, {\it Phys. Rev. Lett.}  {\bf 32 } (1974) 438.
\bibitem{Cho}  Y. M. Cho, {\it J. Math. Phys.} {\bf 16}  (1975) 2029.
\bibitem{ChoFreu} Y. M. Cho, P. G. O.Freund, {\it Phys. Rev.} {\bf D 12} (1975) 1711. 
\bibitem{Zee}  A. Zee, {\it Proceedings of the first Kyoto summer institute on 
grand unified theories and related topics}, Kyoto, Japan, June-July 1981, Ed. by M. Konuma, T. Kaskawa, 
World Scientific Singapore.  
\bibitem{SalStra}  A. Salam, J. Strathdee, {\it Ann. Phys.} (N.Y.) {\bf 141} (1982) 316.
\bibitem{DaeSalStra} S. Randjbar-Daemi, A. Salam, J. Strathdee, {\it Nucl. Phys.} {\bf B 242} (1984) 447.
\bibitem{Mec} W. Mecklenburg, {\it Fortschr. Phys.} {\bf 32} (1984) 207.
\bibitem{HorPalCraSch}
 Z. Horvath, L. Palla, E. Crammer, J. Scherk, {\it Nucl. Phys.} {\bf B 127} (1977) 57.
\bibitem{Asaka}T. Asaka, W. Buchmuller, {\it Phys. Lett.} {\bf B 523} (2001) 199.
\bibitem{ChaSla} G. Chapline, R. Slansky, {\it Nucl. Phys. } {\bf B 209} (1982) 461.
\bibitem{Jackiw} R. Jackiw and K. Johnson, {\it Phys. Rev.} {\bf D 8} (1973) 2386.
\bibitem{Ant} I. Antoniadis, {\it Phys. Lett.}  {\bf B 246} (1990) 377. 
\bibitem{Ramond} P. Ramond, Field Theory, A Modern Primer, Frontier in Physics, Addison-Wesley Pub.,
ISBN 0-201-54611-6.
\bibitem{Horawa}  P. Horawa, E. Witten, {\it Nucl. Phys.}  {\bf B 460} (1966) 506.
\bibitem{KaluzaKlein}  T. Kaluza, ''On the unification problem in Physics'', {\it Sitzungsber. d. Berl. Acad.}
 (1918) 204, O. Klein, ''Quantum theory and five-dimensional relativity'', {\it Zeit. Phys.} {\bf 37}(1926) 895. 
\bibitem{Witten} E. Witten, ”Search for realistic Kaluza-Klein theory”,{\it  Nucl. Phys.} {\bf B 186} (1981) 412.
\bibitem{Duff} M. Duff, B. Nilsson, C. Pope, {\it Phys. Rep.} {\bf C 130} (1984)1,
M. Duff, B. Nilsson, C. Pope, N. Warner, {\it Phys. Lett.} {\bf B 149} (1984) 60.
\bibitem{App} T. Appelquist, H. C. Cheng, B. A. Dobrescu, {\it Phys. Rev.} {\bf D 64}
(2001) 035002.
\bibitem{SapTin}  M. Saposhnikov, P. TinyakovP 2001 {\it Phys. Lett.} {\bf B 515} (2001) 442
 [arXiv:hep-th/0102161v2].
\bibitem{Wetterich} C. Wetterich,{\it Nucl. Phys.} {\bf B 253} (1985) 366.
\bibitem{zelenaknjiga} The authors of the works presented in {\it An introduction to Kaluza-Klein 
theories}, Ed. by H. C. Lee, World Scientific, Singapore 1983.
\bibitem{mil} M. Blagojevi\' c,   {\em Gravitation and gauge symmetries},  IoP Publishing, Bristol 2002.
%
\bibitem{Alvarez}  L. Alvarez-Gaum\' e, "An Introduction to Anomalies", 
Erice School Math. Phys. 1985:0093.
\bibitem{Bilal} A. Bilal, "Lectures on anomalies" [arXiv:0802.0634].
\bibitem{AlvarezBondiaMartin}  L. Alvarez-Gaum\'e, J.M. Gracia-Bond\' ia, C.M. Martin, "Anomaly
Cancellation and the Gauge Group of the Standard Model in NCG" [hep-th/9506115].
%
\bibitem{bereziani} B. Belfatto, R. Beradze, Z. Berezhiani, "The CKM unitarity problem: 
A trace of new physics at the TeV scale?" [arXiv:1906.02714v1].
%
\bibitem{Zurab} Z. Berezhiani, private communication with N.S. Manko\v c Bor\v stnik.
%
\bibitem{EPJC2017}  D. Lukman,  N.S. Manko\v c Bor\v stnik, "Gauge fields with respect to 
$d=(3+1)$ in the Kaluza-Klein   theories  and in the {\it spin-charge-family} theory", 
{\it Eur. Phys. J. C}  (2017) doi: 10.1140/epjc/s10052-017-4804-y.
%
\bibitem{procnh2015}  N.S. Manko\v c Bor\v stnik, H.B.F. Nielsen, "Fermionization in an Arbitrary Number
              of Dimensions", Proceedings to  the $18^{th}$ Workshop "What comes beyond the standard 
              models", Bled, 11-19 of July, 2015, Ed. N.S. Manko\v c Bor\v stnik, H.B. Nielsen,
              D. Lukman, DMFA  Zalo\v zni\v stvo, Ljubljana December 2015, p. 111-128
              [arXiv:1602.03175].
\bibitem{procnh2017} N. S. Manko\v c Bor\v stnik,  H.B. Nielsen, "Fermionization, 
             Number of Families",  Proceedings  to  the $20^{th}$ Workshop "What comes 
             beyond the standard models", Bled, 9-17 of July, 2017, Ed. N.S. Manko\v c Bor\v stnik, H.B. Nielsen,
              D. Lukman, DMFA  Zalo\v zni\v stvo, Ljubljana, December 2017, p.232-257. 
              [arXiv:1703.09699]. 
%
\bibitem{holger} H. Aratyn, H.B. Nielsen, "Constraints On Bosonization In Higher Dimensions",
NBI-HE-83-36,
Conference: C83-10-10.2 (Ahrenshoop Sympos.1983:0260), p.0260 Proceedings.
%
\bibitem{HolMasDSB} H.B. Nielsen, M. Ninomya, "Dirac sea for bosons, I,II", Progress of the theoretical
Physics, {\bf 113}, 606 [arXiv:hep-th/0410218].
%
\bibitem{nd2018} D. Lukman, N.S. Manko\v c Bor\v stnik, "Representations in Grassmann space
and fermion degrees of freedom", [arXiv:1805.06318].
%
\bibitem{nh02}  N.S. Manko\v c Bor\v stnik, H.B.F. Nielsen, {\it J. of Math. Phys.} {\bf 43}, 
5782 (2002) [arXiv:hep-th/0111257].
\bibitem{nh03} N.S. Manko\v c Bor\v stnik, H.B.F. Nielsen,
{\it J. of Math. Phys.} {\bf 44} 4817 (2003) [arXiv:hep-th/0303224].
%
\bibitem{DKhn} N.S. Manko\v c Bor\v stnik and H.B.  Nielsen, 
{\it Phys. Rev.} {\bf D 62}, 044010 (2000) arXiv:[hep-th/9911032].
\bibitem{NH2005} N.S. Manko\v c Bor\v stnik and H.B.  Nielsen, 
"Second quantization of spinors and Clifford algebra objects", 
Proceedings to the $8^{\rm th}$ Workshop ''What Comes Beyond 
                     the Standard Models'', Bled, July 19 - 29, 2005,  Ed. by Norma Manko\v c Bor\v stnik, 
                     Holger Bech Nielsen, Colin Froggatt, Dragan Lukman, DMFA Zalo\v zni\v stvo, 
                     Ljubljana December 2005, p.63-71, hep-ph/0512061.
\bibitem{mdn2006} M. Breskvar, D. Lukman, N. S. Manko\v c Bor\v stnik, 
                     ''On the Origin of Families of Fermions and Their Mass Matrices\,---\,%
Approximate Analyses of Properties of Four Families Within Approach 
Unifying Spins and Charges", 
                     Proceedings to the $9^{\rm th}$ Workshop ''What Comes Beyond the Standard 
                     Models'', Bled, Sept. 16 - 26, 2006,  Ed. by Norma Manko\v c Bor\v stnik, 
		     Holger Bech Nielsen, Colin Froggatt, Dragan Lukman, DMFA Zalo\v zni\v stvo, 
                     Ljubljana December 2006, p.25-50, hep-ph/0612250.
%
\bibitem{gmdn2007} G. Bregar, M. Breskvar, D. Lukman, N.S. Manko\v c Bor\v stnik,
                    "Families of Quarks and Leptons and Their Mass Matrices", 
                    Proceedings to the $10^{th}$ international workshop ''What Comes Beyond 
		    the Standard Model'', 17 -27 of July, 2007, Ed. Norma Manko\v c  
		    Bor\v stnik, Holger Bech Nielsen, Colin Froggatt, Dragan Lukman,
		    DMFA  Zalo\v zni\v stvo, Ljubljana December 2007, 
		    p.53-70, hep-ph/0711.4681.
%
\bibitem{gmdn2008} G. Bregar, M. Breskvar, D. Lukman, N.S. Manko\v c Bor\v stnik, 
                     "Predictions for four families by the Approach unifying spins and charges"
                     {\it New J. of Phys.} {\bf 10} (2008) 093002,
                     hep-ph/0606159, hep/ph-07082846.
%
\bibitem{gn2009} G. Bregar, N.S. Manko\v c Bor\v stnik, "Does dark matter consist of baryons 
	        of new stable family quarks?", {\it Phys. Rev. D } {\bf 80}, 083534 (2009), 1-16
%
\bibitem{gn2013}  G. Bregar, N.S. Manko\v c Bor\v stnik, "Can we predict the fourth family masses 
              for quarks and leptons?", Proceedings (arxiv:1403.4441) to the 16 th Workshop "What comes beyond the 
              standard models", Bled, 14-21 of July, 2013, Ed. N.S. Manko\v c Bor\v stnik, 
              H.B. Nielsen, D. Lukman, DMFA  Zalo\v zni\v stvo, Ljubljana December 2013, p. 31-51, 
              http://arxiv.org/abs/1212.4055.

%
\bibitem{gn2014} G. Bregar, N.S. Manko\v c Bor\v stnik, "The new experimental data for the quarks 
           mixing matrix are in better agreement with the {\it spin-charge-family} theory predictions",               
              Proceedings to 
              the $17^th$ Workshop "What comes beyond the standard models", Bled, 20-28 of July, 2014, 
              Ed. N.S. Manko\v c Bor\v stnik, H.B. Nielsen, D. Lukman, DMFA  Zalo\v zni\v stvo, 
              Ljubljana December 2014, p.20-45 [ arXiv:1502.06786v1] [arxiv:1412.5866].  
%
\bibitem{NA2018}  A. Hernandez-Galeana and N.S. Manko\v c Bor\v stnik, ""The symmetry 
             of $4 \times 4$ mass matrices predicted by the 
             {\it spin-charge-family} theory --- $SU(2) \times SU(2) \times U(1)$ ---  
              remains in all  loop corrections", Proceedings  to  the $21^{st}$ Workshop "What comes 
             beyond the standard models", 23 of June - 1 of July, 2017, Ed. N.S. Manko\v c Bor\v stnik, 
            H.B. Nielsen, D. Lukman, DMFA  Zalo\v zni\v stvo, Ljubljana, December 2018         
           [arXiv:1902.02691, arXiv:1902.10628].
%
\bibitem{NH2017newdata} N.S. Manko\v c Bor\v stnik, H.B.F. Nielsen, "Do the present experiments 
             exclude the 
             existence of the fourth family members?", 
             Proceedings  to  the $19^th$ Workshop "What comes beyond the standard 
              models", Bled, 11-19 of July, 2016, Ed. N.S. Manko\v c Bor\v stnik, H.B. Nielsen,
              D. Lukman, DMFA  Zalo\v zni\v stvo, Ljubljana December 2016, p.128-146 
             [arXiv:1703.09699].
\bibitem{AhmedAli} A. Ali in discussions and in private 
communication at the Singapore
Conference on New Physics at the Large Hadron Collider, 29 February - 4 March 2016.
\bibitem{MatthiasNeubert} M. Neubert, in duscussions at the Singapore
Conference on New Physics at the Large Hadron Collider, 29 February - 4 March 2016.
%
%
\bibitem{nm2015} N.S. Manko\v c Bor\v stnik, M. Rosina, "Are superheavy stable quark clusters viable 
              candidates  for the dark matter?",
              International Journal of Modern Physics D (IJMPD) {\bf 24} (No. 13) (2015) 1545003.
%
\bibitem{Hestenes84} D. Hestenes, G. Sobcyk,  "Clifford algebra to geometric calculus",
Reidel 1984. 
%
\bibitem{Lounesto2001} P. Lounesto, P. Clifford algebras and spinors, Cambridge Univ. Press.2001.
%
\bibitem{MP1707.05695} M. Pav\v si\v c,"Quantized fields a la Clifford and unfication", [arXiv:1707.05695]
%
\bibitem{nhds} N.S. Manko\v c Bor\v stnik and H.B.F. Nielsen,
"Discrete symmetries in the Kaluza-Klein theories",
 {\em JHEP} 04:165,  2014 [arXiv:1212.2362].
%
\bibitem{TDN} T.Troha, D. Lukman, N.S. Manko\v c Bor\v stnik, "Massless and massive 
		                representations in the {\it spinor technique}",
		                {\it .Int. J Mod. Phys.} {\bf A 29}, 1450124 (2014). 
\bibitem{Dirac} P.A.M. Dirac {\it Proc. Roy. Soc. (London)}, {\bf A 117} (1928) 610.
\bibitem{NHD} D. Lukman, N.S. Manko\v c Bor\v stnik and H.B. Nielsen,
"An effective two dimensionality cases bring a new hope to the Kaluza-Klein-like theories", 
{\em New J. Phys.} 13:103027, 2011.
\bibitem{ND012} 
D. Lukman and N.S. Manko\v c Bor\v stnik,  
"Spinor states on a curved infinite disc  with non-zero spin-connection fields",
{\em J. Phys. A:  Math. Theor.} 45:465401, 2012 
[arxiv:1205.1714, arxiv:1312.541, arXiv:hep-ph/0412208 p.64-84].
\bibitem{familiesNDproc}  D. Lukman, N.S. Manko\v c Bor\v stnik and H.B. Nielsen,
  "Families of spinors in $d=(1+5)$ with a zweibein and two kinds of spin connection fields on 
 an almost $S^2$",  Proceedings 
               to the $15^{th}$ Workshop "What comes beyond the standard models", Bled, 
                9-19 of July, 2012, Ed. N.S. Manko\v c Bor\v stnik, 
                H.B. Nielsen, D. Lukman, DMFA  Zalo\v zni\v stvo, Ljubljana December 2012, 157-166,
                [arXiv:1302.4305].
%
\bibitem{pikan2003} A.Bor\v stnik Bra\v ci\v c, N. Manko\v c Bor\v stnik,``The approach Unifying 
                   Spins and Charges and Its Predictions``, 
				Proceedings to the Euroconference on Symmetries Beyond the Standard Model'',
				Portoro\v z, July 12 - 17, 2003,
			        Ed. by Norma Manko\v c Bor\v stnik, Holger Bech Nielsen, Colin Froggatt, Dragan Lukman, DMFA 
				Zalo\v zni\v stvo, Ljubljana December 2003, p. 31-57, [arXiv:hep-ph/0401043, 
                          arXiv:hep-ph/0401055].
%
\bibitem{pikan2006}  A. Bor\v stnik Bra\v ci\v c, N. S. Manko\v c Bor\v stnik, ''On the origin of families 
                     of fermions and their mass matrices'', hep-ph/0512062,  Phys Rev. {\bf D 74} 
                     073013-28  (2006).
%
\bibitem{nh2008} N.S. Manko\v c Bor\v stnik, H.B. Nielsen, ``Particular boundary condition 
                     ensures that a fermion in d=1+5, compactified on a finite disk, manifests 
                     in d=1+3 as massless spinor with a charge 1/2, mass protected and 
                     chirally coupled to the gauge field'',  hep-th/0612126, arxiv:0710.1956, 
                      {\it Phys.  Lett.} {\bf B} 663, Issue 3, 22 May 2008, Pages 265-269.
%
\bibitem{Bethe} H.A. Bethe, "Intermediate quantum mechanics", W.A. Benjamin, 
1964 (New York, Amsterdam).
%
\bibitem{Itzykson} C. Itzykson, J.B. Zuber, "Quantum field theory", McGraw-Hill, 1980 (New York).
%
\bibitem{MP2017} M. Pav\v  si\v c, "Quantized fields \' a la Clifford and unification" [arXiv:1707.05695].
%
\bibitem{NH2007Majorana}  N.S. Manko\v c Bor\v stnik, H. B. Nielsen, ''Fermions with no 
                       fundamental charges call for extra dimensions'',  {\it Phys. Lett.} {\bf B 644}
                      (2007) 198-202 [arXiv:hep-th/0608006].
%
\bibitem{NHD2011} D. Lukman, N.S. Manko\v c Bor\v stnik, H.B. Nielsen, "An effective two 
               dimensionality" cases bring a new hope to the Kaluza-Klein-like theories'', 
               http://arxiv.org/abs/1001.4679v5, 
               {\it New J. Phys.} {\bf 13} (2011) 103027, 1-25

\end{thebibliography}
\end{document}